%% file: thesis.tex
\documentclass[twoside,12pt]{CUEDthesisPSnPDF}

\title{\it A Signature of Higher Dimensions \\ 
						at the Cosmic Singularity}  

  \author{{\sc Paul McFadden }}
  \collegeordept{St.~John's College \\[1ex] Department of Applied Mathematics \& Theoretical Physics}
  \university{University of Cambridge}
  \crest{\includegraphics[bb = 0 0 292 336, width=30mm]{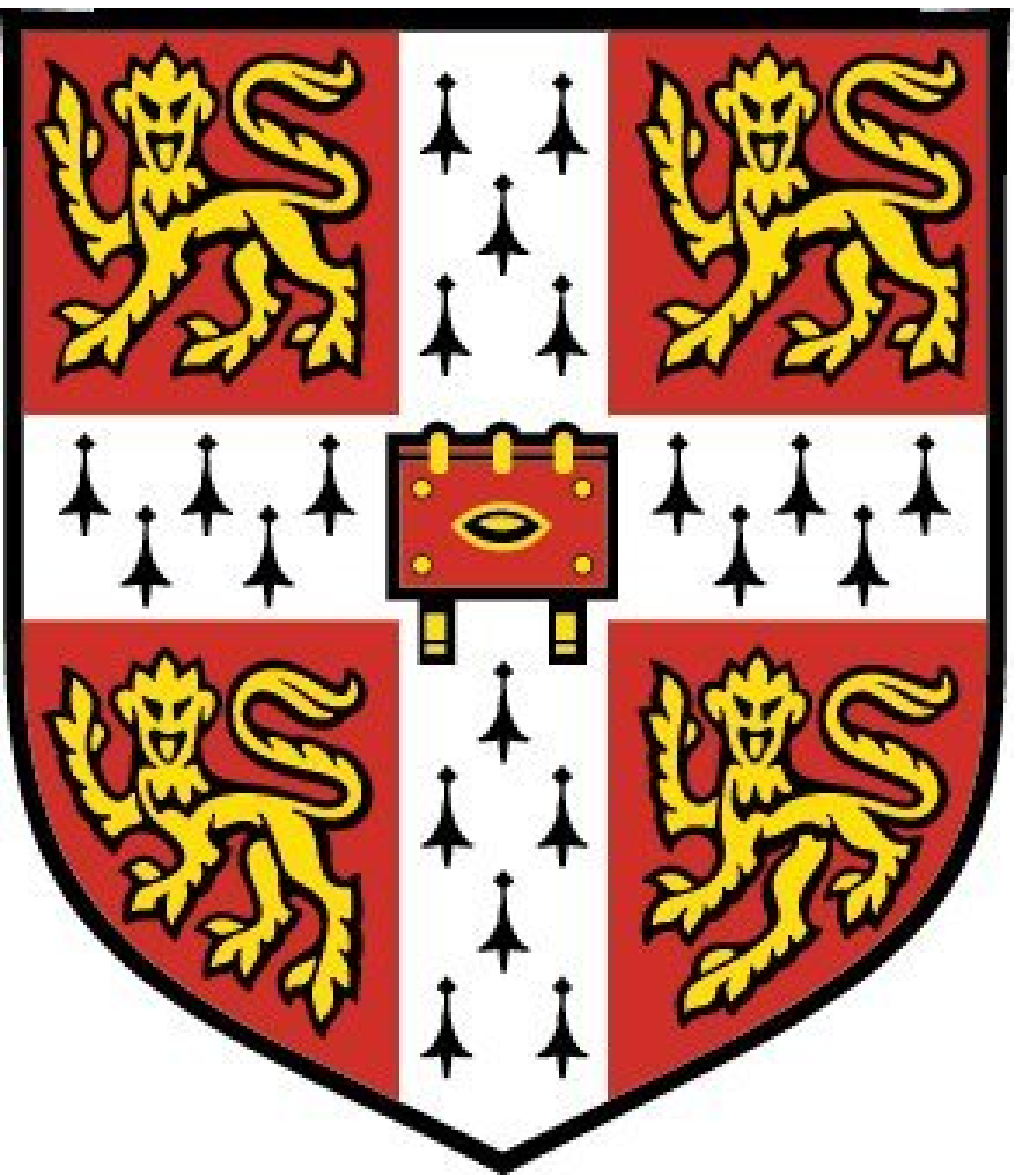}}
\degree{Doctor of Philosophy}
\degreedate{$8^\mathrm{th}$ August 2006}

\usepackage{textcomp}
\usepackage{amssymb}
\usepackage{latexsym}
\usepackage{graphicx}
\usepackage{amsmath}

\newcommand{\nc}{\newcommand}
\nc{\rnc}{\renewcommand}
\nc{\bo}{\raise-0.4mm\hbox{$\Box$}}              
\nc{\Z}{\mathbb{Z}}
\nc{\tz}{\tilde{z}}
\nc{\tg}{\tilde{g}_{\mu \nu}}
\nc{\Tr}{\mathrm{tr}}
\nc{\ep}{\varepsilon}  
\nc{\eg}{\textit{e.g. }}
\nc{\etal}{\textit{et al.}}
\nc{\dx}{\mathrm{d} ^4 x}
\nc{\D}{\partial}
\rnc{\d}{\mathrm{d}}
\nc{\gdxdx}{ g_{\mu \nu}(x) \mathrm{d}x^\mu \mathrm{d}x^\nu}
\nc{\ndxdx}{ \eta _{\mu \nu}(x) \mathrm{d}x^\mu \mathrm{d}x^\nu}
\nc{\tgdxdx}{\tilde{g}_{\mu \nu}(x) \mathrm{d}x^\mu \mathrm{d}x^\nu}
\nc{\dxdx}{\mathrm{d}x^\mu \mathrm{d}x^\nu}
\nc{\rootg}{\sqrt{-g}\,}
\nc{\g}{g_{\mu \nu}}
\nc{\gpm}{g^\pm _{\mu \nu}}
\nc{\gp}{g^+ _{\mu \nu}}
\nc{\gm}{g^- _{\mu \nu}}
\rnc{\[}{\begin{equation}}
\rnc{\]}{\end{equation}}
\rnc{\(}{\left(}
\rnc{\)}{\right)}
\nc{\half}{\frac{1}{2}}
\nc{\vgrad}{\vec{\nabla}}
\nc{\grad}{\nabla}
\nc{\bea}{\begin{eqnarray}}
\nc{\eea}{\end{eqnarray}}
\nc{\ie}{\textit{i.e. }}
\nc{\iec}{\textit{i.e., }}
\nc{\etc}{\textit{etc. }}
\rnc{\tt}{\rightarrow} 
\rnc{\inf}{\infty}
\rnc{\L}{L}
\nc{\w}{\omega}
\nc{\cosech}{\mathrm{cosech}}
\nc{\sech}{\mathrm{sech}}
\rnc{\exp}{\mathrm{exp}}
\nc{\dxsq}{\d \vec{x}^2}
\nc{\chihat}{\hat{\chi}}
\nc{\ahat}{\hat{a}}
\rnc{\k}{\vec{k}}
\nc{\x}{\vec{x}}
\nc{\xp}{\x\,'}
\nc{\ahatk}{\ahat_{\k}}
\nc{\ahatkp}{\ahat_{\k'}}
\nc{\vphi}{\varphi}
\nc{\phih}{\hat{\vphi}}
\rnc{\dag}{\dagger}
\nc{\pihatchi}{\hat{\pi}_\chi}
\rnc{\.}{\cdot}
\rnc{\P}{\mathcal{P}}
\nc{\cH}{\mathcal{H}}  
\nc{\dG}{\delta \mathcal{G}}
\nc{\dT}{\delta \mathcal{T}}
\newcommand{\bra}[1]{\langle #1|}
\newcommand{\ket}[1]{|#1\rangle}

\nc{\eps}{\epsilon}
\nc{\now}{\mathrm{now}}
\nc{\rh}{\mathrm{rh}}
\nc{\GeV}{\mathrm{GeV}}
\nc{\eV}{\mathrm{eV}}

\nc{\inv}{\mathrm{inv}}
\nc{\tN}{\tilde{N}}
\nc{\tB}{\tilde{B}}
\nc{\tA}{\tilde{A}}
\nc{\tk}{\tilde{k}}
\nc{\tl}{\bar{\lambda}}
\nc{\tnu}{\tilde{\nu}}
\nc{\Nbar}{\bar{N}}
\nc{\nn}{\nonumber}
\rnc{\a}{\alpha}

\begin{document}


 \renewcommand\baselinestretch{1.2}
\baselineskip=18pt plus1pt

\maketitle

\clearpage
\thispagestyle{empty}

\setcounter{secnumdepth}{3}
\setcounter{tocdepth}{3}

\frontmatter
\include{dedication}    
\include{blankpage}  

\include{acknowledgement}

\include{blankpage}  

\include{abstract}

\include{blankpage}  

\tableofcontents

\clearpage                      
\thispagestyle{empty}           

\mainmatter

\include{introduction}

\include{chapter1}

\include{chapter2}

\include{chapter3}
\include{chapter4}

\include{chapter5}

\include{chapter6}

\include{conclusions}

\appendix
\include{appendix1}

\include{appendix3}
\include{longappendix}

\bibliographystyle{apsrev} 
\renewcommand{\bibname}{References} 
\bibliography{references} 
\addcontentsline{toc}{chapter}{References} 

\end{document}

%% file: dedication.tex

\begin{dedication} 


{\large \it To my parents}

\end{dedication}




%% file: blankpage.tex

\begin{dedication} 
\end{dedication}

%% file: acknowledgement.tex

\begin{acknowledgements}      

It is a truth universally acknowledged, that a young man, without possession of a good fortune, must be in want of a PhD.

It is a pleasure to thank 
the following people for their help and encouragement along the way:
first and foremost, my supervisor, Neil Turok, for four years of brilliant and inspiring collaboration.
Also Gustavo Niz, comrade-in-arms;
my many other friends and colleagues at DAMTP,  
in particular Anne Davis, Steven Gratton, David Jennings, Jean-Luc
Lehners, Kate Marvel, Gonzalo Palma, Fernando Quevedo, Claudia de Rham, Andrew Tolley, Sam Webster, and Toby Wiseman;
Paul Steinhardt, for an enjoyable visit to Princeton and for collaboration;
the African Institute for Mathematical Sciences, Cape Town, South Africa, where part of this work was completed;
PPARC, Dennis Avery and Stephen Hawking for financial support; St.~John's College for support through tenth term funding and a Benefactor's Scholarship; Wolfson College for an enjoyable Junior Research Fellowship;
Fr{\'e}d{\'e}ric Chopin, Sergei Rachmaninov and Earl Grey, for anaesthetising me to the pain of writing;
my fearless 
cybernetic 
sidekick, Mathematica, for crunching (not nearly enough of) the algebra; 
my many Cambridge friends, especially Laura Protano-Biggs and the
Rusalka Quartet;
my family, for their love and support;
and lastly -- to anyone game enough to actually read this.

\end{acknowledgements}

%% file: abstract.tex

\begin{abstracts}  

In this thesis we study the dynamics  
of higher-dimensional gravity in
a universe emerging from a brane collision. 
We develop a set of powerful analytic methods which, we believe,
render braneworld cosmological perturbation theory solvable.
Our particular concern is to determine the extent to which
the four-dimensional effective theory accurately captures the
higher-dimensional dynamics about the cosmic singularity.

We begin with a simple derivation of the low-energy effective action 
for braneworlds, highlighting the role of conformal invariance, before
showing how the effective action for a positive- and
negative-tension brane pair may be improved using the AdS/CFT correspondence.

We then solve for the cosmological perturbations 
in a five-dimensional background consisting of two separating or colliding 
boundary branes, as an expansion in the collision speed $V$ divided by the 
speed of light $c$.  
Our solution permits a detailed check of the validity of
four-dimensional effective theory in the vicinity of the event
corresponding to the big crunch/big bang singularity. 
We show that the four-dimensional description fails at the first
nontrivial order in $(V/c)^2$. 
At this order, there is nontrivial mixing of the two relevant
four-dimensional perturbation modes (the growing and decaying modes) as the 
boundary branes move from the narrowly-separated limit described by 
Kaluza-Klein theory to the well-separated limit where gravity is confined to
the positive-tension brane. 

We highlight the implications of this result for cosmology,
in particular for the propagation of a scale-invariant spectrum of
density perturbations across the bounce in a big crunch/big bang universe.
The generation of curvature perturbations on the
brane is also examined 
from a five-dimensional perspective.

Finally, as an application of our methods, we develop 
a new colliding-brane solution of the 
Ho{\v r}ava-Witten model of heterotic M-theory.

\end{abstracts}

%% file: introduction.tex
\chapter{Introduction}  


\begin{flushright}
\begin{minipage}{7cm}
\small
{\it \noindent
There is no excellent beauty that hath not \\ some strangeness in the proportion.
}
\begin{flushright}
\noindent 
Francis Bacon
\end{flushright}
\end{minipage}
\end{flushright}
\vspace{0.3cm}


One of the most striking implications of string theory and
M-theory is that there are extra spatial dimensions whose
size and shape determine the particle spectrum and couplings of the
low energy world. 
If the extra dimensions are compact and of fixed size $R$, their
existence results in a tower of Kaluza-Klein massive modes 
whose mass scale is set by $R^{-1}$. Unfortunately, this 
prediction is hard to test if the only energy scales 
accessible to experiment are much lower than $R^{-1}$.
At low energies, the massive modes
decouple from the low energy effective theory and are,
for all practical purposes, invisible. Therefore, we have no means
of checking whether the four-dimensional effective
theory observed to fit particle physics 
experiments is actually the outcome
of a simpler higher-dimensional theory.

The one situation where the extra dimensions seem
bound to reveal themselves is in cosmology. At the big bang,
the four-dimensional effective theory (Einstein gravity 
or its string-theoretic generalisation) breaks down, indicating 
that it must be replaced by an improved description.
Already, there are suggestions of improved behaviour
in higher-dimensional string theory and M-theory.
If matter is localised on two
branes bounding a higher-dimensional bulk, the matter density remains finite
at a brane collision even though this moment is, from the  
perspective of the four-dimensional effective theory,  
the big bang singularity \cite{Cyclicevo,Ekpyrotic,Seiberg}. 
Likewise, the equations of motion for fundamental strings are
actually regular at the collision in string theory, in the relevant
background solutions \cite{perry, Gustavo}.

We will adopt this scenario of colliding branes as our model of the big bang.
Our focus, however, will not be the initial singularity itself, 
but rather, 
the dynamics of higher-dimensional gravity as the universe emerges from a brane collision.
The model we study -- the Randall-Sundrum model \cite{RSI} -- is the simplest possible 
model of braneworld gravity, consisting of two empty $\Z_2$-branes (or orbifold planes)
of opposite tension, separated by a five-dimensional bulk with a negative cosmological
constant.
We develop a solution method for the bulk geometry, for both the background and cosmological perturbations, 
in the form of a perturbative expansion in $(V/c)^2$, where $V$ is the speed of the brane collision and $c$
is the speed of light \cite{long}.
Our solution allows us to track the evolution of the background and cosmological perturbations
from very early times right out to very late times, providing a benchmark against which the predictions of the 
four-dimensional effective theory can be tested.

We will find that the four-dimensional effective theory is accurate in two limits: that of early times,
for which the brane separation is significantly less than the anti-de Sitter (AdS) radius $L$; and that of late times, for which the brane separation is significantly greater than $L$.
In the former limit, the brane tensions and the warping of the bulk become negligible,
and a simple Kaluza-Klein description consisting of four-dimensional gravity and a scalar field applies (the gauge field zero mode having been eliminated by the $\Z_2$ projections).  In the latter limit, however, in which the branes are both widely separated and slowly-moving, the physics is qualitatively very different.  Rather than being uniform across the extra dimension, the low energy gravitational zero modes are now localised around the positive-tension brane, as shown by Randall and Sundrum \cite{RSII}.
Nevertheless, the four-dimensional effective theory describing this limit is identical, consisting of Einstein gravity and a scalar field, the radion, parameterising the separation of the branes.

Surprisingly, however, our five-dimensional solution reveals that in the transition between these two limits -- from Kaluza-Klein to Randall-Sundrum gravity -- the four-dimensional effective theory fails at first nontrivial order in $(V/c)^2$.  
In effect, the separation of the branes at finite velocity serves to excite massive bulk modes, which curb the accuracy of the four-dimensional effective theory until their decay in the late-time asymptotic region.
This process generates a striking signature impossible to forge within any local four-dimensional effective theory;
namely, the mixing of four-dimensional cosmological perturbation modes between early and late times.

This mode-mixing is conveniently described in four-dimensional longitudinal gauge, in which the sole physical degree of freedom associated with adiabatic\footnote{\ie perturbations which do not locally alter the matter history.} scalar perturbations is encoded in the Newtonian potential $\Phi_4$.
For sufficiently long wavelengths such that $|k t_4| \ll 1$, the four-dimensional effective theory predicts that
\[
\label{4dpred}
\Phi_4=A_4-\frac{B_4}{t_4^2},
\]
where $t_4$ is four-dimensional conformal time, and $A_4$ and $B_4$
are constants parameterising the amplitudes of the two perturbation
modes.  The first mode, with amplitude $A_4$, represents a curvature
perturbation on constant energy density or comoving spatial slices,
while the second mode, with amplitude $B_4$, corresponds to a local variation in the time elapsed since the big bang.  (In an expanding universe, the curvature perturbation comes to dominate over the time-delay perturbation at late times, and hence is often referred to as the `growing' mode.  In a collapsing universe, however, the two roles are reversed and the growing mode corresponds instead to the time-delay perturbation).

Now, if the dynamics were truly governed by a local four-dimensional
effective theory, the perturbation amplitudes $A_4$ and $B_4$ would be
constants of the motion.  From our five-dimensional solution, however, 
we can compute the {\it actual} asymptotic behaviour of the
four-dimensional effective Newtonian potential, at early and at late
times, by evaluating the
five-dimensional metric perturbations on the positive-tension
brane\footnote{Explicitly, the four-dimensional effective metric is
  related to the metric on the positive-tension brane via a conformal transformation, and so the anti-conformal part of the four-dimensional effective metric perturbation - namely, $\Phi_4$ - is equal to the anti-conformal part of the induced metric perturbation on the brane.}.
This allows us to identify the four-dimensional effective mode amplitudes $A_4$ and $B_4$ in terms of the  
underlying {\it five-dimensional} mode amplitudes, which are truly constant.  We find that, while the four-dimensional effective theory prediction (\ref{4dpred}) does indeed hold in the limit of both early and late times, the four-dimensional effective mode amplitudes $A_4$ and $B_4$ are mixed in the transition from early to late times.  (For example, if the system starts out purely in the time-delay mode at small $t_4$, then one ends up in a mixture of both the time-delay and the curvature perturbation modes as $t_4\tt\inf$).
This mixing first occurs at order $(V/c)^2$, reflecting the fact that the four-dimensional effective description holds good at leading (zeroth) order.
Equivalently, parameterising the mixing by a matrix relating the dimensionless mode amplitudes $A_4$ and $B_4(L^2/V^2)$ at early times to their counterparts at late times, we find that this matrix differs from the identity at order $(V/c)^2$.

Although this mixing is very small in the limit of a highly non-relativistic brane collision,
it is nonetheless of great significance when considering the origin of the primordial density perturbations.  
Inflation provides one possible for explanation for the origin of these primordial fluctuations by postulating an early phase of quasi-de Sitter expansion, in which quantum fluctuations of the inflaton field are exponentially stretched and amplified into a classical, scale-invariant pattern of large-scale curvature perturbations.

In a scenario such as the present, however, in which the universe undergoes a big crunch to big bang transition described by the collision of two branes, an alternative mechanism is feasible. 
In the ekpyrotic model \cite{Ekpyrotic}, and its cyclic extension
\cite{Steinhardt:2002ih, Cyclicevo}, 
the observed scale-invariant perturbations first arise as ripples on the two branes, 
produced as they attract one another over cosmological time scales.
These ripples are later imprinted on the hot radiation generated as the branes collide.
So far, this process has only been described from a four-dimensional effective point of view,
in which the perturbations are generated during a pre-big bang phase 
in which the four-dimensional Einstein-frame scale factor is contracting \cite{Khoury, Boyle, Gratton2, models}. 

A key difficulty faced by the ekpyrotic and cyclic scenarios is a scale-invariant spectrum of perturbations is generated only in the time-delay mode parameterised by $B_4$, which is the growing mode in a contracting universe.
Naively, this mode is then orthogonal to the growing mode in the present expanding phase of the universe, 
which is the curvature perturbation mode parameterised by $A_4$.  
(In an expanding universe the time-delay perturbation mode $B_4$ decays rapidly to zero).
In order to seed the formation of structure in the present universe, therefore, it is essential that some component of the growing mode time-delay perturbation in the collapsing phase be transmitted, with nonzero amplitude, to the growing mode curvature perturbation post-bang\footnote{This paradox is a recurring theme in the literature, see \eg \cite{gmode4, gmode1, gmode2, gmode3, gmode5, gmode6, gmode7, gmode8, jch2, gmode10, 
gmode11, gmode12, gmode13, Durrer, Copeland, Bozza&Veneziano}.}.

Clearly, our discovery that the four-dimensional perturbation modes mix in the full five-dimensional setup sheds new light on this problem.
The most simple-minded resolution would be for the ekpyrotic mechanism to generate a scale-invariant spectrum in $B_4$ long before the collision (when the branes are relatively far apart and the four-dimensional effective theory is still valid), with a piece of this subsequently being mixed into the curvature perturbation $A_4$ shortly before the collision.
Yet the failure of the four-dimensional effective theory at order $(V/c)^2$, coupled with  
the likely 
five-dimensional nature of any matching rule used to propagate perturbations across the singularity, suggests that instead a reformulation of the problem in terms of purely five-dimensional quantities is necessary. 

To this end, we define the curvature perturbation on the brane directly in five dimensions, showing how it differs from the curvature perturbation in the four-dimensional effective theory at order $(V/c)^2$.
We then proceed to analyse the generation of brane curvature through the action of an additional five-dimensional bulk stress (over and on top of the negative bulk cosmological constant) serving to pull the branes together.
Under the assumption that this additional bulk stress is small, a quantitative estimate of the final brane curvature perturbation prior to the collision can be obtained, although further progress is dependent upon the introduction of a specific five-dimensional model for the additional bulk stress.



Finally, it is worth bearing in mind that 
there is good hope of experimentally distinguishing inflation and the ekpyrotic mechanism
over the coming decades through their very different predictions for the long-wavelength
spectrum of primordial gravitational waves \cite{Boyleinflation}.


In addition to the cosmological ramifications of the work in this thesis, another line of enquiry being actively pursued is the application of our methods to more sophisticated braneworlds stemming from fundamental theory, such as the Ho{\v r}ava-Witten model of heterotic M-theory.
Our efforts to date form the basis of the concluding chapter of the present work.


%

\begin{center}
***
\end{center}

The plan of this thesis is as follows:  
we begin in Chapter \S\,\ref{earlyunivchapter} with an introduction to the early universe, encompassing cosmological perturbation theory and the generation of the primordial density perturbations, via both inflation and the ekpyrotic mechanism.  In Chapter \S\,\ref{branegravitychapter}, we review the Randall-Sundrum model and braneworld cosmology, before proceeding to introduce the low energy four-dimensional effective theory.
The subsequent chapters present original research material: in Chapter \S\,\ref{confsymmchapter}, we investigate the role of conformal invariance in the braneworld construction, showing how the form of the effective action up to quadratic order in derivatives is fully constrained by this symmetry \cite{Conf_sym}.  We also consider how to improve the effective action for a pair of branes of opposite tension through application of the AdS/CFT correspondence.
Chapter \S\,\ref{5dchapter} forms the heart of the present work.  In this chapter, we present our solution methods for the bulk geometry, and apply them to calculate the behaviour of both the background and perturbations in a big crunch/big bang cosmology \cite{long}.  We highlight the failure of the four-dimensional effective theory, and compute explicitly the mixing of four-dimensional effective perturbation modes.
In Chapter \S\,\ref{zetachapter}, we revisit the generation of curvature perturbations from a five-dimensional perspective, considering the action of an additional five-dimensional bulk stress.
Finally, in Chapter \S\,\ref{Mthchapter}, we present work in progress seeking a cosmological solution of the Ho{\v r}ava-Witten model with colliding branes, in which the five-dimensional geometry about the collision is that of a compactified Milne spacetime, and the Calabi-Yau volume at the collision is finite and nonzero.

%% file: chapter1.tex
\chapter{The early universe}
\label{earlyunivchapter}

%
%



\begin{flushright}
\begin{minipage}{7.5cm}
\small
{\it \noindent
 Now entertain conjecture of a time \\
 When creeping murmur and the poring dark \\
 Fills the wide vessel of the universe. }
\begin{flushright}
\noindent 
Henry V, Act IV.  
\end{flushright}
\end{minipage}
\end{flushright}
%




\section{The horizon problem}


Observations of the cosmic microwave background (CMB) allow a precise determination of the nature of the primordial density perturbations.  To date, these observations require the primordial density perturbations to be small-amplitude, 
adiabatic, Gaussian random fluctuations with a nearly scale-invariant spectrum \cite{WMAP3}. 
Even though this is almost the simplest conceivable possibility, to generate density perturbations of this nature demands new physics beyond the Standard Model.

The essence of the problem is simple causality: a universe originating in a big bang has a particle horizon associated with the fact that light has only travelled a finite distance in the finite time elapsed since the big bang.
For a universe with scale factor $a$ given in terms of the proper time $t$ by $a=(t/t_0)^p$, the size of this horizon in comoving coordinates is
\[
\label{d}
d(t) = \int_{t_i}^{t}\frac{\d t'}{a(t')} = \frac{p}{(1-p)}\,(\cH ^{-1}-\cH _i^{-1}), 
\]
where the comoving Hubble radius $\cH^{-1} =  (\d a /\d t)^{-1} 
= p^{-1}\,t_0^p\, t^{1-p}$ and $t_i$ is some initial time.  For standard matter, $p<1$ (\eg $p=1/2$ for radiation, $p=2/3$ for matter), and so the size of the horizon is always increasing.  At sufficiently late times, the integral is dominated by its upper limit; light travels the greatest distance at late times, and the horizon grows as $d\sim \cH^{-1}\sim t^{1-p}$.


Observations of the CMB, however, indicate that the Universe was quasi-homogeneous at the time of last scattering on scales much larger than the size of the causal horizon at that time.  Although the angular scale subtended by the horizon at last scattering is only $\sim 1$\textdegree, the temperature of the microwave sky is uniform to one part in a hundred thousand over much larger angular scales.  
This puzzle of explaining why the universe is nearly homogeneous over regions a priori causally independent is known as the {\it horizon problem}.


\subsection{Inflation}

One possible resolution of the horizon problem is that the early universe underwent a period of inflation, defined as an epoch in which the scale factor is accelerating, \iec $\ddot{a}>0$,
where dots denote differentiation with respect to proper time.  
The comoving Hubble radius $\cH^{-1}$ is therefore shrinking during inflation, since $\dot{\cH}=\ddot{a}>0$. (Note, however, that the converse, shrinking comoving Hubble radius implies inflation, is {\it not} true, as we will see in the next section).  

To see how inflation solves the horizon problem, let us adopt a simple model in which the scale factor expands exponentially, $a=\exp(H(t-t_0))$.  The size of the comoving horizon, equal to the comoving distance travelled by a light ray during inflation, is given by
\[
\label{d_inf}
d_\mathrm{infl.}(t)= \int_{t_i}^t \frac{\d t'}{a(t')} = \cH^{-1}_i - \cH^{-1},
\]
where the comoving Hubble radius $\cH^{-1}=(aH)^{-1}$.  Thus, due to the exponential expansion of the scale factor, after a sufficient number of e-foldings (defined as $N=\ln(a/a_i)$), 
the integral is dominated by the lower limit; light rays travel the greatest distance at early times and we find a near-constant horizon size $d_\mathrm{infl.}\approx \cH_i^{-1}$.
By choosing the time of onset of inflation $t_i$ to be sufficiently small, and the proper Hubble parameter $H$ to be sufficiently large, we can always ensure that the horizon size at the end of inflation is sufficiently large to solve the horizon problem.

As a rough estimate of the number of e-folds of inflation required, we stipulate that the comoving distance travelled by light during the radiation era must equal the comoving distance travelled by light during inflation, \iec the causal horizon at the end of inflation equals the size of our past light-cone at that time, since we live approximately at the end of the radiation era.  This ensures that the portion of the surface of reheating visible to us (and hence any later surface, such as the surface of last scattering) is causally connected.  

From (\ref{d}) with $p=1/2$, we see that the comoving distance travelled by light during the radiation era is approximately equal to the comoving Hubble radius at matter-radiation equality, $\cH^{-1}_\now$.  From (\ref{d_inf}), this must then equal the comoving Hubble radius at the onset of inflation, $\cH^{-1}_i$.  
The shrinking of the comoving Hubble radius during inflation is therefore matched by the growth of the comoving Hubble radius during the radiation era. 
Since during inflation $\cH^{-1}\sim a^{-1}$, whereas during radiation domination $\cH^{-1}\sim a$,
\[
\frac{a_\now}{a_\rh}=\frac{\cH^{-1}_\now}{\cH^{-1}_\rh}=\frac{\cH^{-1}_i}{\cH^{-1}_\rh}=\frac{a_\rh}{a_i}=e^N,
\]
where $a_\rh$ is the scale factor evaluated at reheating, {\it etc}.
Finally, since the scale factor is inversely proportional to the
temperature during radiation domination, we deduce that the number of
e-folds of inflation required is roughly $N\approx \ln(T_\rh/T_\now) \approx
\ln(10^{12}\, \GeV/1\,\mathrm{meV})\approx 55$, assuming that
reheating occurred approximately at the GUT scale\footnote{
Note that if inflation does terminate at the GUT scale, 
observational upper bounds on the density of monopoles constrain the
GUT scale to be less than $10^{12}\,\GeV$ \cite{Peacock}.}.

Having dealt with the horizon problem, we can now ask about the nature of the matter required to sustain a period of inflation.
From the Friedmann equation for a homogeneous and isotropic universe, 
\[
\frac{\ddot{a}}{a} = -\frac{1}{6}\,(\rho+3P)
\]
(where we have set $8\pi G=1$),
we see that inflation ($\ddot{a}>0$) requires matter with density $\rho$ and pressure $P$ such that $\rho+3P<0$. This corresponds to an exotic equation of state with $\w=P/\rho<-1/3$.  Ruling out a cosmological constant with $\w=-1$ on the grounds that inflation must at some point terminate, the next simplest candidate is a massless scalar field $\vphi$, minimally coupled to gravity, and with a potential $V(\vphi)$.  
The stress tensor then takes the form of that for a perfect fluid, with $\rho =\dot{\vphi}^2/2+V$ and $P =\dot{\vphi}^2/2-V$.  Inflation therefore requires that the potential energy is greater than the scalar field kinetic energy, $\dot{\vphi}^2 < V$.

From the action 
\[
\label{scalarfieldaction}
S = \half\int\sqrt{g}\left[R-(\D\vphi)^2-2V(\vphi)\right],
\]
the equations of motion are
\bea
\label{pbgdeom1}
&&3 H^2 =\half\,\dot{\vphi}^2+V, \\
\label{pbgdeom2}
&& 0 =\ddot{\vphi}+3H\dot{\vphi}+V_{,\vphi}, \\
\label{pbgdeom3}
&&\dot{H}=-\half\,\dot{\vphi}^2,
\eea
where the Hubble parameter $H=\dot{a}/a$, and the last equation follows from the first two.
A particularly useful limit in which the dynamics are simplified is the {\it slow-roll} regime, in which 
the scalar field kinetic energy is negligible, $\dot{\vphi}^2\ll V$, and the system is over-damped, $\ddot{\vphi} \ll H\dot{\vphi}$.
This holds provided the slow-roll parameters
\bea
\label{slowrolleps}
\eps &=&\half\,\(\frac{V_{,\vphi}}{V}\)^2 \approx  \half \frac{\dot{\vphi}^2}{H^2}, \\
\label{slowrolleta}
\eta &=& \frac{V_{,\vphi\vphi}}{V} \approx  -\frac{\ddot{\vphi}}{H\dot{\vphi}}+\half\frac{\dot{\vphi}^2}{H^2},
\eea
are much smaller than unity (the approximate relations following in the limit when this is true).
Under conditions of slow roll, the number of e-foldings obtained during inflation is simply
\[
N = \int_{\vphi_i}^{\vphi_f}\frac{\d \vphi}{\dot{\vphi}}H(\vphi) \approx
\int_{\vphi_i}^{\vphi_f} \frac{V}{V_{,\vphi}}\,\d \vphi.
\]

Finally, let us introduce a simple model that will prove useful, obtained by setting the 
equation of state parameter $\w=P/\rho$ to a constant.
In this case the dynamics are exactly solvable (independently of considerations of slow roll), yielding the scaling solution
\[
\label{toymodel}
a=\(\frac{t}{t_0}\)^p, \qquad \vphi = \sqrt{2p}\,\ln\(\frac{t}{t_0}\), 
\qquad V = -V_0 \,\exp(-\sqrt{\frac{2}{p}}\,\vphi),
\]
where $t_0$ and $V_0$ are given in terms of the constant parameter $p$ by $t_0=p-1$ and $V_0 = p\,(1-3p)/t_0^2$.  
The parameter $p$ is in turn related to the equation of state parameter $\w$ by 
$p =2/3(1+\w)$.  
We recover slow-roll inflation in the limit where $p\gg 1$ ($\w \approx -1$), since the slow-roll parameters are $\eps = 1/p\,$ and $\eta=2/p\,$.  In this case the potential takes the form of a positive-valued exponential.


\subsection{A collapsing universe}

An alternative resolution of the horizon problem is that the big bang is not the beginning of the universe.
Instead, the present expanding phase of the universe would be preceded by a collapsing phase, with a concomitant transition from big crunch to big bang.

A concrete example is provided by the simple model with constant $\w$ discussed above.
From (\ref{toymodel}) with $p>1$, we obtain an expanding universe for times $0\le t <\inf$.  If instead we take $0<p<1$, we obtain a contracting universe with $t_0<0$ and the time coordinate 
taking values in the range $-\inf<t\le 0$.  The potential now corresponds to a negative-valued exponential. 
Since $\w>-1/3$, the universe is not inflating and $\ddot{a}<0$.
Nonetheless, the comoving Hubble radius is still shrinking during the collapse: re-expressed in terms in of conformal time $\tau= -[t/(p-1)]^{1-p}$, the scale factor and scalar field in (\ref{toymodel}) are
\[
a=|\tau|^{p/(1-p)}, \qquad \vphi = \frac{\sqrt{2p}}{1-p}\ln|\tau|,
\]
where $-\inf<\tau\le 0$.
(Note that for $0<p<1$, the range $-\inf<\tau\le 0$ is mapped to $-\inf < t \le 0$, whereas for $p>1$, the same range is instead mapped to $0\le t < \inf$). For general $p$, the magnitude of the comoving Hubble radius is therefore 
\[
|\cH|^{-1} = \left|(p^{-1}-1)\,\tau\right|, 
\]
where we have taken the absolute value since in a collapsing universe $\cH<0$. 
With this definition, a shrinking comoving Hubble radius corresponds, in a collapsing universe, to the condition $\dot{\cH}=\ddot{a}<0$. 
In contrast, in an expanding universe, a shrinking Hubble radius implies $\dot{\cH}=\ddot{a}>0$, and hence inflation.

For a universe undergoing a bounce, the resolution of the horizon problem is trivial.
The comoving distance travelled by light rays during the collapsing phase is as given in (\ref{d}), except that now, owing to the shrinking of the comoving Hubble radius during the collapse, the integral is dominated by its lower limit: provided the collapsing phase began at some sufficiently negative initial time $t_i$, the comoving distance travelled scales as $|\cH_i|^{-1}$.  By increasing the duration of the collapsing phase, the horizon size at the big bang can be made arbitrarily large.  A lower bound is given by the size of our past light cone at the big bang, yielding the estimate $|\cH_i|^{-1} \sim \cH^{-1}_\now$, \iec the shrinkage of the comoving Hubble radius during the collapse is matched by its growth in the subsequent expanding phase. 

During both inflation and a collapsing era then, the comoving Hubble radius shrinks inside fixed comoving scales.  The two scenarios differ, however, when interpreted in physical coordinates.  While the proper distance corresponding to a fixed comoving length scales as $a=(t/t_0)^p$, the proper Hubble radius is $H^{-1}=t/p\,$.
In an inflating universe ($p>1$) therefore, the physical wavelength corresponding to a fixed comoving wavevector is stretched outside the Hubble radius as $t\tt \inf$.  In a collapsing universe ($0<p<1$), however, the opposite happens: as $t$ approaches zero from negative values, the shrinking of the Hubble radius outpaces the shrinking of physical wavelengths.

Finally, for completeness,
we note that a stability analysis \cite{Gratton2} shows that the background solution (\ref{toymodel}) is a stable attractor under small perturbations, both in the expanding case where $p>1$ ($\w<-1/3$), and also in the contracting case provided that $p<1/3$ ($\w>1$).
This conclusion remains valid when we include more general forms of matter.
Intuitively, in an inflating universe with $\w\approx-1$, this is because the scalar field potential energy density is nearly constant, whereas the curvature ($\propto a^{-2}$), matter ($\propto a^{-3}$), radiation ($\propto a^{-4}$) and other forms of energy density 
all decrease as the scale factor grows.
Likewise, in a collapsing universe with $\w\gg 1$, the scalar field kinetic energy density increases as $a^{-3(1+\w)}$ as the scale factor shrinks, whereas curvature, matter, radiation and other forms of energy density increase at a slower rate.
Consequently, in each case the dynamics are largely insensitive to the initial conditions: after a few e-folds of expansion or contraction, the solution converges rapidly to the attractor.  This resolves the so-called flatness and homogeneity problems, explaining why the present-day universe is so uniform and geometrically flat, without recourse to fine-tuning of the initial conditions.


\section{Classical cosmological perturbation theory}

We now turn our attention to the behaviour of classical perturbations in linearised general relativity.
Considering only the scalar sector relevant to the density perturbations, the perturbed line element for a flat FRW universe can be expressed in the general form
\[
\d s^2 = a^2(\tau) \( -(1+2\phi)\,\d \tau^2 + 2B_{,i} \d \tau \d x^i+((1-2\psi)\delta_{ij}+2E_{,ij})\,\d x^i \d x^j\),
\]
where the spacetime functions $\phi$, $\psi$, $B$ and $E$ are linear metric perturbations.
Under a gauge transformation $\xi^\mu = (\xi^0,\D^i\xi)$, the perturbations transform as
\bea
\phi & \tt & \phi - {\xi^0}' - \cH\xi ^0, \\
\psi & \tt & \psi+\cH \xi ^0, \\ 
B &\tt& B+\xi^0-\xi', \\
E &\tt& E-\xi,
\eea
where the primes indicate differentiation with respect to the conformal time $\tau$.
We may then construct the gauge-invariant Bardeen potentials \cite{bardeen}
\bea
\Phi &=& \phi+\cH(B-E')+(B-E')' ,\\
\Psi &=& \psi-\cH(B-E').
\eea

Another very useful gauge-invariant quantity is the curvature perturbation on comoving hypersurfaces, $\zeta$, defined in any gauge by
\[
\zeta = \psi-\frac{\cH}{\rho+P}\,q
\]
where the potential $q$ satisfies $q_{,i}=\delta T^0_i$.  (Thus, in comoving gauges characterised by $\delta T^0_i=q_{,i}=0$, $k^2 \zeta$ is proportional to the curvature of spatial slices, given by $R^{(3)}=-4k^2 \psi$).

We can easily relate this definition to an equivalent one in terms of the Bardeen potentials as follows.
Working in longitudinal gauge for convenience ($B_L=E_L=0$, and hence $\phi_L=\Phi$ and $\psi_L=\Psi$, where the subscript $L$ will be used to denote quantities evaluated in this gauge), the perturbed $G^0_i$ Einstein equation yields the momentum constraint equation
\[
\(\delta G^0_i\)_L = -\frac{2}{a^2} \left[\Psi'+\cH \Phi\right]_{,i} = (\delta T^0_i)_L = q_{L,i} 
\]
(where $8\pi G=1$ still).  Making use of the background Friedmann equation in the form $\cH'-\cH^2=-a^2(\rho+P)/2$, we therefore find (in any gauge now)
\[
\label{zetadef}
\zeta = \Psi-\frac{\cH (\Psi'+\cH \Phi)}{\cH'-\cH^2} =\Psi-\frac{H}{\dot{H}}\,(\dot{\Psi}+H\Phi),
\]
where we have reverted to proper time in the last equality.

The remaining components of the perturbed Einstein tensor, in longitudinal gauge, are:
\bea
\(\delta G^0_0\)_L &=& \frac{2}{a^2}\,\left[3\cH^2\Phi+3\cH\Psi'-\grad^2\Psi\right] ,\\
\(\delta G^i_j\)_L &=& \frac{1}{a^2}\,\left[2\Psi''+4\cH\Psi'-\grad^2(\Psi-\Phi)+2\cH\Phi'+2(2\cH'+\cH^2)\Phi\right]\delta^i_j \nonumber \\
&& \qquad +\frac{1}{a^2}\delta^{ik}(\Psi-\Phi)_{,jk}.
\eea
Specialising to the case of a single scalar field, the absence of anisotropic stresses ($\delta T^i_j=0$ for $i\not= j$) forces the two Bardeen potentials to coincide, thus $\Phi=\Psi$.

Evaluating the perturbed stress tensor to linear order and making use of the background equations of motion
in conformal time, 
\bea
\label{bgdeom1}
0 &=& 3\cH^2 - \half \vphi_0'^2-a^2 V, \\
\label{bgdeom2}
0 &=& \vphi_0''+2\cH\vphi_0'+a^2 V_{,\vphi}, \\
\label{bgdeom3}
0 &=& \cH'-\cH^2+\half\vphi_0'^2,
\eea
we find the linearised Einstein equations in longitudinal gauge read
\bea
\label{pert1}
\grad^2\Phi-3\cH\Phi'-(\cH'+2\cH^2)\Phi &=& \half\,\vphi_0'\delta\vphi_L'+\half\,a^2V_{,\vphi}\delta\vphi_L ,\\
\label{pert2}
\Phi'+\cH\Phi &=& \half\,\vphi_0'\delta\vphi_L ,\\
\label{pert3}
\Phi''+3\cH\Phi'+(\cH'+2\cH^2)\Phi &=& \half\,\vphi_0'\delta\vphi_L'-\half\,a^2V_{,\vphi}\delta\vphi_L ,
\eea
where $\delta\vphi_L$ denotes the perturbation of the scalar field in longitudinal gauge.

We now subtract (\ref{pert1}) from (\ref{pert3}). Using the background equation of motion (\ref{bgdeom2}), we can further subtract $(4\cH+2\vphi_0''/\vphi_0')$ times (\ref{pert2}) to find the following second-order differential equation for the Newtonian potential:
\[
\label{phieom}
\Phi''+2\(\cH-\frac{\vphi_0''}{\vphi_0'}\)\Phi'-\grad^2\Phi+2\(\cH'-\cH\frac{\vphi_0''}{\vphi_0'}\)\Phi = 0.
\]
(Note the absence of $\delta\vphi_L$ ensures that this equation is now gauge invariant).

Combining (\ref{phieom}) with (\ref{zetadef}) (recalling $\Phi=\Psi$), we then have
\[
\label{phifromzeta}
\frac{(\cH^2-\cH')}{\cH}\,\zeta' = \frac{\,\vphi_0'^2}{2\cH}\,\zeta'=\grad^2\Phi  .
\]
This tells us that on large scales, where the gradient term is negligible, $\zeta$ is conserved (except possibly if $\vphi_0'= 0$ at some point during the cosmological evolution, \eg during reheating when the inflaton field undergoes small oscillations).  In fact, even if $\vphi_0'$ does vanish at a particular moment in time, it turns out that $\zeta$ is still conserved outside the Hubble radius.  For quite general matter, it can be shown (see \eg \cite{LangloisInflRev}) using the perturbed continuity equation that $\zeta$ is conserved on super-Hubble scales, provided only that the perturbations are {\it adiabatic} (\iec every point in spacetime goes through the same matter history.  Note this automatically holds for single field models).


This accounts for the utility of $\zeta$ in the calculation of inflationary perturbations: once a mode exits the Hubble radius, $\zeta$ for that mode is conserved.  Despite our ignorance of the processes occurring during reheating, the initial conditions for a mode re-entering the Hubble radius in the radiation era are fully specified by the conserved value of $\zeta$ calculated as the mode left the Hubble radius during inflation.

\section{Perturbations in inflation}


Inflation stretches initial vacuum quantum fluctuations into macroscopic cosmological perturbations, thereby providing the initial conditions for the subsequent classical radiation-dominated era.  
Remarkably, the nearly scale-invariant spectrum of perturbations predicted by inflation 
exactly accounts for present-day observations of the CMB.
The computation of the primordial fluctuations arising in inflationary models was first discussed in 
\cite{Mukhanov:1981xt, Starobinsky:1982ee, Hawking:1982cz, Guth:1982ec, Bardeen:1983qw};
the present section comprises a review of these classic calculations.
(Other useful review articles include 
\cite{mukhanov, LangloisInflRev, Brandenberger:2005be, Turok:2002yq}).

\subsection{Massless scalar field in de Sitter}

Before launching into a full calculation of the quantisation of a scalar field including gravitational backreaction, let us first sharpen our claws on a simple toy model; namely, a massless scalar field on a {\it fixed} de Sitter background.
The cosmological scale factor is given by $a \propto \exp(Ht)$, which in conformal time reads $a = -1/H\tau$.  (Note the conformal time takes values in the range $-\inf<\tau<0$).
The action for a massless scalar field in this geometry is 
\[
S = -\half \int \dx \rootg (\D \vphi)^2 = \half \int \d \tau \,\d ^3x\, a^2[\vphi'^2-(\vgrad\vphi)^2],
\]
where differentiation with respect to conformal time is denoted by a prime.
Changing variables to $\chi=a\vphi$ and integrating by parts, we obtain
\[
S = \half\int\d\tau\,\d^3x\,[ \chi'^2-(\vgrad \chi)^2+\frac{a''}{a}\,\chi^2],
\]
where the kinetic term is now canonically normalised.  The fact that the scalar field lives in de Sitter spacetime rather than Minkowski results in a time-dependent effective mass
\[
m_{\mathrm{eff}}^2 = -\frac{a''}{a} = -\frac{2}{\tau^2}.
\]

The quantum field $\chihat$ may then be expanded in a basis of plane waves as
\[
\label{uhat}
\chihat(\tau,\x) = \frac{1}{(2\pi)^3}\,\int\d^3k \,[ \ahatk \,\chi_k (\tau)\, e^{i\k\.\x}+\ahatk^\dag \,\chi_k^*(\tau)\, e^{-i\k\.\x}],
\]
where the creation and annihilation operators $\ahat^\dag$ and $\ahat$ satisfy the usual bosonic commutation relations
\[
\label{acomm}
[\ahatk,\ahatkp]=[\ahatk^\dag,\ahatkp^\dag]=0, \qquad [\ahatk,\ahatkp^\dag]=\delta(\k-\k').
\]
The function $\chi_k(\tau)$ is a complex time-dependent function satisfying the classical equation of motion
\[
\label{ueom}
\chi_k''+(k^2-\frac{a''}{a})\chi_k=0,
\]
representing a simple harmonic oscillator with time-dependent mass.  In the case of de Sitter spacetime, the general solution is given by
\[
\chi_k = A\, e^{-ik\tau}\big(1-\frac{i}{k\tau}\big)+B\, e^{ik\tau}\big(1+\frac{i}{k\tau}\big).
\]
Canonical quantisation consists in imposing the following commutation relations on constant-$\tau$ hypersurfaces:
\[
[\chihat(\tau,\x),\,\chihat(\tau,\xp)]=[\pihatchi(\tau,\x),\,\pihatchi(\tau,\xp)]=0
\]
and
\[
\label{comm}
[\chihat(\tau,\x),\,\pihatchi(\tau,\xp)]=i\hbar \,\delta(\x-\xp),
\]
where the canonical momentum $\pihatchi = \delta S/\delta \chi = \chi'$.
Substituting the mode expansion (\ref{uhat}) into the commutator (\ref{comm}), and making use of (\ref{acomm}),
we obtain the condition
\[
\label{norm}
\chi_k \chi_k'^*-\chi_k^* \chi'_k = i \hbar,
\]
which fixes the normalisation of the Wronskian.

Specifying a particular choice for $\chi_k(\tau)$ corresponds to the choice of a particular physical vacuum $\ket{0}$, defined by $\ahatk \ket{0}=0$.
A different choice for $\chi_k(\tau)$ leads to a different decomposition into creation and annihilation operators, and thus to a different vacuum.
In our case, the most natural physical prescription for the vacuum is to take the particular solution that corresponds to the usual Minkowski vacuum $\chi_k \sim \exp(-ik\tau)$ in the limit $k|\tau|\gg 1$.  This is because the comoving Hubble radius $|\tau|^{-1}$ is shrinking during inflation: provided one travels sufficiently far back in time, then any given mode can always be found within the Hubble radius (\ie we can always choose $|\tau|$ sufficiently large such that $k|\tau|\gg 1$).
Once the wavelength of the mode lies within the Hubble radius, the curvature of spacetime has a negligible effect (\ie the effective mass term in (\ref{ueom}) is negligible) and the mode behaves as if it were in Minkowski spacetime.
Taking account of the normalisation condition (\ref{norm}), we thus arrive at the Bunch-Davies vacuum
\[
\chi_k = \sqrt{\frac{\hbar}{2k}}\,e^{-ik\tau}\big(1-\frac{i}{k\tau}\big).
\]

The power spectrum is defined via the Fourier transform of the correlation function, 
\[
\bra{0}\phih(\x_1)\phih(\x_2)\ket{0} = \int \d^3k\, e^{i\k\.(\x_1-\x_2)} \frac{\P_\vphi(k)}{4\pi k^3},
\]
where 
\[
2\pi^2 k^{-3} \P_\vphi = |\vphi_k|^2=\frac{|\chi_k|^2}{a^2}. 
\]
With this definition, the variance of $\vphi$ in real space is 
\[
\langle\vphi(\x)^2\rangle = \int \P_\vphi \,\d \ln{k}, 
\]
and so a constant $\P_\vphi$ corresponds to a scale-invariant spectrum, \iec one in which the field variance receives an equal contribution from each logarithmic interval of wavenumbers.  

Thus we find that modes well outside the Hubble radius, \iec those for which $k|\tau|\ll 1$, possess the scale-invariant spectrum
\[
\label{dSpowerspec}
\P_\vphi(k) \approx  \hbar \(\frac{H}{2\pi}\)^2.
\]
(Note also that in the opposite limit $k|\tau|\gg 1$, in which the mode is well inside the Hubble radius, we recover the usual result for fluctuations in the Minkowski vacuum, $\P_\vphi=\hbar(k/2\pi a)^2$).

Looking back, we can trace the origin of this scale invariance to the $-2/\tau^2$ effective mass term in the classical equation of motion (\ref{ueom}).  Thanks to this term, once a given mode $\chi_k$ exits the Hubble radius, it ceases to oscillate (\iec the mode `freezes') and instead scales as $\tau^{-1}$ in the limit as $\tau \tt 0$.  Since this amplification sets in at a time $\tau \sim k^{-1}$, the late-time mode amplitude picks up an extra factor of $k^{-1}$ on top of the $k^{-1/2}$ factor from the initial Minkowski vacuum, leading to scale invariance, \iec $|\chi_k|^2 \propto k^{-3}$.

\subsection{Including metric perturbations}
\label{ADMsection}

We now turn to the full calculation of the spectrum of inflationary perturbations including gravitational backreaction.

Recalling our discussion of classical cosmological perturbation theory, there are a total of five scalar modes; four from the metric ($\phi$, $\psi$, $B$ and $E$), and the perturbation of the inflaton itself, $\delta\vphi$.  Gauge invariance allows us to remove two of the five functions 
by fixing the time and spatial reparameterisations $x^\mu\tt x^\mu+\xi^\mu$, where $\xi^\mu=(\xi^0,\D^i\xi)$,  
while the Einstein constraint equations remove a further two.  We are left with a sole physical degree of freedom.

In order to derive the perturbed action up to second order in the metric perturbations, we will work in the ADM formalism, following \cite{Maldacena:2002vr}.
Foliating spacetime into spacelike 3-surfaces of constant time, the metric can be decomposed in terms of the lapse $N(t,x^i)$, the shift $N^i(t,x^i)$ and the spatial 3-metric $h_{ij}$, as
\[
\d s^2 = -N^2 \d t^2 + h_{ij}(\d x^i+N^i\d t)(\d x^j+N^j \d t).
\]
Spatial indices are lowered and raised using the spatial 3-metric.
The 4-curvature can be decomposed in terms of the spatial curvature $R^{(3)}$ and the extrinsic curvature tensor
\[
K_{ij} = \frac{1}{2N}\,(\dot{h}_{ij}-\grad_i N_j-\grad_j N_i)
\]
according to the Gauss-Codacci relation \cite{Wald}
\[
R = R^{(3)}+K_{ij}K^{ij}-K^2,
\]
where $K=K^i_i$ and we have omitted a total derivative term.
The relevant action 
(\ref{scalarfieldaction}) is thus
\bea
\label{ADMaction}
S &=& \half\int \d t \d^3 x \sqrt{h}\,\Big[NR^{(3)}-2NV+N^{-1}(E_{ij}E^{ij}-E^2)\nonumber \\
&& \qquad \qquad +N^{-1}(\dot{\vphi}-N^i\vphi_{,i})^2
		-Nh^{ij}\vphi_{,i}\vphi_{,j}\,\Big],
\eea
where, for convenience, we have defined $E_{ij}=NK_{ij}$. 

The ADM formalism is constructed 
so that $\vphi$ and $h_{ij}$ play the role of dynamical variables, while $N$ and $N^i$ are simply Lagrange multipliers.  Our strategy, after fixing the gauge, will be to solve for these Lagrange multipliers and then backsubstitute them into the action.

Considering only scalar modes, a convenient gauge is
\[
\label{zetagaugedef}
\delta\vphi=0, \qquad h_{ij}=a^2(\tau)\left[(1-2\zeta)\delta_{ij}\right], 
\]
where $a(\tau)$ is the background scale factor but $\zeta$ is a first-order perturbation.  
In this gauge the scalar field is unperturbed, hence $\delta T^0_i=0$ and the gauge is comoving. 
(Note also that the gauge is fully specified by the above).

The equations of motion for $N^i$ and $N$ are the momentum and Hamiltonian constraints respectively, which in this gauge take the form
\bea
&& \grad_i[N^{-1}(E^i_j-\delta^i_jE)]=0, \\
&& R^{(3)}-2V-N^{-2}(E_{ij}E^{ij}-E^2)-N^{-2}\dot{\vphi}_0^2=0.
\eea
At linear order, the solution of these equations is
\[
N=1-\frac{\dot{\zeta}}{H}, \qquad N^i=\delta^{ij}\,\D_j\Big(\frac{\zeta}{a^2H}-\chi\Big), \qquad \D_i^2 \chi = \frac{\dot{\vphi}_0^2}{2H^2}\,\dot{\zeta},
\]
where $\D_i^2$ is a shorthand notation for $\delta^{ij}\,\D_i\D_j$.

Substituting the above back into the action (\ref{ADMaction}), and expanding out to second order, we obtain a quadratic action for $\zeta$.  Note that for this purpose it is unnecessary to compute $N$ or $N^i$ to second order.  This is because these second order terms can only appear multiplying the zeroth order Hamiltonian and momentum constraint equations $\D\mathcal{L}/\D N$ and $\D\mathcal{L}/\D N^i$, which vanish for the background.
Direct replacement in the action gives, up to second order,
\bea
S&=&\half\int \d t\, \d^3 x \,a \,e^{-\zeta}(1-\frac{\dot{\zeta}}{H})[4\D_i^2\zeta-2(\D_i\zeta)^2-2a^2 Ve^{-2\zeta}]\nonumber \\
&&\qquad + a^3 e^{-3\zeta}(1-\frac{\dot{\zeta}}{H})^{-1}[-6(\dot{\zeta}-H)^2+\dot{\vphi}_0^2],
\eea
where we have neglected a total derivative and $(\D_i\zeta)^2 = \delta^{ij}\D_i\zeta\D_j\zeta$.
After integrating by parts, and using the background equations of motion (\ref{pbgdeom1}) to 
(\ref{pbgdeom3}), 
we obtain
\[
S=\half\int\d t\,\d^3x \frac{\dot{\vphi}_0^2}{H^2}\,[a^3 \dot{\zeta}^2-a(\D_i\zeta)^2]=-\half\int\d^4x\sqrt{-g}\,\frac{\dot{\vphi}_0^2}{H^2}\,(g^{\mu\nu}\D_\mu\zeta\D_\nu\zeta),
\]
where $g_{\mu\nu}$ is the background FRW metric.

To connect with our earlier discussion of a scalar field in de Sitter space, let us pass to conformal time and  canonically normalise the kinetic term by introducing the variable $\nu = z\zeta$, where $z=a\vphi_0'/\cH$. 
We find 
\[
S = \half\int\d \tau \d^3x\,[\nu'^2-(\D_i\nu)^2+\frac{z''}{z}\,\nu^2],
\]
the action for a scalar field in Minkowski spacetime with a time-dependent mass $z''/z$.  (Recall that in de Sitter spacetime the effective mass was instead $a''/a$).
Each Fourier mode obeys the equation of motion
\[
\nu_k''+(k^2-\frac{z''}{z})\nu_k=0,
\]
or equivalently,
\[
\label{zetaeom}
\zeta_k''+2\big(\frac{z'}{z}\big)\zeta_k'+k^2\zeta_k=0.
\]
(Since $\zeta$ is gauge-invariant, and $z$ depends only on background quantities, this relation in fact holds in any gauge).

Under conditions of slow-roll, the evolution of $\vphi_0$ and $H$ is much slower than that of the scale factor $a$, and so to leading order in slow-roll one finds
\[
\frac{z''}{z} \approx  \frac{a''}{a}.
\]
Consequently, all the results of the previous section now apply to our variable $\nu$ in the slow-roll approximation.
The correctly normalised expression corresponding to the Bunch-Davies vacuum is approximately
\[
\nu_k \approx  \sqrt{\frac{\hbar}{2k}}\,e^{-ik\tau}\big(1-\frac{i}{k\tau}\big).
\]
In the super-Hubble limit, $k|\tau|\ll 1$, this becomes
\[
\nu_k\approx -\sqrt{\frac{\hbar}{2k}}\frac{i}{k\tau} \approx i\sqrt{\frac{\hbar}{2k}}\frac{a H_*}{k},
\]
where we have used $a\approx  -1/H\tau$ and the asterisk indicates that, to get the most accurate approximation, we should evaluate the value of $H$ at the moment the mode crosses the Hubble radius.

Hence, on scales much larger than the Hubble radius, the power spectrum for $\zeta$ is given by 
\[
\label{zetapowerspec}
\P_\zeta (k)= \frac{k^3}{2\pi^2}\frac{|\nu_k|^2}{z^2} \approx  \frac{\hbar}{4\pi^2}\(\frac{H_*^4}{\,\dot{\vphi}_{0*}^2}\),
\]
where we have reverted once more to physical time, and the quantities labelled with asterisks are to be evaluated when the mode exits the Hubble radius.
This is the celebrated result for the spectrum of scalar cosmological perturbations generated from vacuum fluctuations during slow-roll inflation.

The spectrum is not strictly scale-invariant, however, since the values of the Hubble parameter and the scalar field velocity evolve slowly over time.  This introduces a small additional momentum dependence conveniently parameterised by the spectral index $n_s$, where
\[
n_s(k)-1=\frac{\d \ln\P_\zeta(k)}{\d \ln k} \approx  \frac{1}{H_*}\frac{\d}{\d t_*}\ln\(\frac{H_*^4}{\dot{\vphi}_{0*}^2}\) = -2\(\frac{\ddot{\vphi}_{0*}}{H_*\dot{\vphi}_{0*}}+\frac{\dot{\vphi}_{0*}^2}{H_*^2}\)=2(\eta-3\eps),
\]
using the slow-roll parameters as defined in (\ref{slowrolleps}) and (\ref{slowrolleta}).
A spectral index $n_s=1$ therefore corresponds to an exactly scale-invariant spectrum, whereas $n_s<1$ and $n_s>1$ correspond to a red and a blue spectrum respectively. (Equivalently, we have $k^3 |\zeta_k|^2 \propto k^{n_s-1}$).

A useful heuristic to understand the parametric dependence of the power spectrum (\ref{zetapowerspec}) is as follows.  
Consider a mode just about to cross the Hubble radius and freeze out.  On a constant-time slice the quantum fluctuations in the inflaton field will be of order $\delta\vphi \sim H_*$, approximating the background as de Sitter space and using (\ref{dSpowerspec}).  Comoving surfaces, characterised by $\delta\vphi=0$, are therefore offset from surfaces of constant time by a time delay $\delta t \sim \delta\vphi/\dot{\vphi}\sim H_*/\dot{\vphi}$.
The exponential expansion of the background spacetime as $a\sim \exp(H_*t)$ then leads to a comoving curvature perturbation $\zeta \sim \delta a/a \sim H_* \delta t \sim H_*^2/\dot{\vphi}$, 
consistent with (\ref{zetapowerspec}).

\section{Perturbations in a collapsing universe}

We have seen 
that models formulated in terms of gravity and a single scalar field possess only one physical scalar degree of freedom at linear order.  In the case of inflation, we chose to parameterise this degree of freedom in a gauge-invariant fashion through use of the comoving curvature perturbation $\zeta$.  The utility of this variable lies in its conservation on super-Hubble scales, allowing the amplitudes of modes re-entering the Hubble radius during the radiation era to be calculated independently of the detailed microphysics of re-heating.
In the case of a collapsing universe, however, the situation is more subtle.  The final spectrum of perturbations in the expanding phase is, in general, sensitive to the prescription with which the perturbations are matched across the bounce.  
In particular, it does not necessarily follow that $\zeta$ is conserved across the bounce. 
In this section we will therefore adopt a more generic approach, computing the behaviour of both $\zeta$ and the Newtonian potential $\Phi$ 
during the collapse.  

\subsection{The story of $\zeta$ and $\Phi$}
Let us start by collecting together a number of our earlier results concerning $\zeta$ and $\Phi$.
These variables satisfy the second order linear differential equations
given in (\ref{phieom}) and (\ref{zetaeom}), and are related to each other via (\ref{zetadef}) (recalling that the absence of anisotropic stress for scalar field matter implies that $\Phi=\Psi$) and (\ref{phifromzeta}).
To express these relations more compactly, it is useful to introduce the surrogate variables $u$ and $\nu$, defined as
\[
\label{surrogatevars}
u = \frac{a}{\vphi_0'}\,\Phi, \qquad v=z \zeta,
\]
where the background quantity $z= a\vphi_0'/\cH$ as before.  In particular, $u$ and $\nu$ have the same $k$-dependence as $\Phi$ and $\zeta$, and hence the same spectral properties.
Defining $\theta=1/z$, (\ref{phieom}) and (\ref{zetaeom}) are
\bea
\label{phiueom}
0&=& u''+ (k^2- \frac{\theta''}{\theta})\,u, \\
\label{nueom}
0&=&\nu''+ (k^2-\frac{z''}{z})\,\nu,
\eea
and the relations (\ref{zetadef}) and (\ref{phifromzeta}) become 
\bea
\label{uvrelations}
k\nu &=& 2k \,(u'+\frac{z'}{z}\,u), \\
-ku &=& \frac{1}{2k}\,(\nu'+\frac{\theta'}{\theta}\,\nu).
\eea

Choosing a boundary condition for $u$ and $\nu$ corresponds to specifying a vacuum state for the fluctuations. As usual, we will take this to be the Minkowski vacuum of a comoving observer in the far past, at a time when all comoving scales were far inside the Hubble radius.  Thus, in the limit as $\tau\tt-\inf$,
\bea
\label{ubc}
u&\tt& i\sqrt{\hbar}\,(2k)^{-3/2}e^{-ik\tau}, \\
\label{nubc}
\nu &\tt& \sqrt{\hbar}\,(2k)^{-1/2}e^{-ik\tau}.
\eea
(It is easy to check from (\ref{uvrelations}) that these two boundary conditions are equivalent).

Assuming the equation of state is constant while all wavelengths of cosmological interest leave the Hubble radius, from our scaling solution (\ref{toymodel}) with $\w=2/3p-1$, we find
\bea
\label{thetaduality}
\frac{\,\theta''}{\theta} &=& \frac{p}{(p-1)^2\, \tau^2}\hspace{0.5mm}, \\
\label{zduality}
\frac{\,z''}{z} &=& \frac{p\,(2p-1)}{(p-1)^2 \,\tau^2}\hspace{0.5mm}.
\eea
The solutions of (\ref{phiueom}) and (\ref{nueom}) are then easily obtained in terms of Hankel functions:
\bea
u(x) &=& x^{1/2}\left[A^{(1)}H^{(1)}_\alpha (x)+A^{(2)}H^{(2)}_\alpha (x)\right], \\
\nu(x) &=& x^{1/2}\left[B^{(1)}H^{(1)}_\beta (x)+B^{(2)}H^{(2)}_\beta (x)\right] ,
\eea
where $x=k|\tau|$ is a dimensionless time variable, $A^{(1,2)}$ and $B^{(1,2)}$ are constants,
and the order of the Hankel functions $H^{(1,2)}_s(x)$ is 
\bea
\label{alphadef}
\alpha &=& [(\theta''/\theta)\,\tau^2+1/4]^{1/2} = \half \left|\,\frac{1+p}{1-p}\,\right|, \\
\label{betadef}
\beta &=& [(z''/z)\,\tau^2+1/4]^{1/2} = \half \left|\frac{1-3p}{1-p}\,\right| .
\eea

In the far past ($x\tt \inf$), when comoving scales are well inside the Hubble radius, the asymptotic behaviour of the Hankel functions is
\[
H^{(1,2)}_s (x) \tt \sqrt{\frac{2}{\pi x}}\,\exp\left[\pm i\(x-\frac{s\pi}{2}-\frac{\pi}{4}\)\right]
\]
(where the $+$ sign corresponds to $H^{(1)}_s(x)$).  The boundary conditions (\ref{ubc}) and (\ref{nubc}) then imply
\bea
\label{uresult}
u &=& \frac{\lambda_1}{2k}\,(\hbar\pi x/4k)^{1/2}H^{(1)}_\alpha (x) ,\\
\label{nuresult}
\nu &=& \lambda_2(\hbar\pi x/4k)^{1/2}H^{(1)}_\beta (x),
\eea
where $\lambda_1$ and $\lambda_2$ are the $k$-independent complex phase factors
\bea
\lambda_1 &=& \exp[i(2\alpha+3)\pi/4], \\
\label{lambda2def}
\lambda_2 &=& \exp[i(2\beta+1)\pi/4]. 
\eea

Pausing to catch our breath,  
we notice a curious duality: the equation of motion for $u$ ((\ref{phiueom}), along with (\ref{thetaduality})) is invariant under $p\tt 1/p\,$ \cite{Boyle}.  Moreover, the boundary condition, (\ref{ubc}), is independent of $p$ (as is natural, since the boundary condition is imposed in the far past when comoving scales are well inside the Hubble radius).
Consequently, our expressions (\ref{alphadef}) for $\alpha$, and (\ref{uresult}) for $u$, are invariant under $p\tt 1/p\,$. The spectrum of fluctuations of the Newtonian potential $\Phi$ in an expanding universe with $p>1$ is therefore identical to that in a collapsing universe with $\hat{p}=1/p\,<1$.
(The same cannot be said for $\zeta$, however, as (\ref{zduality}) is not invariant under $p\tt1/p\,$).


\subsection{Power spectra}

Returning now to the calculation of the power spectra for $\zeta$ and $\Phi$, at late times when comoving scales are well outside Hubble radius ($x\tt 0$), the asymptotic behaviour of the Hankel functions is 
\[
\label{H1asymptotics}
H^{(1)}_s (x) \tt -\frac{i}{\pi}\,\Gamma(s)\(\frac{x}{2}\)^{-s}\hspace{-4mm},
\]
where $s>0$ and $\Gamma(s)$ is the Euler gamma function.
On scales much larger than the Hubble radius, the power spectra for $\zeta$ is therefore 
\[
\label{fullzetaspec}
\P_\zeta (k) = \frac{k^3}{2\pi^2}\,\frac{|\nu|^2}{z^2} = \frac{\hbar}{4\pi^2}\(\frac{H_*^4}{\,\dot{\vphi}_{0*}^2}\)
\Lambda(p) \,x^{3-2\beta},
\]
where asterisked quantities are to be evaluated at Hubble radius crossing ($k=aH$), and the $k$-independent numerical factor $\Lambda(p)$ is 
\[
\Lambda(p) = \Gamma^2(\beta)\,\frac{4^\beta(1-p)^2}{2\pi p^2}\,.
\]
We immediately see that the power spectrum for $\zeta$ is only scale-invariant when $\beta=3/2$, cancelling the $x$-dependence on the right-hand side of (\ref{fullzetaspec}).  Equivalently, the spectral index $n_\zeta$ (given by $n_\zeta-1 = 3-2\beta$) must be unity for scale invariance.  From (\ref{betadef}), this requires $p\tt \inf$ ($\w\approx -1$), corresponding to an inflating universe in the slow-roll limit (recall that the slow-roll parameter $\eps=1/p\,$).
In this limit, the numerical factor $\Lambda(p)$ tends to unity and we recover our earlier result (\ref{zetapowerspec}).    

On the other hand, the power spectra on super-Hubble scales for the Newtonian potential $\Phi$ is given by
\[
\label{phipowerspec}
\P_\Phi (k) = \frac{k^3}{2\pi^2}\,\frac{|u|^2 \vphi_0'^2}{a^2}=\frac{1}{(2\pi)^3}\,\dot{\vphi}_{0*}^2\Gamma^2(\alpha)4^{\alpha-1}x^{1-2\alpha}.
\]
The spectral index for $\Phi$ is thus $n_\Phi-1=1-2\alpha$, with scale invariance requiring $\alpha=1/2$.  From (\ref{alphadef}), this is satisfied in only two cases:
firstly, that of an expanding universe undergoing slow-roll inflation ($p\tt\inf$); and secondly, the case of a very slowly collapsing universe with $p\ll 1$ ($\w\gg 1$).  (In this latter scenario, the scalar field is rapidly rolling down a steep negative exponential potential).  
From our earlier remarks about duality, this result is exactly as expected: the power spectrum (\ref{phipowerspec}) is invariant under $p\tt 1/p\,$, since $\alpha$ is invariant.
Nonetheless, it is remarkable that a nearly scale-invariant spectrum can be obtained, without inflation, on a background which is very nearly static Minkowski.  

To sum up, 
while in the case of slow-roll inflation both $\Phi$ and $\zeta$ acquire a scale-invariant spectrum well outside the Hubble radius, in the case of a slowly contracting universe, $\Phi$ is scale-invariant but $\zeta$ is blue ($n_\zeta -1 \approx 2$, indicating greater power at short wavelengths).  

Our analysis has been restricted to the case in which the equation of state is constant.  
More generally, however, when the equation of state is a slowly varying function of time, one can introduce the `fast-roll' parameters $\bar{\eps}$ and $\bar{\eta}$ defined by
\[
\bar{\eps}=\(\frac{V}{V_{,\vphi}}\)^2, \qquad
\bar{\eta}= 1-\frac{V V_{,\vphi\vphi}}{V_{,\vphi}^2}.
\]
Analogously to the case of inflation, one can then express the spectral index in terms of these fast-roll parameters.  A short calculation \cite{Gratton2} yields the result 
\[
n_s-1=-4(\bar{\eps}+\bar{\eta}).
\]

\subsection{The relation between $\zeta$ and $\Phi$}

Since a generic scalar fluctuation can be described using either $\zeta$ or $\Phi$ with equal validity, the fact that one variable possesses a scale-invariant spectrum, while the other does not, is at first sight puzzling.  
To understand this behaviour, it is useful to probe further the relationship between $\zeta$ and $\Phi$.
This is given by
\[
\label{neatzeta}
\zeta= \frac{p}{a}\,\frac{\d}{\d t}\(\frac{a\Phi}{H}\),
\]
which follows directly 
from the definition of $\zeta$ given in (\ref{zetadef}), and the equation of motion $H^2/\dot{H}=-p$.

Expanding once more the Hankel function $H^{(1)}(x)$ as $x\tt 0$, but this time retaining the subleading term as well, we find
\[
H^{(1)}_s (x)= -\frac{i}{\pi}\,\Gamma(s)\(\frac{x}{2}\)^{-s}-\frac{i}{\pi}\,e^{-i\pi s}\,\Gamma(-s)\(\frac{x}{2}\)^s + O(x^{2-s},\,x^{2+s}).
\]
Up to $k$-independent numerical coefficients,
the leading and subleading behaviour of $\Phi$ as $t\tt 0$ is therefore
\[
\label{Phiscale}
\Phi \sim k^{-3/2-p}\,t^{-1-p}+ k^{-1/2+p}\,t_0^{p},
\]
where we have used (\ref{uresult}) and expanded all exponents up to linear order in $p$ (implying $\tau \sim t^{1-p}$). 
The leading term in this expression, the `growing' mode, is responsible for the scale invariance of $\Phi$.  Since $a/H \sim t^{1+p}$, however, we immediately see from (\ref{neatzeta}) that this term makes no contribution to $\zeta$.  Instead, the leading contribution to $\zeta$ comes from the subleading (or `decaying') mode in $\Phi$, which scales as $k^{-1/2+p}$.  The spectrum for $\zeta$ is therefore boosted by a factor of $k^{2+2p}$ relative to scale invariance, yielding a blue spectrum with $n_\zeta-1\approx 2$.

For completeness, from (\ref{nuresult}), the growing and decaying modes for $\zeta$, in the limit as $t\tt 0$, are
\[
\label{zetascale}
\zeta \sim k^{-1/2+p}\,t_0^p+k^{1/2-p}\,t^{1-3p}\,t_0^{2p},
\]
where again we have expanded the exponents to linear order in $p$ and we are neglecting $k$-independent numerical coefficients.
(Incidentally, it should be borne in mind that in a collapsing universe the terms `growing' and `decaying' are gauge-dependent.  The behaviour $\Phi \sim t^{-1-p}$ is the growing mode in Newtonian gauge, yet the same physical fluctuation `decays' as $\zeta\sim t^{1-3p}$ in the gauge specified by (\ref{zetagaugedef})).



\section{Frontiers}


\subsection{Matching at the bounce}

Having seen 
how perturbations are generated in a collapsing universe, we must now
ask how these perturbations relate to observable quantities in the
present-day universe. 

In the expanding phase of the universe $t$ is positive and tending towards $+\inf$, with $p=1/2$ during radiation dominance, followed by $p=2/3$ during matter dominance.  From the analysis leading to (\ref{Phiscale}) and (\ref{zetascale}) (but without expanding the exponents), we see that the dominant contribution to both $\Phi$ and $\zeta$ at late times comes from the time-independent piece, and that both time-dependent pieces decay to zero.
The $\Phi\sim\zeta\sim const.$ mode is thus the `growing' mode in an expanding universe, and
provides the dominant contribution to physical quantities as modes re-enter the Hubble radius.  
Compatibility with observation therefore requires this mode to have a scale-invariant spectrum.

From (\ref{Phiscale}), however, we see that the time-independent term in $\Phi$ does {\it not} have a scale-invariant spectrum in the collapsing phase.  Nevertheless, it can be argued  that the matching rules across the bounce induce a generic mixing between the two solutions, allowing the scale-invariant spectrum of the $\Phi\sim t^{-1-p}$ mode in the collapsing phase to be inherited by the $\Phi\sim const.$ mode in the expanding phase \cite{Ekpyrotic, Gratton2}.  In other words, the mixing at the bounce must match a piece of the scale-invariant growing mode in the collapsing phase to the growing mode in the expanding phase, even though from a naive perspective these modes are orthogonal.

A proposal realising this in the full five-dimensional braneworld setup was made in \cite{TTS}, based on the analytic continuation of variables defined on time slices synchronous with the collision.  
(For subsequent criticism of this proposal, however, see \cite{Creminelli}).  Since the nature of the matching at the bounce continues to be a subject of active research we will not pursue it further here.  
Instead, as explained in the Introduction,
 our chief concern will be to understand how the problem of propagating a scale-invariant spectrum of density perturbations across the bounce is modified, when restored to its true higher-dimensional setting.

\subsection{Breaking the degeneracy: gravitational waves}

As we have seen, 
the power spectrum of the Newtonian potential in a collapsing universe is identical to that of its expanding dual.
Fortunately this degeneracy is broken by tensor perturbations, providing a crucial signature for future observations.

Returning to our calculations in Section $\S\,$\ref{ADMsection} based on the ADM formalism, to compute the action for tensor perturbations we can set  
\[
\delta\vphi=0, \qquad h_{ij}=a^2(\tau)[\delta_{ij}+\gamma_{ij}],
\]
where the tensor perturbation $\gamma_{ij}$ satisfies $\D_j \gamma_{ji}=\gamma_{ii}=0$ and
is automatically gauge-invariant.
Inserting this into the action (\ref{ADMaction}) and expanding to quadratic order, we find
\[
S = \frac{1}{8}\int \d t \,\d^3 x \,[a^3 \dot{\gamma}_{ij}\dot{\gamma}_{ij}-a \,\D_k\gamma_{ij}\D_k\gamma_{ij}].
\]
Expressing $\gamma_{ij}$ in a basis of plane waves with definite polarisation tensors as
\[
\gamma_{ij} = \int \frac{\d^3 k}{(2\pi)^3}\sum_{s=\pm}\eps^s_{ij}(k)\,\gamma^s_{\vec{k}}(t)\,e^{i\vec{k}\cdot\vec{x}},
\]
where 
$\eps_{ii}=k^i\eps_{ij}=0$ and $\eps^s_{ij}(k)\,\eps^{s'}_{ij}(k)=2\delta_{ss'}$,
we see that each polarisation mode obeys essentially the same equation of motion as a massless scalar field:
\[
\gamma''+2\cH\gamma'+k^2\gamma=0.
\]
This can be recast in canonical form by setting $\chi=a \gamma$, yielding
\[
\chi''+(k^2-\frac{a''}{a})\chi=0.
\]
Again, the standard vacuum choice is the Minkowski vacuum of a comoving observer in the far past, corresponding to the boundary condition
\[
\label{chibc}
\chi \tt (\hbar/2k)^{1/2}e^{-ik\tau}
\]
as $\tau \tt -\inf$.
To solve for $\chi$, one need only observe that the equation of motion for $\chi$ is identical to that for $\nu$ given in (\ref{nueom}) (since $z\propto a$ for constant $\w$, hence $z''/z=a''/a$).
The boundary conditions (\ref{chibc}) and (\ref{nubc}) are also identical.  The solution for $\chi$ therefore follows from (\ref{nuresult});
\[
\chi = \nu = \lambda_2(\hbar\pi x/4k)^{1/2}H^{(1)}_\beta (x),
\]
where $x=k|\tau|$, and $\beta$ and $\lambda_2$ are as defined in (\ref{betadef}) and (\ref{lambda2def}) respectively.
Defining the tensor spectral index $n_T$ such that $k^3 |\chi|^2 \propto k^{n_T-1}$, on super-Hubble scales the mode freezes and we have, from (\ref{H1asymptotics}),
\[
n_T-1 = 3-2\beta=3-\left|\frac{1-3p}{1-p}\,\right|.
\]
This expression is \textit{not} invariant under $p\tt 1/p\,$: an expanding universe with a given $p$ produces a tensor spectrum which is much redder than the tensor spectrum produced in a contracting universe with $\hat{p}=1/p\,$ as, from the above, $n_T\le \hat{n}_T-2$.

\section{Summary}

In this chapter we have reviewed two independent scenarios for generating the scale-invariant spectrum of primordial density fluctuations required to fit observation.  In the first scenario, inflation, the early universe undergoes a brief period of accelerated expansion in which physical wavelengths are stretched outside the quasi-static Hubble radius.  The alternative is a slowly collapsing universe followed by a big crunch to big bang transition, as proposed in the cyclic and ekpyrotic models.  In this scenario scale-invariant density perturbations are generated in the growing mode of the Newtonian potential during the collapsing phase, as the Hubble radius shrinks rapidly inside the near-constant physical length scales of the perturbations.  

Thus far, all our considerations have been couched in the framework of four-dimensional effective theory.  In the following chapter we leave the prosaic world of four dimensions, and recast our view of the universe in five dimensions.

%% file: chapter2.tex
\chapter{Braneworld gravity}
\label{branegravitychapter}





\begin{flushright}
\begin{minipage}{9.4cm}
\small
{\it \noindent
Now it is time to explain what was before obscurely said: \\
there was an error in imagining that all the four elements \\
might be generated by and into one another... \\
There was yet a fifth combination which God used \\ in the delineation of the universe.
}
\begin{flushright}
\noindent 
Plato, Timaeus.  
\end{flushright}
\end{minipage}
\end{flushright}
%




One of the most striking ideas to emerge from string theory is that the universe we inhabit may be a brane embedded in, or bounding, a higher-dimensional spacetime \cite{PolchinskiI}.  
The key to this picture is the notion that gravity, being the dynamics of spacetime itself, is free to roam in the full higher-dimensional bulk spacetime, whereas the usual standard model forces are confined to the brane. 
In this section we review the simplest phenomenological model of such a scenario, the Randall-Sundrum model, focusing on its cosmological properties and on its four-dimensional effective description.
For a broader survey of the braneworld literature, see in particular \cite{Brax:2003fv, langlois, maartens}.

\section{The Randall-Sundrum model}

The first Randall-Sundrum model \cite{RSI} comprises a pair of four-dimensional positive- and negative-tension branes embedded in a five-dimensional bulk with a negative cosmological constant.  
For simplicity, a $\Z_2$ symmetry is imposed about each brane, so that the dimension normal to the branes is compactified on an $S^1/\Z_2$ orbifold.  
This scenario provides the simplest possible setup incorporating branes with a nontrivial warp factor in the bulk.

The complete action takes the form:  
\[
\label{5Daction}
S = 2\int_5 \sqrt{-g}\(\half\,m_5^3 R - \Lambda\)-\sum_{\pm} \int_\pm\sqrt{-g^\pm}\(2m_5^3 K^\pm+\sigma^\pm+\mathcal{L}^\pm_\mathrm{matter}\),
\]
where the first group of terms represent the action of the bulk ($m_5$ is the five-dimensional Planck mass, 
$R$ is the five-dimensional bulk Ricci scalar, and $\Lambda<0$ is the bulk cosmological constant), and the second set of terms encode the action of two branes of tension $\sigma^\pm$, each with its own induced metric
 $g^\pm_{ab}$ (where Latin indices run from $0$ to $4$), and (optional) matter content specified by $\mathcal{L}^\pm_\mathrm{matter}$.   

 To deal with the assumed $\Z_2$ symmetry about each brane we have adopted an extended zone scheme in which each brane appears only once, but there are two identical copies of the bulk (hence the factor of two multiplying the bulk action). 
Each brane presents two surfaces to the bulk, and therefore contributes two copies of the Gibbons-Hawking surface term proportional to $K^\pm$, the trace of the extrinsic curvature $K^\pm_{ab}$ (itself defined as the projection onto the brane of the gradient of $n^\pm_a$, the outward-pointing unit normal, 
\iec $K^\pm_{ab}={g^\pm}_a^c \grad_c n^\pm_b$).  
The purpose of these surface terms is to cancel out the second set of surface terms arising from the integration by parts of the Einstein-Hilbert action (required to remove second derivatives of the metric, furnishing an action quadratic in first derivatives only).

\subsection{The Israel matching conditions}

In practice, however, it is usually simplest to adopt a ``cut and paste'' approach when dealing with branes: given a solution of the bulk equations of motion, corresponding to a solution of five-dimensional gravity with a negative cosmological constant, we can create a $\Z_2$-brane simply by gluing together two copies of this geometry.
The total stress-energy $T_{ab}$ induced on the brane is then given in terms of the jump in extrinsic curvature across the brane via the Israel matching conditions \cite{Israel}:
\[
\label{israel}
T_{ab} =-m_5^3\,\([K_{ab}]-h_{ab}[K]\),
\]
where $h_{ab}$ denotes the induced metric on the brane.  
(A simple derivation of this result is provided in Appendix \ref{Isapp}, where we study the Einstein equations in the presence of distributional sources).
To evaluate the jump $[...]$, we introduce a normal vector to the brane and calculate the extrinsic curvature term in the bracket on both sides of the brane using the same normal.  We then subtract the terms calculated on one side (the side from which the normal points away) from the terms calculated on the other side (the side to which the normal is pointing).  It is easy to see that the final result does not depend on the choice of normal: reversing 
the normal changes the order of the subtraction, but also the sign of
the extrinsic curvatures.

Alternatively, we could have evaluated the extrinsic curvatures on each side
using two different normals (pointing away from the brane in each
case).  Their sum is then equal to the jump defined above.  In the
case of a $\Z_2$-symmetric brane, the jump therefore reduces to twice
the extrinsic curvature at the brane, evaluated using the outward-pointing unit normal.
This allows the Israel matching conditions to be re-written as
\[
\label{israel2}
m_5^3\, K_{ab}=-\frac{1}{2}\,(T_{ab}-\frac{1}{3}\,h_{ab}T).
\]
If the stress-energy on the brane is fixed in advance, then these equations constrain the embedding of the brane into the bulk geometry.  (We will see an explicit example of this when we come to consider brane cosmology: the Israel matching conditions will determine the trajectory of the brane through the bulk, and hence the form of the induced Friedmann equation on the brane).


\section{Static solutions}

An obvious solution for the bulk geometry is to take a slice of five-dimensional anti-de Sitter (AdS) spacetime,
\[
\label{staticsoln}
\d s^2 = \d Y^2 + e^{2Y/L}\eta_{\mu\nu}\d x^\mu \d x^\nu,
\]
where $L$ is the AdS radius (defined as $L^2=-6m_5^3 /\Lambda$), and the indices $\mu$ and $\nu$ run from $0$ to $3$.
Let us consider bounding the slice of AdS with flat branes located at constant $Y=Y^\pm$, with $Y^+ > Y^-$.
The coordinates $x^\mu$ then parameterise the four dimensions tangent to the branes, and $Y$ parameterises the direction normal to the branes.
Determining the stress-energy on the branes via the Israel matching conditions (\ref{israel}) as discussed above, a simple calculation shows that $T^\pm_{\mu\nu} = \mp(6m_5^3/L) \,g^\pm_{\mu\nu}$.  To have static branes located at fixed $Y^\pm$ therefore, we see that the branes must be empty apart from a constant tension $\sigma^\pm = \pm 6 m_5^3/L$ (so that $T^\pm_{\mu\nu}=-\sigma^\pm g^\pm_{\mu\nu}$).  

With the tensions tuned to these values there is then a continuous one-parameter family of static solutions, parameterised by the distance between the branes $Y^+ - Y^-$.  (The dependence on the overall position in the warp factor can be removed by a re-scaling of the coordinates tangential to the brane).
Heuristically, the stability of the system derives from a balance between, on the one hand, the cosmological constant and gravitational potential energy of the bulk, and on the other hand, the energy stored in the tensions on the branes.  The warp factor dictates that an increase in the volume of the bulk (rendering the bulk energy more negative) is always accompanied by a corresponding increase in the area (and hence energy) of the positive-tension brane, 
and/or a decrease in the area (hence increase in the energy) of the negative-tension brane.




A useful generalisation of the static solution above is found by taking the bulk line element to be
\[
\label{genstaticsoln}
\d s^2 = \d Y^2 + e^{2Y/L} g_{\mu\nu}(x)\d x^\mu \d x^\nu, 
\]
where $g_{\mu\nu}(x)$ is any Ricci-flat four-dimensional metric.  (In particular, one could choose the Schwarzschild solution, leading to the braneworld black string \cite{BS}).
Again, with the brane tensions appropriately tuned, a solution exists for branes located at arbitrary constant $Y=Y^\pm$.
Later, we will use this solution as the starting point in our derivation of the four-dimensional effective theory via the moduli space approximation.

\section{Cosmological solutions}

Another obvious choice for the bulk metric is the AdS-Schwarzschild black hole.  
This satisfies the five-dimensional Einstein equations with a negative cosmological constant, and moreover, has the interesting property that the horizon is a three-dimensional manifold of constant curvature, which may be either positive, negative, or zero \cite{Birmingham}.  
(This is just one example of the considerably richer phenomenology of black holes in higher dimensions).  
Our suspicions are immediately alerted to the possibility of embedding a 
three-brane of constant spatial curvature into this bulk geometry, thereby obtaining a cosmological solution of the Randall-Sundrum model.

More precisely, the line element for AdS-Schwarzschild takes the form
\[
\label{AdS-Schw}
\d s^2 = -f(r) \d t^2 + f^{-1}(r)\d r^2+r^2 \d\Omega_3^2, 
\]
where the constant-curvature three-geometry, 
\[
\label{omega3}
\d\Omega_3^2=(1-k\chi^2)^{-1}\d\chi^2
+\chi^2\d\Omega_2^2,
\]
is either open, closed or flat according to whether $k=-1$, $+1$ or $0$ respectively,
and where $\d\Omega_2^2$ denotes the usual line element on an $S^2$.
The function $f(r)=k-\mu/r^2+r^2/L^2$, where $L$ is related to the bulk cosmological constant $\Lambda = -6m_5^3/L^2$ as before, and the black hole mass parameter $\mu$ is an arbitrary constant. 
If $\mu$ is positive we have a black hole in AdS, but we can also choose $\mu$ to be zero, yielding pure AdS, or even negative, describing a naked singularity in AdS.  (If $\mu=0$, the different choices for $k$ then correspond respectively to closed, open or flat slicings of AdS).

In the next section bar one, we will show how to embed a brane in this
bulk geometry to obtain an induced metric of FRW form.  The
corresponding Friedmann equations then follow from the Israel 
matching conditions, assuming perfect fluid matter on the branes.
First, however, we will reassure ourselves that AdS-Schwarzschild is
indeed the most general possible choice for the bulk metric, granted
only cosmological symmetry on the branes. 

\subsection{Generalising Birkhoff's theorem} 
\label{Birkhoffsection}

We begin by recalling Birkhoff's theorem in four dimensions: that the
unique spherically symmetric vacuum solution of Einstein's equations
is the static  \linebreak Schwar\-zschild geometry.  
The assumption of spherical symmetry completely determines the form of the metric, up to two arbitrary functions of time and the radial coordinate.  Application of the Einstein equations then fixes these two functions in terms of the radial coordinate alone, yielding the Schwarzschild solution.

Physically, Birkhoff's theorem rules out the emission of gravitational waves by a pulsating spherically symmetric body.
Furthermore, in a configuration consisting of a number of infinitely thin, concentric spherical shells of non-vanishing stress-energy, with everywhere else being held in vacuum, the gravitational field at any point is unaltered by the re-positioning of shells both interior and exterior to this point (provided only that interior shells do not cross to the exterior and vice versa).  Gravity reduced to (1+1) dimensions by means of a symmetry is therefore completely local.

In the present, five-dimensional context (where we also have a bulk cosmological constant), a similar theorem applies: the assumption of cosmological symmetry on the branes, amounting to three-dimensional spatial homogeneity and isotropy, constrains the form of the metric sufficiently that, upon application of the Einstein equations, we obtain a unique solution.  With the right choice of time-slicing, this solution is moreover static.

In analogy with our discussion of concentric shells, the physical import of this generalised Birkhoff theorem is that the bulk geometry seen by a brane is completely unaffected by the presence, and even the motion, of any other branes placed in the same bulk spacetime. (Except, of course, when collisions occur).

To derive the theorem explicitly, following \cite{Gregory} 
(but using different sign conventions), we start by writing the bulk line element, without loss of generality, as
\[
\label{vBansatz}
\d s^2 = e^{2\sigma}B^{-2/3}(-\d \tau^2 +\d y^2)+B^{2/3}\d \Omega_3^2,
\]
where $\sigma(\tau, y)$ and $B(\tau, y)$ are arbitrary functions, 
and $\d\Omega_3^2$ is as defined in (\ref{omega3}).
This represents the most general bulk metric consistent with three-dimensional spatial homogeneity and isotropy, after we have made use of our freedom to write the ($\tau$, $y$) part of the metric in a conformally flat form.
We have further chosen the exponents with the aim of simplifying the bulk Einstein equations as much as possible.
The latter read $G_{ab}=-(\Lambda/m_5^3)\, g_{ab}=(6/L^2)\,g_{ab}$, or equivalently, 
$R_{ab}=-(4/L^2) \,g_{ab}$. 
For the metric (\ref{vBansatz}), we then obtain the following system of partial differential equations:
\bea
B_{,yy}-B_{,\tau\tau} &=& [12L^{-2} B^{1/3}+6kB^{-1/3}]\,e^{2\sigma}, \\
\sigma_{,yy}-\sigma_{,\tau\tau} &=& [2L^{-2}B^{-2/3}-kB^{-4/3}]\,e^{2\sigma},\\
B_{,yy}+B_{\tau\tau} &=& 2\sigma_{,\tau}B_{,\tau}+2\sigma_{,y}B_{,y},\\
B_{,\tau y} &=& \sigma_{,y}B_{,\tau}+\sigma_{,\tau}B_{,y}.
\eea
Switching to the light-cone coordinates
\[
u=\half\,(\tau-y), \qquad v =\half\,(\tau+y),
\]
we find
\bea
\label{PDE1}
-B_{,uv} &=& [12L^{-2} B^{1/3}+6kB^{-1/3}]\,e^{2\sigma}, \\
-\sigma_{,uv} &=& [2L^{-2}B^{-2/3}-kB^{-4/3}]\,e^{2\sigma},\\
B_{,uu} &=& 2\sigma_{,u}B_{,u}, \\
B_{,vv} &=& 2\sigma_{,v}B_{,v}.
\eea
The latter two equations can then be directly integrated to give
\[
e^{2\sigma}=V'(v)B_{,u}=U'(u)B_{,v},
\]
where $U'(u)$ and $V'(v)$ are arbitrary nonzero functions, and we will use primes to denote ordinary differentiation wherever the argument of a function is unique.
It follows that $B$ and $\sigma$ take the form
\[
B=B(U(u)+V(v)), \qquad e^{2\sigma}=B'U'V',
\]
which reduces the sole remaining partial differential equation (\ref{PDE1}) to the ordinary differential equation
\bea
B''+(12L^{-2} B^{1/3}+6kB^{-1/3})B' &=& 0 \\
\label{Beqn}
\Rightarrow B'+9L^{-2}B^{4/3}+9kB^{2/3}&=& 9\mu,
\eea
where $\mu$ is a constant of integration.

The metric then takes the form
\bea
\d s^2 &=& B'U'V' B^{-2/3}(-4\d u\d v)+B^{2/3}\d\Omega_3^2\\
&=&-4B^{-2/3}B'\d U \d V+B^{2/3}\d\Omega_3^2.
\eea
Upon changing coordinates to
\[
r=B^{1/3}, \qquad t=3(V-U),
\]
we recover the AdS-Schwarzschild metric 
\[
\d s^2 = -f(r)\d t^2 +f^{-1}(r)\d r^2 + r^2 \d \Omega_3^2,
\]
where, from (\ref{Beqn}),
\[
\label{f}
f(r)=-\frac{r'}{3}=\frac{1}{9}\,B^{-2/3}B'=k-\frac{\mu}{r^2}+\frac{r^2}{L^2}.
\]

\subsection{The modified Friedmann equations}

The trajectory of a brane embedded in AdS-Schwarzschild can be
specified in parametric form as $t=T(\tau)$ and $r=R(\tau)$, 
where $\tau$ is the proper time.
The functions $T(\tau)$ and $R(\tau)$ are then constrained by
\[
g_{ab}u^a u^b = -f\dot{T}^2+f^{-1}\dot{R}^2 = -1,
\]
where $u^a=(\dot{T},\,\dot{R},\, \vec{0})$ is the brane velocity, dots denote
differentiation with respect to $\tau$, and $f=f(r)$ is as defined in (\ref{f}).
Thus 
\[
\label{Tdotrel}
\dot{T}=f^{-1}(f+\dot{R}^2)^{1/2},
\]
and the components of the
unit normal vector (defined such that $n_a u^a = 0$ and $n_a
n^a=1$) are 
\[
n^\pm_a = \pm\(\dot{R},\,-f^{-1}(f+\dot{R}^2)^{1/2},\, \vec{0}\),
\]
where the choice of sign corresponds to our choice of which side of
the bulk we keep prior to imposing the $\Z_2$ symmetry.
If we keep the side for which $r\le R(\tau)$ (leading to the creation of 
a positive-tension brane), then the normal points
in the direction of decreasing $r$ and we must take the positive sign
in the expression above.
If, instead, we choose to retain the $r\ge R(\tau)$ side of the bulk
(resulting in the formation of a negative-tension brane), then the normal
points in the direction of increasing $r$ and we must take the
negative sign.
Either way, the four-dimensional induced metric on the brane is 
\[
\d s_\pm^2 = -\d \tau^2 + R^2(\tau)\d\Omega_3^2,
\]
and hence the cosmological scale factor on the brane can be
identified with the radial coordinate $R(\tau)$.
The motion of the brane through the bulk
therefore translates into an apparent cosmological evolution on the brane.

To calculate the brane trajectory, we must solve the Israel
matching conditions \cite{KrausFRW, IdaFRW}.  
There are two nontrivial relations; one
deriving from the $\tau\tau$ component, and the other from the
3-spatial components of the extrinsic curvature. 
Beginning with the latter,
writing $\d \Omega_3^2 = \gamma_{ij}\,\d x^i \d x^j$ for $i,\,j=1,\,2,\,3$,
the $ij$ components of the extrinsic curvature tensor are
\[
K^\pm_{ij} = \grad_i n^\pm_j = \half\,n^{\pm a} \D_a g_{ij} = \mp\frac{(f+\dot{R}^2)^{1/2}}{R}\,g_{ij},
\]
where the 3-metric $g_{ij}= r^2 \gamma_{ij}$, and the superscript $\pm$ indicates whether the brane is of positive or negative tension.
Assuming a perfect fluid matter content in addition to the background tension $\sigma^\pm$, the stress-energy tensor on the 
brane takes the form
\[
T^\pm_{ab} = (\rho+p)\, u_a u_b +(p-\sigma^\pm) g^\pm_{ab},
\]
where $\rho$ is the energy density of the fluid, $p$ is its
pressure, and $g^\pm_{ab}$ is the induced metric on the brane.  
The $ij$ components of the matching condition (\ref{israel2}) then yield
the relation
\[
\label{Kij}
\pm m_5^3\,(f+\dot{R}^2)^{1/2} = \frac{1}{6}\,(\rho+\sigma^\pm)\,R .
\]
After squaring this expression, substituting for $f$ using (\ref{f}), replacing the brane tensions by their critical values $\sigma^\pm=\pm 6m_5^3/L$, and relabelling $R(\tau)$ as $a(\tau)$, we find
\[
\label{modFriedmann}
H^2 = \frac{\dot{a}^2}{a^2} = \pm \frac{\rho}{3 m_5^3 L}+\frac{\rho^2}{36 m_5^6}-\frac{k}{a^2}+\frac{\mu}{a^4}.
\]
This is known as the modified Friedmann equation \cite{Binetruy}, and is the braneworld counterpart of the usual Friedmann equation.
Provided the energy density is much less than the critical tension (as is typically the case for standard matter or radiation at sufficiently late times), the linear term in $\rho$ dominates over the quadratic term.  Then, if we set $m_5^3 L = m_4^2$, where $m_4$ is the four-dimensional Planck mass, the modified Friedmann equation is a good approximation to its conventional four-dimensional analogue.
(At least on a positive-tension brane where matter gravitates with the correct sign!)
The additional $\mu a^{-4}$ term is known as the dark radiation term, since it scales like radiation but does not interact with standard matter directly, on account of its gravitational origin.

A second relation follows from the $\tau\tau$ component of the extrinsic curvature, given by
\[
K_{\tau\tau}=K_{ab}u^a u^b = u^a u^b \grad_a n_b = -n_c a^c,
\]
where the acceleration $a^c$ is defined by $a^c=u^b\grad_b u^c$.  Since $a^c u_c=0$, the acceleration can also be written as $a^c = a n^c$, where $a=-K_{\tau\tau}$.
Then, since $\D_t$ is a Killing vector of the static background\footnote{For a Killing vector $\xi^c$, 
$
a n_\xi = an_c \xi^c = a_c \xi^c=\xi^c u^b \grad_b u_c = u^b \grad_b (u_c \xi^c) - u^c u^b \grad_b \xi_c ,
$  
where the last term vanishes by Killing's equation, $\grad_{(b} \xi_{c)}=0$, hence
$a n_\xi = u^b \D_b u_\xi = (\d u_\xi/\d \tau)$.}, we have $a=n_{t}^{-1} (\d u_t/\d \tau)$. 

The $\tau\tau$ component of the Israel matching condition (\ref{israel2}) therefore reads
\[
m_5^3\,K^\pm_{\tau\tau}=\pm \frac{m_5^3}{\dot{R}}\,\frac{\d}{\d \tau}\,(f+\dot{R}^2)^{1/2}
= -\frac{1}{6}\,(2\rho+3p-\sigma^\pm).
\]
Combining this with our first relation (\ref{Kij}) then yields the usual cosmological conservation equation
\[
\dot{\rho}+3H(\rho+p)=0,
\]
where $H=\dot{R}/R=\dot{a}/a$ is the Hubble parameter as above.

\section{A big crunch/big bang spacetime}

Having understood the dynamics of a single brane embedded in AdS-Schwarzschild, 
we turn now to the behaviour of a pair of positive- and negative-tension branes 
when embedded in this same bulk spacetime.  

A useful alternative coordinate system for the 
bulk may be found by
setting $f^{-1}(r)\,\d r^2 = \d Y^2$.  
Restricting our attention to the spatially flat case in which $k=0$, 
the bulk line element (\ref{AdS-Schw}) then takes the form
\[
\label{Bmetric}
\d s^2 = \d Y^2 -N^2(Y) \d T^2 + b^2(Y)\d \vec{x}^2,
\]
where for pure AdS ($\mu=0$),
\[
b^2(Y)=N^2(Y)=e^{2Y/L};
\]
for AdS-Schwarzschild ($\mu>0$) with horizon at $Y=0$,
\[
\label{AdSSsol}
b^2(Y)=\cosh{2Y/L}, \qquad N^2(Y) = \frac{\sinh^2(2Y/L)}{\cosh(2Y/L)};
\]
and for AdS with a naked singularity at $Y=0$ ($\mu <0$),
\[
b^2(Y)=\sinh{2Y/L}, \qquad N^2(Y) = \frac{\cosh^2(2Y/L)}{\sinh(2Y/L)}.
\]
(In the latter two cases, the value of $\mu$ has been absorbed into the definition of $T$ 
and the spatial $x^i$).

For a brane with trajectory $Y=Y^\pm(T)$, the induced metric is
\[
\d s_\pm^2 = -(N^2_\pm-Y^{\pm\,2}_{,T})\d T^2 + b^2_\pm\d \vec{x}^2,
\]
where $N_\pm = N(Y^\pm)$ and similarly $b_\pm = b(Y^\pm)$.
The proper time $\tau$ on the brane then satisfies
\[
\label{taurel}
\d \tau = \sqrt{N_\pm^2-Y^{\pm 2}_{,T}}\,\d T.
\]
In these new coordinates, the spatial components of the Israel matching conditions, (\ref{Kij}), can be put in the form
\[
\label{Newisrael}
\frac{N_\pm}{b_\pm} = \(1\pm \frac{\rho_\pm L}{6 m_5^3}\)\sqrt{1-Y_{,T}^{\pm\,2}/N_\pm^2},
\]
where 
to obtain this expression we have performed the coordinate transformation explicitly on a case by case basis, starting from (\ref{Kij}) with the critical tension, then using (\ref{Tdotrel}) (not confusing the $\dot{T}$ referring to the parameterisation $t=T(\tau)$ with our present coordinate $T$) and (\ref{taurel}).

\subsection{Trajectories of empty branes}

\begin{figure}[!htbp]
  \begin{center}
       \includegraphics[width=10cm]{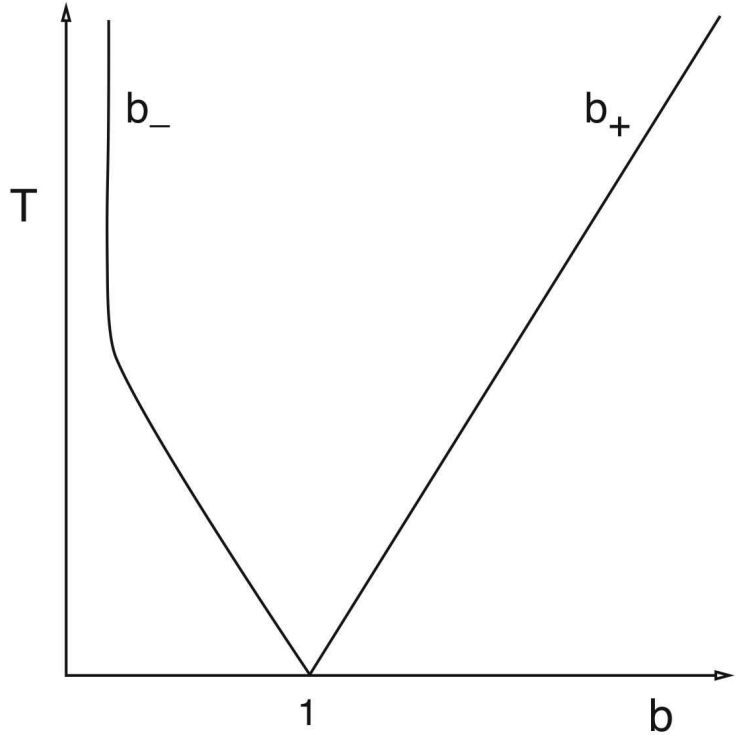}
	\caption{The background brane scale factors $b_\pm$ plotted as a function of the Birkhoff-frame time $T$, 
	where $b_{\pm}$ have 	been normalised to unity at $T=0$.  
	In these coordinates the bulk is AdS-Schwarzschild. The
	brane trajectories are then determined by integrating the Israel
	matching conditions, yielding the result (\ref{traj}).  
	In the limit as $T\tt \inf$, the
	negative-tension brane asymptotes to the event horizon of the black
	hole, while the positive-tension brane asymptotes to the
	boundary of AdS.}
	\label{branesep}
  \end{center}
\end{figure}

In the special case where the branes are empty ($\rho_\pm =0$) as well as flat, the modified Friedmann equations (\ref{modFriedmann}) imply that the bulk must be AdS-Schwarzschild with $\mu>0$ in order to have moving branes (and hence a nontrivial cosmological evolution).  The relevant bulk solution is then (\ref{AdSSsol}),
and the Israel matching condition (\ref{Newisrael}) simplifies to 
\[
Y^\pm_{\,,T} = 
\pm \sinh(2Y^\pm/L)\,\cosh^{-{3/2}}(2Y^\pm/L),
\]
where we have identified the positive root with the positive-tension brane and vice versa.
Equivalently, in terms of the dimensionless brane scale factor $b_\pm$ defined in (\ref{AdSSsol}), we have
\[
L\, b^\pm_{\,,T}=\pm (1-b^{-4}_\pm).
\]
This is easily integrated to yield the brane trajectories:
\[
\label{traj}
\pm \frac{T}{L} = b_\pm -\half\,\tan^{-1}{b_\pm}-\frac{1}{4}\,\ln{\(\frac{b_\pm +1}{b_\pm-1}\)}-C_\pm,
\]
where the $C_\pm$ are constants of integration. 
From the form of these expressions, we see that the branes must inevitably collide at some instant in time.
Translating our time coordinate so that this collision occurs at $T=0$, the constants of integration are then identical;
\[
C_\pm = b_0 -\half\,\tan^{-1}{b_0}-\frac{1}{4}\,\ln{\(\frac{b_0 +1}{b_0-1}\)},
\]
where $b_0$ is the joint scale factor of both branes at the collision.
(If desired, we could of course set $b_0=1$ by a re-scaling of the brane spatial coordinates). 
The brane trajectories are illustrated in Figure \ref{branesep}.  
After emerging from the collision, the positive-tension brane escapes to the boundary of AdS, while the negative-tension brane asymptotes to the event horizon of the bulk black hole.  

\subsection{Brane-static coordinates}

Our analysis of braneworld cosmological dynamics has centred so far on coordinate systems in which the bulk is static and the branes are moving.  Yet it is also possible, however, to find alternative coordinates in which the branes are static, and the bulk evolves dynamically with time \cite{TTS}.  This has the advantage of simplifying the form of the Israel matching conditions, which will be especially useful when we consider cosmological perturbations.  The disadvantage is that the explicit form of the bulk solution in these coordinates is not known.  Nevertheless, it is easy to see that such a transformation from bulk-static to brane-static coordinates must exist, as we now show.

Starting from the Birkhoff-frame metric (\ref{Bmetric}), with $b$ and $N$ given by (\ref{AdSSsol}), we change coordinates from $Y$ to $Z$, where $\d Z = \d Y /N$ and $Z$ is chosen to be zero at the collision event.  Then we have
\[
\d s^2 = N^2 (-\d T^2 + \d Z^2)+b^2\d\vec{x}^2,
\]
where $N$ and $b$ are now functions of $Z$.  Passing to lightcone
coordinates, $T^\pm = T\pm Z$, we have
\[
\d s^2 = N^2(- \d T^+\d T^-) + b^2 \d \vec{x}^2.
\]
Then, under the lightcone coordinate transformation $\bar{T}^\pm = f_\pm(T^\pm)=\bar{T}\pm\bar{Z}$,
\[
\d s^2 = \frac{N^2}{f'_+ f'_-}\,(-\d \bar{T}^2+\d \bar{Z}^2)+b^2 \d \vec{x}^2.
\]
Setting $y=\bar{Z}$ and $t = \pm \,\exp(\pm\bar{T})$ (to describe post- and pre-collision spacetimes respectively), and defining $n^2(t,y)\, t^2 = N^2/f'_+ f'_-$, 
the metric takes the general form
\[
\label{newmetric}
\d s^2 = -n^2(t,y)(-\d t^2 + t^2 \d y^2) + b^2(t,y)\d\vec{x}^2.
\]
Through a suitable choice of the functions $f_\pm$, we can always make
the branes static in the new coordinates.
To see this, observe that the new spatial coordinate $y$ satisfies
\[
\label{yeqn}
y = \half\,[f_+(T+Z)-f_-(T-Z)],
\]
and hence is a solution of the two-dimensional wave equation.  
From the general theory of the wave equation, we can always find a solution $y(T,Z)$ for arbitrarily chosen boundary conditions on the branes (which are themselves described by timelike curves $Z=Z_\pm(T)$).
In particular, we are free to choose $y=y_+$ on the positive-tension brane and $y=y_-$ on the negative-tension brane, for constant $y_\pm$.  (Even after this choice there is additional coordinate freedom, since to completely specify $y(T,Z)$ we need additional Cauchy data, \eg on a constant-time slice).

For the case of empty flat branes discussed in the previous subsection,
\[
Z(b) = \frac{L}{2}\,\left[\tan^{-1}b+\half\,\ln\(\frac{b-1}{b+1}\)\right],
\]
where $b^2=\cosh(2Y/L)$. The brane trajectories $b=b_\pm(T)$ are given implicitly by (\ref{traj}), from which, in principle at least, one could determine the trajectories in the form $Z=Z_\pm(T)$.
The two equations
\[
y_\pm = \half\,[f_+(T+Z_\pm(T))-f_-(T-Z_\pm(T))]
\]
then determine the necessary coordinate transformation.
Equivalently, since the $y_\pm$ are assumed to be constant, differentiating with respect to $T$ gives
\[
f_+(T+Z_\pm(T))(1+V_\pm(T)) = f_-(T-Z_\pm(T))(1-V_\pm(T)),
\]
where $V_\pm(T) = Z^\pm_{\,,T}$ are the brane velocities.
The solution of these equations is not, however, immediately apparent.
Later, when we return to this coordinate system 
in Chapter \S\,\ref{5dchapter}, we will simply construct the metric functions $n$ and $b$ in (\ref{newmetric}) from scratch, by solving the bulk Einstein equations.

\section{Four-dimensional effective theory}

In this section, we will use the moduli space approximation to derive the four-dimensional effective theory describing the Randall-Sundrum model at low energies.
Before we begin, however, it is interesting to consider 
how the nature of 
the dimensional reduction employed in braneworld theories differs from its counterpart in Kaluza-Klein theory.



\subsection{Exact versus inexact truncations}

In a dimensional reduction of the Kaluza-Klein type, 
one expands the higher-dimensional fields in terms of a complete set of harmonics on the compact internal space, before truncating to the massless sector of the resulting four-dimensional theory.
Crucially, this truncation is {\it consistent}, in the sense that all solutions of the lower-dimensional truncated theory are also solutions of the higher-dimensional theory.  In practice, this means that when one considers the equations of motion for the lower-dimensional massive fields prior to truncation, there are no source terms constructed purely from the massless fields that are to be retained.  Thus, if one starts the system off purely in the massless zero mode, it will remain in this mode for all time.  The subsequent dynamics are then exactly described by the four-dimensional effective theory.

Braneworld gravity, in contrast, does not in general possess a consistent truncation down to four dimensions:
the ansatz used in the dimensional reduction, rather than being an exact solution, is typically only an
approximate solution of the higher-dimensional field
equations\footnote{For an interesting counterexample, see however \cite{Koyama}.}.
In effect, the warping of the bulk ensures that the massive higher Kaluza-Klein modes are sourced by the zero mode.  If the branes are moving at nonzero speed, then generically these higher Kaluza-Klein modes will be continuously produced.  The regime of validity of the four-dimensional effective theory is therefore limited to asymptotic regions in which the branes are moving very slowly, or else are very close together (in which case the warp factor and tension on the branes become negligible and the theory reduces to Kaluza-Klein gravity).

\subsection{The moduli space approximation}


The moduli space approximation applies to any field theory whose equations of motion admit a continuous family of static solutions with degenerate action.
This family of static solutions is parameterised by the moduli, which correspond to `flat' directions in configuration space, along which slow dynamical evolution is possible.
During this evolution, the excitation along other directions is consistently small, provided these other directions are stable and are characterised by large oscillatory frequencies.

The action on moduli space may be obtained from the full action by inserting as an ansatz the functional form of the static solutions, but with the moduli promoted from constants to slowly varying functions of spacetime.
Variation with respect to the moduli then yields the equations of motion governing the low energy trajectories of the system along moduli space\footnote{In fact, the moduli space approximation underpins much of our knowledge of the classical and quantum behaviour of solitons, such as magnetic monopoles and vortices \cite{Manton:1981mp, Manton:1985hs}.}.


As we have already discussed, the two brane Randall-Sundrum model does indeed possess a one-parameter family of degenerate static solutions (\ref{staticsoln}), parameterised by a single modulus which is the interbrane separation.
The spectrum of low energy degrees of freedom therefore consists of a single four-dimensional massless scalar field corresponding to this modulus, as well as the four-dimensional graviton zero mode (captured by promoting $\eta_{\mu\nu}$ in (\ref{staticsoln}) to some generic Ricci-flat $g_{\mu\nu}(x)$, as in (\ref{genstaticsoln})).

We will now proceed to calculate the moduli space action for this theory.  
Two approaches are possible: the first is to allow the interbrane separation to slowly fluctuate whilst preserving the form of the bulk metric given in (\ref{genstaticsoln}).  In this approach \cite{Brax:2002nt, Garriga:2001ar}, the kinetic terms in the moduli space action derive from the Gibbons-Hawking boundary terms in (\ref{5Daction}).  The second method \cite{Garriga:2001ar, Ekpyrotic} is to transform to an alternative coordinate system in which the branes are held at fixed locations, with the relevant modulus being instead encoded into the bulk geometry.  The kinetic terms in the moduli space action then derive purely from the bulk Einstein-Hilbert term in (\ref{5Daction}).
For ease of computation we will follow this second method here, using a brane-static coordinate system that permits the bulk Ricci scalar to be evaluated by a standard conformal transformation formula. 

We start with the bulk solution (\ref{genstaticsoln}), re-expressed as
\[
\d s^2 = \frac{L^2}{Z^2}\, (\d Z^2 +  \gdxdx ),
\]
where $Z=L\,\exp{(-Y/L)}$, and the branes are located at constant $Z=Z^\pm$, with $Z^+ < Z^-$.
The coordinate transformation $Z = (Z^- - Z^+)(z /L) + Z^+$
then maps the positive- and negative-tension branes to $z=0$ and $z=L$ respectively. 
The bulk line element now takes the form
\[
\d s^2 = [z/L+c]^{-2}(\d z^2+\gdxdx),
\]
where the dimensionless modulus $c= Z^+/(Z^- - Z^+)$, and we have re-absorbed a factor of $L^2/(Z^- - Z^+)^2$ into $g_{\mu\nu}(x)$.

To apply the moduli space approximation, we simply promote the modulus $c$ to a spacetime function $c(x)$, yielding the variational ansatz
\[
\d s^2 =[z/L+c(x)]^{-2}(\d z^2+\gdxdx).
\]
The five-dimensional Ricci scalar 
 may now be evaluated using the standard conformal transformation formula \cite{Wald}
\[
\label{conftransform}
\tilde{R} = \Omega ^{-2} \left[R - 2(n-1)g^{ab} \nabla _a \nabla _b \ln{\Omega} - (n-2)(n-1)g^{ab} (\nabla _a
\ln{\Omega}) \nabla _b \ln{\Omega} \right], 
\]
where $\tilde{R}$ and $R$ are the Ricci scalars of the two metrics $\tilde{g}_{ab}$ and $g_{ab}$ respectively, which are themselves related by the conformal transformation $\tilde{g}_{ab}=\Omega^2 g_{ab}$, for some smooth strictly positive function $\Omega(x^a)$.  In this formula, $n$ denotes the total number of spacetime dimensions, and it is understood that covariant derivatives are to be evaluated with respect to $g_{ab}$.
In the present case, $n=5$, and we will take $\Omega = [z /L + c(x)]^{-1}$.
The five-dimensional Ricci scalar of the metric $g_{ab}$ (with line element $\d z^2 + \gdxdx$) 
then reduces to the Ricci scalar of the four-dimensional metric $g_{\mu \nu}$, henceforth denoted by $R$. 
Thus
\bea
m_5^3 \int _5 \sqrt{-^5\tilde{g}}\ ^5\tilde{R}  
 &=& m_5^3 \int _5 \sqrt{-g} \,\Omega ^3 \(R + 12g^{ab} (\nabla _a\ln{\Omega}) \nabla _b \ln{\Omega} \) \nonumber \\
 && \qquad -8 m_5^3 \int _4 \sqrt{-g} \left[
  \Omega ^3 \nabla _z \ln{\Omega} \right]^{z=L}_{z=0} ,
\eea
where 
we have integrated by parts yielding a boundary term on the branes.  (The boundary terms
at spatial infinity, $x^\mu \tt \inf$, are assumed to vanish).
Integrating over the $z$-coordinate, we obtain
\[
m_5^3 L \int _4 \sqrt{-g}\left( \half\,( \Omega ^2 _+ - \Omega ^2 _-)R +
(\Omega^4_+-\Omega^4_-)\(3 (\partial{c})^2  -5L^{-2}\)\right) ,
\]
where 
\[
\Omega_+(x) = c^{-1}(x), \qquad \Omega_-(x) = (1+c(x))^{-1},
\]
are the dimensionless scale factors on the $(\pm)$ branes respectively.

Dispensing with the tildes, the remaining terms in the action (\ref{5Daction}) are now easily computed:
\begin{eqnarray}
\int _5 \sqrt{-^5g}\,(-2\Lambda ) &=& 3 m_5 ^3L^{-1} \int _4 \sqrt{-g}\,(\Omega_+^4-\Omega_-^4), \\
-\sum_{\pm} \int _\pm \sqrt{-g^\pm}\,\sigma^\pm &=& -6 m_5^3 L^{-1} \int _4 \sqrt{-g}\,(\Omega_+^4-\Omega_-^4), \\
-\sum_\pm \int _\pm \sqrt{-g^\pm}\,2m_5^3 K^\pm &=& 8 m_5^3 L^{-1} \int _4 \sqrt{-g}\, (\Omega_+^4 - \Omega_-^4),
\end{eqnarray}
where $\Lambda=-6m_5^3/L^2$, and we are assuming the critical tuning $\sigma^\pm = \pm 6m_5^3/L$.
The total moduli space action is thus 
\[
S_{\mathrm{mod}} = m_4^2 \int _4 \sqrt{-g}\,\left(\half\,R(\Omega_+^2-\Omega_-^2)+3(\partial{c})^2(\Omega_+^4-\Omega_-^4)\right),
\]
where we have set $m_5^3 L = m_4^2$ to recover the four-dimensional Planck mass $m_4$.
(Note that the terms proportional to $m_5^3/L$ all sum to zero).
The Einstein frame action is then obtained by a further conformal
transformation (this time setting $n=4$ in (\ref{conftransform})), yielding
\[
S_{\mathrm{mod}} = m_4^2 \int_4\sqrt{-g}\,\left( R-6(\partial{c})^2(1+2c)^{-2}\right).
\]
Setting $\phi = -\sqrt{3/2}\,\ln{(1+2c)}$,
so that $-\inf < \phi \le 0$,
the moduli space action takes the simple form 
\[\label{4Daction}
S_{\mathrm{mod}} =  \half\, m_4^2\int_4\sqrt{-g}\left(R-(\partial{\phi})^2\right) ,
\]
corresponding to gravity and a minimally coupled massless scalar field. 
The induced metrics on the branes (to which matter couples) are given by
\[
g_{\mu\nu}^+ = 
  \cosh^2{(\phi/\sqrt{6})}\,g_{\mu\nu}, \qquad 
g_{\mu\nu}^- = 
\sinh^2{(\phi/\sqrt{6})}\,g_{\mu\nu} ,
\]
and the proper separation of the branes is
\[
d=\int_0^L \d z\, (z/L+c)^{-1} = L \ln{\coth{(-\phi/\sqrt{6})}},
\]
so that the branes collide as $\phi\tt -\inf$.
In this limit, $g^\pm_{\mu\nu}\approx \exp(-\sqrt{2/3}\,\phi)\,g_{\mu\nu}$, whereupon the four-dimensional effective theory reduces to standard Kaluza-Klein theory.

The moduli space approximation is expected to be valid whenever the motion of
the system is sufficiently slow that non-moduli degrees of freedom are
only slightly excited.  In practice, this means the typical scale of
curvature on the branes must be much less than the AdS curvature scale
(\iec four-dimensional derivatives must be small compared to $1/L$).
In the case of cosmology, this implies that the four-dimensional 
Hubble parameter $H$ must be small compared to $1/L$.
Equivalently, 
the matter density on the branes must be much less than the brane tensions,
since 
$\rho \sim m_5^3 L H^2 \ll m_5^3/L \sim \sigma$.

Finally, it should be borne in mind that the above derivation of the four-dimensional effective theory via the moduli space approximation, while possessing the virtue of simplicity, is far from rigorous.  
In particular, performing the variation of 
the full five-dimensional action with a restricted ansatz is not
guaranteed to converge on correct solutions of the field
equations\footnote{In fact, the moduli space ansatz, which may
  equivalently be written $\d s^2 = T(x)^2 \d Y^2 +
  \exp(2T(x)Y/L)\,g_{\mu\nu}(x)\,\d x^\mu \d x^\nu$, is not even sufficiently general to describe linearised perturbations about a flat background \cite{CGR}.}.  

Nevertheless, the validity of the four-dimensional effective action (\ref{4Daction}) up to quadratic order in derivatives can be justified by more rigorous (yet more laborious) techniques in which the bulk field equations are solved in a controlled expansion, before backsubstituting into the action and integrating out the extra dimension \cite{Shiromizu:2002qr, K&S, K&S2, KSValidity}.
Alternatively, one could start out with a more general metric ansatz for the bulk, then attempt to restrict the form of the metric functions via the bulk field equations and the requirement that the dimensionally reduced action be in Einstein frame \cite{GaugesinBulkI}.

In the next chapter we will take a fresh 
approach, 
identifying a hidden symmetry of the four-dimensional effective action that completely constrains its form up to quadratic order in derivatives.

%% file: chapter3.tex
\chapter{Conformal symmetry of braneworld effective actions}
\label{confsymmchapter}

\begin{flushright}
\begin{minipage}{11.5cm}
\small
{\it \noindent
In my work, I have always tried to unite the true with the beautiful;
but when I had to choose one or the other, I usually chose the beautiful. 
}
\begin{flushright}
\noindent 
Hermann Weyl
\end{flushright}
\end{minipage}
\end{flushright}



In this chapter, we show how the brane
construction automatically implies conformal invariance
of the four-dimensional effective theory.
This explains the detailed form of the low energy effective action 
previously found using other methods.  
The AdS/CFT correspondence may then be used to improve
the effective description. 
We show how this works in detail for
a positive- and negative-tension brane pair.


\section{Derivation of the effective action}



For the general, non-static solution to the Randall-Sundrum model (\ref{5Daction}) it is
convenient to choose coordinates in which the bulk metric takes the
form
\[
\label{bulk_metric}
\d s^2 = \d Y^2 + g_{\mu\nu}(x,Y)\dxdx .
\]
The brane loci are now $Y^\pm (x)$ and the metric induced on each brane is 
\[
\label{brane_metric}
\gpm (x) = \D _\mu Y^\pm (x) \D _\nu Y^\pm (x) + \g (x,Y^\pm (x)).
\]
At low energies we expect the configuration to be completely
determined by the metric on one brane and the normal distance to the
other brane, $Y^+ - Y^-$.  That is, we are looking for a
four-dimensional effective theory consisting of gravity plus one
physical scalar degree of freedom.  What we will now show is that this
theory may be determined on symmetry grounds alone.  (See also
 \cite{4d1} for related ideas).

The full five-dimensional theory is diffeomorphism
invariant.  This invariance includes the special set
of transformations
\[
\label{condn1}
Y' = Y + \xi ^5(x), \ \ \ x'^\mu = x^\mu + \xi^\mu (x,Y),
\]
with $\xi^\mu (x,Y)$ satisfying 
\[
\label{condn}
\D_Y \xi ^\mu (x,Y) = -g^{\mu\nu}(x,Y)\D _\nu\xi ^5 (x),
\]
which preserve the form (\ref{bulk_metric}) of the metric. 
The
transformation (\ref{condn1}) displaces the $Y^\pm (x)$ coordinates of
the branes, 
\[
Y^\pm (x) \tt Y^\pm (x) + \xi ^5-\xi^\sigma \D _\sigma Y^\pm (x),
\]
and alters $g^{\mu\nu}(x,Y)$ via the usual Lie derivative.  Using
(\ref{condn}), one finds that the combined effect on each brane metric 
(\ref{brane_metric}) is the four-dimensional diffeomorphism 
\[
x'^\mu = x^\mu + \xi^\mu (x, Y^\pm (x)).
\]  
In fact, by departing from the gauge
(\ref{bulk_metric}) away from the branes, we can construct a five-dimensional
diffeomorphism for which $\xi^\mu$
vanishes on the branes.  To see this, we can set 
\[
\xi^\mu (x,Y)=\xi^\mu _0(x,Y)-f(Y)\xi^\mu _0 (x,Y^+),
\]
where $\xi^\mu _0 (x,Y)$ is the
solution to (\ref{condn}) which vanishes on the negative-tension brane and $f(Y)$
is a function chosen to satisfy $f(Y^-) = 0$, $f(Y^+) = 1$, and
$f'(Y^-)=f'(Y^+)=0$ for all $x$.

We conclude that the four-dimensional theory, in which $Y^\pm
(x)$ are represented as scalar fields, must possess a local symmetry
$\xi ^5(x)$ acting nontrivially on these fields.  The dimensionless 
exponentials $\psi^\pm (x) \equiv \exp( Y^\pm (x)/\L)$ transform as
conformal scalars: 
$\psi ^\pm (x) \tt \exp(\xi ^5 /\L)\,\psi^\pm (x)$, while the induced
brane metrics $\gpm$ remain invariant.
The only local, polynomial, two-derivative action
possessing such a symmetry involves gravity with two conformally coupled
scalar fields.  After diagonalising and re-scaling the fields,
this may be expressed as
\[
m^2\int \dx \sqrt{-g} \left( c_+ \psi^+ \Delta \psi^+ + 
c_- \psi^-\Delta  \psi^-\right) ,
\]
where $\Delta \equiv \Box -\frac{1}{6}R$, $c_\pm = \pm 1$, and $m$ is
a constant with dimensions of mass.  
It should be stressed that the metric $\g$ appearing in this
expression is that of the effective theory, which is in general different to $\gpm$, the
induced metric on the branes.  
Potential terms are excluded by the fact that
flat branes, with arbitrary constant $\psi^\pm$, are solutions of the
five-dimensional theory, \ie the $\psi^\pm$ are moduli.   

By construction, the theory possesses local conformal invariance under
\[
\label{conf_transf}
\psi^\pm \tt \Omega (x)^{-1}\psi^\pm, \ \ \ \g \tt \Omega (x)^2 \g .
\]
For $c_+=-c_-$, without
loss of generality we can set $c_+=-1$. Provided $(\psi^+)^2 - (\psi^-)^2 > 0$,
we obtain the usual sign for the Einstein term, so there are no ghosts
in the gravitational sector. We can then set 
$\psi^+=A\cosh{\phi/\sqrt{6}}$ and
$\psi^- = -A\sinh{\phi/\sqrt{6}}$. The field $A$ has the wrong sign
kinetic term, but it can be set equal to a constant by a choice
of conformal gauge. Therefore, in this case there are no physical
propagating ghost fields. In contrast, a similar analysis reveals that
when $c_+=c_-$ the theory 
possesses physical ghosts, either in the gravitational wave sector
(wrong sign of $R$) or in the scalar sector, no matter how the 
conformal gauge is fixed. 
We conclude that the low energy
effective action must be 
\[
\label{conf_action}
m^2 \int \dx \sqrt{-g} \left(- \psi^+ \Delta  \psi^+ +
 \psi^-\Delta  \psi^-\right).
\]

We know from the above argument that the brane metrics are conformally invariant:
from this, and from general covariance, they must equal $\g$
times homogeneous functions of order two in $\psi^+$ and $\psi^-$.
Yet in the model under consideration, we have static
solutions $\gpm = \exp(2Y^\pm/\L)\, \eta _{\mu\nu}$ for all $Y^+>Y^-$.  The only
choice consistent with this, and with $(\psi^+)^2 - (\psi^-)^2 > 0$, is
\[
\label{g_eqns}
\gpm = \frac{(\psi^\pm)^2}{6}\,\g ,
\]
which is a conformally invariant equation. (The
numerical factor has been introduced for later convenience).

\subsection{Conformal gauges}

It is instructive to fix the conformal gauge in several ways. First,
set $\psi^+=\sqrt{6}$, so that $\g = \gp$ and the metric appearing in
(\ref{conf_action}) is actually the metric on the positive-tension brane.  The
action (\ref{conf_action}) then consists of Einstein gravity (with
Planck mass $m$) plus a conformally invariant
scalar field $\psi^-$ which has to be smaller than
$\sqrt{6}$:  
\[
\label{+action}
m^2 \int \dx \sqrt{-g^+}\left( (1-\frac{1}{6}\,(\psi^-)^2)\,R^+-(\D
  \psi^-)^2 \right).
\]
Changing variables to $\chi = 1 - (\psi^-)^2/6$ produces the
alternative form \cite{K&S, K&S2}
\[
m^2 \int \dx \sqrt{-g^+}\left(\chi R^+ - \frac{3}{2(1-\chi)}(\D \chi)^2
\right).
\]
Conversely, if we set $\psi^-=\sqrt{6}$, then $\g$ is
the metric on the negative-tension brane and $\psi^+$, which has to be larger
than $\sqrt{6}$, is a conformally coupled scalar field.  (The relative 
sign between the gravitational and kinetic terms in the action is now wrong, however,
and so this gauge possesses ghosts).   
If we add matter coupling to the metric on the positive- and negative-tension branes, we find
that matter on the negative-tension brane couples in a
  conformally invariant manner to the positive-tension brane metric and the field
  $\psi^-$, and conversely for matter on the positive-tension brane.
(Note that we are not implying conformal invariance of the matter itself:
it is simply that matter coupled to the brane metrics will be trivially
invariant under the transformation (\ref{conf_transf}), as the brane
metrics are themselves invariant).

A third conformal gauge maps the theory to Einstein gravity with a
minimally coupled scalar field $\phi$, taking values in the range $-\infty <\phi <0$. 
Starting from
(\ref{conf_action}), we can set $\psi^+=A\cosh{\phi/\sqrt{6}}$ and
$\psi^- = -A\sinh{\phi/\sqrt{6}}$, as noted earlier, yielding the action
\[
\label{EFaction1}
-m^2\int \dx \sqrt{-g}\left(A\Delta A+\frac{A^2}{6}(\D \phi
 )^2\right).
\]
Now, choosing the conformal gauge $A=\sqrt{6}$, we find
\[
\label{EFaction2}
m^2 \int \dx \sqrt{-g}\left(R+\phi\Box\phi\right) ,
\]
\iec gravity plus a minimally coupled massless scalar.
In this gauge equations
(\ref{g_eqns}) read:
\[
\label{g_eqns2}
\gp = \cosh ^2 (\phi /\sqrt{6})\,\g, \qquad \gm = \sinh ^2
(\phi/\sqrt{6})\,\g,
\]
in agreement with explicit calculations in the moduli space approach 
 \cite{Ekpyrotic}.  

The present treatment also goes some way towards explaining the
moduli space results.  For example, the fact that the moduli space
metric is flat is seen to be a consequence of conformal invariance.
Specifically, for solutions with cosmological symmetry, one can pick a
conformal gauge in which the metric is static.  The scale factors on
the two branes are determined by $\psi^\pm$.  From
(\ref{conf_action}), the moduli space metric is just two-dimensional
Minkowski space.

A couple of results for conformal gravity follow from the above
discussion.  Firstly, in the $\psi^+ = \sqrt{6}$ gauge, we have
$\psi^-=-\sqrt{6}\tanh{\phi/\sqrt{6}}$. Any solution for a
minimally coupled scalar $\phi$, with metric $\g$, thus yields a
corresponding solution for a conformally coupled scalar $\psi^-$, with
$|\psi^-| < \sqrt{6}$ and metric $\gp$ as in (\ref{g_eqns2}), and vice versa.  
Secondly, in the $\psi^-=\sqrt{6}$ gauge, we have
$\psi^+=-\sqrt{6}\coth{\phi/\sqrt{6}}$ hence we may also obtain a
solution for a conformally coupled scalar $\psi^+$, with
$|\psi^+|>\sqrt{6}$ and metric $\gm$ given in (\ref{g_eqns2}).  
Solutions to conformal scalar gravity therefore
come in pairs: if $\g$ and $\psi$ are a solution, then
$(\psi^2/6)\,\g$ and $\tilde{\psi}=6/\psi$ is another solution.
In terms of branes, this merely states that if $\gp$ and
$\psi^-$ are known in the gauge $\psi^+=\sqrt{6}$, then it is
possible to reconstruct $\gm$ and $\psi^+$ in the gauge
$\psi^-=\sqrt{6}$.

\subsection{Generalisation to other models}

The argument given above establishing the conformal symmetry 
of the effective action is of a very general nature: 
the only step at which we specialised to
the Randall-Sundrum model was in the identification of the brane
metrics in terms of the effective theory variables (\ref{g_eqns}).
This required merely the knowledge of a one-parameter family of solutions. 

To derive the effective theory for other brane models, it is  
only necessary to generalise this last step.
For example, in the case of tensionless branes 
compactified on an $S^1/\Z_2$, the bulk warp
is absent, and so we know that a family of static solutions is given 
by the ground state of Kaluza-Klein theory (in which all fields are
independent of the extra dimension, hence the additional $\Z_2$
orbifolding present in the case of tensionless branes is irrelevant).
Ignoring the gauge fields, the Kaluza-Klein ansatz for the five-dimensional 
metric is 
\[
\d s^2 = e^{2\sqrt{2/3}\,\phi(x)} \d y^2 + e^{-\sqrt{2/3}\,\phi(x)}\gdxdx ,
\]
where $\phi$ and $\g$ extremise an action identical to (\ref{EFaction2}).
For branes located at constant $y$, the induced metrics are
$\exp(-\sqrt{2/3}\,\phi)\,\g$, independent of $y$.

Using the effective action in the form (\ref{EFaction1}), conformal
invariance of the induced brane metrics dictates that
\[
\g ^\pm = A^2 f^\pm (\phi)\, \g ,
\]
for some unknown functions $f^\pm$.  
Upon fixing the conformal gauge to
$A=\sqrt{6}$, one recovers the action (\ref{EFaction2}), which is
just the standard Kaluza-Klein low energy effective action.
The functions $f^\pm$ are thus both equal to
$(1/6)\,\exp(-\sqrt{2/3}\,\phi)$, and we have
\[
\g^\pm = e^{-\sqrt{2/3}\,\phi} \g = \frac{1}{6}\, (\psi^+ + \psi^-)^2 \g.
\]
Note that this is consistent with the $\phi \tt -\inf$ limit of the
Randall-Sundrum theory (\ref{g_eqns2}): as the brane separation goes to
zero, the warping of the bulk becomes negligible and the Randall-Sundrum theory
tends to the Kaluza-Klein limit \cite{TTS}.

Conceivably, arguments similar to the above might also be used to derive the low energy four-dimensional effective action for the Ho{\v r}ava-Witten model (see Chapter \S\,\ref{Mthchapter}), although we will not pursue this connection further here.

\subsection{Cosmological solutions}

We now turn to a discussion of 
the general cosmological solutions representing colliding branes.  
We choose a conformal gauge in which
the metric is static, and all the dynamics are contained in
$\psi^\pm$.  For flat, open and closed spacetimes, the spatial Ricci scalar
is given by $R=6k$, where $k = 0$, $-1$ and $+1$ respectively.
The action (\ref{conf_action}) yields the equations of motion
\bea
\ddot{\psi}^\pm &=& -k\psi^\pm , \\
(\dot{\psi}^+)^2-(\dot{\psi}^-)^2 &=& -k\left(
(\psi^+)^2-(\psi^-)^2\right).
\eea
For $k=0$, we have the solutions
\[
\psi^+ = -At+B, \qquad \psi^-=At+B, \qquad t<0
\]
representing colliding flat branes.  It is natural to match $\psi^+$ to
$\psi^-$ across the collision, and vice versa, to obtain $\psi^\pm =
\pm At+B$ for $t>0$.  This solution then describes two branes which
collide and pass through each other, with the positive-tension brane continuing to
a negative-tension brane, and vice versa \cite{Seiberg, TTS}. 

For $k=-1$, we have the three solutions
\bea
\begin{array}{cccc}
\psi^{(1)}\, = &A\sinh{t}; & A\cosh{t}; & A\,e^t, \\
\psi^{(2)}\, = &A\sinh{(t-t_0)}; & A\cosh{(t-t_0)}; & A\,e^{t-t_0},
\end{array}
\eea
where we set $\psi^+$ equal to the greater, and $\psi^-$
equal to the lesser, of $\psi^{(1)}$ and $\psi^{(2)}$.  For
$k = +1$, we find the bouncing solutions $\psi^{(1)} =
A\sin{t}$, $\psi^{(2)} = A\sin{(t-t_0)}$.  In the absence of matter on
the negative-tension brane, the $\sin$ and $\sinh$ solutions are singular when the negative-tension
brane scale factor $a_-$ vanishes.  Matter on the negative-tension brane
scaling faster than $a_-^{-4}$, however, (\eg scalar kinetic matter) 
causes the solution for $\psi^-$ to bounce smoothly 
at positive $a_-$, because $\psi^-$ has a
positive kinetic term. This bounce is perfectly regular.  
In contrast, the big
crunch/big bang singularity, occurring when the positive- and negative-tension branes collide,
is unavoidable.

The above example illustrates a general feature of the brane pair
effective action. If the positive- and negative-tension brane 
solutions are continued through the collision
without re-labelling
(this means that the orientation of the warp must flip), then the 
four-dimensional 
effective action changes sign.  The re-labelling restores the conventional
sign. The same phenomenon is seen in string theories obtained by
dimensionally reducing eleven-dimensional supergravity, when the 
eleventh dimension collapses and reappears. 
For a discussion of brane world black hole solutions with intersecting branes, see 
\cite{EFT_BH}.

\section{AdS/CFT}

Recently, it has been shown that the AdS/CFT correspondence
 \cite{MaldacenaAdSCFT, WittenADSCFT} provides a powerful approach to the understanding of braneworlds.
For a single positive-tension brane, the four-dimensional effective
description comprises simply Einstein 
gravity plus two copies of the dual CFT \cite{deHaro} (as the $\Z_2$
symmetry implies there are two copies of the bulk).  
Notable successes of this program include reproducing the
$O(1/r^3)$ corrections to Newton's law on the brane \cite{Duff&Liu}, and
reproducing the modified Friedmann equation induced on the
brane \cite{Gubser, Shiromizu&Ida}. 

Consider, for simplicity, a single positive-tension brane containing only radiation.
Taking the trace of the effective Einstein equations, we find 
\[
\label{trace}
-R = 2(8\pi G_4)<T_{CFT}> ,
\]
as the stress tensor of the radiation is traceless.  The trace anomaly
of the dual $\mathcal{N}=4$ $SU(N)$ super-Yang Mills theory must then
be evaluated.  With the help of the AdS/CFT dictionary,
this quantity may be calculated for the case of cosmological symmetry
as shown in \cite{Henningson&Skenderis}, giving 
\[
\label{trace2}
-R=\frac{L^2}{4}\left(R_{\mu\nu}R^{\mu\nu}-\frac{1}{3}R^2\right).
\]
Here, the usual $R^2$ counterterm has been added to the action in
order to eliminate the $\Box R$ term in the trace, thus furnishing second
order equations of motion.  

For a cosmological metric, with scale factor $a$, this becomes
\[
\label{hdoteqn}
2(\ddot{a}a+ka^2)=L^2(k+h^2)\dot{h},
\]
where $h \equiv \dot{a}/a$ and the dot denotes differentiation with
respect to conformal time. 
Re-expressing the left-hand side as $h^{-1}\partial
_t(\dot{a}^2+ka^2)$, we can then integrate to obtain
\[
\label{hinteqn}
h^2+k = \frac{1}{a^2}\,(B-\frac{1}{4}\,k^2 L^2) + \frac{1}{4}\,(h^2+ k)^2 \,{L^2\over a^2},
\]
where $B$ is an integration constant. Now, we can expect to recover 
Einstein gravity on the brane in the limit when $L\rightarrow 0$, with
other physical quantities held fixed.  Expanding all terms in powers of $L$, 
at leading order we must obtain four-dimensional 
Einstein gravity, for which $8\pi G_4 = 8\pi G_5/L$.  We therefore set
$B\sim (8 \pi G_5 \rho_0/3L) +C$,  where $\rho=\rho_0/a^4$ is the  energy
density of conventional radiation, and 
$C$ is a constant independent of $L$ as $L\rightarrow 0$.
From (\ref{hinteqn}), we then obtain the first correction to $h^2+k$,
namely
\[
h^2+k= \frac{8 \pi G_5
  \rho_0}{3La^2} + \frac{C}{a^2}+
  \frac{(8\pi G_5\rho_0)^2}{36a^6} + O(L),
\]
which, thanks to the CFT contribution, now includes 
the well-known dark energy and $\rho^2$ corrections \cite{Binetruy}.

It should come as no surprise that the AdS/CFT correspondence 
only approximates the Randall-Sundrum setup 
up to first nontrivial order in an expansion in $L$. 
The AdS/CFT scenario involves string theory on $AdS_5\times S_5$.  Since 
$\alpha '\sim \ell_s^2 \sim L^2$ at fixed 't Hooft
coupling, and the masses squared of the 
Kaluza-Klein modes on the $S_5$ are of order $1/L^2$,
we expect nontrivial corrections at second order in an expansion in
$L$. Furthermore, 
one can show from the AdS/CFT dictionary 
that, in order
for the $\rho^2$ term to dominate in the modified Friedmann equation,
the temperature
of the conventional radiation must be greater than the 
Hagedorn temperature of the string. 
Clearly, 
the AdS/CFT correspondence
cannot describe this situation.

Let us now consider the extension of the AdS/CFT approach to the case of a pair of
positive- and negative-tension branes,
using the ideas developed in earlier in this chapter.  
The effective action for a single
positive-tension brane is
\[ 
\label{AdS/CFTaction}
\frac{1}{16\pi G_4}\int\dx \sqrt{-g^+}\,R^+  + 2W_{CFT}[g^+] + S_{m}[g^+],
\]
where $\gp$ is the induced metric on the brane, $S_{m}$ is the brane
matter action, and $W_{CFT}$ is the
CFT effective action (including the appropriate $R^2$
counterterms).   
Substituting now for $\gp$ using (\ref{g_eqns}), the Einstein-Hilbert
term $\sqrt{-g^+}\,R^+$ becomes $-\sqrt{-g}\,\psi^+\Delta\psi^+$.
A negative tension brane may then be incorporated as follows: 
\bea
\label{S2}
\frac{1}{16\pi G_4}\int\dx\sqrt{-g}\,(-\psi^+ \Delta \psi^+ +
\psi^- \Delta \psi^-) + 2W_{CFT}[g^+] &&\nonumber  \\
-2W_{CFT}[g^-]+S_{m}[g^+]+S_{m}[g^-]. \ \ \ \ \ \ \ \ \ \ \  
\eea
The action for the positive- and negative-tension brane pair 
must take this form in order to correctly
reproduce the Friedmann equation for each brane. 
To see this, consider again the conformal gauge in which the effective
theory metric is static and all the dynamics are contained in
$\psi^\pm$, which play the role of the brane scale factors.
Variation with respect to the $\psi^\pm$ yields the
scalar field equations 
\[
(\psi^\pm)^{-3}\Delta \psi^\pm = 2(8\pi G_4)<T_{CFT}^\pm >,
\]
where the trace anomaly must be evaluated on the induced brane metric
$\gpm$, but $\Delta$ is evaluated on the effective metric $\g$.
The left-hand side evaluates to
$-(\psi^\pm)^{-3}(\ddot{\psi}^\pm+k(\psi^\pm)^2)$.  After
identifying $\psi^\pm/\sqrt{6}$ with $a_\pm$ according to (\ref{g_eqns}),
we recover equation (\ref{hdoteqn}), upon dropping the plus or minus label.  
From the necessity of recovering the Friedmann equation on each brane,
we may also deduce that cross-terms in the action between $\psi^+$ and
$\psi^-$ are forbidden. 

The signs associated with the gravitational parts of the 
action are required to 
achieve consistency with (\ref{conf_action}).
Consequently, the relative sign between the
gravity plus CFT part of the action, and that of the matter, is reversed for the
negative-tension brane, consistent with the 
modified Friedmann equations \cite{Binetruy}
\[
\label{FRW}
H^2_\pm = \pm \frac{8\pi G_5\rho_\pm}{3\L} + {(8\pi G_5\rho_\pm)^2 \over 36} 
-\frac{k}{a^2}+ \frac{C}{a^4}, 
\]
where plus and minus label the positive- and negative-tension branes,
and $C$ is again a
constant representing the dark radiation.  

In summary, we have elucidated the origin of conformal symmetry in
brane world effective actions, and shown how this 
determines the effective action to lowest
order.  When combined with the 
the AdS/CFT correspondence, our approach also recovers the first
corrections to the brane Friedmann equations.



%% file: chapter4.tex
\chapter{Solution of a braneworld big crunch/big bang cosmology}
\label{5dchapter}

\begin{flushright}
\begin{minipage}{11cm}
\small
{\it \noindent
We can lick gravity, but sometimes  the paperwork is overwhelming.}
\begin{flushright}
\noindent 
Wernher von Braun
\end{flushright}
\end{minipage}
\end{flushright}




In this chapter we solve for the cosmological perturbations in a 
five-dimensional background consisting of
two separating or colliding boundary branes, as
an expansion in the collision speed $V$ divided by
the speed of light $c$.  Our 
solution permits 
a detailed check of the validity of 
four-dimensional effective theory in the 
vicinity of the event corresponding to
the big crunch/big bang singularity. We show that the
four-dimensional description fails at the first
nontrivial order in $(V/c)^2$. 
At this order, there is nontrivial  
mixing of the two relevant four-dimensional perturbation modes  
(the growing and decaying modes) as the 
boundary branes move from the narrowly-separated limit
described by Kaluza-Klein theory to the 
well-separated limit where gravity is confined to
the positive-tension brane. 
We comment on the cosmological
significance of the result and compute other quantities
of interest in five-dimensional cosmological scenarios.



\section{Introduction}

Two limiting regimes can be distinguished 
in which the dynamics of the Randall-Sundrum model simplify: the first is the limit
in which the interbrane separation is much greater than the AdS curvature radius, and the second is the limit in which the interbrane separation is far smaller.
When the two boundary
branes are very close to one another, the warping of the
five-dimensional bulk and the tension of the branes
become irrelevant. In this situation, 
the low energy modes of the system are 
well-described by a simple Kaluza-Klein reduction
from five to four dimensions, \iec gravity plus a
scalar field (the $\Z_2$ projections eliminate the
gauge field zero mode).  
When the two branes are widely 
separated, however, the physics is quite different.
In this regime, the warping of the
bulk plays a key role, causing  
the low energy gravitational modes to be localised 
on the positive-tension brane \cite{RSII,garriga,giddings}.
The four-dimensional effective theory describing this
new situation is nevertheless
identical, consisting of Einstein gravity and 
a scalar field, the radion, describing 
the separation of the two branes. 

In this chapter, we study the transition between these 
two regimes -- from the naive Kaluza-Klein reduction to localised
Randall-Sundrum gravity -- at finite brane speed. 
In the two asymptotic regimes, the narrowly-separated brane limit and
the widely-separated limit, the cosmological 
perturbation modes show precisely the behaviour predicted
by the four-dimensional effective theory. There are
two massless scalar perturbation modes; in 
longitudinal gauge, and in the long 
wavelength ($k\rightarrow 0$) limit, one mode 
is constant and the other
decays as $t_4^{-2}$, where $t_4$ is the  
conformal time.
In the four-dimensional description, 
these two perturbation modes are entirely distinct: 
one is the curvature perturbation mode; the
other is a local time delay to the big bang.
Nonetheless, we shall show that in the five-dimensional
theory, at first nontrivial order in the speed of
the brane collision, the two modes mix.  If,
for example, one starts out in 
the time delay mode at small $t_4$, one ends up 
in a mixture of the time
delay and curvature perturbation modes as 
$t_4 \rightarrow \infty$. 
Thus the two cosmological perturbation modes --
the growing and decaying adiabatic modes -- mix in 
the higher-dimensional braneworld 
setup, a phenomenon which is prohibited
in four dimensions. 

The mode-mixing occurs as a result of a 
qualitative change in the nature of the 
low energy modes of the system. At small
brane separations, the low energy modes are nearly
uniform across the extra dimension. Yet as the brane
separation becomes larger than the bulk warping scale,
the low energy modes
become exponentially localised on
the positive-tension brane. If the branes separate at
finite speed, the localisation process fails to 
keep pace with the brane separation and the 
low energy modes do not evolve adiabatically.
Instead, they evolve into a mixture
involving higher Kaluza-Klein modes,
and the four-dimensional
effective description fails. 

The mixing we see
between the two scalar perturbation modes would
be prohibited in {\it any} local four-dimensional
effective theory consisting of Einstein gravity and matter
fields, no matter what the matter fields were. 
The mixing is therefore a truly five-dimensional phenomenon, 
which cannot be modelled with a local
four-dimensional effective theory.
There is, moreover, an independent argument
against the existence of any local four-dimensional 
description of these phenomena.  In standard 
Kaluza-Klein theory, it is well known that 
the entire spectrum of massive modes is 
actually spin two \cite{duff}. Yet, despite many attempts, no
satisfactory Lagrangian description of massive, purely spin
two fields has ever been found \cite{deser,damour}.
Again, this suggests that one should not expect to 
describe the excitation
of the higher Kaluza-Klein modes
in terms of an improved, local,
four-dimensional effective theory. 

The system we study consists of two branes emerging from
a collision. In this situation, 
there are important simplifications which allow
us to specify initial data rather precisely.
When the brane separation is small, the fluctuation modes 
neatly separate into light Kaluza-Klein zero modes, which are constant along
the extra dimension, and massive modes with nontrivial 
extra-dimensional dependence. Furthermore, the brane tensions
and the bulk cosmological constant become irrelevant 
at short distances. It is thus natural
to specify initial data which map precisely onto 
four-dimensional fields in the naive dimensionally-reduced 
theory describing the limit of  narrowly-separated branes.
 With initial data specified this way, there
are no ambiguities in the system. The two branes
provide boundary conditions for all time and the five-dimensional
Einstein equations yield a unique solution, for arbitrary
four-dimensional initial data.

Our main motivation is the study of cosmologies
in which the big bang was a brane collision, such as the 
cyclic model \cite{Cyclicevo}. Here, a period of 
dark energy domination, followed by slow contraction of the
fifth dimension, renders the branes locally flat and parallel 
at the collision. During the slow contraction phase, growing,
adiabatic, 
scale-invariant
perturbations are imprinted on the branes prior to the 
collision.  Yet if the system is accurately described
by four-dimensional effective theory throughout, 
then, as a number of authors have noted \cite{brand1, brand2, lyth1, jch1, jch2, Creminelli}, there is an apparent roadblock to the passage of 
the scale-invariant perturbations across the bounce.
Namely, it is hard to see how the growing mode
in the contracting phase, usually described as a local time delay, 
could match onto the growing mode in
the expanding phase, usually
described as a curvature perturbation. 
In this chapter, we show that the 
four-dimensional effective theory fails at order $(V/c)^2$,
where $V$ is the collision speed and $c$ is the speed
of light. The four-dimensional description
works well when the branes are close together, or far
apart.  As the branes move from one regime to the
other, however, the two four-dimensional modes mix
in a nontrivial manner.

The mixing we find 
demonstrates that the 
approach of two boundary branes
along a fifth dimension produces physical effects that cannot 
properly be modelled by a local four-dimensional effective theory.  Here, we deal
with the simplest case involving two empty
boundary branes separated by a bulk with a negative
cosmological constant. For the cyclic model, the details are
more complicated. 
In particular, there is 
an additional
bulk stress $\Delta T_5^5$, associated with the interbrane force,
that plays a vital role in
converting a  growing mode 
corresponding to a pure time delay perturbation
into a mixture of time delay and curvature modes on the brane.  
We will present some preliminary considerations of the effects of such a bulk stress in the following chapter.
For now, our main conclusion with 
regard to the cyclic model 
is that, to compute properly the evolution of perturbations before and 
after a brane collision, one must go beyond
the four-dimensional effective theory 
to consider the full five-dimensional theory.
%


The outline of this chapter is as follows.
In Section \S\,\ref{solnmethods}, we provide an overview of 
our three solution methods. In \S\,\ref{seriessoln},
we solve for the background and cosmological
perturbations using a series expansion in time about the
collision.  In \S\,\ref{polysection}, we present an improved method in
which the dependence on the fifth dimension is approximated
using a set of higher-order Dirichlet or Neumann polynomials. 
In \S\,\ref{expaboutscalingsoln}, we develop an expansion about the small-$(V/c)$ scaling
solution, before
comparing our results with those of the four-dimensional effective
theory in \S\,\ref{compwitheft}.  We conclude with a discussion of mode-mixing in
\S\,\ref{mixingsection}.  
Detailed explicit solutions may be found in
Appendix \ref{detailedresults}, and the Mathematica code implementing our calculations 
is available online \cite{Website}.

\section{Three solution methods}
\label{solnmethods}

In this section, we review the three solution methods
employed, noting their comparative merits.
For the model considered here, with no
dynamical bulk fields, as we saw in \S\,\ref{Birkhoffsection}
there is a Birkhoff-like theorem guaranteeing the existence
of coordinates in which the bulk is static. It is easy
to solve for the background in these coordinates. The
motion of the branes 
complicates the
Israel matching conditions, however, rendering
the treatment of perturbations difficult. For this reason, it is preferable
to choose a coordinate system in which the branes
are located at fixed spatial coordinates $y=\pm y_0$,
and the bulk evolves with time. 

We shall employ a coordinate system in which 
the five-dimensional line element for the background 
takes the form
\[
\label{metrica}
\d s^2 = n^2(t,y) (-\d t^2 +t^2 \d y^2) + b^2(t,y) \d \vec{x}^2,
\]
where $y$ parameterises the fifth dimension and $x^i$ (for $i=1,2,3$),
the three noncompact dimensions. 
Cosmological isotropy  excludes $\d t \,\d x^i$ or
$\d y \,\d x^i$ terms, and homogeneity ensures 
$n$ and $b$ are independent of $\vec{x}$. 
The $t,y$ part of the background metric may then
be taken to be conformally flat, and one may further choose to write 
the metric for this two-dimensional Minkowski spacetime in Milne form.
Since we are interested in scenarios with colliding branes in which
the bulk geometry about the collision is Milne, we will assume 
the branes to be located at $y=\pm y_0$, with the collision occurring at
$t=0$. 
By expressing the metric in 
locally Minkowski coordinates, $T=t \cosh{y}$ and
$Y=t \sinh{y}$, one sees that the collision 
speed is $(V/c)= 
\tanh{2 y_0}$ and the 
relative rapidity of the collision
is $2y_0$.
As long as the bulk metric is regular
at the brane collision and possesses cosmological symmetry, 
the line element may
always be put into the 
form (\ref{metrica}). Furthermore,
by suitably re-scaling coordinates one can choose 
$b(0,y)=n(0,y)=1$.

In order to describe  perturbations about this background, one 
needs to specify an 
appropriate gauge choice.
Five-dimensional
longitudinal gauge is particularly convenient \cite{Carsten}:
firstly, it is completely gauge-fixed; 
secondly, the brane trajectories are unperturbed in this gauge \cite{TTS},
so that the Israel matching conditions are relatively simple; and finally,
in the absence of anisotropic stresses, the traceless part
of the Einstein $G^i_j$ (spatial) 
equation yields a constraint among the
perturbation variables, reducing them from four to three. 
In light of these advantages, we will work in
five-dimensional longitudinal gauge throughout.

Our three solution methods are as follows:

\begin{itemize}
\item {\bf Series expansion in \bf\textit{t}}
\end{itemize}

The simplest solution method for the background is to 
solve for the metric functions $n(t,y)$ and
$b(t,y)$ as 
a series in powers of $t$ about $t=0$. At each order,
the bulk Einstein equations yield
a set of ordinary differential equations in $y$,
with the boundary
conditions provided by the Israel matching conditions.
These are straightforwardly solved. 
A similar series approach, involving powers of $t$ and
powers of $t$ times $\ln{t}$ 
suffices for the perturbations.
 
The series approach is useful at small times $(t/L)\ll 1$
since it provides the precise 
solution for the background plus generic perturbations, 
close to the brane collision,
for all $y$ and for any collision rapidity $y_0$.  It allows
one to uniquely specify four-dimensional 
asymptotic data as $t$ tends to
zero. 
Nonetheless, the series thus obtained fails to converge at quite modest times. 
Following the system to long times
requires a more sophisticated method. Instead of taking
$(t/L)$ as our expansion parameter, we want to use the 
dimensionless rapidity of the brane collision $y_0$,
and solve at each order in $y_0$.

\begin{itemize}
\item {\bf Expansion in Dirichlet/Neumann polynomials in \bf\textit{y}}
\end{itemize}

In this approach we represent the spacetime metric
in terms of variables obeying either Dirichlet or Neumann 
boundary conditions
on the branes. We then express these 
variables as series of Dirichlet or Neumann  
polynomials in $y$ and $y_0$, 
bounded at each subsequent order by an increasing power
of the collision rapidity $y_0$.  (Recall that the
range of the $y$ coordinate is bounded by $|y|\le
y_0$).
The coefficients in these expansions are
undetermined functions of $t$. By 
solving the five-dimensional Einstein equations perturbatively
in $y_0$, we obtain a series of ordinary differential equations in
$t$, which can then be solved exactly.
In this Dirichlet/Neumann polynomial expansion, the Israel boundary conditions on the branes
are satisfied automatically at every order in $y_0$, while the initial
data at small $t$ are provided by the previous
series solution method.

The Dirichlet/Neumann polynomial expansion method yields 
simple, explicit solutions for the background and perturbations
as long as $(t/L)$ is smaller than $1/y_0$. Since $y_0 \ll 1$,
this considerably 
improves upon the naive series expansion in $t$. 
For $(t/L)$ of order $1/y_0$, however, the expansion fails because
the growth in the coefficients overwhelms the extra powers of 
$y_0$ at successive orders. Since $(t/L) \sim 1/y_0$ corresponds
to brane separations of order the AdS radius, the Dirichlet/Neumann polynomial
expansion method fails to describe the late-time behaviour of
the system, and a third method is needed.

\begin{itemize}
\item {\bf Expansion about the scaling solution}
\end{itemize}
The idea of our third method is to start by identifying a scaling
solution, whose form is independent of $y_0$ for all
$y_0\ll 1$. This scaling solution is well-behaved for
all times and therefore a perturbation expansion in $y_0$ 
about this solution is similarly well-behaved, even at very late 
times. To find the scaling solution, we first 
change variables from
$t$ and $y$ to an equivalent set of 
dimensionless variables. The characteristic velocity of the system is
the brane speed at the collision, $V=c\tanh 2 y_0 \sim 2 c y_0$
for small $y_0$, where we have temporarily restored the speed of light $c$.    
Thus we have the dimensionless time 
parameter $x = y_0 ct/L \sim V t/L$, of order the time 
for the branes to separate by one AdS radius. We also
re-scale the 
$y$-coordinate by defining $\w = y/y_0$, whose range is 
$-1\leq \w\leq 1$, independent of the characteristic velocity. 

As we shall show, when re-expressed in these variables, 
for small $y_0$, 
the bulk Einstein equations become perturbatively 
\textit{ultralocal}: at each order in $y_0$ one only has to 
solve 
an ordinary differential equation in $\w$, with a source
term determined by time derivatives of lower order terms.
The original partial
differential equations reduce to an infinite series of 
ordinary differential
equations in $\w$ which are then easily solved order by order 
in $y_0$. 

This method, an expansion in $y_0$ about the scaling 
solution, is
the most powerful and may be extended to 
arbitrarily long times $t$ and
for all brane separations. 
In light of the generalised Birkhoff theorem, the bulk in between the two branes 
is just a slice of five-dimensional
AdS-Schwarzschild spacetime, 
within which the two branes move \cite{langlois, maartens, Durrer}.
(The bulk black hole is itself merely virtual, however, as it lies beyond the negative-tension brane and hence is excluded from the physical region).  
As time proceeds, the negative-tension brane becomes closer
and closer to 
the horizon of the virtual AdS-Schwarzschild black hole. Even though 
its location in the Birkhoff-frame (static) coordinates
freezes (see Figure \ref{branesep}), its proper speed grows and the 
$y_0$ expansion fails. 
Nonetheless, by analytic continuation of our solution in $\w$ and $x$, 
we are able to circumvent this temporary breakdown of
the $y_0$ expansion and follow
the positive-tension brane, and the perturbations localised
near it, as they run off to the boundary of anti-de Sitter 
spacetime.

Our expansion about the scaling solution is closely related to 
derivative-expansion techniques developed earlier by a number of authors
\cite{Toby, K&S, Gonzalo}.  In these works, an expansion in terms of
brane curvature over bulk curvature was used. For cosmological
solutions, this  
is equivalent to an expansion in $L \mathcal{H}^+$, where
$\mathcal{H}^+$ is the Hubble constant on the positive-tension brane.  
In the present instance, however, we specifically want to
study the time-dependence of the perturbations for all
times, from the narrowly-separated to the well-separated brane limit.
For this purpose, 
it is better to use a time-independent expansion
parameter ($y_0$), and  to
include all the appropriate time-dependence order by
order in the expansion. 

Moreover, in these earlier works, the goal was to find the
four-dimensional effective description more generally, 
without specifying that the branes emerged from a collision
with perturbations in the lowest Kaluza-Klein modes. 
Consequently, the solutions obtained
contained a number of undetermined functions.  
In the present context, however, the initial conditions along the
extra dimension are completely specified close to the brane collision
by the requirement that only the lowest Kaluza-Klein mode be excited.
The solutions we obtain here are fully determined, with no
arbitrary functions entering our results. 

Returning to the theme of the four-dimensional effective
theory, 
we expect on general grounds that this should be valid
in two particular limits:
firstly, as we have already discussed, a Kaluza-Klein
description will apply at early times near to the collision,
when the separation of the branes is much less than $L$. 
Here, the warping of the bulk geometry and the brane tensions can
be neglected.  Secondly, 
when the branes are separated by many AdS lengths, 
one expects gravity to become localised on the positive-tension
brane, which moves ever more slowly as time proceeds, so
the four-dimensional
effective theory should become more and more accurate.

Equipped with our five-dimensional solution for the background and
perturbations obtained by expanding about the scaling solution, we find ourselves
able to test the four-dimensional effective theory explicitly.  
We will show that the four-dimensional 
effective theory accurately
captures the five-dimensional dynamics to leading order in the
$y_0$-expansion, but fails at the first nontrivial order.
Our calculations reveal that
the four-dimensional perturbation modes undergo a mixing in the
transition between the Kaluza-Klein effective theory at early times
and the brane-localised gravity at late times.
This effect is a consequence of the momentary breakdown of the effective
theory when the brane separation is of the order of an AdS length, and
cannot be seen from four-dimensional effective theory calculations
alone. 

\section{Series expansion in time} 
\label{seriessoln}

As described above,
we find it simplest to work in coordinates
in which the brane locations are fixed but the bulk
evolves.  The bulk metric is therefore given by (\ref{metrica}), with
the brane locations fixed at $y=\pm y_0$ for all time $t$.
The five-dimensional solution then has to satisfy both
the Einstein equations and the Israel matching conditions
on the branes \cite{Israel}. 

The bulk Einstein equations read $G_a^b = -\Lambda \delta_a^b$,
where the bulk cosmological constant
is $\Lambda = -6/L^2$ (we work in units in which the four-dimensional
gravitational coupling $8\pi G_4 = 8\pi G_5/L =1$). 
Evaluating the linear combinations $G^0_0 + G^5_5$
and $G^0_0 + G^5_5 - (1/2)G^i_i$ (where $0$ denotes time, $5$ labels 
the $y$ direction, and $i$ runs over the noncompact directions), 
we find: 
\bea
\label{bgdeqn1}
\beta_{,\tau\tau}-\beta_{,y y} +\beta_{,\tau}^2-\beta_{,y}^2 +
12\,e^{2\nu} &=&0, \\
\label{bgdeqn2}
\nu_{,\tau\tau}- \nu_{,y y} 
+ \frac{1}{3}(\beta_{,y}^2-\beta_{,\tau}^2) - 2\,e^{2\nu} &=& 0,
\eea
where $(t/L)= e^\tau$, $\beta\equiv 3\ln{b}$ and $\nu \equiv  \ln{(nt/L)}$. 
The Israel matching conditions on the branes read \cite{Carsten,
TTS}
\[
\frac{b_{,y}}{b} = \frac{n_{,y}}{n} = \frac{nt}{\L},
\label{ibd}
\]
where all quantities are to be evaluated at the brane locations,
$y=\pm y_0$. 

We will begin our assault on the bulk geometry by constructing a
series expansion in $t$ about the collision, implementing the Israel
matching conditions on the branes at each order in $t$.
This series expansion in $t$ is then exact in both $y$ and the collision
rapidity $y_0$. 
It chief purpose will be to provide
initial data for the more powerful solution
methods that we will develop in the following sections.  

The Taylor series solution in $t$ 
for the background was first presented in \cite{TTS}.  Expanded up to terms of 
$O(t^3/L^3)$,
\bea
n &=& 1 + (\sech\,{y_0}\sinh{y})\,\frac{t}{L}+\frac{1}{4}\,\sech^2\,{y_0}\,(-3+\cosh{2y_0}+2\cosh{2y}) \frac{t^2}{L^2}, \qquad \\ 
\label{tseriesbgd}
b &=& 1 +  (\sech\,{y_0}\sinh{y})\,\frac{t}{L}+\frac{1}{2}\,\sech^2\,{y_0}\,(-\cosh{2y_0}+\cosh{2y}) \frac{t^2}{L^2}. 
\eea
(Note that in the limit as $t\tt 0$ we correctly recover compactified Milne spacetime).

Here, however, we will need the perturbations as well.  
Working in five-dimensional longitudinal gauge for the reasons given
in the previous section, 
the perturbed bulk metric takes the form (see Appendix \ref{appA})
\[
\d s^2 = n^2\left(-(1+2\Phi_L)\,\d t^2-2W_L\,\d t\d y+t^2\,(1-2\Gamma_L)\,\d
  y^2\right)+b^2\left(1-2\Psi_L\right)d\vec{x}^2, 
\]
with $\Gamma_L=\Phi_L-\Psi_L$ being imposed by the five-dimensional
traceless $G^i_j$ equation.
The Israel matching conditions at $y=\pm y_0$ then read
\[
\Psi_{L\, ,y}=\Gamma_L \frac{nt}{L}, \qquad
\Phi_{L\, ,y}=-\Gamma_L \frac{nt}{L}, \qquad
W_L=0.
\label{lgi}
\]
Performing a series expansion, we find
\bea
\Phi_L &=& -\frac{B}{t^2}+\frac{B\,
  \sech{\,y_0}\sinh{y}}{t}+\Big(A-\frac{B}{8}-\frac{Bk^2}{4}+\frac{1}{6}B k^2\ln{|k t|}
\nonumber \\ &&   +\frac{1}{16}B\cosh{2y}\,(-1+6 \,y_0 \coth{2y_0})\,\sech^2{y_0}
 -\frac{3}{8}B\,\sech^2{y_0}\sinh{2y}\Big), 
 \\
\Psi_L
&=&-\frac{B\,\sech{\,y_0}\sinh{y}}{t}+\Big(2A-\frac{B}{4}
+\frac{B k^2}{3}\ln{|kt|} 
+\frac{B}{4}\cosh{2y}\,\sech^2{y_0}\Big), \qquad\,\,\,\,
 \\
\label{tseriesperts}
W_L &=& -\frac{3}{4}B\,\sech^2{y_0}\big(y\cosh{2y}-y_0\cosh{2y_0}\sinh{2y}\big)\,t, 
\eea
where the first two equations are accurate to $O(t)$ and the third is accurate to $O(t^2)$.
We have moreover set $L=1$ for clarity;
except for a few specific instances, we will now adopt this convention
throughout the rest of this chapter.
(To restore $L$, simply replace $t\tt t/L$ and
$k\tt kL$).
The two arbitrary constants $A$ and $B$ (which may themselves be arbitrary 
functions of $\vec{k}$) have been chosen so that,
on the positive-tension brane, to leading order in $y_0$, $\Phi_L$ goes as
\[
\Phi_L = A - \frac{B}{t^2} + O(y_0)+O(k^2)+O(t).
\]

\section{Expansion in Dirichlet and Neumann polynomials}
\label{polysection}

\subsection{Background}

Having solved the relevant five-dimensional Einstein equations as
a series expansion in the time $t$ 
before or after the collision event, we now have
an accurate description of the behaviour of the bulk at small $t$
for arbitrary collision rapidities.
In order to match onto the incoming
and outgoing states, however, 
we really want to study the
long-time behaviour of the solutions, as
the branes become widely separated.  Ultimately, this will
enable us to successfully map the system 
onto an appropriate four-dimensional effective description.
Instead of expanding in powers of the time, 
we approximate the five-dimensional solution
as a power series in the rapidity of the collision, and
determine each metric coefficient for all time
 at each order in the rapidity.

Our main idea is to express
the metric as a series of Dirichlet or Neumann polynomials in 
$y_0$ and $y$, bounded at order $n$ by a constant times
$y_0^n$, 
such that the series satisfies 
the Israel matching
conditions exactly at every order in $y_0$.
To implement this, we first change variables from
$b$ and $n$  to those obeying Neumann
boundary conditions. From 
(\ref{ibd}), $b/n$ 
is Neumann. 
Likewise, if we define $N(t,y)$ by 
\[
nt = \frac{1}{N(t,y) - y},
\label{nt}
\]
then one can easily check that $N(t,y)$ 
is also Neumann on the branes. Notice that if
$N$ and $b/n$ are constant, the metric (\ref{metrica})
is just that for anti-de Sitter spacetime. For fixed $y_0$, $N$ describes the
the proper separation of the
two branes, and $b$ is an additional
modulus describing the three-dimensional
scale-factor of the branes. 

Since $N$ and $b/n$ obey Neumann boundary conditions on
the branes, we
can expand both in a power series
\[
\label{Bgd_ansatz}
N= N_0(t)+\sum_{n=3}^\infty N_n(t) P_n(y), \qquad 
b/n=q_0(t)+\sum_{n=3}^\infty q_n(t) P_n(y), 
\]
where $P_n(y)$ are polynomials
\[
P_n(y)= y^n-\frac{n}{n-2} \,y^{n-2} \,y_0^2, \qquad n=3,4,\dots
\]
satisfying Neumann boundary conditions, 
each bounded by $|P_n(y)|<2y_0^n/(n-2)$  for the relevant
range of $y$. Note that the time-dependent coefficients in this ansatz 
may also be expanded as a power series in $y_0$.
By construction, our ansatz
satisfies the Israel matching conditions exactly at each order
in the expansion.  The bulk
Einstein equations are not satisfied exactly, 
but as the expansion is
continued, the error terms are 
bounded by increasing powers of $y_0$.

Substituting the series ans{\"a}tze (\ref{Bgd_ansatz}) 
into the background Einstein equations (\ref{bgdeqn1}) and (\ref{bgdeqn2}),
we may determine the solution order by order in the rapidity $y_0$.
At each order in $y_0$, one generically obtains a number of linearly
independent algebraic equations, and at most one ordinary differential
equation in $t$.  The solution of the latter introduces a number of
arbitrary 
constants of integration into the solution.

To fix the arbitrary constants, one first applies the remaining
Einstein equations, allowing a small number to be eliminated.  
The rest are then determined using the series expansion in $t$
presented in the previous section: as this solution is
exact to all orders in 
$y_0$, we need only to expand it out to the relevant order in $y_0$,
before 
comparing it term by term with our Dirichlet/Neumann 
polynomial expansion (which is exact
in $t$ but perturbative in $y_0$), taken to a corresponding
order in $t$.
The arbitrary constants are then chosen so as to ensure the
equivalence of the two expansions in the region where both $t$ and
$y_0$ are small. 
This procedure suffices to fix all the remaining arbitrary constants.

The first few terms of the solution are
\bea
N_0 &=& {1\over t}-\frac{1}{2}\,t y_0^2+{1\over 24}\,t(8-9t^2)y_0^4+\dots \\
N_3 &=& -{1\over 6}+\left({5\over 72}-2t^2\right)y_0^2+\dots 
\eea
and
\bea
q_0 &=& 1 -\frac{3}{2}\,t^2 y_0^2+\left(t^2-\frac{7}{8}\,t^4\right)y_0^4+\dots \\
q_3 &=& -2\,t^3 y_0^2+\dots
\eea
The full solution up to $O(y_0^{10})$ may be found in Appendix \ref{appB}.


\subsection{Perturbations}

Following the same principles used in our treatment of the background,
we construct the two linear combinations 
\[
\label{phi4xi4}
\phi_4 = \frac{1}{2}(\Phi_L+\Psi_L), \qquad \xi_4= b^2
(\Psi_L-\Phi_L)=b^2 \Gamma_L,
\]
both of which obey Neumann boundary conditions on the branes, as may
be checked from (\ref{ibd}) and (\ref{lgi}).  In addition, $W_L$ already
obeys simple Dirichlet boundary conditions.

The two Neumann variables, $\phi_4$ and
$\xi_4$, are then expanded in a series
of Neumann polynomials and $W_L$ is 
expanded in a series of
Dirichlet polynomials, 
\[
D_n(y)=y^n-y_0^n,\,\,\,\,\,\, n =2,4,\dots, \,\,\,\,\,\,   D_n(y)=y D_{n-1}(y), \,\,\,\,\,\, n=3,5,\dots,
\]
each bounded by $|D_n(y)|<y_0^n$ for $n$ even
and $y_0^n (n-1)/n^{n/(n-1)}$ for $n$ odd, over the relevant range
of $y$.  As in the case of the background, the time-dependent
coefficients multiplying each of the polynomials should themselves be
expanded in powers of $y_0$.

To solve for the perturbations it is sufficient to use only three of
the perturbed Einstein equations (any solution obtained may then be 
verified against the remainder).  
Setting
\bea
\Phi_L &=& \phi\, e^{-2\nu-\beta/3}, \\
\Psi_L &=& \psi\, e^{-\beta/3}, \\
W_L    &=& w\, e^{\tau-2\nu-\beta/3},
\eea
where $t=e^\tau$, $\beta=3\ln{b}$ and $\nu = \ln{nt}$,
the $G^5_i$, $G^0_i$ and $G^i_i$ equations take the form
\bea
\label{pe1}
w_{,\tau} &=& 2\,\phi_{,y} - 4\,e^{3\nu/2}\,(\psi\, e^{\nu/2})_{,y}, \\
\label{pe2}
\phi_{,\tau} &=& {1\over 2}\,w_{,y} - e^{3\nu}\,(\psi\, e^{-\nu})_{,\tau}, \\
\label{pe3}
(\psi_{,\tau}\, e^{\beta/3})_{,\tau} &=& (\psi_{,y}\, e^{\beta/3})_{,y}+
\psi\,e^{\beta/3}\,
\left(\frac{1}{3}\,\beta_{,\tau}^2-\frac{1}{9}\,\beta_{,y}^2-k^2\,e^{2(\nu-\beta/3)}\right)
\nonumber \\
&& -\frac{2}{9}\,e^{-2\nu+\beta/3}\,\left(\phi
\,(\beta_{,\tau}^2+\beta_{,y}^2)-
w\, \beta_{,\tau}\,\beta_{,y}\right) .
\eea

Using our Neumann and Dirichlet ans{\"a}tze for $\phi_4$, $\xi_4$ and
$W_L$, the Israel matching conditions are automatically satisfied and
it remains only to solve (\ref{pe1}), (\ref{pe2}) and (\ref{pe3})
order by order in the rapidity.
The time-dependent coefficients for $\phi_4$,
$\xi_4$ and  $W_L$ are then found to obey simple
ordinary differential equations, with solutions comprising
Bessel functions in $kt$, given in Appendix \ref{appC}.   
Note that it is not necessary for the set of Neumann or Dirichlet
polynomials we have used to be orthogonal to each other: 
linear independence is perfectly sufficient to determine all the
time-dependent coefficients order by order in $y_0$.

As in the case of the background, the arbitrary constants of
integration remaining in the solution after the application of the
remaining Einstein equations are fixed 
by performing a series
expansion of the solution in $t$.  This expansion can be compared
term 
by term with the series expansion in $t$ given previously, after
this latter series has itself been expanded in $y_0$.  
The arbitrary constants are then chosen so that the two expansions 
coincide in the region where both $t$ and $y_0$ are small.
The results of these calculations, at long wavelengths, are
\bea
\Phi_L &=& A-B\left({1\over  t^2}-{k^2\over 6} {\rm ln}|kt|\right)
 +\left(A t +\frac{B}{t}\right)\,y + 
\dots \\
\Psi_L &=& 2 A +B {k^2\over 3} {\rm ln}|kt| -\left(A t+ \frac{B}{t}\right)\,y + \dots \\
W_L&=& 6 A\, t^2\, (y^2-y_0^2)+\dots 
\label{results}
\eea
where the constants A and B can be arbitrary functions of $k$. The solutions for
all $k$, to fifth order in $y_0$, are given in Appendix \ref{appC}.


\section{Expansion about the scaling solution}
\label{expaboutscalingsoln}

It is illuminating to recast the results of the preceding sections in terms
of a set of dimensionless variables.  Using the relative velocity of
the branes at the moment of collision, $V= 2c \tanh y_0 \simeq 2c y_0$
(where we have temporarily re-introduced the speed of light $c$),
we may construct the dimensionless time parameter $x = y_0 ct/L \sim
Vt/L$ and the dimensionless $y$-coordinate $\w = y/ y_0 \sim y(c/V)$.

Starting from the full Dirichlet/Neumann 
polynomial expansion for the background given in Appendix
\ref{appB}, restoring $c$ to unity and setting $t=x L/y_0$ and $y = \w y_0$, 
we find that
\bea
\label{ad_n_series}
n^{-1} &=& \tN(x)-\w x + O(y_0^2), \\
\label{ad_q}
{b\over n} &=& q(x) + O(y_0^2),
\eea
where 
\bea
\label{N_series}
\tN(x) &=& 1 - \frac{x^2}{2} - \frac{3\,x^4}{8} - \frac{25\,x^6}{48} - \frac{343\,x^8}{384} - 
  \frac{2187\,x^{10}}{1280}+O(x^{12}), \\
\label{q_series}
q(x) &=& 1 - \frac{3\, x^2}{2} - \frac{7\, x^4}{8} - \frac{55\,
  x^6}{48} - \frac{245\, x^8}{128} - \frac{4617\, x^{10}}{1280} +O(x^{12}).
\eea
The single term in (\ref{ad_n_series}) linear in $\w$ is necessary in order that $n^{-1}$
satisfies the correct boundary conditions.  Apart from this 
one term, however, we see that to lowest order in $y_0$ the
metric functions above turn out to be completely independent of $\w$.  
Similar results are additionally found for the perturbations.

Later, we will see how this behaviour leads to the emergence of
a four-dimensional effective theory.  For now, the key point to
notice is that this series expansion 
converges only for $x \ll 1$, corresponding to times $t \ll L/y_0$.
In order to study the behaviour of the theory for all times therefore, we
require a means of effectively resumming the above perturbation expansion to all
orders in $x$.
Remarkably, we will be able to accomplish just this.  
The remainder of this section, divided into five parts, details our method and results:
first, we explain how to find and expand about the scaling solution,
considering only the background for simplicity.  
We then analyse various aspects of the background scaling solution,
namely, the brane geometry and the analytic continuation required to go
to late times, before moving on to discuss higher-order terms in the expansion.  
Finally, we extend our treatment to cover the perturbations.

\subsection{Scaling solution for the background}

The key to the our method is the observation that the 
approximation of small collision rapidity ($y_0\ll 1$) 
leads to a set of equations that are perturbatively ultralocal:
transforming to the dimensionless coordinates $x$ and $\w$, the
Einstein equations for the background
(\ref{bgdeqn1}) and (\ref{bgdeqn2}) become
\bea
\label{e1}
\beta_{,\w\w}+\beta_{,\w}^2-12\,e^{2\tnu}&=& y_0^2 \left(
x(x\beta_{,x})_{,x}+x^2\beta_{,x}^2\right), \\
\label{e2}
\tnu_{,\w\w}-\frac{1}{3}\,\beta_{,\w}^2+2\,e^{2\tnu} &=& y_0^2\big(
x(x\tnu_{,x})_{,x}-\frac{1}{3}\,x^2\beta_{,x}^2\big),
\eea
where we have introduced $\tnu = \nu+\ln{y_0}$.  
Strikingly, all the 
terms involving
$x$-derivatives are 
now suppressed by a factor of $y_0^2$ relative to the remaining terms.
This segregation of $x$- and $\w$-derivatives has profound
consequences: when solving perturbatively in $y_0$,  
the Einstein equations 
(\ref{e1}) and (\ref{e2}) reduce to 
a series of {\it ordinary} differential equations in $\w$,
as opposed to the partial differential equations we started off with.

To see this, consider expanding out both the Einstein equations (\ref{e1}) and (\ref{e2})
as well as the metric functions $\beta$ and $\tnu$ as a series in
positive powers of $y_0$.
At zeroth order in $y_0$, the right-hand sides of (\ref{e1}) and
(\ref{e2}) vanish, and the
left-hand sides can be integrated with respect to $\w$ to yield
anti-de Sitter space.  
(This was our reason for using $\tnu = \nu+\ln{y_0}$ rather than $\nu$:
the former serves to pull the necessary exponential term deriving from
the cosmological constant down to zeroth order in $y_0$, yielding 
anti-de Sitter space as a solution at leading order.  
As we are merely adding a
constant, the derivatives of $\tnu$ and $\nu$ are identical.)
The Israel matching conditions on the branes (\ref{ibd}), which in these coordinates read 
\[
\label{ibd2}
\frac{1}{3}\,\beta_{,\w}=\tnu_{,\w}=e^{\tnu},
\]
are not, however, sufficient to fix all the arbitrary functions of $x$ arising
in the integration with respect to $\w$.  In fact, two arbitrary functions of
$x$ remain in the solution, which may be regarded as time-dependent moduli describing the 
three-dimensional scale factor of the branes and their proper separation.
These moduli may be determined with the help of the $G^5_5$ Einstein equation as we will 
demonstrate shortly.  

Returning to (\ref{e1}) and (\ref{e2}), at $y_0^2$ order now,
the left-hand sides amount to ordinary differential equations in
$\w$ for the $y_0^2$ corrections to $\beta$ and $\tnu$.
The right-hand sides can no longer be neglected, but, because of the
overall factor of $y_0^2$, only the
time-derivatives of $\beta$ and $\tnu$ at {\it zeroth} order in $y_0$
are involved.
Since $\beta$ and $\tnu$ have already been determined to this order,
the right-hand sides therefore act merely as known source terms.
Solving these ordinary differential equations then introduces
two further arbitrary functions of $x$; these serve as $y_0^2$
corrections to the time-dependent moduli and may be fixed in the same
manner as previously.

Our integration scheme therefore proceeds at each order in $y_0$ via a
two-step process: first, we integrate the Einstein 
equations (\ref{e1}) and (\ref{e2}) to determine the $\w$-dependence of the bulk geometry, and then 
secondly, we fix the $x$-dependent moduli pertaining to the brane geometry using the $G^5_5$ equation.
This latter step works as follows: evaluating the $G^5_5$ equation on the branes, we can use the Israel matching
conditions (\ref{ibd2}) to replace the single $\w$-derivatives that appear in this equation, yielding
an ordinary differential equation in time for the geometry on each brane.
Explicitly, we find
\[
\left(\frac{bb_{,x}}{n}\right)_{\hspace{-1mm},x}=0,
\]
where five-dimensional considerations (see Section \S\,\ref{compwitheft}) further allow us to fix the
constants of integration on the ($\pm$) brane as
\[
\label{G55}
\frac{b b_{,x}}{n}=\frac{b b_{,t}}{y_0
  n}=\frac{b_{,t_\pm}}{y_0} 
=\pm\frac{1}{y_0}\tanh{y_0},
\]
where the brane conformal time $t_\pm$ is defined on the branes via $n\d
t = b\d t_\pm$.  
When augmented with the initial conditions that $n$ and
$b$ both tend to unity as $x$ tends to zero (so that we recover
compactified Milne spacetime near the collision), these two equations are fully
sufficient to determine the two $x$-dependent moduli to all orders in $y_0$.


Putting the above into practice, for convenience we will work with
the Neumann variables $\tN$ and $q$, generalising 
(\ref{ad_n_series}) and (\ref{ad_q}) to
\[
n^{-1} = \tN(x,\w)-\w x,  \qquad  \frac{b}{n}=q(x,\w).
\]
Seeking an expansion of the form
\bea
\label{ad_ansatz1}
\tN(x,\w) &=& \tN_0(x,\w) + y_0^2\tN_1(x,\w)+O(y_0^4), \\
\label{ad_ansatz2}
q(x,\w) &=& q_0(x,\w)+y_0^2\, q_1(x,\w)+O(y_0^4),
\eea
the Einstein equations (\ref{e1}) and (\ref{e2}) when expanded to zeroth order in
$y_0$ immediately restrict $\tN_0$ and $q_0$ to be functions of 
$x$ alone.  The bulk geometry to this order is then simply anti-de Sitter space with
time-varying moduli, consistent with (\ref{ad_n_series}) and (\ref{ad_q}).
The moduli $\tN_0(x)$ and $q_0(x)$ may be found
by integrating the brane equations (\ref{G55}), also expanded to lowest order in $y_0$.
In terms of the Lambert W-function \cite{LambertW}, $W(x)$, defined implicitly by
\[
\label{Wdef}
W(x)e^{W(x)}=x,
\]
the solution is
\[
\label{sol1}
\tN_0(x) = e^{\frac{1}{2}W(-x^2)}, \qquad q_0(x) = \left(1+W(-x^2)\right)\,e^{\frac{1}{2}W(-x^2)}.
\]
\begin{figure}[p]
\begin{center}
\hspace{-0.7cm}
      \includegraphics[width=12cm]{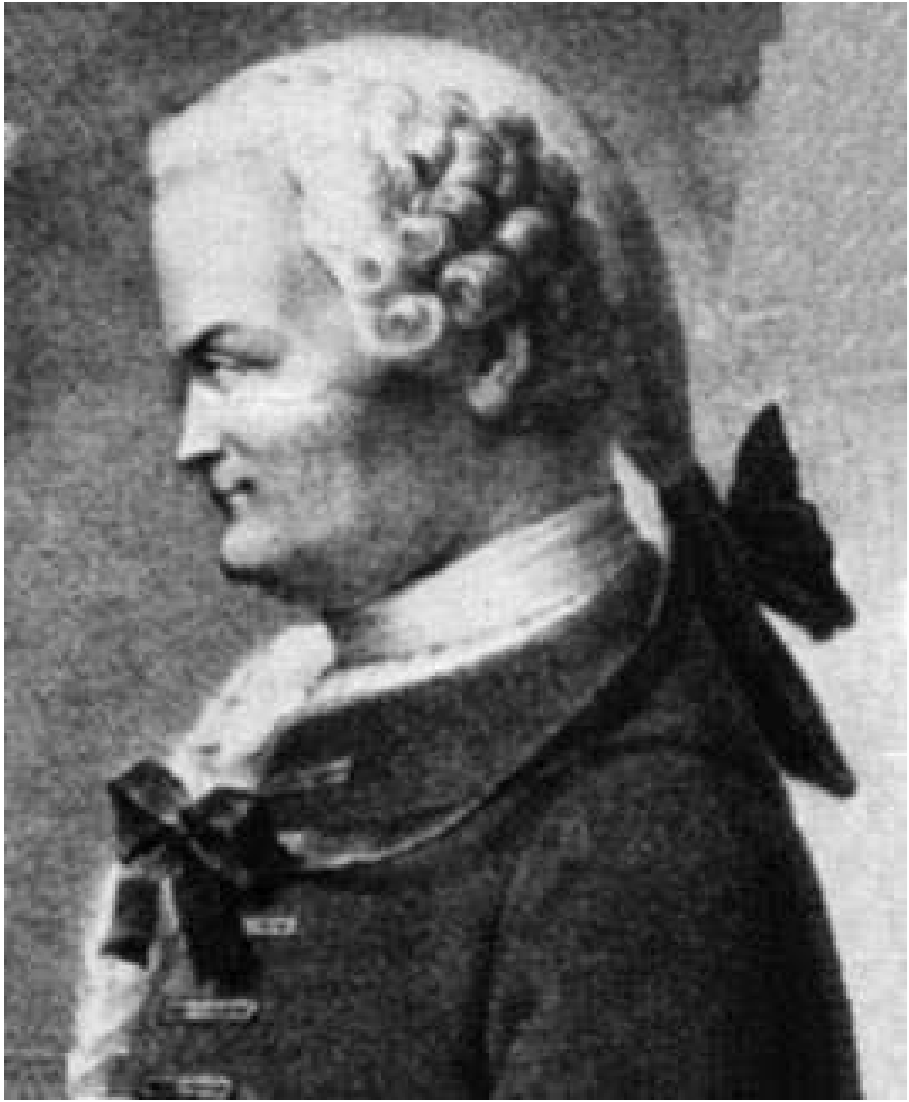}
\caption{The eponymous hero: Johann Heinrich Lambert, 1728-1777. \newline
Famed for his proof of the irrationality of $\pi$, and for his natty dress sense.}
\label{LambertHimself}
\end{center}
\end{figure}
Thus we have found the scaling solution for the background, whose form is
independent of $y_0$, holding for any $y_0\ll 1$.
Using the series expansion for the Lambert W-function about
$x=W(x)=0$, namely\footnote{
Note that the radius of convergence of the series (\ref{W_series}) for
$W(x)$ is $1/e$, and thus it converges for arguments in the range $-1/e\le x
\le 0$ as required.}
\[
\label{W_series}
W(x) = \sum_{m=1}^\inf\frac{(-m)^{m-1}}{m!}x^m,
\]
we can immediately check that the expansion of our solution is in
exact agreement with (\ref{N_series}) and (\ref{q_series}). 
At leading order in $y_0$ then, we have succeeded in resumming  
the Dirichlet/Neumann 
polynomial expansion results for the background to all orders in $x$.

Later, we will return to evaluate the $y_0^2$ corrections in our
expansion about the scaling solution.  
In the next two subsections, however, we will first examine the
scaling solution in greater detail. 

\subsection{Evolution of the brane scale factors}

\begin{figure}[p]
\begin{center}
      \includegraphics[width=12cm]{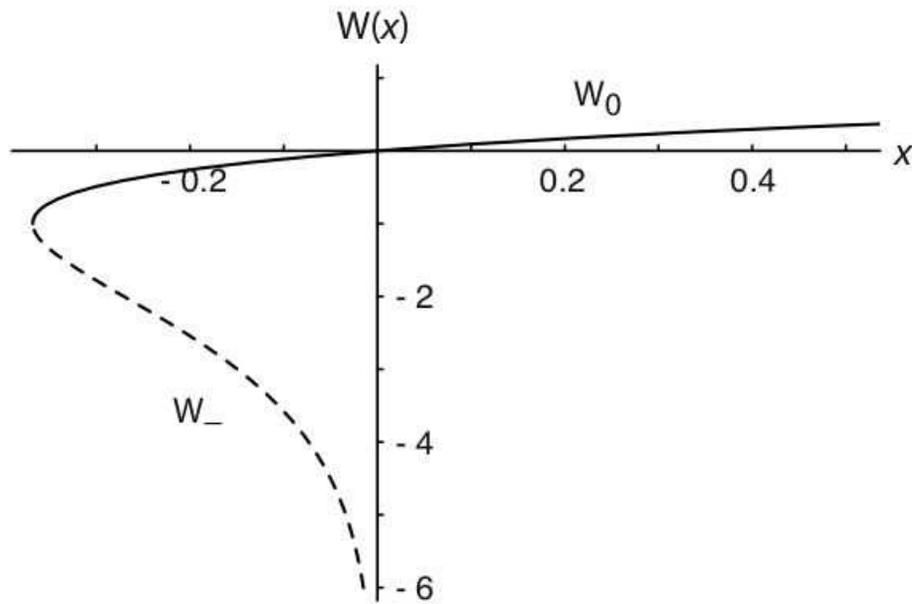}
\caption{
The real values of the Lambert W-function.  The solid line indicates
the principal solution branch, $W_0(x)$, while the dashed line depicts the
$W_{-1}(x)$ branch.  The two branches join smoothly at $x=-1/e$ where 
$W$ attains its negative maximum of $-1$.}
\label{Wfigure}
\end{center}
\end{figure}

Using the scaling solution (\ref{sol1}) to evaluate the scale factors on both
branes, we find to $O(y_0^2)$
\[
b_\pm = 1\pm x e^{-\frac{1}{2}W(-x^2)} = 1\pm\sqrt{-W(-x^2)}.
\]
To follow the evolution of the brane scale factors, it is helpful to
first understand the behaviour of the Lambert W-function, the real
values of which are displayed in Figure \ref{Wfigure}. 
For positive arguments the Lambert W-function is single-valued; 
yet for the negative arguments of interest here, we see that there
are in fact two different real solution branches.
The first branch, denoted $W_0(x)$, satisfies $W_0(x)\ge -1$ and is usually referred
to as the principal branch, while the second branch, $W_{-1}(x)$,
is defined in the range $W_{-1}(x)\le -1$.  The two solution branches
join smoothly at $x=-1/e$, where $W=-1$. 

\begin{figure}[p]
\begin{center}
      \includegraphics[width=12cm]{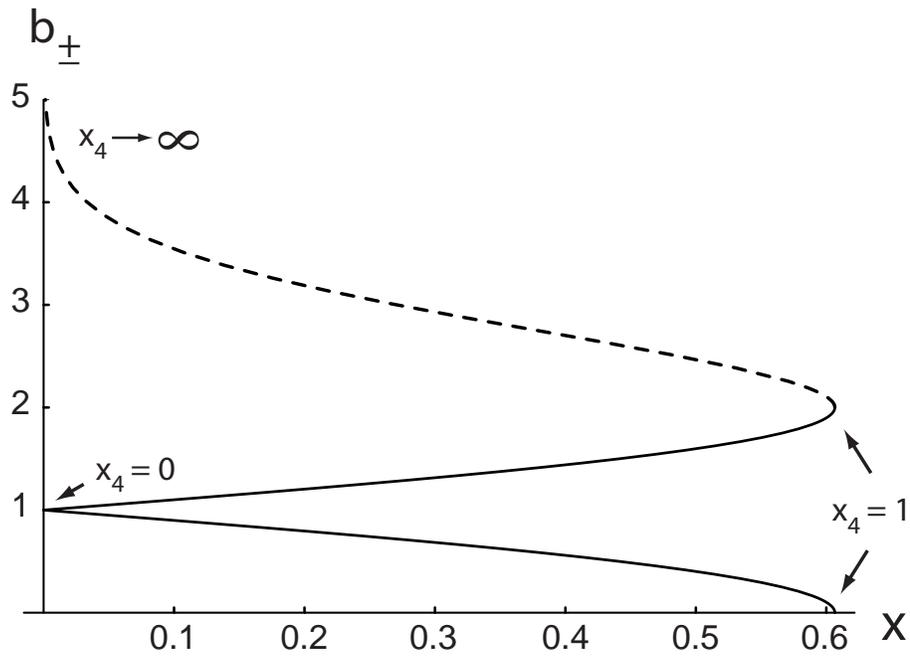}
\caption{
 The scale factors $b_{\pm}$ on the positive-tension brane (rising curve)
and negative-tension brane (falling curve) as a function of the bulk time
 parameter $x$, to zeroth order in $y_0$.  The continuation of the
 positive-tension brane scale factor on to the $W_{-1}$ branch 
of the Lambert W-function is indicated by the dashed line.  }
\label{bfigure}
\end{center}
\end{figure}

Starting at the brane collision where $x=0$, the brane scale factors
are chosen to satisfy $b_\pm=1$, and so we must begin on the
principal branch of the Lambert W-function 
for which $W_0(0)=0$.  Thereafter, as illustrated in Figure \ref{bfigure}, 
$b_+$ increases and $b_-$ decreases monotonically until 
at the critical time $x=x_c$, when $W_0(-x_c^2)=-1$ and $b_-$ shrinks
to zero.  From (\ref{Wdef}), the critical time is therefore
$
x_c = e^{-\frac{1}{2}} = 0.606...,
$
and corresponds physically to the time at which the negative-tension brane
encounters the bulk black hole\footnote{
From the exact solution in bulk-static coordinates, the scale factor
on the negative-tension brane at the horizon obeys $b_-^2 = \sech{2
  Y_0/L} = \tanh{y_0}$, and so $b_- \sim y_0^{1/2}$.}.

At this moment, the scale factor on the positive-tension brane 
has only attained a value of two.  
From the Birkhoff-frame solution, in which the bulk is
AdS-Schwarzschild and the branes are moving, we know that the
positive-tension brane is unaffected by the disappearance 
of the negative-tension brane and simply continues its journey out to the boundary of AdS. 
To reconcile this behaviour with our solution in brane-static
coordinates, it is helpful to pass to $t_+$, the conformal time on the
positive-tension brane.  Working to zeroth order in $y_0$, this
may be converted into the dimensionless form
\[
\label{x4}
x_4= \frac{y_0 t_+}{L} =\frac{y_0}{L}\int \frac{n}{b}\,\d t= \int \frac{\d x}{q_0(x)}=
x e^{-\frac{1}{2}W(-x^2)}=\sqrt{-W(-x^2)} .
\]
Inverting this expression, we find that the bulk time parameter $x=x_4\,e^{-\frac{1}{2}x_4^2}$.
The bulk time $x$ is thus double-valued when expressed as a function
of $x_4$, the conformal time on the positive-tension brane: to continue
forward in $x_4$ beyond $x_4=1$ (where $x=x_c$), the bulk time $x$ must
reverse direction and decrease towards zero.  The metric functions,
expressed in terms of $x$, must then continue back along the other branch
of the Lambert W-function, namely the $W_{-1}$ branch.
In this manner we see that the solution for the scale factor on the positive-tension
brane, when continued on to the $W_{-1}$ branch, tends to infinity as
the bulk time $x$ is reduced back towards zero (see dotted line in Figure \ref{bfigure}),
corresponding to the positive-tension brane approaching the boundary
of AdS as $x_4 \tt \inf$.

For simplicity, in the remainder of this chapter we will work
directly with the brane conformal time $x_4$ itself.  With this choice,
the brane scale factors to zeroth order in $y_0$ are simply
\nopagebreak[1] $b_\pm = 1\pm x_4$.

\subsection{Analytic continuation of the bulk geometry}

In terms of $x_4$, the metric functions $n$ and $b$ are given by
\[
\label{nandb}
n = \frac{e^{\frac{1}{2}x_4^2}}{1-\w x_4}+O(y_0^2), \qquad b = \frac{1-x_4^2}{1-\w x_4}+O(y_0^2).
\]
At $x_4=1$, the three-dimensional scale factor $b$ shrinks to zero at
all values of $\w$ except $\w=1$ (\ie the positive-tension brane).
Since $b$ is a coordinate scalar under transformations of $x_4$ and
$\w$, one might be concerned that that the scaling solution becomes singular at
this point.  When we compute the $y_0^2$ corrections, however, we will 
find that these corrections become large close
to $x_4=1$, precipitating a breakdown of the small-$y_0$
expansion.  
Since it will later turn out that the 
scaling solution
maps directly on to the four-dimensional
effective theory, and that this, like the metric on the
positive-tension brane, is completely regular at $x_4=1$, we are
encouraged to simply analytically continue the scaling solution to times
$x_4>1$.  

When implementing this analytic continuation careful attention must be paid to
the range of the coordinate $\w$.  Thus far, for times $x_4<1$, we
have regarded $\w$ as a coordinate spanning the fifth dimension,
taking values in the range $-1\le \w \le 1$.  The two metric functions
$n$ and $b$ were then expressed in terms of the coordinates $x_4$ and $\w$.
Strictly speaking, however, this parameterisation is redundant: we could
have chosen to eliminate $\w$ by promoting the three-dimensional scale factor $b$ 
from a metric function to an independent coordinate parameterising the
fifth dimension.  Thus we would have only one metric function $n$,
expressed in terms of the coordinates $x_4$ and $b$.   

While this latter parameterisation is more succinct, its disadvantage is
that the locations of the branes are no longer explicit, since the
value of the scale factor $b$ on the branes is time-dependent.
In fact, to track the location of the branes we must re-introduce the
function $\w(x_4,b)=(b+x_4^2-1)/bx_4$ (inverting (\ref{nandb}) at
lowest order in $y_0$).  The trajectories of the branes are then 
given by the contours $\w=\pm 1$.

\begin{figure}[p]
\begin{center}
      \includegraphics[width=13cm]{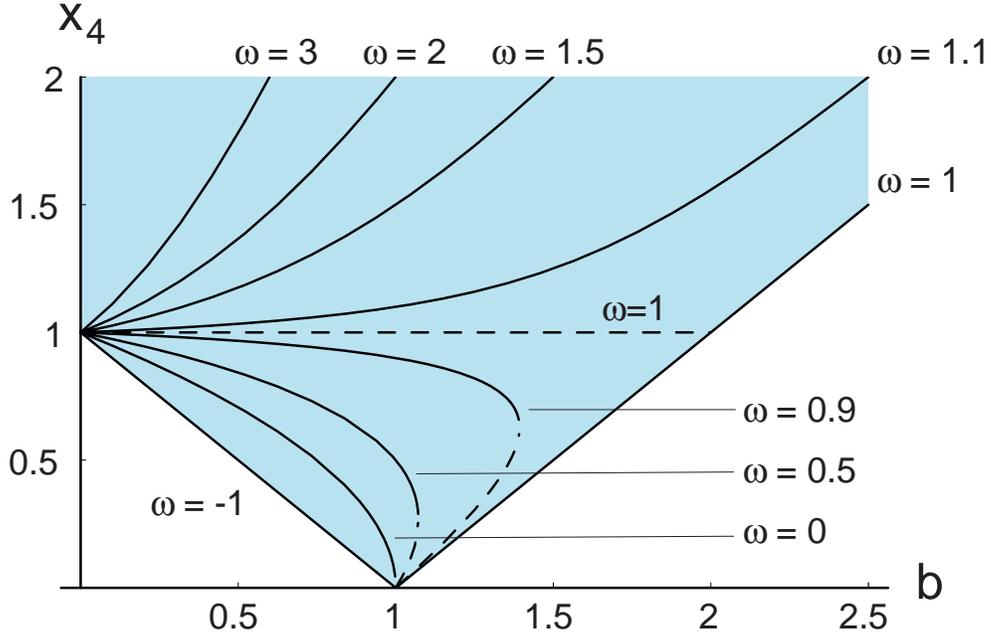}
\caption{
The contours of constant $\w$ in the ($b$, $x_4$) plane.
Working to zeroth order in $y_0$, these are given by $x_4 =
\frac{1}{2}\big(b\w\pm \sqrt{b^2\w^2-4(b-1)}\big)$, where we have
plotted the positive root using a solid line and the negative root using a dashed line.
The negative-tension brane is located at $\w=-1$ for times $x_4<1$,
and the trajectory of the positive-tension brane is given (for all time)
by the positive root solution for $\w=1$.  The region delimited by the
trajectories of the branes (shaded) then corresponds to the bulk.
From the plot we see that, for $0<x_4<1$, the bulk is parameterised by
values of $\w$ in the range $-1\le \w \le 1$.  In contrast, for $x_4>1$, the
bulk is parameterised by values of $\w$ in the range $\w\ge 1$.
} 
\label{coathanger}
\end{center}
\end{figure}

The contours of constant $\w$ as a function of $x_4$ and $b$ are
plotted in Figure \ref{coathanger}.  The analytic continuation to times $x_4>1$ has
been implemented, and the extent of the bulk is indicated by
the shaded region.
From the figure, we see that, if we were to revert to our original
parameterisation of the bulk in terms of $x_4$ and $\w$, the range of
$\w$ required depends on the time coordinate $x_4$: for early times
$x_4<1$, we require only values of $\w$ in the range $-1\le \w \le 1$,
whereas for late times $x_4>1$, we require values in the range $\w\ge
1$.  Thus, while the positive-tension brane remains fixed at $\w=1$
throughout, at early times $x_4<1$ the value of $\w$ {\it decreases} as we
head away from the positive-tension brane along the fifth dimension, whereas
at late times $x_4>1$, the value of $\w$ {\it increases} away from the positive-tension brane.

While this behaviour initially appears paradoxical if $\w$ is
regarded as a coordinate along the fifth dimension, we stress that 
the only variables with meaningful physical content are the brane conformal time $x_4$ and the
three-dimensional scale factor $b$.  These physical variables behave
sensibly under analytic continuation.   
In contrast, $\w$ is simply a convenient
parameterisation introduced to follow the brane trajectories, with the
awkward feature that its range alters under the analytic continuation at $x_4=1$.

\begin{figure}[p]
\begin{center}
\begin{minipage}{10cm}
\vspace{0.3cm}
      \includegraphics[width=10cm]{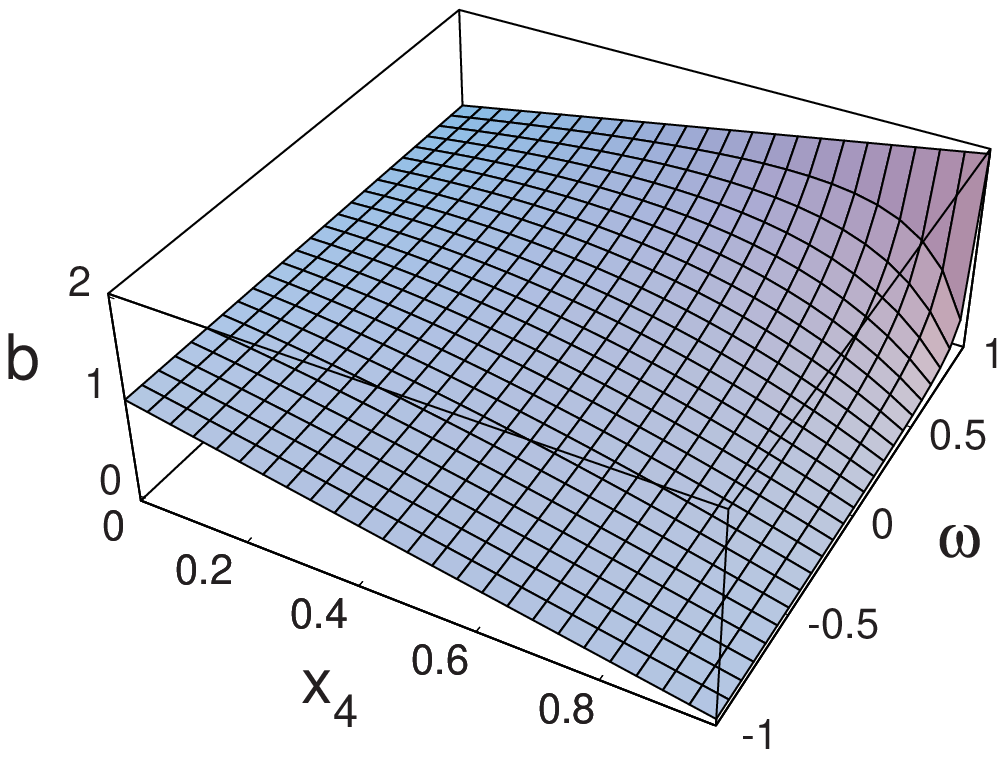}
\vspace{0.4cm}
\end{minipage}
\begin{minipage}{10cm}
      \includegraphics[width=10cm]{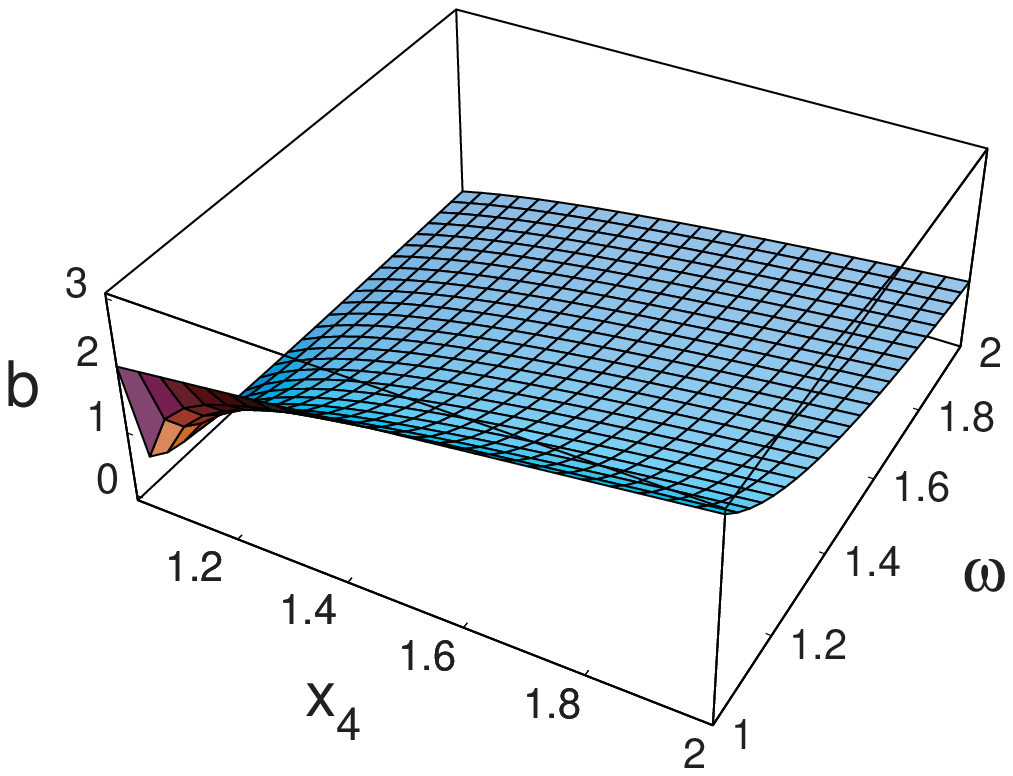}
\vspace{0.5cm}
\end{minipage}
\caption{
The three-dimensional scale factor $b$, plotted to zeroth order in $y_0$ as a
function of $x_4$ and $\w$, for $x_4<1$ (top) and $x_4>1$
(bottom). The positive-tension brane is fixed at $\w=1$ for all time
(note the evolution of its scale factor is smooth and continuous),
and for $x_4<1$, the negative-tension brane is located at $\w=-1$.  
}
\label{bplots}
\end{center}
\end{figure}

For the rest of this chapter, we will find it easiest to continue parameterising the bulk
in terms of $x_4$ and $\w$, adjusting the range of the $\w$ where required.
Figure \ref{bplots} illustrates this approach: at early times $x_4<1$ the
three-dimensional scale factor $b$ is plotted for values of $\w$ in
the range $-1\le \w \le 1$.  At late times $x_4>1$, we must however
plot $b$ for values of $\w$ in the range $\w\ge 1$.  In this fashion,
the three-dimensional scale factor $b$ always decreases along the
fifth dimension away from the brane.  

We have argued that 
the scaling solution for the background, obtained
at lowest order in $y_0$, may 
be analytically continued across $x_4=1$. 
There is a coordinate singularity in the $x_4$, $\w$ 
coordinates but this does not affect the metric on the 
positive-tension brane which remains regular throughout.
The same features will be true when we solve for the
cosmological perturbations. The fact that the
continuation is regular on the positive-tension brane,
and precisely agrees with the predictions of the
four-dimensional effective theory, provides strong evidence for its
correctness. Once the form of the 
the background and the perturbations have been 
determined to lowest order in $y_0$, the 
higher-order corrections are obtained from
differential equations in $y$ with source
terms depending only on the lowest order solutions. 
It is straightforward to obtain these corrections
for $x_4<1$. If we analytically continue them to $x_4>1$ as described, 
we automatically solve the bulk
Einstein equations and the Israel matching 
conditions on the positive
tension brane for all $x_4$. The continued
solution is well behaved 
in the vicinity of the 
positive-tension brane, out to large distances where
the $y_0$ expansion eventually fails. 
%

\subsection{
Higher-order corrections}

In this section we explicitly compute the 
$y_0^2$ corrections.  The size of these
corrections indicates the validity of the expansion about the scaling solution, which
perforce is only valid when the $y_0^2$ corrections are small.

Following the procedure outlined previously, we first evaluate
the Einstein equations (\ref{e1}) and (\ref{e2}) to $O(y_0^2)$ using the ans{\"a}tze
(\ref{ad_ansatz1}) and (\ref{ad_ansatz2}), along with the solutions for
$\tN_0(x)$ and $q_0(x)$ given in (\ref{sol1}). 
The result is two second-order ordinary differential equations in
$\w$, which may straightforwardly be integrated yielding
$\tN_1(x,\w)$ and $q_1(x,\w)$ up to two arbitrary functions of $x_4$. 
These time-dependent moduli are then fixed using the brane equations (\ref{G55}), evaluated
at $O(y_0^2)$ higher than previously.  

To $O(y_0^4)$, we obtain the result:
\bea
\label{ad_n}
n(x_4,\w) &=& \frac{e^{\frac{1}{2}x_4^2}}{1 - \w x_4} + \frac{e^{\frac{1}{2}x_4^2}
  y_0^2}{30 {\left( -1 + \w x_4 \right) }^2 {\left( -1 + x_4^2 \right)
  }^4} \, \big( x_4 \big( 5 \w \left( -3 + \w^2 \right) \nonumber \\ &&  - 5 x_4  +
     40 \w \left( -3 + \w^2 \right)  x_4^2 -  
          5 \left( -14 + 9 \w^2 \left( -2 + \w^2 \right)  \right)
	   x_4^3 \nonumber \\ && + 3 \w^3 \left( -5 + 3 \w^2 \right)  x_4^4  - 19 x_4^5 +  
          5 x_4^7 \big)  - 5 {\left( -1 + x_4^2 \right) }^3 \ln (1 -
	  x_4^2) \big), \nonumber \\  
&& \\
\label{ad_b}
b(x_4,\w) &=&  \frac{1 - x_4^2}{1 - \w x_4} + \frac{x_4 y_0^2}{30  
     {\left( -1 + \w x_4 \right) }^2 {\left( -1 + x_4^2 \right) }^3}\,
 \big( -5 \w \left( -3 +
  \w^2 \right) \nonumber \\ &&  - 20 x_4  + 5 \w \left( -7 + 4 \w^2 \right)  x_4^2 -  
       10 \left( 1 - 12 \w^2 + 3 \w^4 \right)  x_4^3 \nonumber \\ && + 3 \w \left(
       -20 - 5 \w^2 + 2 \w^4 \right)  x_4^4 - 12 x_4^5  + 31 \w x_4^6 -  
       5 \w x_4^8 \nonumber \\ && - 5 {\left( -1 + x_4^2 \right) }^2 \left( \w - 2 x_4 +
       \w x_4^2 \right)  \ln (1 - x_4^2) \big).
\eea

\begin{figure}[p]
\begin{center}
\begin{minipage}{11cm}
\hspace{-1cm}
      \includegraphics[width=11cm]{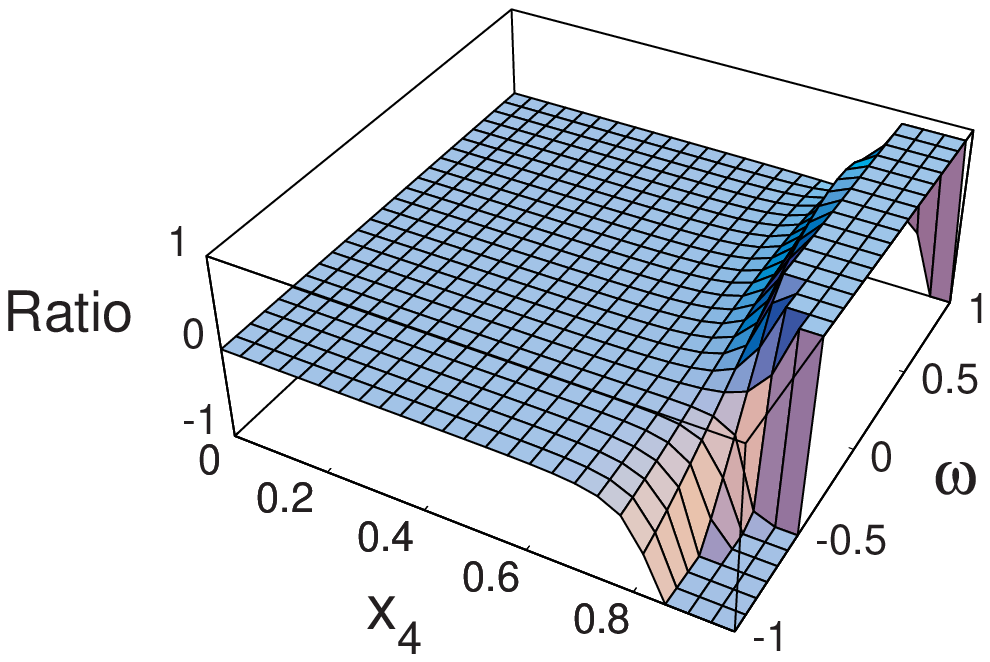}
\vspace{1cm}
\end{minipage}
\begin{minipage}{11cm}
\hspace{-1cm}
      \includegraphics[width=11cm]{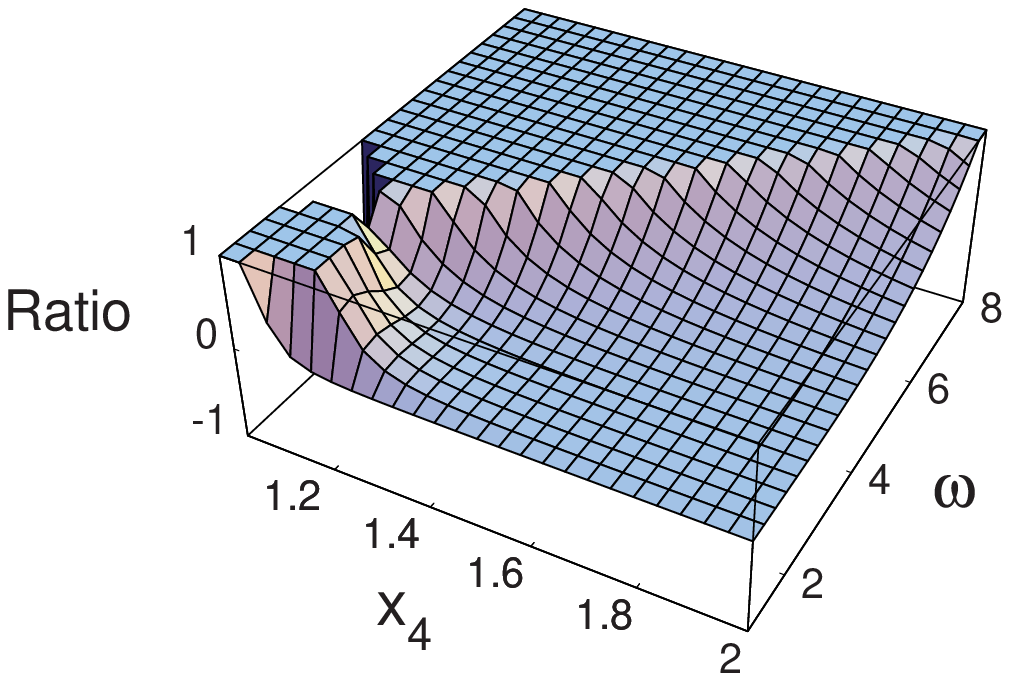}
\vspace{0.5cm}
\end{minipage}
\caption{
The ratio of the $y_0^2$ corrections to the leading term in the
small-$y_0$ expansion for $b$, plotted for $x_4<1$ (top) and $x_4>1$
(bottom), for the case where $y_0=0.1$.   Where this ratio becomes of
order unity the expansion about the scaling solution breaks down.  
The analogous plots for $n$ display similar behaviour.
}
\label{metricplots}
\end{center}
\end{figure}
In Figure \ref{metricplots}, we have plotted the ratio of the $y_0^2$ corrections to
the corresponding terms at leading order: where this ratio becomes of
order unity the expansion about the scaling solution breaks down.  
Inspection shows there are two such regions: the
first is for times close to $x_4=1$, for all $\w$, and the second
occurs at late times $x_4>1$, far away from the positive-tension brane.
In neither case does the failure of the $y_0$ expansion indicate a
singularity of the background metric: from the bulk-static coordinate
system we know the exact solution for the background metric is simply
AdS-Schwarzschild, which is regular everywhere.  
The
exact bulk-static solution in Birkhoff frame  
tells us that the proper speed of the negative-tension brane,
relative to the static bulk, approaches the speed of light 
as it reaches the event horizon
of the bulk black hole.  It therefore seems plausible 
that a small-$y_0$ expansion 
based on slowly moving branes must break
down at this moment, when $x_4=1$ in our chosen
coordinate system.

Analytically continuing our solution in both $x_4$ and $\w$ around
$x_4=1$, the logarithmic terms in the $y_0^2$ corrections now
acquire imaginary pieces for times $x_4>1$.  Since these
imaginary terms are all suppressed by a factor of $y_0^2$, however, they
can only enter the Einstein-brane equations (expanded as a series
to $y_0^2$ order) in a linear fashion.
Hence the real and imaginary parts of the metric necessarily
constitute {\it independent} solutions, permitting us to simply
throw away the imaginary part and work with the real part alone.
As a confirmation of this, it can be checked explicitly that replacing
the $\ln{(1-x_4)}$ terms in (\ref{ad_n}) and (\ref{ad_b}) with
$\ln{|1-x_4|}$ still provides a valid solution to $O(y_0^2)$ of the complete
Einstein-brane equations and boundary conditions.  

Finally, at late times $x_4>1$, note that the extent to which we know the
bulk geometry away from the positive-tension brane is limited by the
$y_0^2$ corrections, which become large at an increasingly large value
of $\w$, away from the positive-tension brane (see Figure \ref{metricplots}).
The expansion about the scaling solution thus breaks down before we
reach the horizon of the bulk black hole, which is located at $\w\tt \inf$ for $x_4>1$.

\subsection{Treatment of the perturbations}

Having determined the background geometry to $O(y_0^2)$ in the
preceding subsections, we now turn our attention to the perturbations.
In this subsection we show how to evaluate the perturbations to
$O(y_0^2)$ by expanding about the scaling solution.  The results will
enable us to perform stringent checks of the four-dimensional
effective theory and moreover to evaluate the mode-mixing between early and
late times.

In addition to the dimensionless variables $x=y_0 ct/L$ and $\w = y/
y_0$, when we consider the metric perturbations we must further introduce 
the dimensionless perturbation amplitude $\tB = B y_0^2 c^2/L^2 \sim B
V^2/L^2$ and the
dimensionless wavevector $\tk = kL/y_0\sim ckL/V$.  In this fashion, 
to lowest order in $y_0$ and $k$, we then find $\Phi_L = A - B/t^2
= A - \tB /x^2$ and similarly $kct=\tk x$.  (Note that
the perturbation amplitude A is already dimensionless however).

Following the treatment of the perturbations in the Dirichlet/Neumann 
polynomial expansion,
we will again express the metric perturbations in terms of $W_L$
(obeying Dirichlet boundary conditions), and the Neumann variables
$\phi_4$ and $\xi_4$, defined in (\ref{phi4xi4}).  
Hence we seek an expansion of the form
\bea
\label{adperts}
\phi_4(x_4,\w)&=&\phi_{40}(x_4,\w)+y_0^2\,\phi_{41}(x_4,\w)+O(y_0^4), \\
\xi_4(x_4,\w) &=& \xi_{40}(x_4,\w)+y_0^2\,\xi_{41}(x_4,\w)+O(y_0^4),\\
W_L(x_4,\w)&=& W_{L0}(x_4,\w)+y_0^2\,W_{L1}(x_4,\w)+O(y_0^4).
\eea

As in the case of the background, we will use the $G^5_5$ equation
evaluated on the brane to fix the arbitrary functions of
$x_4$ arising from integration of the Einstein equations with respect to
$\w$.  By substituting
the Israel matching conditions into the $G^5_5$ equation, along with
the boundary conditions for the perturbations, it is possible to 
remove the single $\w$-derivatives that appear.  
We arrive at the following 
second-order ordinary differential equation, valid on both branes, 
\bea
\label{breq}
0&=&
 2n(x_4^2-1)(2b^2\phi_4+\xi_4)\dot{b}^2+b^2\left(nx_4(x_4^2-3)-(x_4^2-1)\dot{n}\right)
(2b^2\dot{\phi}_4-\dot{\xi}_4)\nonumber 
\\ &&
+b\dot{b}\left(4nx_4(x_4^2-3)(b^2\phi_4+\xi_4)-(x_4^2-1)(4(b^2\phi_4+\xi_4)\dot{n}
-n(10b^2\dot{\phi}_4+\dot{\xi}_4))\right)\nonumber 
\\ &&+bn(x_4^2-1)\left(4(b^2\phi_4+\xi_4)\ddot{b}+2b^3\ddot{\phi}_4-b\ddot{\xi}_4\right),
\eea
where dots indicate differentiation with respect to $x_4$, 
and where, in the interests of clarity, we have omitted terms of $O(\tk^2)$. 

\begin{figure}[p]
\begin{center}
\begin{minipage}{11cm}
\vspace{0.4cm}
\hspace{-0.8cm}
      \includegraphics[width=11cm]{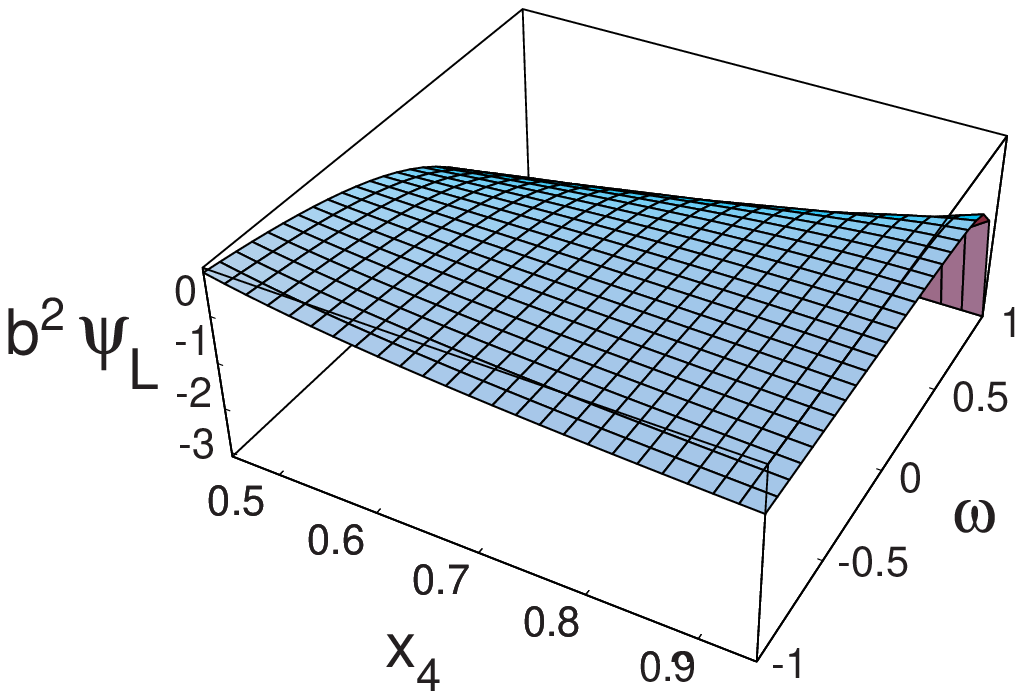}
\vspace{1cm}
\end{minipage}
\begin{minipage}{11cm}
\hspace{-0.8cm}
      \includegraphics[width=11cm]{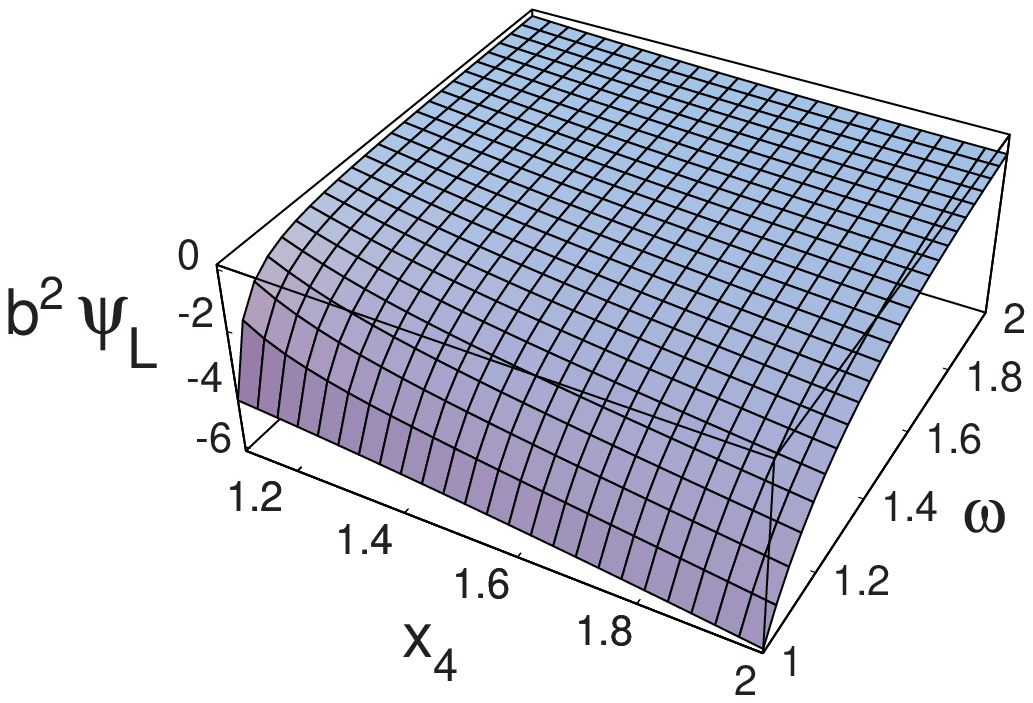}
\vspace{0.5cm}
\end{minipage}
\caption{
The perturbation to the three-dimensional scale factor, $b^2 \Psi_L$, plotted on long
wavelengths to zeroth order in $y_0$ for early times (top) and late times (bottom).
Only the $\tB$ mode is displayed (\ie $A=0$ and $\tB=1$).  Note
how the perturbations are localised on the positive-tension brane
(located at $\w =1$), and decay away from the brane.
}
\label{pertplots}
\end{center}
\end{figure}

Beginning our computation, the $G^5_i$ and $G^5_5$ Einstein  
equations when evaluated to lowest order in $y_0$ immediately restrict
$\phi_{40}$ and $\xi_{40}$ to be functions of $x_4$ only.
Integrating the $G^0_i$ equation with respect to $\w$ then gives $W_{L0}$
in terms of $\phi_{40}$ and $\xi_{40}$, up to an arbitrary function of
$x_4$.  Requiring that $W_{L0}$ vanishes on both branes allows us to
both fix this arbitrary function, and also to solve for $\xi_{40}$ in
terms of $\phi_{40}$ alone.  Finally, evaluating (\ref{breq}) on both branes 
to lowest order in $y_0$ and solving simultaneously yields a
second-order ordinary differential equation for $\phi_{40}$, with
solution 
\[
\label{phi40soln}
\phi_{40} =
\Big(\frac{3A}{2}-\frac{9\tB}{16}\Big)-\frac{\tB}{2x_4^2} +
O(\tk^2),  
\]
where the two arbitrary constants have been chosen to match the
small-$t$ series expansion given in Section \S\,\ref{seriessoln}.
With this choice, dropping terms of $O(\tk^2)$ and higher,
\bea
\label{xi4soln}
\xi_{40} &=&
-A+\frac{11\tB}{8}-\frac{\tB}{x_4^2}+\Big(A-\frac{3\tB}{8}\Big)x_4^2 \\
W_{L0} &=& \frac{(1-\w^2)}{(1-x_4^2)^2}\Big(3
Ax_4^2(-2+\w 
  x_4)+\tB x_4\big(\frac{9}{4}x_4+\w (1-\frac{9}{8}x_4^2)\big)\Big)\, e^{-\frac{1}{2}x_4^2}.\qquad 
\eea
The resulting behaviour for the perturbation to the three-dimensional
scale factor, $b^2 \Psi_L$, is plotted in Figure \ref{pertplots}.

In terms of the original Newtonian gauge variables, an 
identical calculation (working now to all orders in $\tk$ but dropping terms of $O(y_0^2)$) yields,
{\allowdisplaybreaks
\bea
\label{PhiLallk}
\Phi_L &=& \frac{2\,\tk \,(1-\w x_4)^2}{3\, (x_4^2-1)}\,\big(A_0 J_0(\tk
x_4)+B_0 Y_0(\tk x_4)\big)\nonumber \\ && \qquad +\frac{1}{x_4}\Big(1+\frac{(1-\w
  x_4)^2}{1-x_4^2}\Big)\big(A_0 J_1(\tk x_4)+B_0 Y_1(\tk x_4)\big), \\
\label{PsiLallk}
\Psi_L &=& \frac{1}{3\,(1-x_4^2)}\Big(2\tk (1-\w x_4)^2 \big(A_0 J_0(\tk
x_4)+B_0 Y_0(\tk x_4)\big) \nonumber \\ && \qquad - 3\,(x_4+\w(-2+\w x_4))\big(A_0 J_1(\tk
x_4)+B_0 Y_1(\tk x_4)\big)\Big), \\
 W_{L} &=&
2\,x_4^2
\,e^{-\frac{1}{2}x_4^2}\,\frac{(\w^2-1)}{(1-x_4^2)^2}\,\Big(\tk\, (1-\w
 x_4)\big(A_0 J_0(\tk x_4)+B_0
 Y_0(\tk x_4)\big)\nonumber \\
 &&\qquad  +\w\,\big(A_0 J_1(\tk
 x_4)+B_0 Y_1(\tk x_4)\big) \Big), 
\eea
}
where the constants $A_0$ and $B_0$ are given by
\[
\label{A0B0}
A_0 = \frac{3A}{\tk}-\frac{9\tB}{8\tk}+\frac{1}{2}\tB \tk (\ln{2}-\gamma)+O(y_0^2), \qquad B_0 =
  \frac{\tB \tk \pi}{4} +O(y_0^2).
\]

To evaluate the $y_0^2$ corrections, we repeat the same sequence of
steps: integrating the $G^5_i$ and $G^5_5$ Einstein equations (at
$y_0^2$ higher order) gives us $\phi_{41}$ and $\xi_{41}$ up to two
arbitrary functions of $x_4$, and integrating the $G^0_i$ equation
then gives us $W_{L1}$ in terms of these two arbitrary functions plus
one more.   Two of the three arbitrary functions are then determined
by imposing the Dirichlet boundary conditions on $W_{L1}$, and the
third is found to satisfy a second-order ordinary differential
equation after making use of (\ref{breq}) on both branes.
Solving this differential equation, the constants of integration appearing in the solution are again chosen
so as to match the small-$t$ series expansion of Section \S\,\ref{seriessoln}.

Converting back to the original longitudinal gauge variables, the
results to $O(y_0^4)$ and to $O(k^2)$ take the schematic form 
\bea
\Phi_L &=&  f^\Phi_0+y_0^2 (f^\Phi_1+f^\Phi_2
\ln{(1+x_4)} + f^\Phi_3 \ln{(1-x_4)}+f^\phi_4 \ln{(1-\w x_4)}, \qquad \\
\Psi_L &=&  f^\Psi_0+y_0^2 (f^\Psi_1+f^\Psi_2
\ln{(1+x_4)} + f^\Psi_3 \ln{(1-x_4)}+f^\Psi_4 \ln{(1-\w x_4)}, \qquad  \\
W_L&=&e^{-\frac{1}{2}\,x_4^2} \,\big( f^W_0+y_0^2 (f^W_1+f^W_2
\ln{(1+x_4)} \nonumber \\ && \qquad\qquad\qquad\qquad
+ f^W_3 \ln{(1-x_4)} +f^W_4 \ln{(1-\w x_4)})\big), 
\eea
where the $f$ are rational functions of $x_4$ and $\w$ which, due to
their length, have been listed separated separately in Appendix \ref{appD}.
(If desired, more detailed results including the $O(k^2)$ corrections
are available \cite{Website}). 

It is easy to check that the results obtained by expanding about the
scaling solution are consistent with those obtained using our previous
method based upon Dirichlet/Neumann 
polynomials.  Taking the results
from the polynomial expansion given in Appendix \ref{appC}, 
substituting $t=(x_4/y_0) e^{-\frac{1}{2}x_4^2}$ and $y=\w y_0$,
retaining only terms of $O(y_0^2)$ or less, one finds agreement with
the results listed in Appendix \ref{appD} after these have been re-expressed as a
series in $x_4$.  This has been checked explicitly, both for the
background and the perturbations.

Just as in the case of the background, the small-$y_0$ expansion breaks down
for times close to $x_4=1$, when the $y_0^2$ corrections to the
perturbations become larger than the corresponding zeroth order terms.
Again, we will simply analytically continue the solution in $x_4$ and
$\w$ around this point.  In support of this, the induced metric on the
positive-tension brane is, to zeroth order in $y_0$, completely
regular across $x_4=1$, even including the perturbations as
can be seen from (\ref{PhiLallk}) and (\ref{PsiLallk}).

As in the case of the background, any imaginary pieces acquired from
analytically continuing logarithmic terms are all suppressed by order
$y_0^2$.  Thus they may only enter the Einstein-brane equations
(when these are expanded to order $y_0^2$) in a linear fashion, and
hence the real and imaginary parts of the metric constitute independent
solutions.  We can therefore simply drop the imaginary parts, or
equivalently replace the $\ln{(1-x_4)}$ and $\ln{(1-\w x_4)}$ terms
with $\ln{|1-x_4|}$ and $\ln{|1-\w x_4|}$ respectively.  
We have checked explicitly that this still satisfies the Einstein-brane
equations and boundary conditions.

\section{Comparison with the four-dimensional effective theory}
\label{compwitheft}

We have now arrived at a vantage point from which we may scrutinise
the predictions of the four-dimensional effective theory using our
expansion of the bulk geometry about the scaling solution.
We will find that the four-dimensional effective theory is in exact
agreement with the scaling solution.
Beyond this, the $y_0^2$ corrections lead to effects 
that cannot be described within a four-dimensional effective framework.
Nonetheless, the higher-order corrections are automatically small at
very early and very late times, restoring the accuracy of the
four-dimensional effective theory in these limits.

In the near-static limit, the mapping from four to five dimensions
may be calculated from the moduli space approach \cite{Terning, 
Ekpyrotic,KhouryZ}: putting the four-dimensional effective theory metric
$g^4_{\mu\nu}$ into Einstein frame, the mapping
reads
\[
g_{\mu \nu}^{+}= \cosh^2 (\phi/\sqrt{6})g_{\mu \nu}^{4} \qquad 
g_{\mu \nu}^{-}= \sinh^2 (\phi/\sqrt{6})g_{\mu \nu}^{4},
\label{map4}
\]
where $g_{\mu \nu}^{+}$ and $g_{\mu \nu}^{-}$ are the metrics
on the positive- and negative-tension branes respectively, 
and $\phi$ is the radion.
As we showed in Chapter \S\,\ref{confsymmchapter}, on 
symmetry grounds this is the unique local mapping
involving no derivatives \cite{Conf_sym}, and that to leading order,
the action for $g_{\mu \nu}^{4}$ and $\phi$ is that for
Einstein gravity with a 
minimally coupled 
scalar field. 

Solving the four-dimensional effective theory is trivial: 
the background is conformally flat, $g^{4}_{\mu \nu}
= b_4^2(t_4)\,\eta_{\mu \nu}$, and the Einstein-scalar
equations yield the following solution, unique up to a 
sign choice for
$\phi_0$, of the form
\[
b_4^2=\bar{C}_4 t_4, \qquad e^{\sqrt{2\over 3}\phi_0} = \bar{A}_4 t_4,
\label{back4i}
\]
with $\phi_0$ the background scalar field, and
$\bar{A}_4$ and $\bar{C}_4$ arbitrary constants. 
(Throughout this chapter we adopt units where $8 \pi G_4 =1$).

According to the map (\ref{map4}), the brane scale factors
are then predicted to be 
\[
b_\pm= {1\over 2}  b_4
e^{-{\phi_0\over \sqrt{6}}} \left( 1 \pm e^{\sqrt{2\over 3} 
\phi_0}\right) = 1\pm \bar{A}_4 t_4,
\label{back4}
\]
where we have chosen $\bar{C}_4= 4 \bar{A}_4$, so that the brane scale
factors 
are unity at the brane collision. As emphasised
in \cite{TTS}, the result (\ref{back4}) 
is actually exact for the
induced brane metrics, when $t_4$ is 
identified with the conformal time on the branes. 
From this correspondence, one can read off the five-dimensional
meaning of the 
parameter $\bar{A}_4$: it equals $L^{-1} \tanh{y_0}$ (our definition
of $y_0$ differs from that of \cite{TTS} by a factor of 2). 

With regard to the perturbations, in longitudinal gauge (see \eg 
\cite{mukhanov,bardeen}) the perturbed line element of the
four-dimensional effective theory reads
\[
\d s_4^2= b_4^2(t_4)\left[-(1+2\Phi_4)\d t_4^2+
(1-2\Phi_4)\d\vec{x}^2\right],
\label{4dmet}
\]
and the general solution to the perturbation
equations is \cite{TTS, Khoury} 
\bea
\label{phi4eft}
\Phi_4 &=& \frac{1}{t_4}\,\left(\tA_0 J_1(k t_4)+\tB_0 Y_1(k t_4)\right),\qquad\\ 
\label{deltaradioneft}
\frac{\delta \phi}{\sqrt{6}} &=&\frac{2}{3}\,k\,\left(\tA_0
J_0(kt_4)+\tB_0Y_0(kt_4)\right) -\frac{1}{t_4}\,\left(\tA_0 J_1(kt_4)+\tB_0
Y_1(kt_4)\right),\qquad
\eea
with $\tA_0$ and $\tB_0$ being the amplitudes of the 
two linearly independent perturbation modes. 

\subsection{Background}

In the case of the background, we require 
only the result that the scale factors on the positive- and
negative-tension branes are given by
\[
b_\pm = 1\pm \bar{A}_4 t_4,
\]
where the constant $\bar{A}_4=L^{-1}\tanh{y_0}$ and $t_4$ denotes conformal
time in the four-dimensional effective theory.  (Note this solution has been
normalised so as to set the brane scale factors at the collision to unity).
Consequently, the four-dimensional effective theory restricts $b_+ + b_-=2$.  In comparison, our results from the expansion about the scaling solution
(\ref{ad_b}) give
\[
b_+ + b_- = 2 + \frac{2\,x_4^4\,\left( x_4^2-3 \right)
  \,y_0^2}{3\,{\left(1- x_4^2 \right) }^3} + O(y_0^4).
\]
Thus, the four-dimensional effective theory captures the behaviour of the
full theory only in the limit in which the $y_0^2$ corrections
are small, \iec when the scaling solution is an accurate description
of the higher-dimensional dynamics.  
At small times such that $x_4\ll 1$, the $y_0^2$ corrections will
additionally be suppressed by $O(x_4^2)$, and so the effective theory
becomes increasingly accurate in the Kaluza-Klein limit near to the
collision.  Close to $x_4=1$, the small-$y_0$ expansion fails, hence
our results for the bulk geometry are no longer reliable.
For late times $x_4>1$, the negative-tension brane no longer exists and the above
expression is not defined.

We can also ask what the physical counterpart of $t_4$, conformal time in the 
four-dimensional effective theory, is:
from (\ref{ad_b}), to $O(y_0^3)$, we find
\[
t_4 = \frac{b_+-b_-}{2 \bar{A}_4} = \frac{x_4}{y_0}-
\frac{y_0}{30 
     {( 1 - x_4^2) }^3}\,\Big( x_4^3 ( 5 - 14 x_4^2 + 5 x_4^4 )  - 
       5 x_4{( -1 + x_4^2 ) }^2 \ln (1 - x_4^2)\Big).
\]
In comparison, the physical conformal times on the positive- and
negative-tension branes, defined via $b\, \d t_\pm = n\,\d t =
(n/y_0)(1-x_4^2)\,e^{-x_4^2/2}\,\d x_4 $, are, to $O(y_0^3)$,
\[
t_+ = \frac{x_4}{y_0} + \frac{y_0}{30 
     {( 1 - x_4^2 ) }^3}\Big( 10 - 30x_4^2- x_4^3( 5 - 14 x_4^2 + 5
x_4^4 ) +  5 x_4 {( 1 - x_4^2 ) }^2 \ln (1 - x_4^2) \Big)  
\]
and
\[
t_- = \frac{x_4}{y_0} - \frac{y_0}{30 
     {( 1 - x_4^2 ) }^3}\Big( 10 -30 x_4^2+x_4^3(5 - 14 x_4^2 + 5
x_4^5 ) -  5 x_4 {( 1 - x_4^2 ) }^2 \ln (1 - x_4^2) \Big), 
\]
where we have used (\ref{ad_n}) and (\ref{ad_b}).

Remarkably, to lowest order in $y_0$, the two brane conformal times are
in agreement not only with each other, but also with the four-dimensional effective theory
conformal time.  Hence, in the limit in which $y_0^2$
corrections are negligible, there exists a universal four-dimensional time.
In this limit, $t_4=x_4/y_0$ and the brane scale factors are simply
given by $b_\pm = 1\pm\bar{A}_4 t_4 = 1\pm x_4$.
The four-dimensional effective scale factor, $b_4$, is given by
\[
\label{b4def}
(b_4)^2=b_+^2-b_-^2=4 \bar{A}_4 t_4 = 4 x_4 \qquad \Rightarrow b_4 = 2 x_4^{1/2}.
\]


In order to describe the full five-dimensional geometry, 
one must specify the distance between
the branes $d$ as well as the metrics induced upon them. 
The distance between the branes is of particular interest
in the cyclic scenario, where an interbrane force depending
on the interbrane distance 
$d$ is postulated. In the lowest approximation,
where the branes are static, the four-dimensional effective theory
predicts that 
\[
\label{lncoth}
d =L \ln \coth\left({|\phi|\over\sqrt{6}}\right)=
L\ln\left(\frac{b_+}{b_-}\right) = L \ln \left({1+\bar{A}_4 t_4\over
  1-\bar{A}_4 t_4} \right). 
\]
Substituting our scaling solution and evaluating to leading order in 
$y_0$, we find
\[
\label{4d_pred}
d = L \ln \left(\frac{1+x_4}{1-x_4}\right)+O(y_0^2).
\]
(Again, this quantity is ill-defined for $x_4>1$).

In the full five-dimensional setup, 
a number of different measures of the interbrane distance are
conceivable, and the interbrane force could depend upon each 
of these, according to the precise higher-dimensional physics. 
One option would be to take the metric distance along
the extra dimension
\[
d_m =L \int_{-y_0}^{y_0} \sqrt{g_{yy}} \d y = L \int_{-y_0}^{y_0} nt
\d y = L \int_{-1}^{1} n x_4 e^{-\frac{1}{2}x_4^2}\d \w.
\]
Using (\ref{ad_n}), we obtain
\[
\label{d_m}
d_m = L \int_{-1}^{1}\frac{x_4}{1-\w x_4}\d \w+ O(y_0^2) = L \ln \left(\frac{1+x_4}{1-x_4}\right)+O(y_0^2),
\]
in agreement with (\ref{4d_pred}).

An alternative measure of the interbrane distance is provided by
considering affinely parameterised spacelike geodesics running
from one brane to the other 
at constant Birkhoff-frame time
and noncompact coordinates $x^i$. 
The background interbrane distance
is just the affine parameter distance along
the geodesic, and the fluctuation in distance
is obtained by integrating the metric fluctuations
along the geodesic, as 
discussed in Appendix \ref{appE}.
One finds that, to leading order in $y_0$ only, the geodesic trajectories lie
purely in the $y$-direction.  Hence the affine distance
$d_a$ is trivially equal to the metric distance $d_m$ at leading
order, since
\[
d_a = L \int \sqrt{g_{ab}\dot{x}^a\dot{x}^b}\,\d \lambda = L \int nt\dot{y} \d \lambda = d_m,
\]
where the dots denote differentiation with respect to the affine parameter $\lambda$.
Both measures of the interbrane distance therefore
coincide and are moreover in agreement with the four-dimensional
effective theory prediction, but only at leading order in $y_0$.

\subsection{Perturbations}

Since the four-dimensional Newtonian potential $\Phi_4$ represents the
\textit{anticonformal} part of the perturbed four-dimensional effective metric 
(see (\ref{4dmet})), it is unaffected by the conformal
factors in (\ref{map4}) relating the four-dimensional effective metric to the induced
brane metrics.  Hence we can directly compare the anticonformal part
of the perturbations of the induced metric on the branes, as calculated in five dimensions,
with $2\Phi_4$ in the four-dimensional effective theory.  The induced metric on the
branes is given by
\bea
\d s^2 &=& b^2\, \left( -(1+2 \Phi_L)\, \d t_\pm^2 + (1-2 \Psi_L )\, \d
\vec{x}^2 \right) \nonumber \\
 &=& b^2\, (1+\Phi_L-\Psi_L) \left( -(1+\Phi_L+\Psi_L)\, \d t_\pm^2 + (1-
\Psi_L-\Phi_L )\, \d \vec{x}^2 \right),\qquad
\eea
where the background brane conformal time, $t_\pm$, is related to the bulk time via $b\, \d t_\pm = n\,\d t$. 
The anticonformal part of the metric perturbation is thus simply
$\Phi_L+\Psi_L$.
It is this quantity, evaluated on the branes to leading order in $y_0$, that we expect to
correspond to $2\Phi_4$ in the four-dimensional effective theory.

\begin{figure}[p]
\begin{center}
\begin{minipage}{10cm}
      \includegraphics[width=10cm]{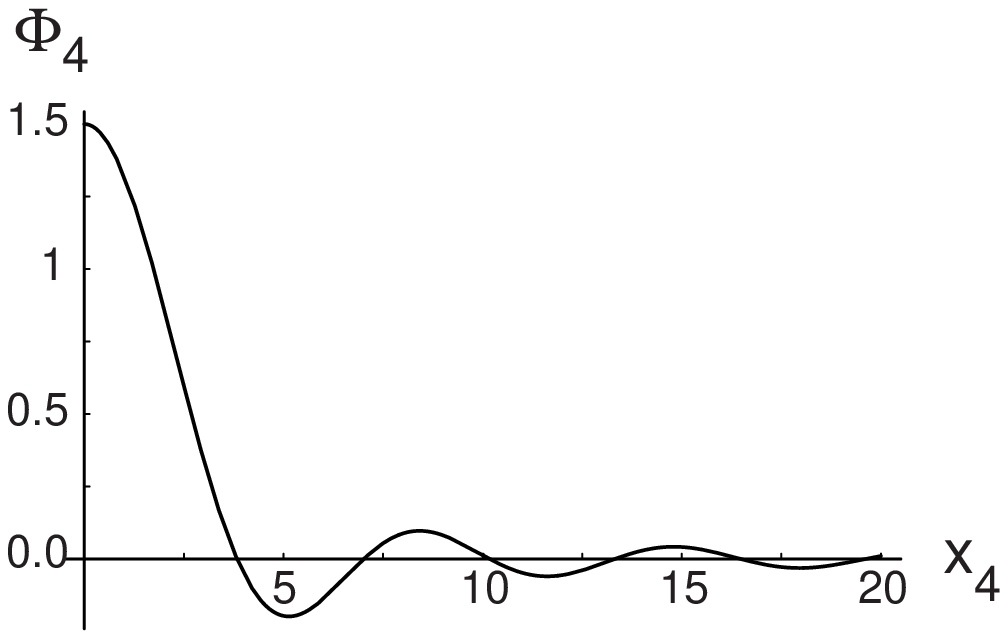}
\vspace{1cm}
\end{minipage}
\begin{minipage}{10cm}
\hspace{-0.32cm}
      \includegraphics[width=10cm]{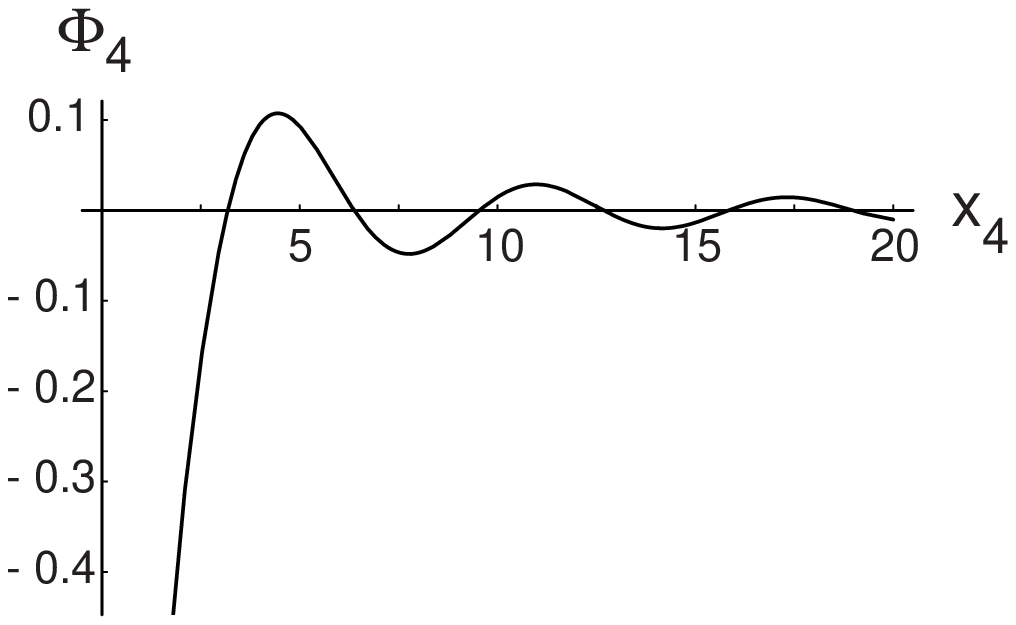}
\vspace{1cm}
\end{minipage}
\caption{
The four-dimensional Newtonian potential $\Phi_4$ on the
positive-tension brane, plotted to zeroth order in $y_0$ as a function
of the time $x_4$ for wavelength $\tk = 
1$.  The upper plot illustrates the mode with $A=1$ and $\tB=0$,
while the lower plot has $A=0$ and $\tB=1$.  
}
\label{phifourplots}
\end{center}
\end{figure}

Using our results (\ref{adperts})
and (\ref{phi40soln}) from expanding about the scaling solution, we have to $O(y_0^2)$, 
\[
\label{thiseqn}
\frac{1}{2}(\Phi_L+\Psi_L)_+= \frac{1}{2}(\Phi_L+\Psi_L)_-=\phi_{40}(x_4)
= \frac{1}{x_4}\left(A_0 J_1(\tk x_4)+B_0 Y_1(\tk x_4)\right),
\]
with $A_0$ and $B_0$ as given in (\ref{A0B0}).
On the other hand, the Newtonian potential of the four-dimensional effective theory
is given by (\ref{phi4eft}).  Since $t_4$, the conformal time in the
four-dimensional effective theory, is related to the physical dimensionless brane
conformal time $x_4$ by $t_4=x_4/y_0$ (to lowest order in $y_0$), and moreover
$\tk=k/y_0$, we have $\tk x_4 = k t_4$.  Hence, the four-dimensional effective theory
prediction for the Newtonian potential is in exact agreement with the
scaling solution holding at leading order in $y_0$, upon identifying
$\tA_0$ with $A_0/y_0$ and $\tB_0$ with $B_0/y_0$. 
The behaviour of the Newtonian potential is
illustrated in Figure \ref{phifourplots}. 

Turning our attention now to the radion perturbation, $\delta \phi$, we know from our
earlier considerations that this quantity is related to the
perturbation $\delta d$ in the interbrane separation.  Specifically,
from varying (\ref{lncoth}), we find
\[
\delta d = 2 L\, \cosech\Big(\sqrt{\frac{2}{3}}\,\phi\Big)\,\frac{\delta
  \phi}{\sqrt{6}}.
\]   
Inserting the four-dimensional effective theory predictions for $\phi$ and $\delta
\phi$, we obtain
\[
\label{deltad}
\frac{\delta d}{L}=\Big(\frac{4 \bar{A}_4 t_4}{\bar{A}^2_4 t^2_4
  -1}\Big)\Big(\frac{2}{3}\,k\,\big(\tA_0
J_0(kt_4)+\tB_0Y_0(kt_4)\big) -\frac{1}{t_4}\,\big(\tA_0 J_1(kt_4)+\tB_0
Y_1(kt_4)\big) \Big),  
\]
where to lowest order $\bar{A}_4 = y_0+O(y_0^3)$.

In comparison, the perturbation in the metric distance
between the branes is
\[
\label{deltad_m}
\frac{\delta d_m}{L} = \int_{-y_0}^{y_0}nt\Gamma_L \d y =
\int_{-1}^{1} \frac{n x_4 \xi_4}{b^2}\, e^{-\frac{1}{2}x_4^2} \d \w,
\]
where we have used (\ref{phi4xi4}).  Evaluating the integral using
(\ref{xi4soln}), to an accuracy of $O(y_0^2)$ we obtain
\bea
\frac{\delta d_m}{L}&=&\int_{-1}^{1} \frac{n x_4 \xi_{40}(x_4)}{b^2}\,
e^{-\frac{1}{2}x_4^2} \d \w
= \frac{2 x_4 \xi_{40}(x_4)}{(1-x_4^2)^2}
\qquad\qquad\qquad\qquad \nonumber \\ &=& 
\frac{1}{(x_4^2-1)}\Big(\frac{8}{3}\,\tk x_4 (A_0 J_0(\tk x_4)+B_0
Y_0(\tk x_4)) \qquad \qquad\qquad\qquad \nonumber \\ && \qquad \qquad\qquad\qquad \qquad -4 (A_0 J_1(\tk x_4)+B_0 Y_1(\tk x_4))\Big),
\eea
which is in agreement with (\ref{deltad}) when we set $\bar{A}_4 t_4
\sim y_0 t_4 = x_4$, along with $\tA_0=A_0/y_0$, $\tB_0=B_0/y_0$ and $k=\tk y_0$. 
The calculations in Appendix \ref{appE} show moreover that the perturbation
in the affine distance between the branes, $\delta d_a$, is 
identical to the perturbation in the metric distance $\delta d_m$,
to lowest order in $y_0$.  

The four-dimensional effective theory thus correctly predicts the Newtonian
potential $\Phi_4$ and the radion perturbation $\delta\phi$, but only
in the limit in which the $y_0^2$ corrections are
negligible and the bulk geometry is described by the scaling solution.  
While these corrections are automatically small at very
early or very late times, at intermediate times they cannot be
ignored and introduce effects that cannot be described by four-dimensional
effective theory.
The only five-dimensional longitudinal gauge metric perturbation we
have not used in any of the above is $W_L$: 
this component is effectively invisible to the four-dimensional effective
theory, since it vanishes on both branes and has no effect on the
interbrane separation.  

\section{Mixing of growing and decaying modes}
\label{mixingsection}

Regardless of the rapidity of the brane collision
$y_0$, one expects a four-dimensional effective description to hold
both near to the collision, when the brane separation is much less than an AdS
length, and also when the branes are widely separated over many AdS lengths.  In the
former case, the warping of the bulk geometry is negligible and a
Kaluza-Klein type reduction is feasible, and in the latter case, one
expects to recover brane-localised gravity.  At the transition between
these two regions, when the brane separation is of order one AdS
length, one might anticipate a breakdown of the four-dimensional effective description.

When the brane separation is of order a few AdS
lengths, however, the negative-tension brane reaches the horizon of the bulk
black hole and the small-$y_0$ expansion fails.  This failure hampers any efforts to
probe the breakdown of the four-dimensional effective theory at $x_4=1$ directly;
instead, we will look for evidence of mixing between the four-dimensional perturbation modes 
in the transition from Kaluza-Klein to brane-localised gravity.

To see this in action we have to compare the behaviour of the
perturbations at very small times with that at very late times: in
both of these limits a four-dimensional effective description should apply,
regardless of the collision rapidity $y_0$, in which the four-dimensional Newtonian
potential $\Phi_4$ satisfies
\[
\Phi_4 = \frac{1}{x_4}\left(A_0 J_1(\tk x_4)+B_0 Y_1(\tk x_4)\right).
\]
Expanding this out on long wavelengths $\tk\ll 1$, taking in addition $\tk x_4 \ll 1$, we find
\[
\label{phi4exp}
\Phi_4 = -\,\frac{\tB_4^0}{x_4^2}+A_4^0+\frac{1}{2}\,\tB_4^0 \tk^2 \ln{\tk
  x_4}-\frac{1}{8}\,A_4^0 \tk^2 x_4^2+O(k^4),
\]
where the dimensionless constants $A_4^0$ and $\tB_4^0$ are given in terms of the five-dimensional 
perturbation amplitudes $A$ and $\tB$ by
\[
\label{A4B4}
A_4^0 = \frac{3}{2}\,A-\frac{9}{16}\,\tB-\frac{1}{8}\tB\tk^2, \qquad \tB_4^0
= \frac{1}{2}\tB,
\]
where we have used (\ref{A0B0}) and (\ref{thiseqn}), recalling that
$\Phi_4 = \phi_{40}$ at leading order in $y_0$.

In comparison, using our results from expanding about the scaling solution,
we find the Newtonian potential on the positive-tension brane at small
times $x_4 \ll 1$ is given by
\bea
\label{phi4early}
\Phi_4 &=&
\(-\,\frac{\tB}{2 x_4^2}+
\frac{3}{2}\,A-\frac{9}{16}\,\tB-\frac{1}{8}\,
\tB \tk^2+\frac{1}{4}\, \tB \tk^2 \ln{\tk x_4} -\frac{3}{16}\,A\tk^2 x_4^2+\frac{9}{128}\,\tB \tk^2
x_4^2\) \nonumber \\ &&
+y_0^2 \,\Bigg(\frac{11}{120}\,\tB-3 A x_4-\frac{47}{8}\,\tB
x_4-\frac{1}{2}\,\tB\tk^2 x_4 \ln{\tk x_4}+6Ax_4^2\nonumber \\ &&
\qquad +\frac{1084}{105}\,\tB x_4^2 -\frac{211}{960}\,\tB
\tk^2x_4^2+\tB\tk^2 x_4^2 \ln{\tk
  x_4}\Bigg) +O(x_4^3)+O(y_0^4).
\eea
Examining this, we see that to zeroth order in $y_0$ the result is in
exact agreement with (\ref{phi4exp}) and (\ref{A4B4}).  At $y_0^2$
order, however, extra terms appear that are not present in
(\ref{phi4exp}).  Nonetheless, at sufficiently small times the
effective theory is still valid as these `extra' terms are subleading
in $x_4$: in this limit we find
\[
\Phi_4 =
-\,\frac{\tB^E_4}{x_4^2}+A_4^E+\frac{1}{2}\,\tB_4^E\,\tk^2\ln{\tk
    x_4}+O(x_4) +O(\tk^4)
\]
(the superscript $E$ indicating early times), in accordance with the
four-dimensional effective theory, where  
\[
\label{A4B4E}
A_4^E = A_4^0 + \frac{11}{120}\tB y_0^2, \qquad
\tB_4^E =\tB_4^0.
\]

At late times such that $x_4\gg 1$ (but still on sufficiently long
wavelengths that $\tk x_4 \ll 1$),
we find on the positive-tension brane
\bea
\label{phi4late}
\Phi_4 &=&
\left(-\,\frac{\tB}{2 x_4^2}+
\frac{3}{2}\,A-\frac{9}{16}\,\tB-\frac{1}{8}\,
\tB \tk^2+\frac{1}{4}\, \tB \tk^2 \ln{\tk x_4} -\frac{3}{16}\,A\tk^2 x_4^2+\frac{9}{128}\,\tB \tk^2
x_4^2\right) \nonumber \\ &&
+y_0^2 \,\Bigg(-\,\frac{A}{3 x_4^2}-\frac{\tB}{24
  x_4^2}-\frac{A\tk^2}{8x_4^2} +\frac{173\tB\tk^2}{960
  x_4^2}-\frac{\tB \tk^2 \ln{\tk}}{18x_4^2}-\frac{\tB \tk^2 \ln{
  x_4}}{12 x_4^2}\nonumber \\ &&\qquad
-\frac{3}{8}\,\tB+\frac{2}{9}\tB\tk^2+\frac{1}{6}\tB\tk^2\ln{x_4}
+\frac{3}{64}\tB\tk^2 x_4^2\Bigg)+O\(\frac{1}{x_4^3}\)+O(y_0^4).
\eea
To zeroth order in $y_0$, the results again coincide with the
effective theory prediction (\ref{phi4exp}) and (\ref{A4B4}).
At $y_0^2$ order, however, extra terms not present in the
four-dimensional effective description once more appear.  In spite of
this, at sufficiently late times the effective description still holds as these `extra'
terms are suppressed by inverse powers of $x_4$ relative to the leading
terms, which are 
\[
\Phi_4 = A_4^L-\,\frac{\tB^L_4}{x_4^2} -\,\frac{1}{8}\,A_4^L\, \tk^2
x_4^2
+ O(\tk^2 \ln{\tk x_4})
\]
(where the superscript $L$ indicates late times),
in agreement with the four-dimen\-sional effective theory.  (Since $x_4\gg 1$, we find $\tk \ll \tk x_4 \ll 1$, and we have chosen
to retain terms of $O(\tk^2 x_4^2)$ but to drop terms of $O(\tk^2)$.
The term of $O(x_4^{-2})$ is much larger than $O(\tk^2)$ and
so is similarly retained).
Fitting this to (\ref{phi4late}), we find
\[
\label{A4Ldef}
A_4^L = A_4^0 -\frac{3}{8}\,\tB y_0^2, \qquad \tB_4^L =
\tB_4^0+\Big(\frac{A}{3}+\frac{\tB}{24}\Big)y_0^2.
\]

Comparing the amplitudes of the two four-dimensional modes
at early times, $A_4^E$ and $\tB_4^E$, 
with their counterparts $A_4^L$ and $\tB_4^L$ at late times, we see
clearly that the amplitudes differ at $y_0^2$ order.
Using (\ref{A4B4}), we find:
\[
\label{mixingmatrix}
\left(\begin{array}{c} A_4^L \\[1ex] \tB_4^L \end{array}\right) =
\left(\begin{array}{cc} 1 \  \ \ \ & -\frac{14}{15} y_0^2 \\[1ex] 
  \frac{2}{9}y_0^2 \ \ \ \ &
  1+\big(\frac{1}{3}+\frac{\tk^2}{18}\big)y_0^2\end{array}\right)\left(\begin{array}{c} A_4^E \\[1ex] \tB_4^E \end{array}\right) .
\]
Hence 
the four-dimensional perturbation modes (as defined at very
early or very late times) undergo mixing.   

\section{Summary}

In this chapter we have developed a set of powerful analytical methods
which, we believe, render braneworld cosmological perturbation
theory solvable. 

Considering the simplest possible cosmological scenario, consisting of slowly moving,
flat, empty branes emerging from a collision, 
we have found a striking example of how the four-dimensional effective
theory breaks down at first nontrivial order in the brane speed.  
As the branes separate, a qualitative change in
the nature of the low energy modes occurs, from being nearly uniform
across the extra dimension when the brane separation is small, to
being exponentially localised on the positive-tension brane when the branes
are widely separated.  
If the branes separate at finite speed, the 
localisation process fails to keep up with the brane separation and
the low energy modes do not evolve adiabatically.
Instead, a given Kaluza-Klein zero mode at early times will
generically evolve into a mixture of both brane-localised zero modes and excited modes
in the late-time theory.  
From the perspective of the four-dimensional theory, this is
manifested in the mixing of the four-dimensional effective
perturbation modes between early and late times, as we have calculated explicitly.
Such a mixing would be
impossible were a local four-dimensional effective theory to remain valid
throughout cosmic history: mode-mixing is literally a signature of
higher-dimensional physics, writ large across the sky.



The strength of our expansion about the scaling
solution 
lies in its ability to interpolate between very early and very
late time behaviours, spanning the gap in 
which the effective theory fails.
Not only can we solve for the full five-dimensional
background and perturbations of a colliding braneworld, 
but our solution takes 
us beyond the four-dimensional
effective theory and into the domain of intrinsically
higher-dimensional physics.

%% file: chapter5.tex
\chapter{Generating brane curvature}
\label{zetachapter}

\begin{flushright}
\begin{minipage}{8.5cm}
\small
{\it \noindent
If you can look into the seeds of time, \\
And say which grain will grow, and which will not... 
}
\begin{flushright}
\noindent 
MacBeth, Act I.
\end{flushright}
\end{minipage}
\end{flushright}

\section{Introduction}

A key quantity of cosmological interest is $\zeta$, the 
curvature perturbation on comoving slices. 
Well away from the collision,
$\zeta_4$, the curvature perturbation in the four-dimensional effective theory,
agrees well with the same quantity $\zeta$ defined
on the branes. Yet whereas the brane curvature perturbation 
is precisely conserved (in the absence of additional 
bulk stresses), the curvature perturbation in the four-dimensional 
effective theory changes as the branes approach 
at order $(V/c)^2$, due to the mixing of four-dimensional perturbation modes.

In light of this failure of the four-dimensional effective theory, it is interesting 
to revisit the generation of curvature on the branes from a five-dimensional perspective.
In the ekpyrotic and cyclic models, 
this takes place through the action of
an additional bulk stress, $\Delta T^5_5$, on top of the background negative cosmological constant \cite{Ekpyrotic, Steinhardt:2002ih, Cyclicevo}.
In this chapter, we show how $\zeta$ on the branes is generated as the two branes approach by an effect 
proportional to $\Delta T_5^5$ times an `entropy' perturbation.
The latter measures the relative time delay between the interbrane distance
and a quantity measuring the time delay on the branes. The entropy
perturbation is nonzero even in the model with no $\Delta T_5^5$,
for which we have solved the five-dimensional Einstein equations.
We are thus able to give an expression for the final brane curvature
which is accurate to leading order in $\Delta T_5^5$.
The $(V/c)^2$ violations of the four-dimensional effective theory
cause the entropy perturbation to acquire a scale-invariant spectrum, 
which is then converted, under the influence of $\Delta T_5^5$, 
into a scale-invariant brane curvature perturbation.

\subsection{$\zeta$ in the four-dimensional effective theory}

Working in longitudinal gauge, from (\ref{zetadef}) the comoving curvature perturbation in the four-dimensional effective theory is given by
\[
\zeta_4 = \Phi_4 - \frac{\cH(\Phi_4'+\cH \Phi_4)}{\cH'-\cH^2} = \frac{4}{3}\,A_4,
\]
where primes denote differentiation with respect to the four-dimensional conformal time $t_4$. 
(Recall from (\ref{b4def}) that the comoving Hubble parameter $\cH = b_4'/b_4 = 1/2 t_4$, and that 
$\Phi_4 = A_4 -B_4/t_4^2$ on long wavelengths).
From the four-dimensional effective theory, we expect that $\zeta_4$ is constant on long wavelengths.
As a consequence of the mode-mixing in (\ref{mixingmatrix}), however, the value of $A_4$ (and hence $\zeta_4$) differs at $O(y_0^2)$ between early and late times.

\subsection{$\zeta$ on the branes}

To find a quantity that is exactly conserved on long wavelengths, let us instead
define the comoving curvature perturbation on the branes directly in five dimensions.
The following linear combination of the five-dimensional isotropic spatial metric
perturbation $\Psi$ and the lapse perturbation $\Phi$ 
is gauge-invariant, when evaluated on either brane:
\[
\zeta \equiv \Psi - {H\over H_{,T}}\left( \Psi_{,T}+H\Phi\right),
\label{zeta}
\]
where $T$ is now the proper time on the brane and $H=b_{,T}/b$. 
In four-dimensional cosmological perturbation
theory, the three-curvature of spatial slices is 
proportional to $k^2 \Psi$, and the second term in 
(\ref{zeta}) is proportional to the matter velocity with
respect to those slices \cite{mukhanov}. The latter may be gauged away in 
a suitable comoving time-slicing. Thus, $k^2 \zeta$ may be interpreted 
as the curvature perturbation of spatial slices which are comoving
with respect to the matter. In our case, no matter is present
on the branes. Nevertheless, 
$\zeta$ is still a well-defined gauge-invariant quantity. 
For empty branes and an empty 
bulk (apart from the background negative cosmological 
constant), as we shall see below, 
$\zeta$ is a conserved quantity at long 
wavelengths. We can therefore calculate $\zeta$ at small $t$ from 
the results (\ref{tseriesbgd}) and (\ref{tseriesperts}) from the previous chapter.
Evaluating (\ref{zeta}) on each brane, we find that, to order $y_0^4$,
\[
\zeta_+=\zeta_- = 2A-\frac{3}{4}\,B\,y_0^2-\frac{1}{2}\,B\,y_0^4  
= \frac{4}{3}\,A_4^L 
= \frac{4}{3}\,A_4^E-\frac{56}{45}\,y_0^2 \tB_4^E ,
\label{qzeta}
\]
where the latter two equalities follow from (\ref{A4Ldef}), (\ref{A4B4}) and (\ref{mixingmatrix}).
Thus, in the limit of late times, $\zeta_\pm$ and $\zeta_4$ coincide.  At early times, however, 
$\zeta_4$ differs from $\zeta_\pm$ at $O(y_0^2)$.



\section{Inducing brane curvature from a five-dimen\-sional perspective}

In the ekpyrotic and cyclic models, the primordial ripples
on the branes are generated by a force between the branes \cite{Ekpyrotic,
Khoury}.
In the four-dimensional effective theory, this is described by an
assumed potential $V(\phi)$ for the radion field $\phi$. 
In the five-dimensional description, the force between 
the branes is associated with an 
additional bulk stress $\Delta T^5_5$, on top of the 
background negative cosmological constant.
This extra stress enters the background evolution
equations as follows. The background $G_5^5$ Einstein equation,
when evaluated on each brane with the appropriate boundary
conditions \cite{Carsten},
yields an equation for the Hubble constant $H$ on each brane \cite{Binetruy}:
\[
H_{,T}+ 2H^2 = -{1 \over 3 m_5^3}\, \Delta T^5_5,
\label{back5}
\]
where  $\Delta T^5_5$ is the five-dimensional stress evaluated on
the brane, and the five-dimensional Planck mass $m_5$ is 
defined so that coefficient of the Ricci scalar in the five-dimensional Einstein-Hilbert action is $m_5^3/2$. 
Likewise for the linearised perturbations,
the $G_5^5$ equation, when supplemented with the appropriate boundary conditions, 
yields \cite{Carsten}
\[
\Psi_{,TT}+ H\,(\Phi_{,T} +4 \Psi_{,T})
+2 \Phi\,(H_{,T} +2 H^2) = {k^2 \over 3 b^2}\, (\Phi-2 \Psi)
+{1\over 3 m_5^3} \,\delta T^5_5,
\label{psieq}
\]
where $\delta T^5_5$ is the linearised perturbation of the 
bulk stress \cite{Carsten}.

From equations  (\ref{zeta}),  (\ref{back5}) and (\ref{psieq}), one 
can derive the following evolution equation for $\zeta$:
\[
\zeta_{,T} = -{1\over 3 m_5^3}\, {1\over H_{,T}}\, \left[(\Delta T^5_5)_{,T} \,
{\cal S} - k^2\, b^{-2} H\, (\Phi-2 \Psi) \right],
\label{zetaeq}
\]
where we define a gauge-invariant `entropy' perturbation 
\[
{\cal S}\equiv {H\over H_{,T}}\,\left( \Psi_{,T}+H\Phi\right) 
+H \,{\delta T^5_5\over (\Delta T^5_5)_{,T}},
\label{seq}
\]
which measures the difference between the dimensionless `time delay'
in $\delta T^5_5$ and a quantity transforming in an identical
manner constructed from the four-dimensional metric perturbations.
Therefore ${\cal S}$ can be thought of as a measure of the extent
to which $\delta T^5_5$ is synchronised with the metric perturbations on
the brane. For adiabatic perturbations in four dimensions, 
${\cal S}$ is zero on long wavelengths for all scalar variables such as 
$\Delta T^5_5$. 


In a fully five-dimensional description of the 
attractive interbrane  force, (\ref{zetaeq}) 
and (\ref{seq}) could be used to study how  
$\zeta$ is generated on the branes. Such
a calculation is beyond the scope of the present work.
Instead, we will content ourselves with 
an estimate of the induced  $\zeta$ on the
branes, under the assumption that 
$\Delta T_5^5$ is a function of the 
interbrane separation $d$.  Since
${\cal S}$ does not depend on the magnitude of
$\Delta T_5^5$ (nor indeed on its functional dependence 
upon $d$), as long as $\Delta T_5^5$ is small we can 
still use our solution from the preceding chapter, in which 
the effects of $\Delta T_5^5$ have been ignored,
to reliably calculate ${\cal S}$ to lowest order
in $\Delta T_5^5$. Nonetheless, some ambiguity remains in 
our result because it is not obvious {\it a priori}
precisely which measure of the interbrane distance
$d$ should be used. In fact, this will depend on
the detailed microphysics generating the
interbrane force, which will be model-dependent.
We will investigate two possible cases here and
use the difference between the answers as a 
measure of the uncertainty in the final result. 

As a first choice, let us assume that the bulk stress $\Delta T_5^5$ is
a function of the 
geodesic separation between the
two branes, along spacelike geodesics chosen to be 
orthogonal to the
four translational Killing vectors of 
the static background (corresponding to shifts in
$\vec{x}$ and Birkhoff-frame time). 
The relevant distance $d$ and its perturbation are calculated in the final section of Appendix D. 
We find that, as the collision
approaches, 
${\cal S}$ tends to a constant on both branes: 
\[
{\cal S}\equiv {H\over H_{,T}}\,\left( \Psi_{,T}+H\Phi\right) 
+H \,{\delta d\over d_{,T}} \approx C_S B_4\,(V^2/L^2)\, (V/c)^2 +O(t^2),
\label{sdeq}
\]
where $V$ is the relative speed of the branes as they collide,
$B_4$ is the growing mode amplitude in the four-dimensional effective theory\footnote{
From (\ref{A4B4}) and (\ref{A4B4E}), to leading order in $(V/c)^2$ we have $B_4= B/2$ as the collision approaches, where $B$ is the five-dimensional growing mode amplitude.}, and $C_S=7$ for this choice of $d$.
We have also calculated ${\cal S}$ under the assumption that
the relevant $d$ is just the naive metric separation between
the two branes, $\int dy \sqrt{g_{55}}$, in which case we find $C_S=1$. 

The absence of any $k$-dependence in this
formula means that in the cyclic model, where the coefficient
$B_4$ has a scale-invariant spectrum in the incoming state, 
the entropy perturbation is scale-invariant
as the two branes collide. We emphasise that ${\cal S}$ 
is a fully gauge-invariant
quantity, measuring the relative time delay between the
perturbations on the branes, and the time to the collision
as measured by the distance $d$, upon which the interbrane force
depends.

We can see from (\ref{zetaeq}) how a bulk stress $\Delta T_5^5$,
acting on the entropy perturbation ${\cal S}$, leads directly
to a scale-invariant $\zeta$ on the branes. If we work 
perturbatively in $\Delta T_5^5$, then using (\ref{back5}) and
(\ref{zetaeq}), to leading order in $k$ we find
\[
\zeta \approx {1\over 6 m_5^3} \,\int dT \,{(\Delta T_5^5)_{,T}\over
H^2}\, {\cal S}.
\label{pertz}
\]
When the branes are close, ${\cal S}$ is approximately
constant. Since the integrand is not a total derivative, 
and therefore does not cancel,
it follows that $\zeta$ acquires the same $k$-dependence as
${\cal S}$. For example, taking $\Delta T_5^5$ to be 
a narrow negative square well of strength $\Delta$ and width $\tau$, 
we find 
\[
\zeta \approx \,\frac{{\cal S}_0 \tau \Delta }{3 m_5^3 H}
\]
to leading order in $\tau$,
where ${\cal S}_0$ is the constant value in (\ref{sdeq}). 

In the cyclic model, however, $\Delta T_5^5$ is not small. 
Its presence, as discussed in \cite{Cyclicevo}, 
qualitatively alters the evolution of the background
solution so that 
the negative tension brane, initially contracting after the
collision,  
turns around and starts expanding so that the
interbrane distance never becomes large. An accurate
calculation of the amplitude of the final perturbation
spectrum will therefore  
require the effect of
$\Delta T_5^5$ on the background, and on the perturbations, to be included.

It is nevertheless not hard to estimate the parametric dependence
of the result. In the scaling solution of the four-dimensional effective
theory, one has $H \sim T^{-1}$ and $m_5^{-3} \Delta T_5^5 \sim -
T^{-2}$ from (\ref{back5}).
Equation (\ref{zetaeq}) then reads $T (d \zeta /dT)  \sim {\cal S}$.
Well before the collision, when the branes are moving 
slowly with respect both to each other and the bulk, one
expects the four-dimensional effective description to be
accurate and hence ${\cal S}$ should be zero. The brane
curvature is then directly related to the  four-dimensional effective
theory curvature perturbation $\zeta_4$, which is
zero on long wavelengths (hence $A_4=0$). As the branes approach,
their speed picks up and  
${\cal S}$ rises to a value of order $\sim  B_4(V^2/L^2) (V/c)^2$.
Finally, as the branes become very close, $\Delta T_5^5$ turns off
and $\zeta$ is, from (\ref{zetaeq}), conserved once more. 

On long wavelengths, therefore, we obtain the following equation
for the brane curvature as the collision approaches: up to a factor of
order unity, we have
\[
\zeta_+(0^-) \approx \zeta_-(0^-) \sim B_4 (V^2/L^2) (V/c)^2.
\label{zap}
\]
In the ekpyrotic and cyclic models $B_4$ is scale-invariant in the incoming state,
and so the final-state spectrum of curvature perturbations on the brane is also scale-invariant.

%% file: chapter6.tex
\chapter{Colliding branes in heterotic M-theory}
\label{Mthchapter}


\begin{flushright}
\begin{minipage}{7.5cm}
\small
{\it \noindent
What is it that breathes fire into the equations \\
and makes a universe for them to describe?
}
\begin{flushright}
\noindent 
Hawking, A Brief History of Time.
\end{flushright}
\end{minipage}
\end{flushright}
\vspace{0.2cm}


Our goal in this chapter is to derive a cosmological solution of the Ho{\v r}ava-Witten model with colliding branes, in which the five-dimensional geometry about the collision is that of a compactified Milne spacetime, and the Calabi-Yau volume at the collision is finite and nonzero.
We construct this solution as a perturbation expansion in the rapidity of the brane collision, applying the formalism developed in Chapter \S\,\ref{5dchapter}.
As this work is currently still in progress, we will present here only the leading terms in this expansion, corresponding to the scaling solution for the background geometry. 

\section{Heterotic M-theory}

One of the most successful models relating M-theory to phenomenology is that of Ho{\v r}ava and Witten \cite{HW}, in which an eleven-dimensional spacetime is the product of a compact Calabi-Yau manifold with a five-dimensional spacetime consisting of two parallel 3-branes or domain walls, one with negative and the other with positive tension.
As in the case of the Randall-Sundrum model,
the dimension normal to the 3-branes is compactified on an $S^1/\Z_2$ orbifold.

Performing a generalised dimensional reduction of the eleven-dimensional theory, 
such that the 4-form has non-vanishing flux on the Calabi-Yau internal space \cite{Lavrinenko}, one obtains a five-dimensional supergravity theory as shown in \cite{Lukas}\footnote{Note in particular that the truncation from eleven to five dimensions is \textit{consistent}, \iec a solution of the five-dimensional theory is an exact solution of the full eleven-dimensional theory.}.
The five-dimensional action for the bulk metric and the scalar field $\phi$, representing the volume of the internal Calabi-Yau manifold, then takes the form
\[
S = \int_{5} \sqrt{-g}\, [R - \half (\D \phi)^2 - 6\a^2 e^{-2 \phi}] 
-\sum_{\pm}\int_\pm \sqrt{g^\pm}\,\sigma^\pm e^{-\phi},
\]
where the sum runs over the two branes, with induced metrics $g^\pm_{\mu\nu}$, and we have dropped the Gibbons-Hawking term.  The constants $\sigma^\pm$ are given by $\sigma^\pm = \pm 12 \a$, where $\a$ is an arbitrary constant parameterising the amount of 4-form flux threading the Calabi-Yau.
The exponential potential in the bulk action is the remnant of the 4-form field strength, and plays a role analogous to that of 
the bulk cosmological constant in the Randall-Sundrum model.
In the above, and for the remainder of this chapter, we have adopted
units in which the five-dimensional Planck mass is set to unity.  

A static solution of this model \cite{Lukas}, comprising two domain walls located at $y=\pm y_0$, 
is given by
\bea
\label{domainwall}
\d s^2 &=& H(y)\eta_{\mu\nu}\dxdx + H^4(y)\d y^2, \\
e^\phi &=& H^3(y), \\
H(y) &=& 1+2\a|y+y_0| , 
\eea
where, to create the second domain wall, we must additionally impose a $\Z_2$ reflection symmetry about $y=+y_0$.
Recently, exact solutions of the Ho{\v r}ava-Witten model with colliding branes were found in \cite{gibbons}.
An analysis of these solutions from a four-dimensional effective
perspective was given in \cite{Koyama}. 
In these solutions, however, 
the volume of the internal Calabi-Yau manifold
shrinks to zero at the collision.  
For the purposes of constructing an M-theory model of a big crunch/big bang transition, 
it seems preferable to seek solutions in which the volume of the Calabi-Yau manifold at the collision is nonzero.
This way, the only singular behaviour at the collision is that of the five-dimensional spacetime geometry, 
which we will take to be Milne about the collision.

\section{A cosmological solution with colliding branes}

\subsection{Equations of motion}

As in the case of the Randall-Sundrum model\footnote{See Section \S\,\ref{solnmethods}.}, cosmological symmetry on the branes permits the bulk line element to be written in the form
\[
\label{goodoldmetric}
\d s^2 = n^2(t,y)(-\d t^2 + t^2 \d y^2)+b^2(t,y)\d \vec{x}^2,
\]
where the $x^i$ coordinates span the flat, three-dimensional spatial worldvolume of the branes.  Moreover, in this coordinate system, the branes are held fixed at $y=\pm y_0$.  (Recall also that, physically, $y_0$ represents the rapidity of the brane collision occurring at $t=0$). 
As in Chapter \S\,\ref{5dchapter}, it is useful to introduce the variables $\w=y/y_0$ and $x=y_0 t$, along with the  metric functions $\beta(x, \w) = 3\ln b$ and $\nu(x, \w) = \ln (nx)$.  

The junction conditions, evaluated at $\w=\pm 1$, are then
\[
\label{Mjn}
\nu' = \a\, e^{\nu-\phi}, \qquad
\beta' = 3\a\, e^{\nu-\phi}, \qquad
\phi' = 6\a\, e^{\nu-\phi},
\]
where primes denote differentiation with respect to $\w$.
Evaluating the $\phi$, $G^0_0+G^5_5$, $G^5_5$, $G^0_0+G^5_5-(1/2)G^i_i$, 
and $G_{05}$ equations in the bulk, 
we obtain the following set of equations:
\bea
\label{Mphi}
\phi''+\beta'\phi'+12\a^2 e^{2\nu-2\phi} &=& y_0^2\,(\ddot{\phi}+\dot{\beta}\dot{\phi}),  \\
\label{MG00p55}
\beta''+\beta'^2+6\a^2 e^{2\nu-2\phi} &=& y_0^2 \,(\ddot{\beta}+\dot{\beta}^2), \\
\label{MG55}
\frac{1}{3}\,\beta'^2+\beta'\nu'-\frac{1}{4}\,\phi'^2+3\a^2 e^{2\nu-2\phi} &=& y_0^2\,(\ddot{\beta}+\frac{2}{3}\dot{\beta}^2 -\dot{\beta}\dot{\nu}+\frac{1}{4}\,\dot{\phi}^2), \\
\label{MG0055ii}
\nu''-\frac{1}{3}\,\beta'^2+\frac{1}{4}\,\phi'^2-\a^2 e^{2\nu-2\phi} &=& y_0^2 \,(\ddot{\nu}-\frac{1}{3}\,\dot{\beta}^2+\frac{1}{4}\,\dot{\phi}^2), \\
\label{MG05}
\dot{\beta}'+\frac{1}{3}\,\dot{\beta}\beta'-\dot{\nu}\beta'-\nu'\dot{\beta}+\half\,\dot{\phi}\phi' &=& 0,
\eea
where, 
throughout this chapter, we will use a dot as a shorthand notation for $x\D_x$.

In the above, both the $G^5_5$ equation (\ref{MG55}) and the $G_{05}$ equation (\ref{MG05}) involve only single derivatives with respect to $\w$.
Applying the junction conditions, we find that both left-hand sides vanish when evaluated on the branes.
The $G_{05}$ equation is then trivially satisfied, while the $G^5_5$ equation yields the relation
\[
\frac{b_{,xx}}{b}+\frac{b_{,x}^2}{b^2}-\frac{b_{,x}\,n_{,x}}{bn}+\frac{1}{12}\,\phi_{,x}^2 = 0,
\]
valid on both branes.  Introducing the brane conformal time $\tau$, defined on either brane via $b\,\d \tau = n\, \d x$, this relation can be re-expressed as
\[
\label{Mbraneeq}
\frac{b_{,\tau\tau}}{b}+\frac{1}{12}\,\phi_{,\tau}^2=0.
\]

\subsection{Initial conditions}

Through a suitable rescaling of coordinates, we can always arrange for the metric functions $n$ and $b$ to tend to unity as $t\tt 0$, so that the geometry about the collision is that of a Milne spacetime.
Furthermore, we are seeking a solution in which the Calabi-Yau volume $e^\phi$ tends to a finite, nonzero value at the collision.  Since a constant shift in $\phi$ is a symmetry of the system (given that $\a$ is arbitrary), without loss of generality we can set $e^\phi \tt 1$ as $t\tt 0$.

To check that such a solution is possible, and to help ourselves later with fixing the initial conditions, it is useful to solve for the bulk geometry about the collision as a series expansion in $t$.  Up to terms of order $t^3$, the solution corresponding to the Kaluza-Klein zero mode is:
\bea
n &=& 1 + \a\,(\sech{y_0}\sinh{y})\,t + \frac{\a^2}{8}\,\sech^2 y_0\,(9-\cosh{2y_0}-8\cosh{2y})\,t^2, \qquad\\
b &=& 1 + \a\,(\sech{y_0}\sinh{y})\,t+\frac{\a^2}{4}\,\sech^2 y_0 \,(3+\cosh{2y_0}-4\cosh{2y})\, t^2, \qquad\\
e^\phi &=& 1+6\a\,(\sech{y_0}\sinh{y})\,t-\frac{3\a^2}{2}\,\sech^2 y_0 \,(2-\cosh{2y_0}-\cosh{2y})\, t^2, \qquad
\eea
where we have used the junction conditions to fix the arbitrary constants arising in the integration of the bulk equations with respect to $y$.

The brane conformal times are then given by
\[
\tau_\pm = \int \frac{n_\pm}{b_\pm}\,\d x  =x + O(x^3),
\]
in terms of which the brane scale factors $b_\pm$ are
\[
\label{btau}
b_\pm = 1 \pm \a\tanh{y_0}\,\Big(\frac{\tau_\pm}{y_0}\Big)-\frac{3}{2}\,\a^2 \tanh^2{y_0}\,\Big(\frac{\tau_\pm}{y_0}\Big)^2 + O\Big(\frac{\tau_\pm}{y_0}\Big)^3.
\]

\subsection{A conserved quantity}

In the case of the Randall-Sundrum model, where there were only two time-dependent moduli, the $G^5_5$ equation evaluated on both branes provided sufficient information to determine both moduli.
In the present setup, however, we have an extra time-dependent modulus due to the presence of the scalar field $\phi$.
We therefore require a further equation to fix this.

Introducing the variable $\chi=\phi-2\beta$, upon subtracting twice (\ref{MG00p55}) from (\ref{Mphi}) we find
\[
\label{chiwave}
(\chi'e^\beta)' = y_0^2 (\dot{\chi}e^\beta)\dot{\,},
\]
which is simply the two-dimensional massless wave equation $\bo\chi=0$. 
Since the junction conditions imply $\chi'=0$ on the branes, the left-hand side vanishes upon integrating over $\w$.
A second integration over $x$ then yields
\[
\label{charge}
\int_{-1}^{+1} \d \w \dot{\chi}e^\beta = \gamma,
\]
for some constant $\gamma$, which we can set to zero
since our initial conditions are such that $\dot{\chi}e^\beta \tt 0$ as $x\tt 0$
Let us now consider solving (\ref{chiwave}) for $\chi$, as a perturbation expansion in $y_0^2$.  
Setting $\chi=\chi_0+y_0^2 \chi_1+O(y_0^4)$, and similarly for $\beta$, at zeroth order we have $(\chi_0'e^{\beta_0})'=0$.
Integrating with respect to $\w$ introduces an arbitrary function of $x$ which we can immediately set to zero using the boundary condition on the branes, which, when evaluated to this order, read $\chi_0'=0$.  This tells us that $\chi_0'=0$ throughout the bulk; $\chi_0$ is then a function of $x$ only, and can be taken outside the integral in (\ref{charge}).  Since $\gamma=0$, yet the integral of $e^\beta$ across the bulk cannot vanish, it follows that $\chi_0$ must be a constant.

At order $y_0^2$, the right-hand side of (\ref{chiwave}) evaluates to $y_0^2\, (\dot{\chi}_0e^{\beta_0})\dot{\,}$, which vanishes.  Evaluating the left-hand side, we have $(\chi_1'e^{\beta_0})'=0$, and hence, by a sequence of steps analogous to those above, we find that $\chi_1$ must also be constant.  It is easy to see that this behaviour continues to all orders in $y_0^2$.  We therefore deduce that $\chi = \phi-2\beta$ is exactly constant. 
Since both $\phi$ and $\beta$ tend to zero as $x\tt 0$, this constant must be zero, and so we find 
$\phi=2\beta$.

The essence of this result is that the scaling solution, and consequently any expansion about it in powers of $y_0^2$, exists only when $\chi$ is in the Kaluza-Klein zero mode.  
This may be seen from (\ref{chiwave}), which reduces, in the limit where $x\tt 0$ and $\beta\tt 0$, to $\chi''=y_0^2\, \ddot{\chi}$.  
For a perturbative expansion in $y_0^2$ to exist, the right-hand side of this equation must vanish at leading order.
This is only the case, however, for the Kaluza-Klein zero mode: all the higher modes have a rapid oscillatory time dependence such that the right-hand side does contribute at leading order.  
This means that for the higher Kaluza-Klein modes a gradient expansion does not exist.

Setting $\phi=2\beta$ from now on, returning to (\ref{Mbraneeq}) and recalling that $\beta=3\ln b$, we immediately obtain the equivalent of the Friedmann equation on branes;
\[
\Big(\frac{b_{,\tau}}{b}\Big)_{,\tau}+4\,\Big(\frac{b_{,\tau}}{b}\Big)^2=0.
\]
Integrating, the brane scale factors are given by
\[
b_\pm = \bar{A}_\pm (\tau_\pm - c_\pm)^{1/4}.
\]
To fix the arbitrary constants $\bar{A}_\pm$ and $c_\pm$, we need only to expand the above in powers of $\tau_\pm$ and compare with (\ref{btau}).  We find
\[
\label{Mb}
b_\pm = (1\pm A \tau_\pm)^{1/4},
\]
where $A = (4\a/y_0)\tanh{y_0}$.  This equation determines the brane scale factors to all orders in $y_0^2$ in terms of the conformal time on each brane, and is one of our main results.  (As a straightforward check, it is easy to confirm that the $O(\tau_\pm^2)$ terms in (\ref{btau}) are correctly reproduced). 

\subsection{The scaling solution}

Having set up the necessary prerequisites, we are now in a position to
solve the bulk equations of motion perturbatively in $y_0^2$.  
Setting
\[
\beta = \beta_0+y_0^2 \beta_1 + O(y_0^4), \qquad \nu =
\nu_0+y_0^2\nu_1+O(y_0^4),
\]
the scaling solution corresponds to the leading terms in this
expansion, \iec to $\beta_0$ and $\nu_0$.
To determine the $\w$-dependence of these functions, we must solve the bulk equations of
motion at zeroth order in $y_0^2$.
Evaluating the linear combination $(\ref{MG00p55})-3[(\ref{MG55})+(\ref{MG0055ii})]$,
noting that at this order the right-hand sides all vanish
automatically, we find
\[
((\beta'_0-3\nu'_0)e^{\beta_0})'=0.
\]
Using the boundary condition $\beta'_0-3\nu'_0=0$ on the branes,
we then obtain $\beta_0=3\nu_0+f(x)$, for some arbitrary function $f(x)$.
Backsubstituting into (\ref{MG55}) and taking the square root then yields
\[
\nu'_0=\a\,e^{-5\nu_0-2f}, 
\]
where consistency with the junction conditions (\ref{Mjn}) forced us to take the positive root.
Integrating a second time, and re-writing $e^f=B^{-5}$, we find
\[
\label{wdep}
e^{\nu_0} = B^2(x)h^{1/5}, \qquad e^{\beta_0} = B(x)h^{3/5}, 
\]
where $h = 5 \a\w+C(x)$ and $C(x)$ is arbitrary.
  
In the special case where $B$ and $C$ are both constant, we recover the exact 
static domain wall solution (\ref{domainwall}), up to a coordinate transformation.  
In general, however, these two moduli will be time dependent.
Inverting the relation $b_\pm^5 = B^{5/3}(\pm 5\a+C)$ to re-express $B$ and $C$ in terms of the brane scale factors $b_\pm(x)$, we find
\[
B^{5/3} = \frac{1}{10\a}\,(b_+^5-b_-^5), \qquad C=5\a\(
\frac{b_+^5+b_-^5}{b_+^5-b_-^5}\).
\]
Furthermore, at zeroth order in $y_0^2$, the conformal times on
both branes are equal, since
\[
\d \tau = \frac{n}{b}\,\d x = e^{\nu_0-\beta_0/3}\,\frac{\d
  x}{x}=B^{5/3}(x)\frac{\d x}{x}
\]
is independent of $\w$.  To this order then, (\ref{Mb}) reduces to 
\[
b_\pm = (1\pm 4\a \tau)^{1/4},
\]
allowing us to express the moduli in terms of $\tau$ as
\bea
\label{Bmod}
B^{5/3}(\tau) &=& \frac{1}{10\a}\,[(1+4\a\tau)^{5/4}-(1-4\a\tau)^{5/4}], \\\nn \\
\label{Cmod}
C(\tau) &=& 5\a\left[\frac{(1+4\a\tau)^{5/4}+(1-4\a\tau)^{5/4}}{(1+4\a\tau)^{5/4}-(1-4\a\tau)^{5/4}}\right].
\eea

The brane conformal time $\tau$ and the Milne time $x$ are then related by
\[
\ln x = 10 \a \int [(1+4\a\tau)^{5/4}-(1-4\a\tau)^{5/4}]^{-1}\d\tau.
\]
As we do not know how to evaluate this integral analytically, it seems that the best approach is simply to adopt $\tau$ as our time coordinate\footnote{In support of this, note that the relation between $x$ and $\tau$ is monotonic.  At small times $x = \tau+(\a^2/4)\tau^3+O(\tau^5)$.} in place of $x$.
The complete scaling solution is then given by (\ref{wdep}), (\ref{Bmod}) and (\ref{Cmod}).

\section{Discussion}

In this chapter we have taken some first steps towards a cosmological solution of the Ho{\v r}ava-Witten model with colliding branes, in which the Calabi-Yau volume at the collision is finite and nonzero, and the five-dimensional spacetime geometry about the collision is Milne.
Employing the techniques developed in Chapter \S\,\ref{5dchapter}, we have shown how to construct the scaling solution
for the background geometry.  All the relevant tools are now in place to compute the corrections at $O(y_0^2)$.
Only two steps are required: firstly, integrating the bulk equations of motion at $O(y_0^2)$ to find the $\w$-dependence of the solution; and secondly, fixing the time dependence using our explicit knowledge of the scale factors on the branes, according to (\ref{Mb}).   With this accomplished, it is then feasible to consider solving for cosmological perturbations about this background, along the lines of the work presented in Chapter \S\,\ref{5dchapter}.

An interesting feature of our solution is the vanishing of the scale factor (and also the volume of the Calabi-Yau manifold) on the negative-tension brane at $\tau=(1/4\a)$.  
In more general models incorporating matter on the negative-tension brane, four-dimensional effective theory arguments suggest that under similar circumstances the scale factor on the negative-tension brane, rather than shrinking to zero, would instead undergo a bounce at some small nonzero value.  Since this behaviour persists even in the limit of only infinitesimal amounts of matter on the negative-tension brane, one might consider implementing a bounce in the present model by replacing the factors of $(1-4\a\tau)$ in (\ref{Bmod}) and (\ref{Cmod}) with $|1-4\a\tau|$.  The scale factor on the positive-tension brane, along with the bulk time $x$, 
would then continue smoothly across the bounce and out to late times as $\tau\tt \inf$.  The implications of this continuation remain to be explored.

%% file: conclusions.tex
\def\baselinestretch{1}
\chapter{Conclusions}

\def\baselinestretch{1.66}

%


\begin{flushright}
\begin{minipage}{10cm}
\small
{\it \noindent
All of this danced up and down, like a company of gnats, 
each separate, but all marvellously controlled in an invisible 
elastic net - danced up and down in Lily's mind...
}
\begin{flushright}
\noindent 
Virginia Woolf, To the Lighthouse.
\end{flushright}
\end{minipage}
\end{flushright}
\vspace{0.2cm}

In this thesis we have explored the impact of
higher-dimensional physics in the vicinity of the cosmic singularity.
Taking a simple Randall-Sundrum braneworld with colliding branes as
our model of a big crunch/big bang universe, we have explicitly
calculated the full five-dimensional gravitational dynamics, both for
the background solution, and for cosmological perturbations. 
Our solution method, a perturbative expansion in $(V/c)^2$ (or equivalently, in the collision rapidity),
provides a powerful analytical tool enabling us to assess the
validity of the low energy four-dimensional effective theory.

While this four-dimensional effective theory holds good
at leading order in our expansion of the bulk geometry, at order
$(V/c)^2$ new effects appear that cannot be accounted for within any 
local four-dimensional effective theory.
Our principal example is the mixing of four-dimensional effective
perturbation modes in the transition from early to late times (and
vice versa).

Despite the small size of this effect in the case of a
highly non-relativistic collision, 
mode-mixing is nonetheless an important piece 
of the puzzle concerning the origin of the primordial density perturbations.
If these primordial fluctuations were indeed generated via the ekpyrotic
mechanism in a prior contracting phase of the universe, then
mode-mixing suggests a resolution to the problem of 
matching the scale-invariant spectrum of growing mode time-delay perturbations
in the collapsing phase, to the growing mode curvature perturbations post-bang.
In such a scenario, higher-dimensional physics,  rather than being insignificant, 
is instead responsible for all the structure we see in the universe today.

The failure of the four-dimensional effective theory, however, reminds us
that in order to address this issue satisfactorily, a fully
five-dimensional description of the generation of curvature
perturbations on the brane is required.
Initiating this programme, 
we have shown how brane curvature is generated through the action of an 
additional five-dimensional bulk stress pulling the branes together.
Our understanding is sufficient to provide a quantitative estimate of the 
final brane curvature prior to the collision, but to progress further
it will be necessary to construct a specific five-dimensional model of the additional bulk stress.
Work along these lines is already underway.

Another direction of active research lies in the adaptation of our solution methods
to other braneworld scenarios.  Of particular interest are the cosmological colliding-brane
solutions of the Ho{\v r}ava-Witten model.  Here, we have constructed the leading terms of a background solution in which the five-dimensional geometry about the collision is Milne,
and the Calabi-Yau volume tends to unity at the collision.
Evaluating the corrections to the background at subleading order,
as well as solving for cosmological perturbations, are both the subject of current research. 
Our methods should also extend to scenarios in which matter is present on the brane.
A more challenging extension would be to probe the evolution of a black string \cite{BS} in an expanding cosmological background consisting of two separating branes. As the branes separate, the gradual stretching of the black string will eventually trigger the onset of the Gregory-Laflamme instability \cite{GL}.

We conclude in the only fashion possible -- with a call-to-arms.  
Higher-dimensional physics, beyond the four-dimensional effective theory, necessarily holds the key to the decisive experimental signature over which braneworld cosmologies will one day stand or fall.

%% file: appendix1.tex
\chapter{Israel matching conditions}
\label{Isapp}

In this appendix we derive the Israel matching conditions \cite{Israel} from scratch, based on the treatment presented in \cite{Poisson}.

We begin with a timelike (or spacelike) hypersurface $\Sigma$, pierced by a congruence of geodesics which intersect it orthogonally.  Denoting the proper distance (or proper time) along these geodesics by $\ell$, we can always adjust our parameterisation to set $\ell=0$ on $\Sigma$.  (One side of the hypersurface is therefore parameterised by values of $\ell<0$, and the other by values $\ell>0$).  
Introducing $n^\alpha$, the unit normal to $\Sigma$ in the direction of increasing $\ell$,  the
displacement away from the hypersurface along one of the geodesics is given by $\d x^\alpha = n^\alpha \d \ell$.
From this it follows that $n^\alpha n_\alpha=\ep$, where $\ep=+1$ for spacelike $\Sigma$, and $\ep=-1$ for timelike $\Sigma$, and that
\[
n_\alpha = \ep \,\D_\alpha \ell.
\]

It is also useful to introduce a continuous coordinate system $x^\alpha$ spanning both sides of the hypersurface,
along with a second set of coordinates $y^a$ installed on the hypersurface itself.  
(Henceforth Latin indices will be used for hypersurface coordinates and Greek indices for coordinates in the embedding spacetime).  
The hypersurface may then be parameterised as $x^\alpha=x^\alpha(y^a)$, 
and the vectors (in a Greek sense)
\[
e^\alpha_a=\frac{\D x^\alpha}{\D y^a}
\]
are tangent to curves contained in $\Sigma$ (hence $e^\alpha_a n_\alpha=0$).
For displacements within $\Sigma$, 
\bea
\d s^2_\Sigma &=& g_{\alpha\beta}\,\d x^\alpha \d x^\beta \\
&=& g_{\alpha\beta}\,\Big(\frac{\D x^\alpha}{\D y^a}\,\d y^a \Big)\Big(\frac{\D x^\beta}{\D y^b}\,\d y^b \Big) \\
&=& h_{ab}\,\d y^a \d y^b,
\eea
and so the induced metric $h_{ab}$ on $\Sigma$ is given by
\[
h_{ab} = g_{\alpha\beta}\,e^\alpha_a e^\beta_b.
\]

To denote the jump in an arbitrary tensorial quantity $A$ across the hypersurface, we will use the notation
\[
[A] = \lim_{\ell\tt0_+}(A) - \lim_{\ell\tt 0^-}(A).
\]
The continuity of $x^\alpha$ and $\ell$ across $\Sigma$  immediately imply $[n^\alpha]=[e^\alpha_a]=0$.

To decompose the bulk metric in terms of the different metrics on either side of $\Sigma$, we will require the services of the Heaviside distribution $\Theta(\ell)$, equal to $+1$ if $\ell>0$; $0$ if $\ell<0$; and indeterminate if $\ell=0$.  Note in particular the properties:
\[
\Theta^2(\ell)=\Theta(\ell), \qquad \Theta(\ell)\Theta(-\ell)=0, \qquad \frac{\d}{\d \ell}\,\Theta(\ell) = \delta(\ell),
\]
where $\delta(\ell)$ is the Dirac distribution.

We can now express the metric $g_{\alpha\beta}$ in terms of the coordinates $x^\alpha$ as a distribution-valued tensor:
\[
g_{\alpha\beta}=\Theta(\ell)\,g^+_{\alpha\beta}+\Theta(-\ell)\,g^-_{\alpha\beta},
\]
where $g^+_{\alpha\beta}$ denotes the metric on the $\ell>0$ side of $\Sigma$, and $g^-_{\alpha\beta}$ the metric on the $\ell<0$ side.
Differentiating, we find
\[
g_{\alpha\beta , \gamma}=\Theta(\ell)\,g^+_{\alpha\beta , \gamma}+\Theta(-\ell)\,g^-_{\alpha\beta , \gamma} + \ep\, \delta(\ell)[g_{\alpha\beta}]n_\gamma.
\]
The last term is singular; moreover, this term creates problems when we compute the Christoffel symbols by generating the product $\Theta(\ell)\delta(\ell)$, which is not defined as a distribution. 
In order for the connection to exist as a distribution, we are forced to impose the continuity of the metric across the hypersurface: $[g_{\alpha\beta}]=0$.  This statement can be reformulated in terms of hypersurface coordinates alone as $0=[g_{\alpha\beta}]\,e^\alpha_a e^\beta_b = [g_{\alpha\beta}\,e^\alpha_a e^\beta_b] = [h_{ab}]$, \iec the induced metric $h_{ab}$ must be the same on both sides of $\Sigma$.  
This condition is often referred to as the `first' junction condition.

To derive the `second' junction condition (the Israel matching condition), we must calculate the distribution-valued Riemann tensor.  Beginning with the Christoffel symbols, we obtain
\[
\Gamma^\alpha_{\beta\gamma} = \Theta(\ell) \Gamma^{+\alpha}_{\beta\gamma}+\Theta(-\ell)\Gamma^{-\alpha}_{\beta\gamma},
\]
where $\Gamma^{\pm\alpha}_{\beta\gamma}$ are the Christoffel symbols constructed from $g^\pm_{\alpha\beta}$.  Thus
\[
\Gamma^\alpha_{\beta\gamma , \delta} = \Theta(\ell)\Gamma^{+\alpha}_{\beta\gamma , \delta} +\Theta(-\ell)\Gamma^{-\alpha}_{\beta\gamma , \delta} + \ep \delta(\ell)[\Gamma^\alpha_{\beta\gamma}]n_\delta,
\]
and the Riemann tensor is
\[
R^\alpha_{\beta\gamma\delta}=\Theta(\ell)R^{+\alpha}_{\beta\gamma\delta}+\Theta(-\ell)R^{-\alpha}_{\beta\gamma\delta}
+\delta(\ell)A^\alpha_{\beta\gamma\delta},
\]
where
\[
A^\alpha_{\beta\gamma\delta}=\ep\([\Gamma^\alpha_{\beta\delta}]n_\gamma-[\Gamma^\alpha_{\beta\gamma}]n_\delta\).
\]

The quantities $A^\alpha_{\beta\gamma\delta}$ transform as a tensor since they are the difference of two sets of Christoffel symbols.  We will now try to find an explicit expression for this tensor.

Observe that the continuity of the metric across $\Sigma$ in the coordinates $x^\alpha$ implies that its tangential derivatives must also be continuous.  Thus, if $g_{\alpha\beta , \gamma}$ is to be discontinuous, the discontinuity must be directed along the normal vector $n^\alpha$.  We can therefore write
\[
[g_{\alpha\beta,\gamma}]=\kappa_{\alpha\beta}n_\gamma,
\]
for some tensor $\kappa_{\alpha\beta}$ (given explicitly by $\kappa_{\alpha\beta}=\ep [g_{\alpha\beta,\gamma}]n^\gamma$).  We then find
\[
[\Gamma^\alpha_{\beta\gamma}] = \half\,(\kappa^\alpha_\beta n_\gamma+\kappa^\alpha_\gamma n_\beta - \kappa_{\beta\gamma} n^\alpha),
\]
and hence
\[
A^\alpha_{\beta\gamma\delta}=\frac{\ep}{2}\,(\kappa^\alpha_\delta n_\beta n_\gamma - \kappa^\alpha_\gamma n_\beta n_\delta - \kappa_{\beta\delta} n^\alpha n_\gamma +\kappa_{\beta\gamma}n^\alpha n_\delta).
\]

A few lines of calculation show us that the $\delta$-function part of the Einstein tensor is 
\[
\label{Seq}
S_{\alpha\beta} = \frac{\ep}{2}\,\(\kappa_{\mu\alpha}n^\mu n_\beta+\kappa_{\mu\beta}n^\mu n_\alpha-\kappa n_\alpha n_\beta-\ep \kappa_{\alpha\beta}-(\kappa_{\mu\nu}n^\mu n^\nu - \ep \kappa)g_{\alpha\beta}\).
\]
On the other hand, the total stress-energy tensor is of the form
\[
T^\mathrm{\,total}_{\alpha\beta}=\Theta(\ell)T^+_{\alpha\beta}+\Theta(-\ell)T^-_{\alpha\beta}+\delta(\ell)T_{\alpha\beta},
\]
where $T^+_{\alpha\beta}$ and $T^-_{\alpha\beta}$ represent the bulk stress-energy in the regions where $\ell> 0$ and $\ell < 0$ respectively, while $T_{\alpha\beta}$ denotes the stress-energy localised on the hypersurface $\Sigma$
itself.  From the Einstein equations, we find $T_{\alpha\beta}=(8\pi G)^{-1} S_{\alpha\beta}$.
It then follows that $T_{\alpha\beta}n^\beta=0$, implying $T_{\alpha\beta}$ is tangent to $\Sigma$.  This allows us to decompose $T^{\alpha\beta}$ as
\[
T^{\alpha\beta}=T^{ab}e^\alpha_a e^\beta_b,
\]
where $T_{ab}=T_{\alpha\beta}e^\alpha_a e^\beta_b$ re-expresses the hypersurface stress-energy tensor in terms of coordinates intrinsic to $\Sigma$.
From (\ref{Seq}), we have
\bea
16\pi G T_{ab} &=& -\kappa_{\alpha\beta}e^\alpha_a e^\beta_b-\ep(\kappa_{\mu\nu}n^\mu n^\nu-\ep\kappa)h_{ab} \nonumber\\
&=& -\kappa_{\alpha\beta}e^\alpha_a e^\beta_b-\kappa_{\mu\nu}(g^{\mu\nu}-h^{mn}e^\mu_m e^\nu_n)h_{ab}+\kappa h_{ab} \nonumber\\
&=&-\kappa_{\alpha\beta}e^\alpha_a e^\beta_b+h^{mn}\kappa_{\mu\nu}e^\mu_m e^\nu_n h_{ab}.
\eea

Finally, we can relate $T_{ab}$ to the jump in extrinsic curvature across $\Sigma$.
From
\[
[\grad_{\alpha}n_\beta]= - [\Gamma^\gamma_{\alpha\beta}]n_\gamma = \half\,(\ep \kappa_{\alpha\beta}-\kappa_{\gamma\alpha}n_\beta n^\gamma-\kappa_{\gamma\beta}n_\alpha n^\gamma),
\]
we deduce that
\[
[K_{ab}] = [\grad_\alpha n_\beta]e^\alpha_a e^\beta_b = \frac{\ep}{2}\,\kappa_{\alpha\beta}e^\alpha_a e^\beta_b.
\]
This leads us to our goal; 
\[
8\pi G T_{ab}=-\ep\([K_{ab}]-[K]h_{ab}\),
\]
which is the Israel matching condition (or `second' junction condition).


%% file: appendix3.tex
\chapter{Five-dimensional longitudinal gauge}
\label{appA}




Starting with the background metric in the form (\ref{metrica}), 
the most general scalar metric perturbation can be written as \cite{Carsten}
\bea
\d s^2 &=& n^2\left(-(1+2\Phi)\,\d t^2 -2W\,\d t\, \d y + t^2 (1-2\Gamma)\,\d
  y^2 \right) - 2\nabla_i \alpha \,\d x^i \,\d t \nonumber \\ && +2t^2\,\nabla_i\beta \,\d y \,\d
  x^i+ b^2\left( (1-2\Psi)\,\delta_{ij}-2\nabla_i\nabla_j\chi\right)\d
  x^i\,\d x^j .
\eea
Under a gauge transformation $x^A\tt x^A+\xi^A$, these variables transform as
\bea
	\nonumber & &
	\Phi \rightarrow \Phi-\dot{\xi}^t-\xi^t \frac{\dot{n}}{n}-\xi^y \frac{n'}{n}, \\
	\nonumber & &
	\Gamma \rightarrow \Gamma+\xi'^y+\frac{1}{t}\xi^t+\xi^t 
	\frac{\dot{n}}{n}+\xi^y \frac{n'}{n}, \\
	\nonumber & &
	W \rightarrow W-\xi'^t+t^2 \dot{\xi}^y,  \\
	\nonumber & &
	\alpha \rightarrow \alpha-\xi^t+\frac{b^2}{n^2} \dot{\xi}^s, \\
	\nonumber & &
	\beta \rightarrow \beta-\xi^y-\frac{b^2}{n^2 t^2} \xi'^s,  \\
	\nonumber & &
	\Psi \rightarrow \Psi+\xi^t \frac{\dot{b}}{b}+\xi^y \frac{b'}{b}, \\
	& &
	\chi \rightarrow \chi+\xi^s , 
\eea
where dots and primes indicate differentiation with respect to $t$ and $y$ respectively. 
Since a five-vector $\xi^A$ has three scalar degrees of freedom $\xi^t$,
$\xi^y$ and $\xi^i=\nabla_i \xi^{s}$,  only four of the seven functions
$(\Phi,\Gamma,W,\alpha,\beta,\Psi,\chi)$ are physical. 
We can therefore construct four gauge-invariant variables, which are
\bea
\nonumber & &
\Phi_{\inv}=\Phi-\dot{\tilde{\alpha}}-\tilde{\alpha}\, \frac{\dot{n}}{n}-\tilde{\beta}\, \frac{n'}{n}, \\
\nonumber & &
\Gamma_{\inv}=\Gamma+\tilde{\beta}'+\frac{1}{t}\,\tilde{\alpha}
+\tilde{\alpha}\, \frac{\dot{n}}{n}+\tilde{\beta}\, \frac{n'}{n}, \\
\nonumber & &
W_{\inv}=W-\tilde{\alpha}'+t^2\, \dot{\tilde{\beta}}, \\
 & &
\Psi_{\inv}=\Psi+\frac{\dot{b}}{b}\,\tilde{\alpha}+\frac{b'}{b}\,
\tilde{\beta},
\label{gauges}
\eea
where $\tilde{\alpha}=\alpha-(b^2/n^2)\, \dot{\chi}$ and
$\tilde{\beta}=\beta+(b^2/n^2 t^2)\, \chi'$. 

In analogy with the four-dimensional case, we then define
five-dimensional longitudinal gauge by $\chi=\alpha=\beta=0$, giving
\begin{eqnarray}
	\Phi_{\inv}&=&\Phi_L, \qquad	\,\,\Gamma_{\inv}=\Gamma_L, \nonumber \\
	W_{\inv}&=&W_L, \qquad 	\Psi_{\inv}=\Psi_L ,
\end{eqnarray}
\ie the gauge-invariant variables are equal to the values of the metric perturbations
in longitudinal gauge.
This gauge is spatially
isotropic in the $x^i$ coordinates, although in general there will be a
non-zero $t, y$ component of the metric.



As for the locations of the branes, this will in general be different for
different choices of gauge.  
In the case where the brane matter has no
anisotropic stresses, the location of the branes is easy to establish.  
Working out the
Israel matching conditions, we find that $\beta$ on the branes is
related to the anisotropic part of the brane stress-energy.
If we consider only perfect fluids, for which the shear vanishes, then
the Israel matching conditions give $\beta(y=\pm y_0)=0$.  

From the gauge transformations above, we can transform into the gauge $\alpha=\chi=0$
using only a $\xi^s$ and a $\xi^t$ transformation.  We may then pass to
longitudinal gauge ($\alpha=\beta=\chi=0$) with the transformation
$\xi^y=\tilde{\beta}$ alone.  Since $\beta$ (and hence $\tilde{\beta}$)
vanishes on the branes, $\xi^y$ must also vanish leaving the brane
trajectories unperturbed.  Hence, in longitudinal gauge the brane 
locations remain at their unperturbed values $y=\pm y_0$.
Transforming to a completely arbitrary gauge, we see that in general
the brane locations are given by $y = \pm y_0 - \tilde{\beta}$.



%% file: longappendix.tex
\chapter{Detailed results}
\label{detailedresults}


\begin{flushright}
\begin{minipage}{5.7cm}
\small
\noindent
{\it By the pricking of my thumbs, \\
Something wicked this way comes.}
\begin{flushright}
\noindent 
MacBeth, Act IV.
\end{flushright}
\end{minipage}
\end{flushright}

\vspace{0.2cm}

In this appendix, we list the results of detailed calculations of the background geometry and
cosmological perturbations in a big crunch/big bang universe, obtained using a variety of methods.
For clarity, the AdS radius $L$ has been set to unity throughout. 
(To restore $L$, simply replace $t \tt t/L$ and $k \tt kL$).

\section{Polynomial expansion for background}
\label{appB}

Using the expansion in Dirichlet/Neumann
polynomials presented in Section \S\,\ref{polysection} to solve for the
background geometry, we find
{\allowdisplaybreaks \bea
N_0 &=& {1\over t}-{1\over 2}\,t y_0^2+{1\over 24}\,t \(8-9 \,t^2\) y_0^4-{1\over
  720}\, t \(136+900 \,t^2+375 \,t^4\) y_0^6 \nonumber \\
&& +{1\over 40320}\,t \(3968+354816 \,t^2-348544 \,t^4-36015 \,t^6\) y_0^8+O(y_0^{10}) \qquad\\
N_3 &=& -\frac{1}{6} + \left(\frac{5}{72} - 2 \,t^2 \right)y_0^2 - 
  {1\over 2160}\,\( 61 - 20880 \,t^2 + 19440 \,t^4 \) y_0^4 \nonumber \\
&& + \left( \frac{277}{24192} - \frac{743 \,t^2}{20} + \frac{677 \,t^4}{6}
  - \frac{101 \,t^6}{3}\right)y_0^6+O(y_0^{8}) \\
N_4 &=& \frac{3}{2}\, t^3 y_0^2 + \frac{1}{4}\,t^3\( -28 + 33 \,t^2\)  y_0^4 \nonumber \\ && 
    + \frac{1}{80}\,t^3 \( 1984 - 6776 \,t^2 + 2715 \,t^4 \)
    y_0^6 +O(y_0^8) \\
N_5 &=& -\frac{1}{120} + \frac{1}{1800}\,\( 7 - 1800 \,t^2 - 540 \,t^4\)  y_0^2 \nonumber \\
&& - \frac{1}{201600}\,\( 323 - 990528 \,t^2 + 2207520 \,t^4 + 362880 \,t^6 \)
     y_0^4+O(y_0^6)  \qquad\\
N_6 &=& t^3 y_0^2 + \frac{1}{30}\,t^3 \( -142 + 371 \,t^2 \)
 y_0^4+O(y_0^6) \\
N_7 &=& -\frac{1}{5040}- \frac{1}{94080}\,\( -9 + 20384 \,t^2 + 23520 \,t^4\)  y_0^2+O(y_0^4) \\
N_8 &=& \frac{3}{10}\,t^3 y_0^2+O(y_0^4) \\
N_9 &=& \frac{1}{362880} +O(y_0^2) \\
N_{10} &=& O(y_0^2) 
\eea}
and
\bea
q_0 &=& 1 - \frac{3}{2}\,t^2 y_0^2 + \left( t^2 -
\frac{7\,t^4}{8} \right) y_0^4 +  
  \left( \frac{-17\,t^2}{30} + \frac{17\,t^4}{12} - \frac{55\,t^6}{48}
  \right) y_0^6\qquad \qquad \nonumber \\
&& + \left( \frac{31\,t^2}{105} - \frac{9\,t^4}{5} + \frac{233\,t^6}{90} -
  \frac{245\,t^8}{128} \right) y_0^8 +  O(y_0^{10}) \qquad
\\
q_3 &=&-2\,t^3 y_0^2 + \left( \frac{29\, t^3}{3} - 8\, t^5 \right) y_0^4 \nonumber \\ && 
  - \left( \frac{743\, t^3}{20} - \frac{322\, t^5}{3} + 27\, t^7 \right)
  y_0^6  +O(y_0^8) 
\\ 
q_4 &=& \frac{1}{2}\,t^4 y_0^2 + \left( \frac{-5\,t^4}{3} +
\frac{9\,t^6}{4} \right) y_0^4+O(y_0^6)
\\
q_5&=& - t^3 y_0^2  + \left( \frac{737\,t^3}{150} - \frac{58\,t^5}{5}
\right) y_0^4+O(y_0^6)
\\
q_6 &=& \frac{1}{3}\,t^4 y_0^2+O(y_0^4)
\\
q_7 &=& -\frac{13}{60}\,t^3 y_0^2+O(y_0^4)   
\\
q_8 &=& O(y_0^2)
\\
q_9 &=& O(y_0^2),
\eea
This solution explicit satisfies all the Einstein equations up to $O(y_0^{10})$.


\section{Polynomial expansion for perturbations}
\label{appC}

\subsection{All wavelengths}


Using the Dirichlet/Neumann polynomial expansion 
to solve for the perturbations, the
solution may be expressed in terms of the original longitudinal gauge
variables as
\[
\Phi_L = \mathcal{P}_\Phi^{(0)}(y,t)F^{(0)}(t)+ \mathcal{P}_\Phi^{(1)}(y,t)F^{(1)}(t),
\]
where
\[
F^{(n)}(t) =
\bar{A} J_n(kt)+\bar{B} Y_n(kt),
\]
for $n=0,\,1$ and $\gamma = 0.577\dots$ is the Euler-Mascheroni
constant.  The constants $\bar{A}$ and $\bar{B}$ are arbitrary functions of
$k$.  In order to be consistent with the series expansion in $t$
presented in Section \S\,\ref{seriessoln}, we must set
\bea
\bar{A} &=& 12\,A+2\,B\,k^2\,(\ln{2}-\gamma ) -
\frac{9\,B\,y_0^2}{2} + \frac{233\,B\,y_0^4}{45}+O(y_0^6),
 \\
\bar{B} &=& B\, k^2\,\pi+O(y_0^6).
\eea

The polynomials $\mathcal{P}_\Phi^{(n)}$ are then given (for all $k$
and $t$) by
\bea
 \mathcal{P}_\Phi^{(0)}(t,y) &= & -\frac{1}{6} + \frac{1}{3}\,t y 
      + \frac{1}{12}\,t^2 \left(-2 y^2 + y_0^2 \right) +
  \frac{1}{36}\,t 
      y \left(11 y^2 + 3 (-11 + 3 t^2) 
          y_0^2 \right)  \nonumber \\
&&   +\frac{t^2}{2160}\, \left(-525 y^4 + 90 (19 - 5t^2 ) y^2 
          y_0^2 +
            (511 - 180 t^2 + 45 k^2 t^4 )
          y_0^4 \right)  \nonumber \\
&& - \frac{t\,y}{2160}\,\Bigg(3\(-92 + (9 + 4 k^2) t^2 \) 
          y^4 \nonumber \\ &&  
          + 30 \(92 - (219 + 4 k^2) t^2 + k^2 
                t^4 \) y^2 y_0^2 \nonumber \\ &&  
           - \big(6900 - (20087 + 300 k^2) 
                  t^2  + 90\, (13 + k^2)\, t^4 - 90\, k^2\, 
                  t^6 \big)\, y_0^4\Bigg) \nonumber \\ &&
		  +O(y_0^6),\nonumber \\
\eea
and by
\bea
\mathcal{P}_\Phi^{(1)}(t,y) &=& \frac{1}{k t}\,\Bigg[\frac{1}{2} -
  \frac{t y}{2} + \frac{t^2}{12}\left(3  y^2 + (-3 + k^2  t^2)\,  
          y_0^2 \right)  \nonumber \\
&&  -\frac{t y}{36} \left( (3 + k^2  t^2 ) \, 
          y^2 + 3  \left(-3 + (3 - k^2 )\,  t^2 + 2  k^2  
                t^4 \right)  y_0^2 \right)  \nonumber \\
&&+\frac{t^2}{2160}\,\Bigg(75 \, (6 + k^2  
                t^2 ) \, 
          y^4 + 90 \left(-12 + 3\, (2 - k^2)\, t^2 + 2 k^2 
                t^4 \right) y^2 y_0^2  \nonumber \\
&&           + \left(718 + k^2  t^2  \,(-101 + 225  t^2) \right)  
          y_0^4 \Bigg) \nonumber \\
&& -\frac{t y}{2160}\, \Bigg(3 \left(3 + 2  \,(-9 + 31  k^2)\,  t^2 + 2  
                k^2  t^4 \right)  
          y^4 \nonumber \\
&& +30 \left(-3 + (219 - 62  k^2) \, t^2 + 16  
                k^2  t^4 \right)  y^2  y_0^2  \nonumber \\
&&          +  \left(225 -(20104 - 4650 k^2)\,  
                  t^2 + (1215 - 1822 k^2) \, t^4 + 765 k^2 
                  t^6 \right)
          y_0^4 \Bigg) \Bigg]\nonumber \\ && +O(y_0^6) . 
\eea
Since the $F^{(n)}$ are of zeroth order in $y_0$, the solution for
$\Phi_L$ to a given order less than $O(y_0^6)$ is found simply by truncating the
polynomials above.  (Should they be needed, results up to
$O(y_0^{14})$ can in addition be found at \cite{Website}). 

In a similar fashion we may express the solution for $\Psi_L$ as
\[
\Psi_L = \mathcal{P}_\Psi^{(0)}(y,t)F^{(0)}(t)+ \mathcal{P}_\Psi^{(1)}(y,t)F^{(1)}(t),
\]
where $F^{(n)}$ is defined as above and
\bea
\mathcal{P}_\Psi^{(0)}(t,y) &=& 
\frac{1}{6} - \frac{t  y}{3} + \frac{t^2}{6}\, (y^2 + y_0^2 ) +
  \frac{t  
      y}{36} \left(-2  y^2 + 3 \,(2 - 3  t^2 )\,  
          y_0^2 \right) \nonumber \\
&& + \frac{t^2}{2160}\, \left(120 y^4 + 450 \,(-2 + t^2 )\, y^2 y_0^2 +
            (644 + 450 t^2 - 45 k^2 t^4 ) \,
          y_0^4 \right)  \nonumber \\
&& + \frac{t y}{2160} \Bigg(3 \left(-2 + (9 + 4 k^2)\, t^2 \right) 
          y^4 + 30 \left(2 + 4 \,(9 - k^2) \,t^2 + k^2 
                t^4 \right) y^2 y_0^2  \nonumber \\
&&         -   \left(150 + (2863 - 300  k^2 ) \, 
                  t^2 + 90 \, (13 + k^2) \, t^4 - 90  k^2  
                  t^6 \right)  y_0^4 \Bigg)\nonumber \\&& +O(y_0^6),
\eea
\bea
\mathcal{P}_\Psi^{(1)}(t,y) &=& \frac{1}{k t}\Bigg[
\frac{t y}{2} - \frac{t^2}{12} \left(3  y^2 + (3 + k^2  t^2)\,  
          y_0^2 \right)  \nonumber \\
&&  +\frac{t y}{36} \left( (3 + k^2  t^2 )\,  
          y^2 + 3 \left(-3 + (3 - k^2 )\, t^2 + 2 k^2 
                t^4 \right) 
          y_0^2 \right)  \nonumber \\
&&- \frac{t^2}{2160}\,\Bigg(15\, (12 + 5 k^2 
                t^2 ) \,
          y^4 - 90 \left(6 - 3\,(2 - k^2)\, t^2 - 2 k^2 
                t^4 \right) y^2 y_0^2 \nonumber \\
&&            -\left(752 - (540 - 101  k^2) \, t^2 - 360  k^2  
                  t^4 \right)  y_0^4 \Bigg)  \nonumber \\
&&  +\frac{t y}{2160}\,\Bigg( \left(9 + 6 \,(-9 + k^2 )\,  t^2 + 6  k^2  
                t^4 \right)  
          y^4  \nonumber \\
&& -30  \left(3 + (33 + 2  k^2 )\,  t^2 - 7  k^2  
                t^4 \right)  y^2  y_0^2  \nonumber \\
&&           + \left(225 + 2 \,(1288 + 75 k^2 )\, 
                  t^2 + (1215 - 1012 k^2 )\, t^4 + 765 k^2 
                  t^6 \right) y_0^4 \Bigg)
\Bigg]\nonumber \\ && +O(y_0^6) . \nonumber \\ 
\eea

Finally, writing
\[
W_L = \mathcal{P}_W^{(0)}(y,t)F^{(0)}(t)+ \mathcal{P}_W^{(1)}(y,t)F^{(1)}(t),
\]
we find
\bea
\mathcal{P}_W^{(0)}(t,y) &=&-\frac{1}{60}\,t^2\, (y^2 - y_0^2)\,
\Big(-30 + 30 t 
    y - 25 \left(y^2 + (-5 + 3 t^2)\, y_0^2 \right)
    \nonumber \\
&& +t  y \left(21 y^2 +(-149 + 75 t^2)\, y_0^2 \right)
    \Big)+O(y_0^6), \\
\nonumber \\
\mathcal{P}_W^{(1)}(t,y) &=&-\frac{1}{60 k}\,t^2\, (y^2 - y_0^2)\,\Big(30 y - 5 
k^2 t \left(2 y^2 + (-10 + 3 t^2)\, y_0^2 \right)
 \nonumber \\
&& + y\left( (12 + 11 k^2 t^2)\, 
    y^2 + \left(-38 + (60 - 69 k^2)\, t^2 + 15 k^2 
            t^4 \right)y_0^2\right)\Big) \nonumber \\ && +O(y_0^6)
	    .\nonumber \\ 
\eea

\subsection{Long wavelengths}

On long wavelengths, $F^{(n)}$ reduces to
\bea
F^{(0)}(t) &=& 12\,A - \frac{9\,B\,y_0^2}{2} + \frac{233\,B\,y_0^4}{45} +O(k^2)+O(y_0^6), \\
F^{(1)}(t) &=& \left( 6At - \left( \frac{2}{t} +
\frac{9t y_0^2}{4} - \frac{233t y_0^4}{90}\right)B  \right)k
+O(k^2)+O(y_0^6).\qquad
\eea
For convenience, we list below 
the metric perturbations truncated at $O(k^2)$.  
\bea
\Phi_L &=& \big(A - \frac{B}{t^2}\big) + \big( \frac{B y}{t} + A t y \big)    + 
  \frac{1}{8}\left( B ( y_0^2 - 4 y^2 )  - 4 A t^2 ( y_0^2 + y^2 )
  \right) \nonumber \\ && + 
  \frac{y}{24 t}\big( B ( 3 y_0^2 ( -4 + t^2 )  + 4 y^2 )  + 
       4 A t^2 ( 3 y_0^2 ( -19 + 3 t^2 )  + 19 y^2 \big)  \nonumber \\
       && + 
  \Big( \frac{1}{6} A t^2 ( y_0^4 ( 29 - 6 t^2 )  +
    3 y_0^2 ( 13 - 2 t^2 )  y^2 - 10 y^4 )\nonumber \\ &&  - 
     \frac{1}{240}(B ( y_0^4 ( 56 - 45 t^2 )  +
       15 y_0^2 ( -16 + 5 t^2 )  y^2 + 100 y^4 )
     ) \Big) \nonumber \\ &&  
     + \frac{y}{240 t} \big( 2 A t^2 ( 5 y_0^4 ( 905 - 1338 t^2 + 75 t^4 )  + 
          10 y_0^2 ( -181 + 219 t^2 )  y^2 + 181 y^4 )  \nonumber \\
	  && + 
       B ( y_0^4 ( 50 - 3509 t^2 + 135 t^4 )  +
       5 y_0^2 ( -4 + 235 t^2 )  y^2 +  
          ( 2 - 12 t^2 )  y^4 )\big) \nonumber \\ && +  O(y_0^6), \\ 
\Psi_L &=& 2 A - \frac{y}{t}( B + A t^2 ) +
\frac{1}{4}\big( 2 A t^2 ( y_0^2 + y^2 )  +  
       B ( -y_0^2 + 2 y^2 )  \big)  \nonumber \\ && - 
  \frac{y}{24 t} \big( 4 A t^2 ( 3 y_0^2 ( -1 + 3 t^2 )  + y^2 )  + 
       B ( 3 y_0^2 ( -4 + t^2 )  + 4 y^2 )  \big) \nonumber \\ && + 
  \frac{1}{48}\Big( 8 A t^2 ( 2 y_0^4 ( 17 + 3 t^2 )  +
    3 y_0^2 ( -7 + 2 t^2 )  y^2 + y^4 )  \nonumber \\ && +  
       B ( y_0^4 ( 8 + 15 t^2 )  + 3 y_0^2 ( -8
       + 5 t^2 )  y^2 + 8 y^4 )  \Big)\nonumber \\ &&  -  
  \frac{y}{240 t} \Big( 2 A t^2 ( 25 y_0^4 ( 1 + 42 t^2 +
    15 t^4 )  - 10 y_0^2 ( 1 + 39 t^2 )  y^2 + y^4 
          )  \nonumber \\ && + B ( y_0^4 ( 50 + 721 t^2 + 135 t^4
	  )  - 5 y_0^2 ( 4 + 47 t^2 )  y^2 +  
          ( 2 - 12 t^2 )  y^4 )  \Big)  \nonumber \\ &&+  O(y_0^6), \\ 
W_L &=& 
6 A t^2 ( -y_0^2 + y^2 )  - t ( B + 3 A t^2
)  y ( -y_0^2 + y^2 )  \nonumber \\ &&  +  
  \frac{1}{4}t^2 ( -y_0^2 + y^2 )  ( -9 B y_0^2 +
    20 A ( y_0^2 ( -5 + 3 t^2 )  + y^2 )
    )\nonumber \\ &&  - \frac{1}{120} t y ( -y_0^2 + y^2 )  (
     120 A t^2 ( y_0^2 ( -26 + 9 t^2 )  + 3 y^2
     ) \nonumber \\ &&  +  
       B ( y_0^2 ( -152 + 105 t^2 )  + 48 y^2 )  ) +  O(y_0^6).
\eea


\section{Perturbations from expansion about the scaling solution}
\label{appD}


Following the method of expanding about the scaling solution 
presented in Section \S\,\ref{expaboutscalingsoln}, we have computed the
perturbations to $O(y_0^4$). On long wavelengths,
the five-dimensional longitudinal gauge variables 
take the form
\bea
\Phi_L &=&  f^\Phi_0+y_0^2 (f^\Phi_1+f^\Phi_2
\ln{(1+x_4)} + f^\Phi_3 \ln{(1-x_4)}+f^\phi_4 \ln{(1-\w x_4)},\qquad \qquad \\
\Psi_L &=&  f^\Psi_0+y_0^2 (f^\Psi_1+f^\Psi_2
\ln{(1+x_4)} + f^\Psi_3 \ln{(1-x_4)}+f^\Psi_4 \ln{(1-\w x_4)}, \qquad\qquad\\
W_L&=& e^{-\frac{1}{2}x_4^2}\,\big( f^W_0+y_0^2 (f^W_1+f^W_2
\ln{(1+x_4)} 
+ f^W_3 \ln{(1-x_4)} \nonumber \\ && \qquad 
+f^W_4 \ln{(1-\w x_4)}\big), 
\eea
where the $f$ are rational functions of $x_4$ and $\w$.  For $\Phi_L$,
we have
\bea
f^\Phi_0 &=& \frac{1}{16 x_4^2 ( -1 + x_4^2 ) }\Big(16 \tB - 16 \tB \w x_4 - 2 ( 8 A + \tB -
  4 \tB \w^2 )  x_4^2 \nonumber \\ && \qquad 
  + 2 ( -8 A + 3 \tB )  \w x_4^3 +  
    ( 8 A - 3 \tB )  ( 3 + \w^2 )
     x_4^4\Big),\\
f^\Phi_1 &=& \frac{1}{960 x_4^4 {( -1 + x_4^2 ) }^5}\Big(8 A x_4^5 \big(
-580 x_4 + 95 x_4^3 - 576 \w^5 x_4^4 + 281 x_4^5 \nonumber \\ && + 96 \w^6 x_4^5 
	- 60 x_4^7  + 
       5 \w^4 x_4 ( 40 + 167 x_4^2 - 39 x_4^4 )  + 20 \w^3 ( -19 - 5
       x_4^2 + 48 x_4^4 ) \nonumber \\ && -  
       10 \w^2 x_4 ( 78 + 117 x_4^2 - x_4^4 - 2 x_4^6 )  +
       4 \w ( 285 + 100 x_4^2 - 25 x_4^4 - 29 x_4^6 + 5 x_4^8
       )  \big)  \nonumber \\ && +  
    \tB \big( -1920 + x_4 ( 480 \w + 80 ( 91 + 15 \w^2
    )  x_4 \nonumber \\ && - 160 \w ( 5 + 4 \w^2 )  x_4^2 +  
          40 ( -273 - 116 \w^2 + 7 \w^4 )  x_4^3 + 20 \w ( 649 - 127
	  \w^2 )  x_4^4 \nonumber \\ && + 
          4 ( 1231 - 3285 \w^2 + 2060 \w^4 )  x_4^5 -
	  4 \w ( 2036 - 2115 \w^2 + 1152 \w^4 )  x_4^6 \nonumber \\ &&
	  +  
          ( 1107 + 8102 \w^2 - 4905 \w^4 + 768 \w^6 )  x_4^7 + 
          12 \w ( 233 + 48 \w^2 ( -5 + 3 \w^2 )  )  x_4^8 \nonumber \\
	  && - 
          ( 3131 + 1582 \w^2 - 585 \w^4 + 288 \w^6 )  x_4^9
	  - 772 \w x_4^{10} + 20 ( 85 + 29 \w^2 )  x_4^{11}
	 \nonumber \\ && +  
          180 \w x_4^{12}  - 120 ( 3 + \w^2 )  x_4^{13}
	  )  \big) \Big), \\ 
f^\Phi_2 &=& \frac{1}{48 x_4^5 {( -1 + x_4^2 ) }^3}\Big(\tB\big( 48 +
( 36 - 48 \w )  x_4 + 12 ( -7 - 6 \w + 2 \w^2 )
 x_4^2 \nonumber \\ && +  
       4 ( -13 + 6 \w + 9 \w^2 )  x_4^3 + ( 60 + 80 \w
       - 12 \w^2 )  x_4^4 + 4 ( 8 + 6 \w - 9 \w^2 )
        x_4^5 \nonumber \\ && -  
       3 ( 8 + 7 \w + 4 \w^2 )  x_4^6 + ( -11 + 5 \w^2 )  x_4^7 + 3 \w x_4^8 \big)  + 
    8 A x_4^6 ( x_4 + \w^2 x_4 \nonumber \\ && - \w ( 1 + x_4^2 )  )
     \Big), 
\eea

\bea
f^\Phi_3 &=& \frac{1}{48 x_4^5 {( -1 + x_4^2 )
  }^3}\Big(\tB\big( -48 + 12( 3 + 4\w )  x_4 + 12 ( 7 - 6 \w - 2 \w^2
)  x_4^2 \nonumber \\ && - 
       4 ( 13 + 6 \w - 9 \w^2 )  x_4^3 - 4 ( 15 - 20 \w - 3 \w^2 )  x_4^4 + 
       4 ( 8 - 6 \w - 9 \w^2 )  x_4^5 \nonumber \\ && + 3 ( 8 - 7 \w
       + 4 \w^2 )  x_4^6 + ( -11 + 5 \w^2 )  x_4^7 +
       3 \w x_4^8 
       )  \nonumber \\ && + 8 A x_4^6 ( x_4 + \w^2 x_4 - \w ( 1 + x_4^2
       )  ) \Big), \\
f^\phi_4 &=& \frac{3 \tB {( -1 + \w x_4 )
  }^2}{2 x_4^4 {( -1 + x_4^2 ) }^2} .
\eea
For $\Psi_L$, we find
{\allowdisplaybreaks
\bea
f^\Psi_0 &=& \frac{1}{16 x_4 ( -1 + x_4^2 ) }\Big(16 \tB \w - 4 ( 8 A + \tB ( -1 + 2 \w^2
  )  )  x_4 + 2 ( 8 A - 3 \tB )  \w x_4^2 \nonumber \\ && \qquad +  
    ( -8 A + 3 \tB )  ( -3 + \w^2 )
     x_4^3\Big), \\
f^\Psi_1 &=& \frac{1}{960 x_4^3 {( -1 +
    x_4^2 ) }^5}\Big(-480 \tB \w - 240 \tB ( -7 + 5 \w^2 )  x_4
  + 160 \tB \w ( 5 + 4 \w^2 )  x_4^2 \nonumber \\ && -  
    40 \tB ( 143 - 104 \w^2 + \w^4 )  x_4^3 + 20 \w 
     ( \tB ( 197 - 155 \w^2 )  + 8 A ( -3 + \w^2 )  )  x_4^4 \nonumber
    \\ && + 
    20 ( -8 A ( 34 - 21 \w^2 + \w^4 )  + \tB (
    98 - 369 \w^2 + 101 \w^4 )  )  x_4^5\nonumber \\ && -  
    4 \w ( 200 A ( -14 + 5 \w^2 )  +  \tB ( 34 -
    975 \w^2 + 288 \w^4 )  )  x_4^6 \nonumber \\ && +  
    ( 40 A ( 55 - 234 \w^2 + 67 \w^4 )  + \tB (
    2455 + 838 \w^2 - 1005 \w^4 + 192 \w^6 )  )  x_4^7  \nonumber \\ && -  
    4 \w ( 3 \tB ( 53 + 120 \w^2 - 36 \w^4 )  +
    8 A ( 155 - 120 \w^2 + 36 \w^4 )  )  x_4^8 \nonumber \\ && -  
    ( \tB ( 3515 - 1042 \w^2 - 225 \w^4 + 72 \w^6 )   -
    8 A ( 89 + 170 \w^2 - 75 \w^4 + 24 \w^6 )  )
     x_4^9 \nonumber \\ && +  
    4 ( 232 A + 193 \tB )  \w x_4^{10} - 20 (
    8 A ( 1 + \w^2 )  \nonumber \\ && + \tB ( -91 + 29 \w^2 )
    )   
     x_4^{11} - 20 ( 8 A + 9 \tB )  \w x_4^{12} +
     120 \tB ( -3 + \w^2 )  x_4^{13}\Big), \\
f^\Psi_2 &=& \frac{-1}{48 x_4^4  
    {( -1 + x_4^2 ) }^3}\Big( 8 A x_4^5 ( x_4 + \w^2 x_4 - \w ( 1 + x_4^2 )  )  + 
      \tB ( 36 + 60 x_4 - 36 x_4^2 \nonumber \\ && - 84 x_4^3 +  24 x_4^5 + 5 x_4^6 +
      \w^2 x_4 ( 24 + 36 x_4 - 12 x_4^2 - 36 x_4^3 - 12 x_4^4 + 5 x_4^5
      ) \nonumber \\ && +  
         \w ( -48 - 72 x_4 +  24 x_4^2 + 80 x_4^3 + 24 x_4^4 - 21 x_4^5
      + 3 x_4^7 )  )   \Big),
\\
f^\Psi_3 &=& \frac{1}{48 x_4^4 {( -1 + x_4^2 ) }^3}\Big(8 A x_4^5 ( - x_4 - \w^2 x_4 + \w (1+x_4^2) )
   + 
    \tB ( -36 + 60 x_4 + 36 x_4^2 \nonumber \\ && - 84 x_4^3 +  24 x_4^5 - 5 x_4^6 -
      \w^2 x_4 ( -24 + 36 x_4 + 12 x_4^2 - 36 x_4^3 + 12 x_4^4 +
      5 x_4^5 ) \nonumber \\ && +  
       \w ( -48 +  72 x_4 + 24 x_4^2 - 80 x_4^3 + 24 x_4^4 + 21 x_4^5 -
      3 x_4^7 )  ) \Big),
 \\
f^\Psi_4 &=& \frac{-3 \tB {( -1 + \w x_4 )
  }^2}{2 x_4^4 {( -1 + x_4^2 ) }^2} .
\eea
}
Finally, for $W_L$, we have
\bea
f^W_0 &=&\frac{( -1 + \w^2 )  x_4 ( -24 A x_4 (
  -2 + \w x_4 )  + \tB ( -18 x_4 + \w ( -8 + 9 x_4^2
  )  ) 
      ) }{8 {( -1 + x_4^2 ) }^2},\\
f^W_1 &=&\frac{(-1 + \w^2)}{480 x_4^2 {( -1 + x_4^2 ) }^7}  \Big( 8 A x_4^4 \big( 1500
  + 84 \w^4 x_4^4 ( -12 + x_4^2 ) \nonumber \\
	 &&  -  
         6 \w^2 ( 50 + 100 x_4^2 - 427 x_4^4 + x_4^6 )  +
	 3 \w^3 x_4 ( 60 + 585 x_4^2 - 160 x_4^4 + 3 x_4^6 )
	  \nonumber \\ &&+  
         \w^5 ( 168 x_4^5 - 36 x_4^7 )  - 6 x_4^2 ( -590
	 + 243 x_4^2 + 201 x_4^4 - 4 x_4^6 + 10 x_4^8 )  \nonumber \\
	 && +  
         \w x_4 ( -1560 - 4935 x_4^2 + 2000 x_4^4 - 265 x_4^6 + 92
	 x_4^8 )  \big)  \nonumber \\ && + 
      \tB \big( 1440 + x_4 ( -84 \w^4 x_4^5 ( 24 + 28 x_4^2 +
      3 x_4^4 )  \nonumber \\ && + 12 \w^5 x_4^6 ( 40 + 6 x_4^2 + 9 x_4^4
      )  +  
            6 \w^2 x_4^3 ( -450 + 964 x_4^2 + 863 x_4^4 + 3 x_4^6 )
	    \nonumber \\ && - 
            3 \w^3 x_4^2 ( -40 - 748 x_4^2 - 2333 x_4^4 + 672 x_4^6 +
	    9 x_4^8 )  \nonumber \\ && - 
            6 x_4 ( 2160 - 5490 x_4^2 + 3770 x_4^4 - 2249 x_4^6 +
	    597 x_4^8 - 588 x_4^{10} + 90 x_4^{12} ) \nonumber \\ &&  +  
            \w ( -2160 + 18920 x_4^2 - 53216 x_4^4 + 20629 x_4^6 -
	    11216 x_4^8 + 5579 x_4^{10} \nonumber \\ && - 2236 x_4^{12} + 360 x_4^{14}
	    )  )  
         \big)  \Big), \\
f^W_2 &=& \frac{( 1 - \w )}{24 {( x_4 - x_4^3 ) }^4}  \Big( \tB \big( -144 + 108 ( -1 + 2 \w )  x_4 - 
           24 ( -5 + \w )  ( 3 + 2 \w )  x_4^2 \nonumber \\ && -
	   36 ( -11 + \w ( 16 + \w )  )  x_4^3 +  
           24 \w ( -29 + 7 \w )  x_4^4 \nonumber \\ && + 36 ( -2 +
	   \w ( -2 + 7 \w )  )  x_4^5 +  
           3 ( 1 + \w )  ( 3 + 32 \w )  x_4^6 +
	   7 \w ( 1 + \w )  x_4^7 \nonumber \\ &&+ 27 ( 1 + \w )
	    x_4^8  -  
           9 \w ( 1 + \w )  x_4^9 \big)  + 24 A ( 1 + \w )  x_4^6 
         ( -1 - 3 x_4^2 + \w ( x_4 + x_4^3 )  )
	  \Big),  \nonumber \\ &&  \\
f^W_3 &=& \frac{( 1 + \w )}{24 x_4^4 {( -1 + x_4^2 ) }^4}  \Big( \tB \big( -144 + 108 ( 1 + 2 \w )  x_4 - 
         24 ( 5 + \w )  ( -3 + 2 \w )  x_4^2 \nonumber \\ && +
	 36 ( -11 + ( -16 + \w )  \w )  x_4^3 +  
         24 \w ( 29 + 7 \w )  x_4^4 - 36 ( -2 + \w ( 2 + 7 \w )  )  x_4^5 \nonumber \\ && + 
         3 ( -1 + \w )  ( -3 + 32 \w )  x_4^6 -
	 7 ( -1 + \w )  \w x_4^7  - 27 ( -1 + \w )
	  x_4^8 \nonumber \\ &&+  
         9 ( -1 + \w )  \w x_4^9 \big)  - 24 A ( -1 + \w )  x_4^6 
       ( -1 - 3 x_4^2 + \w ( x_4 + x_4^3 )  )
       \Big),   \\
f^W_4 &=& \frac{3 \tB {( -1 + \w x_4 ) }^2 ( 4 -
  10 x_4^2 + \w x_4 ( -1 + 7 x_4^2 )  ) }{x_4^4 {(
    -1 + x_4^2 ) }^4} .
\eea

Results including the corrections at $O(\tk^2)$ can be found at \cite{Website}.


\section{Bulk geodesics}
\label{appE}

To calculate the affine distance between the branes along a spacelike
geodesic we must solve the geodesic equations in the bulk.
Let us first consider the situation in Birkhoff-frame coordinates for which the
bulk metric is static and the branes are moving.  The Birkhoff-frame
metric takes the form (see Chapter \S\,\ref{branegravitychapter}) 
\[
\d s^2 = \d Y^2 - N^2(Y)\, \d T^2 + A^2(Y)\, \d \vec{x}^2,
\]
where for AdS-Schwarzschild with a horizon at $Y=0$,
\[
A^2(Y) = \frac{\cosh(2 Y/L)}{\cosh(2Y_0/L)}, \qquad N^2(Y) =
  \frac{\cosh(2 Y_0/L)}{\cosh(2Y/L)}\left(\frac{\sinh{(2Y/L)}}{\sinh{(2Y_0/L)}}\right)^2.
\]
At $T=0$, the $Y$-coordinate of the branes is represented by the parameter $Y_0$.
The subsequent brane trajectories $Y_\pm(T)$ can then be determined by integrating 
the Israel matching conditions, which read $\tanh{(2Y_\pm/L)}= \pm
\sqrt{1-V_\pm^2}\,$, 
where $V_\pm = (\d Y_\pm/\d T)/N(Y_\pm)$
are the proper speeds of the positive- and negative-tension branes respectively.
From this, it further follows that $Y_0$
is related to the rapidity $y_0$ of the collision
by $\tanh y_0 =\sech(2Y_0/L)$.

For the purpose of measuring the distance between the branes, a
natural choice is to use spacelike geodesics that are orthogonal to the four
translational Killing vectors of the static bulk, corresponding to
shifts in $\vec{x}$ and $T$. 
Taking the $\vec{x}$ and $T$ coordinates to be fixed along the
geodesic, 
we find that $Y_{,\lambda}$ is constant for an affine parameter $\lambda$ along the geodesic.

To make the connection to our original brane-static coordinate system,
recall that the metric function $b^2(t,y) = A^2(Y)$, and thus
\[
Y_{,\lambda}^2 = \frac{(bb_{,t}t_{,\lambda}+b b_{,y} y_{,\lambda})^2}{b^4 -
  \theta^2} = n^2 (-t_{,\lambda}^2+t^2 y_{,\lambda}^2),
\]
where we have introduced the constant $\theta=\tanh{y_0}=V/c$.
Adopting $y$ now as the affine parameter, 
we have
\[
0 = (b_{,t}^2 b^2+n^2(b^4-\theta^2))t_{,y}^2 + 2 b_{,t}b_{,y}b^2 t_{,y}+(b_{,y}^2
b^2-n^2t^2(b^4-\theta^2)),
\]
where $t$ is to be regarded now as a function of $y$.

We can solve this equation order by order in $y_0$ using the series ansatz
\[
t(y) = \sum_{n=0}^\inf c_n y^n,
\]
where the constants $c_n$ are themselves
series in $y_0$.
Using the series solution for the background geometry given in Section \ref{appB}, 
and imposing the boundary condition that $t(y_0)=t_0$, we obtain
\bea
c_0 &=&
t_0 + \frac{t_0\,y_0^2}{2} - 2\,t_0^2\,y_0^3 + \frac{\left( t_0 + 36\,t_0^3 \right) y_0^4}{24} - 
  t_0^2\left( 1 + 5\,t_0^2 \right) y_0^5 \nonumber \\ && + \left( \frac{t_0}{720}
  + \frac{17\,t_0^3}{4} + 4\,t_0^5 \right) y_0^6 
 - \frac{t_0^2\left( 13 + 250\,t_0^2 + 795\,t_0^4 \right) y_0^7}{60}\qquad \nonumber \\ &&  + O(y_0^8), \\
c_1 &=& 
2\,t_0^2\,y_0^2 + \left( \frac{5\,t_0^2}{3} + 5\,t_0^4 \right) \,y_0^4 - 8\,t_0^3\,y_0^5 \nonumber \\ &&
+ \left( \frac{91\,t_0^2}{180} + \frac{23\,t_0^4}{6} + \frac{53\,t_0^6}{4}
  \right) \,y_0^6  + O(y_0^7), \\
c_2 &=& 
-\frac{t_0}{2} - \frac{t_0\left( 1 + 6\,t_0^2 \right) y_0^2}{4} + t_0^2\,y_0^3 - 
  \left( \frac{t_0}{48} - 2t_0^3 + 4t_0^5 \right)y_0^4 \nonumber \\ &&+ \frac{\left( t_0^2 + 23\,t_0^4 \right) y_0^5}{2} + 
  O(y_0^6),  \\
c_3 &=& 
-\frac{5\,t_0^2\,y_0^2}{3} - \frac{t_0^2\,\left( 25 + 201\,t_0^2 \right)
  y_0^4}{18} + O(y_0^5), \\
c_4 &=&
\frac{5\,t_0}{24} + \left( \frac{5\,t_0}{48} + \frac{7\,t_0^3}{4} \right) y_0^2 - \frac{5\,t_0^2\,y_0^3}{12} + 
  O(y_0^4), \\
c_5 &=&
\frac{61\,t_0^2\,y_0^2}{60} + O(y_0^3), \\
c_6 &=& 
-\frac{61\,t_0}{720} + O(y_0^2), \\
c_7 &=& 0 + O(y_0) .
\eea
Substituting $t_0=x_0/y_0$ and $y=\w y_0$, we find $x(\w)=x_0/y_0+O(y_0)$,
\ie to lowest order in $y_0$, 
the geodesics are trajectories of constant time lying solely along the
$\w$ direction.
Hence in this limit, the affine and metric separation of the branes
(defined in (\ref{d_m})) must necessarily agree.  
To check this, the affine distance between the branes is given by
\bea
\frac{d_a}{L} &=& \int_{-y_0}^{y_0}n\sqrt{t^2-t'^2}\,\d y \nonumber \\
&=& 2\,t_0\,y_0 + \frac{\left( t_0 + 5\,t_0^3 \right) \,y_0^3}{3} - 4\,t_0^2\,y_0^4 + 
  \frac{\left( t_0 - 10\,t_0^3 + 159\,t_0^5 \right) \,y_0^5}{60} \nonumber \\[1ex] && -
  \frac{2\,\left( t_0^2 + 30\,t_0^4 \right) \,y_0^6}{3}  +  
  \frac{\left( t_0 + 31115\,t_0^3 - 5523\,t_0^5 + 12795\,t_0^7 \right) \,y_0^7}{2520} \nonumber \\ && + O(y_0^8),
\eea
which to lowest order in $y_0$ reduces to
\[
\frac{d_a}{L} = 2\,x_0 + \frac{5\,x_0^3}{3} + \frac{53\,x_0^5}{20} +
\frac{853\,x_0^7}{168} + O(x_0^8) + O(y_0^2),
\]
in agreement with the series expansion of (\ref{d_m}).
(Note however that the two distance measures differ nontrivially at order
$y_0^2$).

To evaluate the perturbation $\delta d_a$ in the affine distance
between the branes, consider 
\bea
\delta \int \sqrt{\g \dot{x}^\mu \dot{x}^\nu} \d \lambda &=&
\frac{1}{2}\int \frac{\d \lambda}{\sqrt{g_{\rho\sigma} \dot{x}^\rho \dot{x}^\sigma}} \left(\delta\g
\dot{x}^\mu\dot{x}^\nu+g_{\mu\nu ,\kappa}\delta x^\kappa\dot{x}^\mu
\dot{x}^\nu +2 \g \dot{x}^\mu \delta \dot{x}^\nu\right) \nonumber \\
&=&\left[\frac{\dot{x}_\nu\delta x^\nu}{\sqrt{g_{\rho\sigma} \dot{x}^\rho
      \dot{x}^\sigma}}\right]+\frac{1}{2}\int\frac{\delta\g\dot{x}^\mu
  \dot{x}^\nu}{\sqrt{g_{\rho\sigma} \dot{x}^\rho \dot{x}^\sigma}}\,\d \lambda, 
\eea
where dots indicate differentiation with respect to the affine parameter
$\lambda$, and in going to the second line we have integrated by parts
and made use of the background geodesic equation
$\ddot{x}_\sigma=\frac{1}{2} g_{\mu\nu ,\sigma}\dot{x}^\mu \dot{x}^\nu$ and
the constraint $\g \dot{x}^\mu \dot{x}^\nu=1$. 
If the endpoints of the geodesics on the branes are unperturbed, this
expression is further simplified by the vanishing of the surface term.
Converting to coordinates where $t_0= x_0/y_0$
and $y= \w y_0$, to lowest order in $y_0$ the unperturbed geodesics
lie purely in the $\w$ direction, and so the perturbed affine distance
is once again identical to the perturbed metric distance (\ref{deltad_m}).

Explicitly, we find
\bea
\frac{\delta d_a}{L} &=&  
 -\frac{2\,\left( B + A\,t_0^2 \right) \,y_0}{t_0} \nonumber \\ && -
\left( \frac{B\,\left( 4 + 3\,t_0^2 \right)}{12\,t_0} +  
     \frac{A\,\left( t_0+ 9\,t_0^3 \right) }{3} \right) y_0^3 +
     \left( -4\,B + 4\,A\,t_0^2 \right) y_0^4\nonumber \\
&& - \left(\frac{B\,\left( 2 + 2169\,t_0^2 + 135\,t_0^4 \right)  + 
       2\,A\,t_0^2\,\left( 1 + 1110\,t_0^2 + 375\,t_0^4 \right)}
       {120\,t_0}\right)y_0^5 \nonumber \\
&& +  \left(\frac{4\,A\,t_0^2\,\left( 1 + 42\,t_0^2 \right)  -
       B\,\left( 4 + 57\,t_0^2 \right)}{6}\right) y_0^6 \nonumber
     \\
&& -  \frac{1}{10080 t_0}\, \Big(B\left( 4 + 88885t_0^2 + 952866t_0^4 + 28875t_0^6 \right)  \nonumber \\ && + 
       4At_0^2\left( 1 - 152481 t_0^2 + 293517 t_0^4 +
       36015 t_0^6 \right)\Big)\,y_0^7 \nonumber \\
&& +  O(y_0^8),
\eea
which, substituting $t_0=x_0/y_0$ and dropping terms of $O(y_0^2)$,  reduces to 
\bea
\frac{\delta d_a}{L} &=& -\frac{2\,\tB}{x_0} - 2\,A\,x_0 - \frac{\tB}{4}\,x_0
  - 3\,A\,x_0^3 - \frac{9}{8}\,\tB\,x_0^3 - \frac{25}{4}\,A\,x_0^5 \nonumber \\ &&  -  
  \frac{275}{96}\,\tB\,x_0^5 - \frac{343}{24}\,A\,x_0^7 + O(x_0^8),\qquad
\eea
where $\tB=B y_0^2$.
Once again, this expression is in accordance with the series expansion
of (\ref{deltad_m}).  At $O(y_0^2)$, however, the perturbed affine and metric
distances do not agree.

%% file: thesis.bbl
\begin{thebibliography}{100}
\expandafter\ifx\csname natexlab\endcsname\relax\def\natexlab#1{#1}\fi
\expandafter\ifx\csname bibnamefont\endcsname\relax
  \def\bibnamefont#1{#1}\fi
\expandafter\ifx\csname bibfnamefont\endcsname\relax
  \def\bibfnamefont#1{#1}\fi
\expandafter\ifx\csname citenamefont\endcsname\relax
  \def\citenamefont#1{#1}\fi
\expandafter\ifx\csname url\endcsname\relax
  \def\url#1{\texttt{#1}}\fi
\expandafter\ifx\csname urlprefix\endcsname\relax\def\urlprefix{URL }\fi
\providecommand{\bibinfo}[2]{#2}
\providecommand{\eprint}[2][]{\url{#2}}

\bibitem[{\citenamefont{Steinhardt and Turok}(2002{\natexlab{a}})}]{Cyclicevo}
\bibinfo{author}{\bibfnamefont{P.~J.} \bibnamefont{Steinhardt}}
  \bibnamefont{and} \bibinfo{author}{\bibfnamefont{N.}~\bibnamefont{Turok}},
  \bibinfo{journal}{Phys. Rev.} \textbf{\bibinfo{volume}{D65}},
  \bibinfo{pages}{126003} (\bibinfo{year}{2002}{\natexlab{a}}),
  \eprint{hep-th/0111098}.

\bibitem[{\citenamefont{Khoury et~al.}(2001)\citenamefont{Khoury, Ovrut,
  Steinhardt, and Turok}}]{Ekpyrotic}
\bibinfo{author}{\bibfnamefont{J.}~\bibnamefont{Khoury}},
  \bibinfo{author}{\bibfnamefont{B.~A.} \bibnamefont{Ovrut}},
  \bibinfo{author}{\bibfnamefont{P.~J.} \bibnamefont{Steinhardt}},
  \bibnamefont{and} \bibinfo{author}{\bibfnamefont{N.}~\bibnamefont{Turok}},
  \bibinfo{journal}{Phys. Rev.} \textbf{\bibinfo{volume}{D64}},
  \bibinfo{pages}{123522} (\bibinfo{year}{2001}), \eprint{hep-th/0103239}.

\bibitem[{\citenamefont{Khoury et~al.}(2002{\natexlab{a}})\citenamefont{Khoury,
  Ovrut, Seiberg, Steinhardt, and Turok}}]{Seiberg}
\bibinfo{author}{\bibfnamefont{J.}~\bibnamefont{Khoury}},
  \bibinfo{author}{\bibfnamefont{B.~A.} \bibnamefont{Ovrut}},
  \bibinfo{author}{\bibfnamefont{N.}~\bibnamefont{Seiberg}},
  \bibinfo{author}{\bibfnamefont{P.~J.} \bibnamefont{Steinhardt}},
  \bibnamefont{and} \bibinfo{author}{\bibfnamefont{N.}~\bibnamefont{Turok}},
  \bibinfo{journal}{Phys. Rev.} \textbf{\bibinfo{volume}{D65}},
  \bibinfo{pages}{086007} (\bibinfo{year}{2002}{\natexlab{a}}),
  \eprint{hep-th/0108187}.

\bibitem[{\citenamefont{Turok et~al.}(2004)\citenamefont{Turok, Perry, and
  Steinhardt}}]{perry}
\bibinfo{author}{\bibfnamefont{N.}~\bibnamefont{Turok}},
  \bibinfo{author}{\bibfnamefont{M.}~\bibnamefont{Perry}}, \bibnamefont{and}
  \bibinfo{author}{\bibfnamefont{P.~J.} \bibnamefont{Steinhardt}},
  \bibinfo{journal}{Phys. Rev.} \textbf{\bibinfo{volume}{D70}},
  \bibinfo{pages}{106004} (\bibinfo{year}{2004}), \eprint{hep-th/0408083}.

\bibitem[{\citenamefont{Niz and Turok}(2006)}]{Gustavo}
\bibinfo{author}{\bibfnamefont{G.}~\bibnamefont{Niz}} \bibnamefont{and}
  \bibinfo{author}{\bibfnamefont{N.}~\bibnamefont{Turok}}
  (\bibinfo{year}{2006}), \eprint{hep-th/0601007}.

\bibitem[{\citenamefont{Randall and Sundrum}(1999{\natexlab{a}})}]{RSI}
\bibinfo{author}{\bibfnamefont{L.}~\bibnamefont{Randall}} \bibnamefont{and}
  \bibinfo{author}{\bibfnamefont{R.}~\bibnamefont{Sundrum}},
  \bibinfo{journal}{Phys. Rev. Lett.} \textbf{\bibinfo{volume}{83}},
  \bibinfo{pages}{3370} (\bibinfo{year}{1999}{\natexlab{a}}),
  \eprint{hep-ph/9905221}.

\bibitem[{\citenamefont{McFadden et~al.}(2005)\citenamefont{McFadden, Turok,
  and Steinhardt}}]{long}
\bibinfo{author}{\bibfnamefont{P.~L.} \bibnamefont{McFadden}},
  \bibinfo{author}{\bibfnamefont{N.}~\bibnamefont{Turok}}, \bibnamefont{and}
  \bibinfo{author}{\bibfnamefont{P.}~\bibnamefont{Steinhardt}}
  (\bibinfo{year}{2005}), \eprint{hep-th/0512123}.

\bibitem[{\citenamefont{Randall and Sundrum}(1999{\natexlab{b}})}]{RSII}
\bibinfo{author}{\bibfnamefont{L.}~\bibnamefont{Randall}} \bibnamefont{and}
  \bibinfo{author}{\bibfnamefont{R.}~\bibnamefont{Sundrum}},
  \bibinfo{journal}{Phys. Rev. Lett.} \textbf{\bibinfo{volume}{83}},
  \bibinfo{pages}{4690} (\bibinfo{year}{1999}{\natexlab{b}}),
  \eprint{hep-th/9906064}.

\bibitem[{\citenamefont{Steinhardt and
  Turok}(2002{\natexlab{b}})}]{Steinhardt:2002ih}
\bibinfo{author}{\bibfnamefont{P.~J.} \bibnamefont{Steinhardt}}
  \bibnamefont{and} \bibinfo{author}{\bibfnamefont{N.}~\bibnamefont{Turok}},
  \bibinfo{journal}{Science} \textbf{\bibinfo{volume}{296}},
  \bibinfo{pages}{1436} (\bibinfo{year}{2002}{\natexlab{b}}).

\bibitem[{\citenamefont{Khoury et~al.}(2002{\natexlab{b}})\citenamefont{Khoury,
  Ovrut, Steinhardt, and Turok}}]{Khoury}
\bibinfo{author}{\bibfnamefont{J.}~\bibnamefont{Khoury}},
  \bibinfo{author}{\bibfnamefont{B.~A.} \bibnamefont{Ovrut}},
  \bibinfo{author}{\bibfnamefont{P.~J.} \bibnamefont{Steinhardt}},
  \bibnamefont{and} \bibinfo{author}{\bibfnamefont{N.}~\bibnamefont{Turok}},
  \bibinfo{journal}{Phys. Rev.} \textbf{\bibinfo{volume}{D66}},
  \bibinfo{pages}{046005} (\bibinfo{year}{2002}{\natexlab{b}}),
  \eprint{hep-th/0109050}.

\bibitem[{\citenamefont{Boyle et~al.}(2004)\citenamefont{Boyle, Steinhardt, and
  Turok}}]{Boyle}
\bibinfo{author}{\bibfnamefont{L.~A.} \bibnamefont{Boyle}},
  \bibinfo{author}{\bibfnamefont{P.~J.} \bibnamefont{Steinhardt}},
  \bibnamefont{and} \bibinfo{author}{\bibfnamefont{N.}~\bibnamefont{Turok}},
  \bibinfo{journal}{Phys. Rev.} \textbf{\bibinfo{volume}{D70}},
  \bibinfo{pages}{023504} (\bibinfo{year}{2004}), \eprint{hep-th/0403026}.

\bibitem[{\citenamefont{Gratton et~al.}(2004)\citenamefont{Gratton, Khoury,
  Steinhardt, and Turok}}]{Gratton2}
\bibinfo{author}{\bibfnamefont{S.}~\bibnamefont{Gratton}},
  \bibinfo{author}{\bibfnamefont{J.}~\bibnamefont{Khoury}},
  \bibinfo{author}{\bibfnamefont{P.~J.} \bibnamefont{Steinhardt}},
  \bibnamefont{and} \bibinfo{author}{\bibfnamefont{N.}~\bibnamefont{Turok}},
  \bibinfo{journal}{Phys. Rev.} \textbf{\bibinfo{volume}{D69}},
  \bibinfo{pages}{103505} (\bibinfo{year}{2004}), \eprint{astro-ph/0301395}.

\bibitem[{\citenamefont{Khoury et~al.}(2004)\citenamefont{Khoury, Steinhardt,
  and Turok}}]{models}
\bibinfo{author}{\bibfnamefont{J.}~\bibnamefont{Khoury}},
  \bibinfo{author}{\bibfnamefont{P.~J.} \bibnamefont{Steinhardt}},
  \bibnamefont{and} \bibinfo{author}{\bibfnamefont{N.}~\bibnamefont{Turok}},
  \bibinfo{journal}{Phys. Rev. Lett.} \textbf{\bibinfo{volume}{92}},
  \bibinfo{pages}{031302} (\bibinfo{year}{2004}), \eprint{hep-th/0307132}.

\bibitem[{\citenamefont{Lyth}(2002{\natexlab{a}})}]{gmode4}
\bibinfo{author}{\bibfnamefont{D.~H.} \bibnamefont{Lyth}},
  \bibinfo{journal}{Phys. Lett.} \textbf{\bibinfo{volume}{B524}},
  \bibinfo{pages}{1} (\bibinfo{year}{2002}{\natexlab{a}}),
  \eprint{hep-ph/0106153}.

\bibitem[{\citenamefont{Martin et~al.}(2002)\citenamefont{Martin, Peter,
  Pinto~Neto, and Schwarz}}]{gmode1}
\bibinfo{author}{\bibfnamefont{J.}~\bibnamefont{Martin}},
  \bibinfo{author}{\bibfnamefont{P.}~\bibnamefont{Peter}},
  \bibinfo{author}{\bibfnamefont{N.}~\bibnamefont{Pinto~Neto}},
  \bibnamefont{and} \bibinfo{author}{\bibfnamefont{D.~J.}
  \bibnamefont{Schwarz}}, \bibinfo{journal}{Phys. Rev.}
  \textbf{\bibinfo{volume}{D65}}, \bibinfo{pages}{123513}
  (\bibinfo{year}{2002}), \eprint{hep-th/0112128}.

\bibitem[{\citenamefont{Martin et~al.}(2003)\citenamefont{Martin, Peter,
  Pinto-Neto, and Schwarz}}]{gmode2}
\bibinfo{author}{\bibfnamefont{J.}~\bibnamefont{Martin}},
  \bibinfo{author}{\bibfnamefont{P.}~\bibnamefont{Peter}},
  \bibinfo{author}{\bibfnamefont{N.}~\bibnamefont{Pinto-Neto}},
  \bibnamefont{and} \bibinfo{author}{\bibfnamefont{D.~J.}
  \bibnamefont{Schwarz}}, \bibinfo{journal}{Phys. Rev.}
  \textbf{\bibinfo{volume}{D67}}, \bibinfo{pages}{028301}
  (\bibinfo{year}{2003}), \eprint{hep-th/0204222}.

\bibitem[{\citenamefont{Lyth}(2002{\natexlab{b}})}]{gmode3}
\bibinfo{author}{\bibfnamefont{D.~H.} \bibnamefont{Lyth}},
  \bibinfo{journal}{Phys. Lett.} \textbf{\bibinfo{volume}{B526}},
  \bibinfo{pages}{173} (\bibinfo{year}{2002}{\natexlab{b}}),
  \eprint{hep-ph/0110007}.

\bibitem[{\citenamefont{Brandenberger and
  Finelli}(2001{\natexlab{a}})}]{gmode5}
\bibinfo{author}{\bibfnamefont{R.}~\bibnamefont{Brandenberger}}
  \bibnamefont{and} \bibinfo{author}{\bibfnamefont{F.}~\bibnamefont{Finelli}},
  \bibinfo{journal}{JHEP} \textbf{\bibinfo{volume}{11}}, \bibinfo{pages}{056}
  (\bibinfo{year}{2001}{\natexlab{a}}), \eprint{hep-th/0109004}.

\bibitem[{\citenamefont{Finelli and
  Brandenberger}(2002{\natexlab{a}})}]{gmode6}
\bibinfo{author}{\bibfnamefont{F.}~\bibnamefont{Finelli}} \bibnamefont{and}
  \bibinfo{author}{\bibfnamefont{R.}~\bibnamefont{Brandenberger}},
  \bibinfo{journal}{Phys. Rev.} \textbf{\bibinfo{volume}{D65}},
  \bibinfo{pages}{103522} (\bibinfo{year}{2002}{\natexlab{a}}),
  \eprint{hep-th/0112249}.

\bibitem[{\citenamefont{Hwang}(2002{\natexlab{a}})}]{gmode7}
\bibinfo{author}{\bibfnamefont{J.-c.} \bibnamefont{Hwang}},
  \bibinfo{journal}{Phys. Rev.} \textbf{\bibinfo{volume}{D65}},
  \bibinfo{pages}{063514} (\bibinfo{year}{2002}{\natexlab{a}}),
  \eprint{astro-ph/0109045}.

\bibitem[{\citenamefont{Hwang and Noh}(2002{\natexlab{a}})}]{gmode8}
\bibinfo{author}{\bibfnamefont{J.-c.} \bibnamefont{Hwang}} \bibnamefont{and}
  \bibinfo{author}{\bibfnamefont{H.}~\bibnamefont{Noh}},
  \bibinfo{journal}{Phys. Rev.} \textbf{\bibinfo{volume}{D65}},
  \bibinfo{pages}{124010} (\bibinfo{year}{2002}{\natexlab{a}}),
  \eprint{astro-ph/0112079}.

\bibitem[{\citenamefont{Hwang and Noh}(2002{\natexlab{b}})}]{jch2}
\bibinfo{author}{\bibfnamefont{J.}~\bibnamefont{Hwang}} \bibnamefont{and}
  \bibinfo{author}{\bibfnamefont{H.}~\bibnamefont{Noh}},
  \bibinfo{journal}{Phys. Lett.} \textbf{\bibinfo{volume}{B545}},
  \bibinfo{pages}{207} (\bibinfo{year}{2002}{\natexlab{b}}),
  \eprint{hep-th/0203193}.

\bibitem[{\citenamefont{Peter and Pinto-Neto}(2002)}]{gmode10}
\bibinfo{author}{\bibfnamefont{P.}~\bibnamefont{Peter}} \bibnamefont{and}
  \bibinfo{author}{\bibfnamefont{N.}~\bibnamefont{Pinto-Neto}},
  \bibinfo{journal}{Phys. Rev.} \textbf{\bibinfo{volume}{D66}},
  \bibinfo{pages}{063509} (\bibinfo{year}{2002}), \eprint{hep-th/0203013}.

\bibitem[{\citenamefont{Peter et~al.}(2003)\citenamefont{Peter, Pinto-Neto, and
  Gonzalez}}]{gmode11}
\bibinfo{author}{\bibfnamefont{P.}~\bibnamefont{Peter}},
  \bibinfo{author}{\bibfnamefont{N.}~\bibnamefont{Pinto-Neto}},
  \bibnamefont{and} \bibinfo{author}{\bibfnamefont{D.~A.}
  \bibnamefont{Gonzalez}}, \bibinfo{journal}{JCAP}
  \textbf{\bibinfo{volume}{0312}}, \bibinfo{pages}{003} (\bibinfo{year}{2003}),
  \eprint{hep-th/0306005}.

\bibitem[{\citenamefont{Gasperini et~al.}(2004)\citenamefont{Gasperini,
  Giovannini, and Veneziano}}]{gmode12}
\bibinfo{author}{\bibfnamefont{M.}~\bibnamefont{Gasperini}},
  \bibinfo{author}{\bibfnamefont{M.}~\bibnamefont{Giovannini}},
  \bibnamefont{and}
  \bibinfo{author}{\bibfnamefont{G.}~\bibnamefont{Veneziano}},
  \bibinfo{journal}{Nucl. Phys.} \textbf{\bibinfo{volume}{B694}},
  \bibinfo{pages}{206} (\bibinfo{year}{2004}), \eprint{hep-th/0401112}.

\bibitem[{\citenamefont{Gasperini et~al.}(2003)\citenamefont{Gasperini,
  Giovannini, and Veneziano}}]{gmode13}
\bibinfo{author}{\bibfnamefont{M.}~\bibnamefont{Gasperini}},
  \bibinfo{author}{\bibfnamefont{M.}~\bibnamefont{Giovannini}},
  \bibnamefont{and}
  \bibinfo{author}{\bibfnamefont{G.}~\bibnamefont{Veneziano}},
  \bibinfo{journal}{Phys. Lett.} \textbf{\bibinfo{volume}{B569}},
  \bibinfo{pages}{113} (\bibinfo{year}{2003}), \eprint{hep-th/0306113}.

\bibitem[{\citenamefont{Durrer and Vernizzi}(2002)}]{Durrer}
\bibinfo{author}{\bibfnamefont{R.}~\bibnamefont{Durrer}} \bibnamefont{and}
  \bibinfo{author}{\bibfnamefont{F.}~\bibnamefont{Vernizzi}},
  \bibinfo{journal}{Phys. Rev.} \textbf{\bibinfo{volume}{D66}},
  \bibinfo{pages}{083503} (\bibinfo{year}{2002}), \eprint{hep-ph/0203275}.

\bibitem[{\citenamefont{Cartier et~al.}(2003)\citenamefont{Cartier, Durrer, and
  Copeland}}]{Copeland}
\bibinfo{author}{\bibfnamefont{C.}~\bibnamefont{Cartier}},
  \bibinfo{author}{\bibfnamefont{R.}~\bibnamefont{Durrer}}, \bibnamefont{and}
  \bibinfo{author}{\bibfnamefont{E.~J.} \bibnamefont{Copeland}},
  \bibinfo{journal}{Phys. Rev.} \textbf{\bibinfo{volume}{D67}},
  \bibinfo{pages}{103517} (\bibinfo{year}{2003}), \eprint{hep-th/0301198}.

\bibitem[{\citenamefont{Bozza and Veneziano}(2005)}]{Bozza&Veneziano}
\bibinfo{author}{\bibfnamefont{V.}~\bibnamefont{Bozza}} \bibnamefont{and}
  \bibinfo{author}{\bibfnamefont{G.}~\bibnamefont{Veneziano}},
  \bibinfo{journal}{Phys. Lett.} \textbf{\bibinfo{volume}{B625}},
  \bibinfo{pages}{177} (\bibinfo{year}{2005}), \eprint{hep-th/0502047}.

\bibitem[{\citenamefont{Boyle et~al.}(2006)\citenamefont{Boyle, Steinhardt, and
  Turok}}]{Boyleinflation}
\bibinfo{author}{\bibfnamefont{L.~A.} \bibnamefont{Boyle}},
  \bibinfo{author}{\bibfnamefont{P.~J.} \bibnamefont{Steinhardt}},
  \bibnamefont{and} \bibinfo{author}{\bibfnamefont{N.}~\bibnamefont{Turok}},
  \bibinfo{journal}{Phys. Rev. Lett.} \textbf{\bibinfo{volume}{96}},
  \bibinfo{pages}{111301} (\bibinfo{year}{2006}), \eprint{astro-ph/0507455}.

\bibitem[{\citenamefont{McFadden and Turok}(2005{\natexlab{a}})}]{Conf_sym}
\bibinfo{author}{\bibfnamefont{P.~L.} \bibnamefont{McFadden}} \bibnamefont{and}
  \bibinfo{author}{\bibfnamefont{N.}~\bibnamefont{Turok}},
  \bibinfo{journal}{Phys. Rev.} \textbf{\bibinfo{volume}{D71}},
  \bibinfo{pages}{021901} (\bibinfo{year}{2005}{\natexlab{a}}),
  \eprint{hep-th/0409122}.

\bibitem[{\citenamefont{Spergel et~al.}(2006)}]{WMAP3}
\bibinfo{author}{\bibfnamefont{D.~N.} \bibnamefont{Spergel}}
  \bibnamefont{et~al.} (\bibinfo{year}{2006}), \eprint{astro-ph/0603449}.

\bibitem[{\citenamefont{Peacock}(1999)}]{Peacock}
\bibinfo{author}{\bibfnamefont{J.~A.} \bibnamefont{Peacock}},
  \emph{\bibinfo{title}{Cosmological physics}} (\bibinfo{publisher}{Cambridge
  University Press}, \bibinfo{year}{1999}).

\bibitem[{\citenamefont{Bardeen}(1980)}]{bardeen}
\bibinfo{author}{\bibfnamefont{J.~M.} \bibnamefont{Bardeen}},
  \bibinfo{journal}{Phys. Rev.} \textbf{\bibinfo{volume}{D22}},
  \bibinfo{pages}{1882} (\bibinfo{year}{1980}).

\bibitem[{\citenamefont{Langlois}(2004)}]{LangloisInflRev}
\bibinfo{author}{\bibfnamefont{D.}~\bibnamefont{Langlois}}
  (\bibinfo{year}{2004}), \eprint{hep-th/0405053}.

\bibitem[{\citenamefont{Mukhanov and Chibisov}(1981)}]{Mukhanov:1981xt}
\bibinfo{author}{\bibfnamefont{V.~F.} \bibnamefont{Mukhanov}} \bibnamefont{and}
  \bibinfo{author}{\bibfnamefont{G.~V.} \bibnamefont{Chibisov}},
  \bibinfo{journal}{JETP Lett.} \textbf{\bibinfo{volume}{33}},
  \bibinfo{pages}{532} (\bibinfo{year}{1981}).

\bibitem[{\citenamefont{Starobinsky}(1982)}]{Starobinsky:1982ee}
\bibinfo{author}{\bibfnamefont{A.~A.} \bibnamefont{Starobinsky}},
  \bibinfo{journal}{Phys. Lett.} \textbf{\bibinfo{volume}{B117}},
  \bibinfo{pages}{175} (\bibinfo{year}{1982}).

\bibitem[{\citenamefont{Hawking}(1982)}]{Hawking:1982cz}
\bibinfo{author}{\bibfnamefont{S.~W.} \bibnamefont{Hawking}},
  \bibinfo{journal}{Phys. Lett.} \textbf{\bibinfo{volume}{B115}},
  \bibinfo{pages}{295} (\bibinfo{year}{1982}).

\bibitem[{\citenamefont{Guth and Pi}(1982)}]{Guth:1982ec}
\bibinfo{author}{\bibfnamefont{A.~H.} \bibnamefont{Guth}} \bibnamefont{and}
  \bibinfo{author}{\bibfnamefont{S.~Y.} \bibnamefont{Pi}},
  \bibinfo{journal}{Phys. Rev. Lett.} \textbf{\bibinfo{volume}{49}},
  \bibinfo{pages}{1110} (\bibinfo{year}{1982}).

\bibitem[{\citenamefont{Bardeen et~al.}(1983)\citenamefont{Bardeen, Steinhardt,
  and Turner}}]{Bardeen:1983qw}
\bibinfo{author}{\bibfnamefont{J.~M.} \bibnamefont{Bardeen}},
  \bibinfo{author}{\bibfnamefont{P.~J.} \bibnamefont{Steinhardt}},
  \bibnamefont{and} \bibinfo{author}{\bibfnamefont{M.~S.}
  \bibnamefont{Turner}}, \bibinfo{journal}{Phys. Rev.}
  \textbf{\bibinfo{volume}{D28}}, \bibinfo{pages}{679} (\bibinfo{year}{1983}).

\bibitem[{\citenamefont{Mukhanov et~al.}(1992)\citenamefont{Mukhanov, Feldman,
  and Brandenberger}}]{mukhanov}
\bibinfo{author}{\bibfnamefont{V.~F.} \bibnamefont{Mukhanov}},
  \bibinfo{author}{\bibfnamefont{H.~A.} \bibnamefont{Feldman}},
  \bibnamefont{and} \bibinfo{author}{\bibfnamefont{R.~H.}
  \bibnamefont{Brandenberger}}, \bibinfo{journal}{Phys. Rept.}
  \textbf{\bibinfo{volume}{215}}, \bibinfo{pages}{203} (\bibinfo{year}{1992}).

\bibitem[{\citenamefont{Brandenberger}(2005)}]{Brandenberger:2005be}
\bibinfo{author}{\bibfnamefont{R.~H.} \bibnamefont{Brandenberger}}
  (\bibinfo{year}{2005}), \eprint{hep-th/0501033}.

\bibitem[{\citenamefont{Turok}(2002)}]{Turok:2002yq}
\bibinfo{author}{\bibfnamefont{N.}~\bibnamefont{Turok}},
  \bibinfo{journal}{Class. Quant. Grav.} \textbf{\bibinfo{volume}{19}},
  \bibinfo{pages}{3449} (\bibinfo{year}{2002}).

\bibitem[{\citenamefont{Maldacena}(2003)}]{Maldacena:2002vr}
\bibinfo{author}{\bibfnamefont{J.~M.} \bibnamefont{Maldacena}},
  \bibinfo{journal}{JHEP} \textbf{\bibinfo{volume}{05}}, \bibinfo{pages}{013}
  (\bibinfo{year}{2003}), \eprint{astro-ph/0210603}.

\bibitem[{\citenamefont{Wald}(1984)}]{Wald}
\bibinfo{author}{\bibfnamefont{R.~M.} \bibnamefont{Wald}},
  \emph{\bibinfo{title}{General Relativity}} (\bibinfo{publisher}{Chicago
  University Press}, \bibinfo{year}{1984}).

\bibitem[{\citenamefont{Tolley et~al.}(2004)\citenamefont{Tolley, Turok, and
  Steinhardt}}]{TTS}
\bibinfo{author}{\bibfnamefont{A.~J.} \bibnamefont{Tolley}},
  \bibinfo{author}{\bibfnamefont{N.}~\bibnamefont{Turok}}, \bibnamefont{and}
  \bibinfo{author}{\bibfnamefont{P.~J.} \bibnamefont{Steinhardt}},
  \bibinfo{journal}{Phys. Rev.} \textbf{\bibinfo{volume}{D69}},
  \bibinfo{pages}{106005} (\bibinfo{year}{2004}), \eprint{hep-th/0306109}.

\bibitem[{\citenamefont{Creminelli et~al.}(2005)\citenamefont{Creminelli,
  Nicolis, and Zaldarriaga}}]{Creminelli}
\bibinfo{author}{\bibfnamefont{P.}~\bibnamefont{Creminelli}},
  \bibinfo{author}{\bibfnamefont{A.}~\bibnamefont{Nicolis}}, \bibnamefont{and}
  \bibinfo{author}{\bibfnamefont{M.}~\bibnamefont{Zaldarriaga}},
  \bibinfo{journal}{Phys. Rev.} \textbf{\bibinfo{volume}{D71}},
  \bibinfo{pages}{063505} (\bibinfo{year}{2005}), \eprint{hep-th/0411270}.

\bibitem[{\citenamefont{Polchinski}(1998)}]{PolchinskiI}
\bibinfo{author}{\bibfnamefont{J.}~\bibnamefont{Polchinski}},
  \emph{\bibinfo{title}{String theory. Vol.~1: An introduction to the bosonic
  string}} (\bibinfo{publisher}{Cambridge University Press},
  \bibinfo{year}{1998}).

\bibitem[{\citenamefont{Brax and van~de Bruck}(2003)}]{Brax:2003fv}
\bibinfo{author}{\bibfnamefont{P.}~\bibnamefont{Brax}} \bibnamefont{and}
  \bibinfo{author}{\bibfnamefont{C.}~\bibnamefont{van~de Bruck}},
  \bibinfo{journal}{Class. Quant. Grav.} \textbf{\bibinfo{volume}{20}},
  \bibinfo{pages}{R201} (\bibinfo{year}{2003}), \eprint{hep-th/0303095}.

\bibitem[{\citenamefont{Langlois}(2003)}]{langlois}
\bibinfo{author}{\bibfnamefont{D.}~\bibnamefont{Langlois}},
  \bibinfo{journal}{Prog. Theor. Phys. Suppl.} \textbf{\bibinfo{volume}{148}},
  \bibinfo{pages}{181} (\bibinfo{year}{2003}), \eprint{hep-th/0209261}.

\bibitem[{\citenamefont{Maartens}(2004)}]{maartens}
\bibinfo{author}{\bibfnamefont{R.}~\bibnamefont{Maartens}},
  \bibinfo{journal}{Living Rev. Rel.} \textbf{\bibinfo{volume}{7}},
  \bibinfo{pages}{7} (\bibinfo{year}{2004}), \eprint{gr-qc/0312059}.

\bibitem[{\citenamefont{Israel}(1966)}]{Israel}
\bibinfo{author}{\bibfnamefont{W.}~\bibnamefont{Israel}},
  \bibinfo{journal}{Nuovo Cim.} \textbf{\bibinfo{volume}{B44S10}},
  \bibinfo{pages}{1} (\bibinfo{year}{1966}).

\bibitem[{\citenamefont{Chamblin et~al.}(2000)\citenamefont{Chamblin, Hawking,
  and Reall}}]{BS}
\bibinfo{author}{\bibfnamefont{A.}~\bibnamefont{Chamblin}},
  \bibinfo{author}{\bibfnamefont{S.~W.} \bibnamefont{Hawking}},
  \bibnamefont{and} \bibinfo{author}{\bibfnamefont{H.~S.} \bibnamefont{Reall}},
  \bibinfo{journal}{Phys. Rev.} \textbf{\bibinfo{volume}{D61}},
  \bibinfo{pages}{065007} (\bibinfo{year}{2000}), \eprint{hep-th/9909205}.

\bibitem[{\citenamefont{Birmingham}(1999)}]{Birmingham}
\bibinfo{author}{\bibfnamefont{D.}~\bibnamefont{Birmingham}},
  \bibinfo{journal}{Class. Quant. Grav.} \textbf{\bibinfo{volume}{16}},
  \bibinfo{pages}{1197} (\bibinfo{year}{1999}), \eprint{hep-th/9808032}.

\bibitem[{\citenamefont{Bowcock et~al.}(2000)\citenamefont{Bowcock, Charmousis,
  and Gregory}}]{Gregory}
\bibinfo{author}{\bibfnamefont{P.}~\bibnamefont{Bowcock}},
  \bibinfo{author}{\bibfnamefont{C.}~\bibnamefont{Charmousis}},
  \bibnamefont{and} \bibinfo{author}{\bibfnamefont{R.}~\bibnamefont{Gregory}},
  \bibinfo{journal}{Class. Quant. Grav.} \textbf{\bibinfo{volume}{17}},
  \bibinfo{pages}{4745} (\bibinfo{year}{2000}), \eprint{hep-th/0007177}.

\bibitem[{\citenamefont{Kraus}(1999)}]{KrausFRW}
\bibinfo{author}{\bibfnamefont{P.}~\bibnamefont{Kraus}},
  \bibinfo{journal}{JHEP} \textbf{\bibinfo{volume}{12}}, \bibinfo{pages}{011}
  (\bibinfo{year}{1999}), \eprint{hep-th/9910149}.

\bibitem[{\citenamefont{Ida}(2000)}]{IdaFRW}
\bibinfo{author}{\bibfnamefont{D.}~\bibnamefont{Ida}}, \bibinfo{journal}{JHEP}
  \textbf{\bibinfo{volume}{09}}, \bibinfo{pages}{014} (\bibinfo{year}{2000}),
  \eprint{gr-qc/9912002}.

\bibitem[{\citenamefont{Binetruy et~al.}(2000)\citenamefont{Binetruy, Deffayet,
  Ellwanger, and Langlois}}]{Binetruy}
\bibinfo{author}{\bibfnamefont{P.}~\bibnamefont{Binetruy}},
  \bibinfo{author}{\bibfnamefont{C.}~\bibnamefont{Deffayet}},
  \bibinfo{author}{\bibfnamefont{U.}~\bibnamefont{Ellwanger}},
  \bibnamefont{and} \bibinfo{author}{\bibfnamefont{D.}~\bibnamefont{Langlois}},
  \bibinfo{journal}{Phys. Lett.} \textbf{\bibinfo{volume}{B477}},
  \bibinfo{pages}{285} (\bibinfo{year}{2000}), \eprint{hep-th/9910219}.

\bibitem[{\citenamefont{Arroja and Koyama}(2006)}]{Koyama}
\bibinfo{author}{\bibfnamefont{F.}~\bibnamefont{Arroja}} \bibnamefont{and}
  \bibinfo{author}{\bibfnamefont{K.}~\bibnamefont{Koyama}},
  \bibinfo{journal}{Class. Quant. Grav.} \textbf{\bibinfo{volume}{23}},
  \bibinfo{pages}{4249} (\bibinfo{year}{2006}), \eprint{hep-th/0602068}.

\bibitem[{\citenamefont{Manton}(1982)}]{Manton:1981mp}
\bibinfo{author}{\bibfnamefont{N.~S.} \bibnamefont{Manton}},
  \bibinfo{journal}{Phys. Lett.} \textbf{\bibinfo{volume}{B110}},
  \bibinfo{pages}{54} (\bibinfo{year}{1982}).

\bibitem[{\citenamefont{Manton}(1985)}]{Manton:1985hs}
\bibinfo{author}{\bibfnamefont{N.~S.} \bibnamefont{Manton}},
  \bibinfo{journal}{Phys. Lett.} \textbf{\bibinfo{volume}{B154}},
  \bibinfo{pages}{397} (\bibinfo{year}{1985}).

\bibitem[{\citenamefont{Brax et~al.}(2003)\citenamefont{Brax, van~de Bruck,
  Davis, and Rhodes}}]{Brax:2002nt}
\bibinfo{author}{\bibfnamefont{P.}~\bibnamefont{Brax}},
  \bibinfo{author}{\bibfnamefont{C.}~\bibnamefont{van~de Bruck}},
  \bibinfo{author}{\bibfnamefont{A.~C.} \bibnamefont{Davis}}, \bibnamefont{and}
  \bibinfo{author}{\bibfnamefont{C.~S.} \bibnamefont{Rhodes}},
  \bibinfo{journal}{Phys. Rev.} \textbf{\bibinfo{volume}{D67}},
  \bibinfo{pages}{023512} (\bibinfo{year}{2003}), \eprint{hep-th/0209158}.

\bibitem[{\citenamefont{Garriga et~al.}(2003)\citenamefont{Garriga, Pujolas,
  and Tanaka}}]{Garriga:2001ar}
\bibinfo{author}{\bibfnamefont{J.}~\bibnamefont{Garriga}},
  \bibinfo{author}{\bibfnamefont{O.}~\bibnamefont{Pujolas}}, \bibnamefont{and}
  \bibinfo{author}{\bibfnamefont{T.}~\bibnamefont{Tanaka}},
  \bibinfo{journal}{Nucl. Phys.} \textbf{\bibinfo{volume}{B655}},
  \bibinfo{pages}{127} (\bibinfo{year}{2003}), \eprint{hep-th/0111277}.

\bibitem[{\citenamefont{Charmousis et~al.}(2000)\citenamefont{Charmousis,
  Gregory, and Rubakov}}]{CGR}
\bibinfo{author}{\bibfnamefont{C.}~\bibnamefont{Charmousis}},
  \bibinfo{author}{\bibfnamefont{R.}~\bibnamefont{Gregory}}, \bibnamefont{and}
  \bibinfo{author}{\bibfnamefont{V.~A.} \bibnamefont{Rubakov}},
  \bibinfo{journal}{Phys. Rev.} \textbf{\bibinfo{volume}{D62}},
  \bibinfo{pages}{067505} (\bibinfo{year}{2000}), \eprint{hep-th/9912160}.

\bibitem[{\citenamefont{Shiromizu and Koyama}(2003)}]{Shiromizu:2002qr}
\bibinfo{author}{\bibfnamefont{T.}~\bibnamefont{Shiromizu}} \bibnamefont{and}
  \bibinfo{author}{\bibfnamefont{K.}~\bibnamefont{Koyama}},
  \bibinfo{journal}{Phys. Rev.} \textbf{\bibinfo{volume}{D67}},
  \bibinfo{pages}{084022} (\bibinfo{year}{2003}), \eprint{hep-th/0210066}.

\bibitem[{\citenamefont{Kanno and Soda}(2002{\natexlab{a}})}]{K&S}
\bibinfo{author}{\bibfnamefont{S.}~\bibnamefont{Kanno}} \bibnamefont{and}
  \bibinfo{author}{\bibfnamefont{J.}~\bibnamefont{Soda}},
  \bibinfo{journal}{Phys. Rev.} \textbf{\bibinfo{volume}{D66}},
  \bibinfo{pages}{083506} (\bibinfo{year}{2002}{\natexlab{a}}),
  \eprint{hep-th/0207029}.

\bibitem[{\citenamefont{Kanno and Soda}(2002{\natexlab{b}})}]{K&S2}
\bibinfo{author}{\bibfnamefont{S.}~\bibnamefont{Kanno}} \bibnamefont{and}
  \bibinfo{author}{\bibfnamefont{J.}~\bibnamefont{Soda}},
  \bibinfo{journal}{Phys. Rev.} \textbf{\bibinfo{volume}{D66}},
  \bibinfo{pages}{043526} (\bibinfo{year}{2002}{\natexlab{b}}),
  \eprint{hep-th/0205188}.

\bibitem[{\citenamefont{Kanno and Soda}(2005)}]{KSValidity}
\bibinfo{author}{\bibfnamefont{S.}~\bibnamefont{Kanno}} \bibnamefont{and}
  \bibinfo{author}{\bibfnamefont{J.}~\bibnamefont{Soda}},
  \bibinfo{journal}{Phys. Rev.} \textbf{\bibinfo{volume}{D71}},
  \bibinfo{pages}{044031} (\bibinfo{year}{2005}), \eprint{hep-th/0410061}.

\bibitem[{\citenamefont{Kim et~al.}(2004)\citenamefont{Kim, Tupper, and
  Viollier}}]{GaugesinBulkI}
\bibinfo{author}{\bibfnamefont{J.~E.} \bibnamefont{Kim}},
  \bibinfo{author}{\bibfnamefont{G.~B.} \bibnamefont{Tupper}},
  \bibnamefont{and} \bibinfo{author}{\bibfnamefont{R.~D.}
  \bibnamefont{Viollier}}, \bibinfo{journal}{Phys. Lett.}
  \textbf{\bibinfo{volume}{B593}}, \bibinfo{pages}{209} (\bibinfo{year}{2004}),
  \eprint{hep-th/0404180}.

\bibitem[{\citenamefont{Kanno and Soda}(2004)}]{4d1}
\bibinfo{author}{\bibfnamefont{S.}~\bibnamefont{Kanno}} \bibnamefont{and}
  \bibinfo{author}{\bibfnamefont{J.}~\bibnamefont{Soda}},
  \bibinfo{journal}{Phys. Lett.} \textbf{\bibinfo{volume}{B588}},
  \bibinfo{pages}{203} (\bibinfo{year}{2004}), \eprint{hep-th/0312106}.

\bibitem[{\citenamefont{McFadden and Turok}(2005{\natexlab{b}})}]{EFT_BH}
\bibinfo{author}{\bibfnamefont{P.~L.} \bibnamefont{McFadden}} \bibnamefont{and}
  \bibinfo{author}{\bibfnamefont{N.~G.} \bibnamefont{Turok}},
  \bibinfo{journal}{Phys. Rev.} \textbf{\bibinfo{volume}{D71}},
  \bibinfo{pages}{086004} (\bibinfo{year}{2005}{\natexlab{b}}),
  \eprint{hep-th/0412109}.

\bibitem[{\citenamefont{Maldacena}(1998)}]{MaldacenaAdSCFT}
\bibinfo{author}{\bibfnamefont{J.~M.} \bibnamefont{Maldacena}},
  \bibinfo{journal}{Adv. Theor. Math. Phys.} \textbf{\bibinfo{volume}{2}},
  \bibinfo{pages}{231} (\bibinfo{year}{1998}), \eprint{hep-th/9711200}.

\bibitem[{\citenamefont{Witten}(1998)}]{WittenADSCFT}
\bibinfo{author}{\bibfnamefont{E.}~\bibnamefont{Witten}},
  \bibinfo{journal}{Adv. Theor. Math. Phys.} \textbf{\bibinfo{volume}{2}},
  \bibinfo{pages}{253} (\bibinfo{year}{1998}), \eprint{hep-th/9802150}.

\bibitem[{\citenamefont{de~Haro et~al.}(2001)\citenamefont{de~Haro, Skenderis,
  and Solodukhin}}]{deHaro}
\bibinfo{author}{\bibfnamefont{S.}~\bibnamefont{de~Haro}},
  \bibinfo{author}{\bibfnamefont{K.}~\bibnamefont{Skenderis}},
  \bibnamefont{and} \bibinfo{author}{\bibfnamefont{S.~N.}
  \bibnamefont{Solodukhin}}, \bibinfo{journal}{Class. Quant. Grav.}
  \textbf{\bibinfo{volume}{18}}, \bibinfo{pages}{3171} (\bibinfo{year}{2001}),
  \eprint{hep-th/0011230}.

\bibitem[{\citenamefont{Duff and Liu}(2000)}]{Duff&Liu}
\bibinfo{author}{\bibfnamefont{M.~J.} \bibnamefont{Duff}} \bibnamefont{and}
  \bibinfo{author}{\bibfnamefont{J.~T.} \bibnamefont{Liu}},
  \bibinfo{journal}{Phys. Rev. Lett.} \textbf{\bibinfo{volume}{85}},
  \bibinfo{pages}{2052} (\bibinfo{year}{2000}), \eprint{hep-th/0003237}.

\bibitem[{\citenamefont{Gubser}(2001)}]{Gubser}
\bibinfo{author}{\bibfnamefont{S.~S.} \bibnamefont{Gubser}},
  \bibinfo{journal}{Phys. Rev.} \textbf{\bibinfo{volume}{D63}},
  \bibinfo{pages}{084017} (\bibinfo{year}{2001}), \eprint{hep-th/9912001}.

\bibitem[{\citenamefont{Shiromizu and Ida}(2001)}]{Shiromizu&Ida}
\bibinfo{author}{\bibfnamefont{T.}~\bibnamefont{Shiromizu}} \bibnamefont{and}
  \bibinfo{author}{\bibfnamefont{D.}~\bibnamefont{Ida}},
  \bibinfo{journal}{Phys. Rev.} \textbf{\bibinfo{volume}{D64}},
  \bibinfo{pages}{044015} (\bibinfo{year}{2001}), \eprint{hep-th/0102035}.

\bibitem[{\citenamefont{Henningson and Skenderis}(1998)}]{Henningson&Skenderis}
\bibinfo{author}{\bibfnamefont{M.}~\bibnamefont{Henningson}} \bibnamefont{and}
  \bibinfo{author}{\bibfnamefont{K.}~\bibnamefont{Skenderis}},
  \bibinfo{journal}{JHEP} \textbf{\bibinfo{volume}{07}}, \bibinfo{pages}{023}
  (\bibinfo{year}{1998}), \eprint{hep-th/9806087}.

\bibitem[{\citenamefont{Garriga and Tanaka}(2000)}]{garriga}
\bibinfo{author}{\bibfnamefont{J.}~\bibnamefont{Garriga}} \bibnamefont{and}
  \bibinfo{author}{\bibfnamefont{T.}~\bibnamefont{Tanaka}},
  \bibinfo{journal}{Phys. Rev. Lett.} \textbf{\bibinfo{volume}{84}},
  \bibinfo{pages}{2778} (\bibinfo{year}{2000}), \eprint{hep-th/9911055}.

\bibitem[{\citenamefont{Giddings et~al.}(2000)\citenamefont{Giddings, Katz, and
  Randall}}]{giddings}
\bibinfo{author}{\bibfnamefont{S.~B.} \bibnamefont{Giddings}},
  \bibinfo{author}{\bibfnamefont{E.}~\bibnamefont{Katz}}, \bibnamefont{and}
  \bibinfo{author}{\bibfnamefont{L.}~\bibnamefont{Randall}},
  \bibinfo{journal}{JHEP} \textbf{\bibinfo{volume}{03}}, \bibinfo{pages}{023}
  (\bibinfo{year}{2000}), \eprint{hep-th/0002091}.

\bibitem[{\citenamefont{Duff et~al.}(1984)\citenamefont{Duff, Nilsson, Pope,
  and Warner}}]{duff}
\bibinfo{author}{\bibfnamefont{M.~J.} \bibnamefont{Duff}},
  \bibinfo{author}{\bibfnamefont{B.~E.~W.} \bibnamefont{Nilsson}},
  \bibinfo{author}{\bibfnamefont{C.~N.} \bibnamefont{Pope}}, \bibnamefont{and}
  \bibinfo{author}{\bibfnamefont{N.~P.} \bibnamefont{Warner}},
  \bibinfo{journal}{Phys. Lett.} \textbf{\bibinfo{volume}{B149}},
  \bibinfo{pages}{90} (\bibinfo{year}{1984}).

\bibitem[{\citenamefont{Boulware and Deser}(1972)}]{deser}
\bibinfo{author}{\bibfnamefont{D.~G.} \bibnamefont{Boulware}} \bibnamefont{and}
  \bibinfo{author}{\bibfnamefont{S.}~\bibnamefont{Deser}},
  \bibinfo{journal}{Phys. Rev.} \textbf{\bibinfo{volume}{D6}},
  \bibinfo{pages}{3368} (\bibinfo{year}{1972}).

\bibitem[{\citenamefont{Damour and Kogan}(2002)}]{damour}
\bibinfo{author}{\bibfnamefont{T.}~\bibnamefont{Damour}} \bibnamefont{and}
  \bibinfo{author}{\bibfnamefont{I.~I.} \bibnamefont{Kogan}},
  \bibinfo{journal}{Phys. Rev.} \textbf{\bibinfo{volume}{D66}},
  \bibinfo{pages}{104024} (\bibinfo{year}{2002}), \eprint{hep-th/0206042}.

\bibitem[{\citenamefont{Brandenberger and
  Finelli}(2001{\natexlab{b}})}]{brand1}
\bibinfo{author}{\bibfnamefont{R.}~\bibnamefont{Brandenberger}}
  \bibnamefont{and} \bibinfo{author}{\bibfnamefont{F.}~\bibnamefont{Finelli}},
  \bibinfo{journal}{JHEP} \textbf{\bibinfo{volume}{11}}, \bibinfo{pages}{056}
  (\bibinfo{year}{2001}{\natexlab{b}}), \eprint{hep-th/0109004}.

\bibitem[{\citenamefont{Finelli and
  Brandenberger}(2002{\natexlab{b}})}]{brand2}
\bibinfo{author}{\bibfnamefont{F.}~\bibnamefont{Finelli}} \bibnamefont{and}
  \bibinfo{author}{\bibfnamefont{R.}~\bibnamefont{Brandenberger}},
  \bibinfo{journal}{Phys. Rev.} \textbf{\bibinfo{volume}{D65}},
  \bibinfo{pages}{103522} (\bibinfo{year}{2002}{\natexlab{b}}),
  \eprint{hep-th/0112249}.

\bibitem[{\citenamefont{Lyth}(2002{\natexlab{c}})}]{lyth1}
\bibinfo{author}{\bibfnamefont{D.~H.} \bibnamefont{Lyth}},
  \bibinfo{journal}{Phys. Lett.} \textbf{\bibinfo{volume}{B524}},
  \bibinfo{pages}{1} (\bibinfo{year}{2002}{\natexlab{c}}),
  \eprint{hep-ph/0106153}.

\bibitem[{\citenamefont{Hwang}(2002{\natexlab{b}})}]{jch1}
\bibinfo{author}{\bibfnamefont{J.-c.} \bibnamefont{Hwang}},
  \bibinfo{journal}{Phys. Rev.} \textbf{\bibinfo{volume}{D65}},
  \bibinfo{pages}{063514} (\bibinfo{year}{2002}{\natexlab{b}}),
  \eprint{astro-ph/0109045}.

\bibitem[{Web()}]{Website}
\bibinfo{note}{Please follow link from {\it
  http://www.damtp.cam.ac.uk/user/ngt1000/}}.

\bibitem[{\citenamefont{van~de Bruck et~al.}(2000)\citenamefont{van~de Bruck,
  Dorca, Brandenberger, and Lukas}}]{Carsten}
\bibinfo{author}{\bibfnamefont{C.}~\bibnamefont{van~de Bruck}},
  \bibinfo{author}{\bibfnamefont{M.}~\bibnamefont{Dorca}},
  \bibinfo{author}{\bibfnamefont{R.~H.} \bibnamefont{Brandenberger}},
  \bibnamefont{and} \bibinfo{author}{\bibfnamefont{A.}~\bibnamefont{Lukas}},
  \bibinfo{journal}{Phys. Rev.} \textbf{\bibinfo{volume}{D62}},
  \bibinfo{pages}{123515} (\bibinfo{year}{2000}), \eprint{hep-th/0005032}.

\bibitem[{\citenamefont{Wiseman}(2002)}]{Toby}
\bibinfo{author}{\bibfnamefont{T.}~\bibnamefont{Wiseman}},
  \bibinfo{journal}{Class. Quant. Grav.} \textbf{\bibinfo{volume}{19}},
  \bibinfo{pages}{3083} (\bibinfo{year}{2002}), \eprint{hep-th/0201127}.

\bibitem[{\citenamefont{Palma and Davis}(2004)}]{Gonzalo}
\bibinfo{author}{\bibfnamefont{G.~A.} \bibnamefont{Palma}} \bibnamefont{and}
  \bibinfo{author}{\bibfnamefont{A.-C.} \bibnamefont{Davis}},
  \bibinfo{journal}{Phys. Rev.} \textbf{\bibinfo{volume}{D70}},
  \bibinfo{pages}{064021} (\bibinfo{year}{2004}), \eprint{hep-th/0406091}.

\bibitem[{\citenamefont{Corless et~al.}(1996)\citenamefont{Corless, Gonnet,
  Hare, Jeffrey, and Knuth}}]{LambertW}
\bibinfo{author}{\bibfnamefont{R.~M.} \bibnamefont{Corless}},
  \bibinfo{author}{\bibfnamefont{G.~H.} \bibnamefont{Gonnet}},
  \bibinfo{author}{\bibfnamefont{D.~E.~G.} \bibnamefont{Hare}},
  \bibinfo{author}{\bibfnamefont{D.~J.} \bibnamefont{Jeffrey}},
  \bibnamefont{and} \bibinfo{author}{\bibfnamefont{D.~E.} \bibnamefont{Knuth}},
  \bibinfo{journal}{Adv. Comp. Math.} pp. \bibinfo{pages}{329--359}
  (\bibinfo{year}{1996}), \bibinfo{note}{see also {\it
  http://mathworld.wolfram.com/LambertW-Function.html}}.

\bibitem[{\citenamefont{Csaki et~al.}(2000)\citenamefont{Csaki, Graesser,
  Randall, and Terning}}]{Terning}
\bibinfo{author}{\bibfnamefont{C.}~\bibnamefont{Csaki}},
  \bibinfo{author}{\bibfnamefont{M.}~\bibnamefont{Graesser}},
  \bibinfo{author}{\bibfnamefont{L.}~\bibnamefont{Randall}}, \bibnamefont{and}
  \bibinfo{author}{\bibfnamefont{J.}~\bibnamefont{Terning}},
  \bibinfo{journal}{Phys. Rev.} \textbf{\bibinfo{volume}{D62}},
  \bibinfo{pages}{045015} (\bibinfo{year}{2000}), \eprint{hep-ph/9911406}.

\bibitem[{\citenamefont{Khoury and Zhang}(2002)}]{KhouryZ}
\bibinfo{author}{\bibfnamefont{J.}~\bibnamefont{Khoury}} \bibnamefont{and}
  \bibinfo{author}{\bibfnamefont{R.-J.} \bibnamefont{Zhang}},
  \bibinfo{journal}{Phys. Rev. Lett.} \textbf{\bibinfo{volume}{89}},
  \bibinfo{pages}{061302} (\bibinfo{year}{2002}), \eprint{hep-th/0203274}.

\bibitem[{\citenamefont{Horava and Witten}(1996)}]{HW}
\bibinfo{author}{\bibfnamefont{P.}~\bibnamefont{Horava}} \bibnamefont{and}
  \bibinfo{author}{\bibfnamefont{E.}~\bibnamefont{Witten}},
  \bibinfo{journal}{Nucl. Phys.} \textbf{\bibinfo{volume}{B460}},
  \bibinfo{pages}{506} (\bibinfo{year}{1996}), \eprint{hep-th/9510209}.

\bibitem[{\citenamefont{Lavrinenko et~al.}(1997)\citenamefont{Lavrinenko, Lu,
  and Pope}}]{Lavrinenko}
\bibinfo{author}{\bibfnamefont{I.~V.} \bibnamefont{Lavrinenko}},
  \bibinfo{author}{\bibfnamefont{H.}~\bibnamefont{Lu}}, \bibnamefont{and}
  \bibinfo{author}{\bibfnamefont{C.~N.} \bibnamefont{Pope}},
  \bibinfo{journal}{Nucl. Phys.} \textbf{\bibinfo{volume}{B492}},
  \bibinfo{pages}{278} (\bibinfo{year}{1997}), \eprint{hep-th/9611134}.

\bibitem[{\citenamefont{Lukas et~al.}(1999)\citenamefont{Lukas, Ovrut, Stelle,
  and Waldram}}]{Lukas}
\bibinfo{author}{\bibfnamefont{A.}~\bibnamefont{Lukas}},
  \bibinfo{author}{\bibfnamefont{B.~A.} \bibnamefont{Ovrut}},
  \bibinfo{author}{\bibfnamefont{K.~S.} \bibnamefont{Stelle}},
  \bibnamefont{and} \bibinfo{author}{\bibfnamefont{D.}~\bibnamefont{Waldram}},
  \bibinfo{journal}{Phys. Rev.} \textbf{\bibinfo{volume}{D59}},
  \bibinfo{pages}{086001} (\bibinfo{year}{1999}), \eprint{hep-th/9803235}.

\bibitem[{\citenamefont{Chen et~al.}(2006)\citenamefont{Chen, Chong, Gibbons,
  Lu, and Pope}}]{gibbons}
\bibinfo{author}{\bibfnamefont{W.}~\bibnamefont{Chen}},
  \bibinfo{author}{\bibfnamefont{Z.~W.} \bibnamefont{Chong}},
  \bibinfo{author}{\bibfnamefont{G.~W.} \bibnamefont{Gibbons}},
  \bibinfo{author}{\bibfnamefont{H.}~\bibnamefont{Lu}}, \bibnamefont{and}
  \bibinfo{author}{\bibfnamefont{C.~N.} \bibnamefont{Pope}},
  \bibinfo{journal}{Nucl. Phys.} \textbf{\bibinfo{volume}{B732}},
  \bibinfo{pages}{118} (\bibinfo{year}{2006}), \eprint{hep-th/0502077}.

\bibitem[{\citenamefont{Gregory and Laflamme}(1993)}]{GL}
\bibinfo{author}{\bibfnamefont{R.}~\bibnamefont{Gregory}} \bibnamefont{and}
  \bibinfo{author}{\bibfnamefont{R.}~\bibnamefont{Laflamme}},
  \bibinfo{journal}{Phys. Rev. Lett.} \textbf{\bibinfo{volume}{70}},
  \bibinfo{pages}{2837} (\bibinfo{year}{1993}), \eprint{hep-th/9301052}.

\bibitem[{\citenamefont{Poisson}(2004)}]{Poisson}
\bibinfo{author}{\bibfnamefont{E.}~\bibnamefont{Poisson}},
  \emph{\bibinfo{title}{A Relativist's Toolkit}} (\bibinfo{publisher}{Cambridge
  University Press}, \bibinfo{year}{2004}).

\end{thebibliography}
